\documentclass[
        aps,
        prd,
        onecolumn,
        a4paper,
        10pt,
        amsmath,
        amssymb,
        preprintnumbers,
        tightenlines,
        notitlepage,
        floatfix,
        nofootinbib,
        longbibliography,
        superscriptaddress
	]{revtex4-2}

\usepackage{natbib}
\usepackage{graphicx}
\usepackage{amsfonts}
\usepackage{amsmath,amssymb,amsthm} 
\usepackage{bm}
\usepackage[dvipsnames]{xcolor}
\usepackage[pdftex,breaklinks,colorlinks,
anchorcolor=red,
urlcolor=cyan,
pdfencoding=auto]{hyperref}
\usepackage{caption}
\usepackage{subcaption}
\usepackage{tabularx}
\usepackage[normalem]{ulem}
\usepackage{mathtools}
\usepackage{tensor}
\usepackage{etoolbox}
\usepackage{upgreek}
\usepackage{float}
\usepackage{orcidlink}

\input{CQGformat.input}

\newcommand{\nhat}{{\hat{n}}}
\renewcommand{\O}{\mathcal{O}}

\definecolor{CiteColor}{rgb}{0,0.5,0}
\hypersetup{citecolor=CiteColor}
\definecolor{RefColor}{rgb}{0.55,0,0}
\hypersetup{linkcolor=RefColor}

\allowdisplaybreaks

\makeatletter
\renewcommand*{\fnum@figure}{{\normalfont\bfseries \figurename~\thefigure}}
\renewcommand*{\@caption@fignum@sep}{\textbf{ : }}
\makeatother

\makeatletter
\def\@mkboth#1#2{}
\newlength\appendixwidth
\preto\appendix{\addtocontents{toc}{\protect\patchl@section}}
\newcommand{\patchl@section}{%
  \settowidth{\appendixwidth}{\textbf{Appendix }}%
  \addtolength{\appendixwidth}{1.5em}%
  \patchcmd{\l@section}{1.5em}{\appendixwidth}{}{\ddt}%
}
\makeatother

\theoremstyle{plain}

\newtheorem*{theorem*}{Theorem}

\newcommand{\dd}{\text{d}}

\newcommand{\e}{\epsilon}						

\DeclareMathOperator{\bigO}{\mathcal{O}}					

\newcommand{\beq}{\begin{equation}}
\newcommand{\eeq}{\end{equation}}

\newcommand{\cU}{\mathcal{U}}
\newcommand{\cV}{\mathcal{V}}
\newcommand{\cM}{\mathcal{M}}
\newcommand{\cS}{\mathcal{S}}
\newcommand{\cI}{\mathcal{I}}
\newcommand{\cJ}{\mathcal{J}}
\newcommand{\cW}{\mathcal{W}}
\newcommand{\cX}{\mathcal{X}}
\newcommand{\cY}{\mathcal{Y}}
\newcommand{\cZ}{\mathcal{Z}}

\newcommand{\FP}{\mathop{\mathrm{FP}}_{B=0}}
\newcommand{\<}{\left\langle}
\renewcommand{\>}{\right\rangle}

\begin{document}

\title{Gravitational memory: new results from post-Newtonian and self-force theory}

\author{Kevin~Cunningham \orcidlink{0000-0002-1094-2267},\textsuperscript{1} Chris~Kavanagh \orcidlink{0000-0002-2874-9780},\textsuperscript{1} Adam~Pound \orcidlink{0000-0001-9446-0638},\textsuperscript{2} David~Trestini \orcidlink{0000-0002-4140-0591},\textsuperscript{2} Niels~Warburton \orcidlink{0000-0003-0914-8645},\textsuperscript{1} Jakob~Neef \orcidlink{0000-0002-3215-5694}\textsuperscript{1}
}
\address{
	\textsuperscript{1} School of Mathematics \& Statistics, University College Dublin, Belfield, Dublin 4, Ireland\\
	\textsuperscript{2} School of Mathematical Sciences and STAG Research Centre, University of Southampton, Southampton SO17 1BJ, United Kingdom
}

%
%
%
%

\begin{abstract}
	We compute the (displacement) gravitational wave memory due to a quasi-circular inspiral of two black holes using a variety of perturbative techniques.
	Within post-Newtonian theory, we extend previous results to 3.5PN order, both for non-spinning compact binaries and for non-precessing, spinning binary black holes (including horizon flux contributions).
		Using the gravitational self-force approach, we compute the memory at first order in the mass ratio for inspirals into a Kerr black hole.
	We do this both numerically and via a double post-Newtonian--self-force expansion which we carry out to 5PN order.
	At second order in the self-force approach, near-zone calculations encounter an infrared divergence  associated with memory, which is resolved through matching the near-zone solution to a post-Minkowskian expansion in the far zone. We describe that matching procedure for the first time and show how it introduces nonlocal-in-time memory effects into the two-body dynamics at second order in the mass ratio, as was also predicted by recent 5PN calculations within the effective field theory approach.
	We then compute the gravitational-wave memory through second order in the mass ratio (excluding certain possible memory distortion effects) and find that it agrees well with recent results from numerical relativity simulations for near-comparable-mass binaries.
\end{abstract}
\maketitle

\tableofcontents

\newpage
\section{Introduction}

The direct detection of gravitational waves (GWs) sourced by compact binary coalescences has confirmed many predictions of the theory of general relativity (GR)~\cite{LIGOScientific:2021sio}.
One prediction that has yet to be experimentally established is the permanent displacement of test masses after the oscillatory part of a GW has passed.
This imprint of a passing wave is known as GW memory.

There are a variety of effects that fall under the umbrella of GW memory.
Linear (displacement) memory is sourced by the motion of bodies that may initially be bound but end up unbound in their final state~\cite{ZeldovichCluster}, e.g., supernova explosions~\cite{Richardson:2024wpp, Bhattacharya:2023wzl} and hyperbolic scatterings~\cite{DeVittori:2014psa,Hait:2022ukn}.
Compact binary coalescences, the only type of source observed so far by GW detectors, cannot generate linear memory, but instead can source the so-called \textit{nonlinear} (displacement) memory, which arises from nonlinear interactions between different multipole modes of the linear piece of the GWs~\cite{Blanchet:1990twn,Christodoulou,Blanchet:1992br, ThorneMemoryII, PhysRevD.44.R2945, Blanchet:1997ji, Favata:2008yd, Favata:2011qi, Faye:2014fra, Trestini:2023wwg}.
Recently, other higher-order memory effects have also been studied, though in general these have a smaller influence and are thus more difficult to observe~\cite{Nichols:2017rqr,Flanagan:2018yzh,Grant:2023jhd,Grant:2023ged,Siddhant:2024nft}. 
The influence of memory on test gyroscopes has also been explored~\cite{Seraj:2021rxd,Faye:2024utu}.


In addition to being an interesting physical prediction of GR, GW memory  is of significant theoretical interest. Memory represents the set of conserved charges associated with the Bondi-Metzner-Sachs (BMS) group~\cite{Barnich:2011mi,Compere:2019gft}, the infinite-dimensional symmetry group of future null infinity~\cite{Bondi:1962px,Sachs:1962wk} (see \cite{Madler:2016xju} for a review of the Bondi-Sachs formalism). Memory is also closely related to the soft graviton theorem~\cite{Strominger:2014pwa, He:2014laa, Pasterski:2015tva}. Together, memory, BMS symmetries, and soft theorems form an ``infrared triangle''~\cite{Strominger:2017zoo}, which plays a fundamental role in the celestial holography program~\cite{Pasterski:2021raf,Pasterski:2021rjz,Raclariu:2021zjz}. Memory effects (BMS charges and analogous charges on black hole horizons~\cite{Rahman:2019bmk,Bhattacharjee:2020vfb,Galoppo:2024vww}) can also be interpreted as soft hair on black holes~\cite{Hawking:2016msc,Donnay:2016ejv,Chandrasekaran:2018aop,Donnay:2019jiz,Chandrasekaran:2023vzb,Bhat:2024cyq}, suggesting they could be significant for the black hole information paradox.

Detecting GW memory is challenging due to its low frequency and low amplitude compared to the oscillatory part of the GW.  
A variety of methods have thus been proposed to detect it with different GW detectors.
Current ground-based detectors are not sufficiently sensitive to detect memory from a single binary coalescence, but
by detecting hundreds or thousands of events,  they could achieve a cumulative signal-to-noise ratio large enough to confirm the presence of the memory effect~\cite{Lasky:2016knh, Hubner:2019sly}.
Future low-frequency GW detectors are more promising for detecting GW memory~\cite{Islo:2019qht,Sesana:2019vho,Inchauspe:2024ibs}.
Pulsar timing arrays have already searched for GW memory in the nano-hertz regime~\cite{Agazie:2023eig}, and the recently adopted space-based Laser Interferometer Space Antenna, LISA~\cite{Colpi:2024xhw}, will observe massive black hole binary coalescences in the millihertz band;
with signal-to-noise ratios in the thousands, these sources should provide a direct detection of GW memory from a single event~\cite{Favata:2009ii}.
In addition to the above, there have also been proposals to detect the effect of GW memory via astrometric observations of distant light sources~\cite{Madison:2020xhh,Boybeyi:2024aax}.

The nature of GW astronomy means that, in order to detect and characterise GW memory, we need theoretical predictions of its effect that we can compare with the data from the detectors.
By comparing with GW template models computed with and without memory, we can decide which is a better fit to the data, and thus further test GR. 
The inclusion of GW memory in template models may also bolster parameter estimation efforts by breaking degeneracies between parameters, e.g., inclination and luminosity distance~\cite{Gasparotto:2023fcg,Xu:2024ybt,Goncharov:2023woe}.
If we have models that include corrections to GR, then these can be used to search for new physics~\cite{Kilicarslan:2018bia}. 
For example, in alternative theories of gravity such as scalar-tensor theories, memory effects can enter a lower PN order than in~GR~\cite{Lang:2013fna, Sennett:2016klh, Bernard:2022noq}.

There has been much progress in modelling the memory effects in different regimes of the compact binary parameter space.
When the two bodies are widely separated, the post-Newtonian (PN) expansion can be employed.
In that context, GW memory effects have been studied at various PN order for both circular and eccentric binaries, featuring both non-spinning and spinning black holes \cite{Favata:2008yd, Favata:2011qi, Ebersold:2019kdc, Mitman:2022kwt}. This effort has recently culminated in the completion of the memory piece of the waveform at 3PN order for spinning black holes (with aligned or anti-aligned spins) on eccentric orbits~\cite{Henry:2023tka}. In the case on eccentric orbits, memory effects do not only enter the waveform, but also the angular momentum flux~\cite{Arun:2009mc,Loutrel:2016cdw,Trestini:2024mfs}, which arises at 2.5PN beyond the leading quadrupolar expression.
In the deep strong-field regime, the PN approximation breaks down, so we instead turn to numerical relativity (NR) simulations~\cite{baumgarte2010numerical}.
This approach solves the full non-linear Einstein field equations on supercomputers.
Until recently, the methods used to extrapolate the waveform from the edge of the computational domain to future null infinity made it challenging to directly compute memory effects from NR simulations (though post-processing methods were developed for displacement memory~\cite{Mitman:2020bjf,Talbot:2018sgr}).
Recently developed Cauchy-characteristic evolution methods have enabled the memory to be directly calculated from NR simulations~\cite{Mitman:2020pbt, Mitman:2024uss}.
However, the already high computational cost of NR simulations grows as the mass ratio becomes more extreme due to the high resolution requirements and the need to resolve many more orbits.
This places systematic studies of binaries with mass ratios beyond $\sim15:1$ out of reach of current NR technology. 

By combining PN and NR results, fast-to-generate waveform models have been developed that include GW memory contributions.
This was achieved \textit{via} a variety of methods including surrogate~\cite{Yoo:2023spi}, phenomenological models~\cite{Rossello-Sastre:2024zlr} and effective-one-body models \cite{Grilli:2024lfh}.
Black hole perturbation theory (BHPT), specifically gravitational self-force (GSF) theory~\cite{Barack:2018yvs,PoundAndWardell}, can also assist in building these models. 
Traditional BHPT sources like extreme-mass-ratio inspirals are not ideal for measuring memory (though intermediate mass ratio inspirals hold more promise~\cite{Islam:2021old}).
Nonetheless, BHPT results can be useful in developing global models that span a wide portion of the binary configuration parameter space.
Concurrently with the development of the present work, two other authors have taken this approach~\cite{Elhashash:2024thm}.
In that work, they computed the memory effect at first order in the mass ratio to a very high PN order for a circular orbit around a non-spinning black hole and used the result to inform a more global model for non-spinning binaries developed by fitting to NR data.

While GW memory can be viewed purely as a feature of the waveform at future null infinity, it is also directly associated with hereditary, nonlocal-in-time effects within the two-body systems that generate the GWs. Such effects have a storied history in PN theory~\cite{Blanchet:1987wq}, where they are generated both by tails describing the backscattering of waves off of spacetime curvature and by memory effects associated with GW fluxes that re-radiate soft, zero-frequency waves back into the near zone. Tail effects, which enter the waveform at relative 1.5PN order, were critical in completing the 4PN two-body dynamics~\cite{Damour:2014jta}. Memory effects, which enter the waveform at relative 0PN order, are at the center of disagreements in the 5PN two-body dynamics~\cite{Blumlein:2020pyo,Blumlein:2021txe,Almeida:2023yia,Porto:2024cwd,Almeida:2024klb}. These near-zone memory effects in the 5PN dynamics enter at second order in the mass ratio~\cite{Porto:2024cwd}, suggesting they should enter GSF theory at second order. They have also been loosely studied directly within GSF theory in the context of a nonlinear scalar toy model~\cite{Pound:2015fma}, where it was found they do enter at second order, in agreement with the recent PN results.

In this work, we study GW memory at future null infinity, as well as memory effects in the near zone, for quasi-circular inspirals \textit{via} a variety of approaches.
The overarching approach we use to compute the GW memory is described in section~\ref{sec:theory}. 
In section~\ref{sec:3.5PN}, we first apply it to the case of  widely separated compact objects using PN techniques. Making use of the recent completion of the oscillatory piece of the 3.5PN waveform~\cite{Henry:2022ccf}, we are able to obtain the GW memory piece at 3.5PN, both for non-spinning compact objects and for spinning, non-precessing,  binary black holes, which completes the waveform of Ref.~\cite{Henry:2022ccf} and extends the 3PN results of Favata~\cite{Favata:2008yd}. We also take advantage of this computation to present in one place a certain number of important quantities (energy, energy flux at infinity, horizon energy flux, and phasing) in the case of spinning, non-precessing, binary black hole systems on circular orbits, at 3.5PN and to all orders in spin. These results were for the most part known, but were scattered throughout the literature. 

Next, we consider the case where the mass ratio is small (without restrictions on the separation), and overview the required GSF theory in section~\ref{sec:SF_theory}.
At second order in the mass ratio, there is considerable subtlety in extracting the asymptotic amplitudes of the metric perturbation at null infinity due to an infrared divergence, which arises in the multiscale expansion underpinning second-order self-force (2SF) calculations~\cite{Pound:2015fma}. We give a detailed derivation of our waveform extraction in section \ref{far-zone}, where we develop a post-Minkowskian approximation to handle the far-zone behavior, following reference~\cite{Pound:2015fma}. This method was used in all 2SF calculations to date~\cite{Pound:2019lzj,Warburton:2021kwk,Wardell:2021fyy} but was never detailed before now. We explore for the first time how it introduces memory effects into GSF theory, in both the near and far zones. We use it to show that memory enters into the near-zone two-body dynamics at second order in the mass ratio, consistent with the 5PN effects recently highlighted in PN theory~\cite{Porto:2024cwd,Almeida:2024klb}. The appearance at second order is in contrast with far-zone memory effects, which enter the waveform at first (rather than second) order in the mass ratio; this difference between near- and far-zone orders is analogous to the difference (of 2.5PN orders) between near and far zone orders in PN theory. Our post-Minkowskian analysis also reveals possible new second-order contributions to the far-zone metric, `memory distortion' terms generated by an interaction between memory and emitted waves.

For spinning black holes, we employ BHPT to calculate GW memory for circular orbits at first order in the mass ratio in section~\ref{sec:KerrMemoryFirstOrder}.
We do this both numerically and analytically \textit{via} a double PN-BHPT expansion, which we carry through to 5PN order.
Returning to non-spinning black holes, we make use of the recent results extending GSF theory to second order~\cite{Wardell:2021fyy,Warburton:2021kwk,Pound:2019lzj} to compute the GW memory through second order in the mass ratio for circular orbits around a non-spinning black hole --- see section~\ref{Sec:2ndOrderResults}. Our calculation omits the possible contribution from new memory distortion effects mentioned above, but we nevertheless find that our results agree remarkably well with NR simulations for mass ratios as small as 10:1.

Except in section~\ref{sec:3.5PN}, we use geometrized units where $G = c = 1$, and we take the metric signature $(-,+,+,+)$. We denote Euler's constant as $\gamma_\mathrm{E}$. The index $l$ is used to denote spherical harmonic modes, while $\ell$ denotes spheroidal harmonic modes. We also define $\epsilon = m_2 / m_1$ as the small mass ratio, $q = m_1/ m_2$ as the large mass ratio, $M=m_1 + m_2$ as the total mass and $\nu = {m_1 m_2}/{M^2}$ as the symmetric mass ratio. In an asymptotic frame and in a set of radiative coordinates, the position of a distant observer will be denoted $\bm{X} = R \bm{N}$, where $R$ is the Euclidean norm of $\bm{X}$, and $\bm{N}$ is the unit normal~$\bm{N} = (\sin\Theta\cos\Phi,\sin\Theta\sin\Phi,\cos\Theta)$ associated to angles $\Theta^A=(\Theta, \Phi)$. Time is denoted~$t$ whereas retarded time is defined as $U =  t - R$.  The components of spatial vectors decomposed onto a Cartesian basis are denoted with Latin indices, denoted equivalently $N_i$ or $N^i$, since, asymptotically, indices are raised and lowered with the flat background metric $\eta_{\mu\nu}$, whose restriction to the spatial sector is the Kronecker~$\delta_{ij}$, associated with the Levi-Civita symbol $\epsilon_{ijk}$. A multi-index of length~$l$ will be denoted $L = i_1\cdots i_l$, whereas $L-1 = i_1 \cdots i_{l-1}$, etc. We use uppercase Latin indices from the middle of the alphabet ($L$, $M$, $N$,...) for multi-indices, reserving indices from the beginning of the alphabet ($A$, $B$, $C$, ...) for angular components.  These multi-indices are compatible with the Einstein summation conventions, and we write, e.g., $N_L = N_{i_1} \cdots N_{i_l}$. The symmetric projection is denoted with parentheses, as in  $\alpha_{(i}\beta_{j)} = \frac{1}{2}(\alpha_i \beta_j + \alpha_j \beta_i)$, whereas the symmetric trace-free (STF) projection is denoted with angled brackets, as in $\alpha_{\langle i}\beta_{j \rangle} = \alpha_{(i} \beta_{j)} - \frac{1}{3}\delta_{ij} \alpha_k \beta_k $. A common abbreviation will be to denote the STF projection with a hat, e.g., $\hat{N}_L \equiv N_{\langle i_1} N_{i_2} \cdots N_{i_{l-1}} N _{i_l \rangle}$. Angled brackets on  angular indices refer to the STF projection with respect to the metric $\Omega_{AB}$ on the unit 2-sphere, as in $h_{\langle AB\rangle} = h_{AB} - \frac{1}{2}\Omega_{AB}\Omega^{CD}h_{CD}$, and angular indices are lowered and raised with $\Omega_{AB}$ and its inverse $\Omega^{AB}$.

\section{Calculation of gravitational wave memory}\label{sec:theory}

The non-linear displacement memory was first discovered independently by Blanchet~\cite{Blanchet:1990twn} and by Christodoulou~\cite{Christodoulou} and subsequently studied by different groups~\cite{Blanchet:1992br, ThorneMemoryII, PhysRevD.44.R2945, Blanchet:1997ji}. 
Later on, Favata~\cite{Favata:2008yd, Favata:2011qi}  developed a systematic formalism for obtaining the memory contribution to the waveform, which we will follow. 
The displacement memory can be interpreted as the non-oscillatory component of the waveform arising from the re-radiation of GWs by massless gravitons~\cite{ThorneMemoryII, PhysRevD.44.R2945}. Decomposing the asymptotic metric~$g_{\mu\nu}$ into background and perturbation as $g_{\mu\nu} = \eta_{\mu\nu} + h_{\mu\nu} + \mathcal{O}(1/R^2)$, the dynamical piece of the metric perturbation is expressed by the elegant formula%
\begin{equation}
    \label{eq:NonlinearMemory}
    \left[h_{ij}^\text{TT}\right]^\text{mem} = \frac{4}{R} \perp^{\text{TT}}_{ij,ab}\int_{-\infty}^{U} \dd t' \int \dd \Omega'_2 \, \frac{\dd E^\text{GW}}{\dd t'\dd \Omega'_2}\frac{N_a' N_b'}{1- \bm{N'}\cdot\bm{N}}\;,
\end{equation}
where the angular integration element is defined as $\dd \Omega'_2 = \sin(\Theta) \dd \Theta' \dd \Phi'$ and is associated with $\bm{N}'$, whereas $R$, $\bm{N}$ and $U$ refer to the field point.   The angular flux distribution of GWs is denoted $\frac{\dd E^\text{GW}}{\dd t\dd \Omega_2}$, and the algebraic transverse-traceless (TT) projector for asymptotic metrics is defined by~\cite{Blanchet:2024mnz}
\begin{equation}\label{eq:TT projector}
\perp^{\text{TT}}_{ij,ab} = \perp_{a(i} \perp_{j) b} - \,\frac{1}{2}\! \perp_{ij} \perp_{ab} \,,
\end{equation}
where $\perp_{ij} = \delta_{ij} - N_i N_j$.
 In particular, this means that it is possible to predict the non-oscillatory memory contribution to the waveform solely based on the asymptotic expression of the oscillatory waveform at future null infinity. 
 Despite its elegance, it will prove more useful to decompose  equation~\eqref{eq:NonlinearMemory} as a multipolar expansion, which reads~\cite{Blanchet:2024mnz}
\begin{equation}\label{eq:NonlinearMemoryMultipolar}
\left[h_{ij}^\text{TT}\right]^\text{mem} =  \frac{8}{R} \perp^{\text{TT}}_{ij,ab} \sum_{l = 2}^{\infty} \frac{(2l +1)!!}{(l+2)!} N_{L-2} \int_{-\infty}^{U} \dd t' \int \dd \Omega' _2\, N'_{ab L-2} \frac{\dd E^\text{GW}}{\dd t'\dd \Omega'_2} \,.
\end{equation}
 
We now need to control the oscillatory piece of the waveform before we can derive the non-oscillatory piece. Regardless of the approximation method used to solve the Einstein equations, the TT asymptotic metric can be written as a multipolar expansion in terms of so-called \textit{radiative moments}, decomposed either in a basis of spin-weighted spherical harmonics or on a basis of symmetric trace-free tensor harmonics~\cite{Thorne:1980ru, Blanchet:2024mnz}:
\begin{align}\label{eq:metricRadiativeMoments}
h_{ij}^\text{TT} &=  \frac{4}{R} \perp^{\text{TT}}_{ij,ab} \sum_{l = 2}^{\infty} \frac{1}{\ell!}  \Bigg[N_{L-2}\, \cU_{ab L-2} - \frac{2l}{l+1} N_{cL-2} \,\epsilon_{cd(a} \cV_{b)dL-2}\Bigg] + \mathcal{O}\left(\frac{1}{R^2}\right) \nonumber\\
& = \frac{1}{R}\sum_{l =2}^{\infty} \sum_{m = -l}^{l} \Bigg[ \cU_ {lm} T_{ij}^{E2,lm} + \cV_ {lm} T_{ij}^{B2,lm}\Bigg] + \mathcal{O}\left(\frac{1}{R^2}\right)  \,.
\end{align}
In the first line, we have introduced the STF radiative moments $\cU_L$ and $\cV_L$, which are STF in their $l$ indices and pure functions of retarded time $U$. In the second line, we have introduce the spherical-harmonic radiative moments $\cU_{lm}$ and $\cV_{lm}$, which are pure functions of retarded time $U$, as well as the basis of ``pure-spin tensor harmonics''~\cite{Mathews1962, Zerilli1970TensorHI, Thorne:1980ru} denoted $T_{ij}^{E2,lm}$ and $T_{ij}^{B2,lm}$, which are given explicitly in (2.30) of~\cite{Thorne:1980ru}.
Note that the two types of radiative moments are straightforwardly related by the algebraic expressions
\begin{subequations}\label{eq:UlmVlm_vs_ULVL}\begin{align}
\cU_{lm} &= \frac{4}{\ell!}\sqrt{\frac{(l+1)(l+2)}{2l(l-1)}} \alpha_L^{\ell m} \cU_L \,, \\*
\cV_{lm} &= -\frac{8}{\ell!}\sqrt{\frac{l(l+2)}{2(l+1)(l-1)}} \alpha_L^{\ell m} \cV_L \, ,
\end{align}\end{subequations}
where the change-of-basis matrix reads~\cite{Henry:2021cek}
\begin{equation}
\alpha_L^{\ell m} = \frac{ \sqrt{4\pi} (-1)^{m} 2^{m/2} l! }{\sqrt{(2l+1)(l+m)!(l-m)!}} \overline{\frak{m}}_0^{\langle M} l_0^{L-M \rangle}.
\end{equation}
We have introduced some non-corotating reference triad $(\bm{n}_0, \bm{\lambda}_0, \bm{l}_0)$, with $\bm{l}_0$ orthogonal to the orbital plane, and $\overline{\bm{\frak{m}}}_0$ is the complex conjugate of $\bm{\frak{m}}_0 = (\bm{n}_0 + i \bm{\lambda}_0)/\sqrt{2}$.

Another useful formulation of the waveform is to express it in terms of the usual ``plus'' and ``cross'' polarizations (see p.~80 and footnote 90 of~\cite{Blanchet:2024mnz} for the exact conventions), and recast them into a single complex number that can be expanded onto a basis of spin-weighted spherical harmonics~\cite{Newman:1966ub}:  
\begin{equation} \label{eq:h_sphericalModes}
    h_+ - i h_\times = \frac{1}{R}\sum_{l=2}^\infty\sum_{m=-l}^{l} h_{lm} {\ }_{-2} Y_{lm}(\Theta,\Phi).
\end{equation}
The strain amplitudes are straightforwardly expressed in terms of radiative moments\footnote{This corresponds to the Blanchet \textit{et al.} convention as described by Favata~\cite{Favata:2008yd}, namely $\alpha = -1$. This sign difference originates from the passage from the ``gothic'' metric perturbation $\frak{h}^{\mu\nu} = \sqrt{g} g^{\mu\nu} - \eta^{\mu\nu}$ to the usual metric perturbation $h_{\mu\nu} = g_{\mu\nu} - \eta_{\mu\nu}$. See reference~\cite{Blanchet:2024mnz} for details.}:
\begin{equation}
    \label{eq:strainModeMultipoles}
    h_{lm} = -  \frac{1}{\sqrt{2}}\left(\cU_{lm} - i \cV_{lm}\right).
\end{equation}
We can also write the polar components of the strain in terms of the Bondi shear $C_{AB}$ and tensor harmonics in  Bondi-Sachs coordinates $(U,R,\Theta^A)$~\cite{Madler:2016xju}:
\begin{equation} \label{eq:shear}
  h_{AB}   = R\, C_{AB}(U,\Theta^C) = -R\sum_{l=2}^\infty\sum_{m=-l}^{l}\sqrt{\frac{2(l-2)!}{(l+2)!}}(\cU_{lm} Y^{lm}_{AB}+\cV_{lm} X^{lm}_{AB}),
\end{equation}
where $Y^{lm}_{AB}$ and $X^{lm}_{AB}$ are even- and odd-parity tensor harmonics, respectively~\cite{MartelPoisson}. The two representations are related by $h_+ - i h_\times = \bar m^A  \bar m^B h_{AB}$, where $m^A=\frac{1}{\sqrt{2}R}(1,i\csc\Theta )$ and its complex conjugate $\bar m^A$ form a (Newman-Penrose) complex dyad on the celestial sphere of radius $R$~\cite{Goldberg:1966uu}.
 We refer to~\cite{Compere:2019gft,Blanchet:2020ngx,Blanchet:2023pce} for detailed translations between the two common treatments of radiative spacetime asymptotics: Bondi-type formulations and the Cartesian multipolar formulations used in the multipolar post-Minkowskian (MPM) approach. For Bondi quantities, we follow the conventions of references~\cite{Flanagan:2015pxa,Compere:2019gft}.

By comparing the memory piece of the waveform \eqref{eq:NonlinearMemoryMultipolar} with the general ansatz \eqref{eq:metricRadiativeMoments}, we can then identify the memory contribution to the radiative moments. The contribution to the mass-type radiative moment is given by~\cite{Blanchet:1992br, Favata:2008yd}
\begin{subequations}    \label{Umem}\begin{align}  
\cU_{L}^\text{(mem)} &= \frac{2(2l+1)!!}{(l+1)(l+2)} \int_{-\infty}^U \dd t \int \dd \Omega_2 \frac{\dd E^\text{GW}}{\dd t\dd \Omega_2} n_{\langle L \rangle}  \label{seq:UL_mem} \,, \\
    \cU_{lm}^\text{(mem)} &= 32 \pi \sqrt{\frac{(l-2)!}{2(l+2)!}}\int_{-\infty}^{U}  \dd t \int \dd \Omega_2 \frac{\dd E^\text{GW}}{\dd t\dd \Omega_2} {\ }_{0} Y_{lm}^*  \, , \label{seq:Ulm_mem}
\end{align}\end{subequations}
whereas the memory contribution to the current-type moments always vanishes, namely $\cV_L^\text{(mem)} = 0$ and $\cV_{lm}^\text{(mem)} = 0$.

We now use the  expression of the angular energy flux distribution $\frac{\dd E^\text{GW}}{\dd t\dd \Omega_2}$ in terms of the waveform, which reads
\begin{equation}
    \label{modeflux}
    \frac{\dd E^\text{GW}}{\dd t\dd \Omega_2} = \frac{R^2}{16\pi} \left( \dot h_+^2  + \dot h_\times^2 \right) .
\end{equation}
This can equivalently be written in terms of the Bondi news tensor, $\dot C_{AB}$, as
\begin{equation}
    \label{modeflux - news}
    \frac{\dd E^\text{GW}}{\dd t\dd \Omega_2} = \frac{1}{32\pi} \dot C_{AB} \dot C^{AB}.
\end{equation}
Injecting equation \eqref{eq:h_sphericalModes} into equation \eqref{modeflux}, one obtains the angular energy flux distribution in terms of the spherical harmonic $(l,m)$ modes~\cite{Favata:2008yd}:
\begin{equation}
\frac{\dd E^\text{GW}}{\dd t\dd \Omega_2} = \frac{1}{16\pi}\sum_{l'=2}^{\infty}\sum_{l''=2}^{\infty}\sum_{m'=-l'}^{l'}\sum_{m''=-l''}^{l''} \dot{h}_{l'm'}\, \dot{h}_{l''m''}^* {\ }_{-2} Y_{l'm'}  {\ }_{-2} Y_{l''m''}^* \,.
\end{equation}
Plugging \eqref{modeflux} into \eqref{seq:Ulm_mem} and applying a time derivative (so as to remove the hereditary time integral), we finally obtain~\cite{Favata:2008yd}
 %
\begin{equation}
    \label{Umemdot}
    \dot{\cU}_{lm}^\text{(mem)} =  \sqrt{\frac{2(l-2)!}{(l+2)!}}\sum_{l'=2}^{\infty}\sum_{l''=2}^{\infty}\sum_{m'=-l'}^{l'}\sum_{m''=-l''}^{l''} (-1)^{m+m''} \dot{h}_{l'm'}\, \dot{h}_{l''m''}^* \,G^{2,-2,0}_{l' \!,\, l'' \!,\, m' \!,\, -m'' \!,\, -m}   \,,
\end{equation}
where we have introduced the numerical constant
\begin{equation}\label{eq:G_def}
G^{s_1,s_2,s_3}_{l_1,l_2,l_3, m_1,m_2,m_3} \equiv \int \dd \Omega_2\, {\ }_{-s_1} Y_{l_1m_1}\!\left(\Theta,\Phi\right)  {\ }_{-s_2} Y_{l_2m_2}\!\left(\Theta,\Phi\right) {\ }_{-s_3} Y_{l_3m_3}\!\left(\Theta,\Phi\right)  \,.
\end{equation}
This is related to the quantity $C^{\,l_1,m_1,s_1}_{\,l_2,m_2,s_2, l_3,m_3,s_3}$, used in 2SF literature and the \texttt{PerturbationEquations} package~\cite{PerturbationEquations}, by  \mbox{$C^{\,l_1,m_1,s_1}_{\,l_2,m_2,s_2, l_3,m_3,s_3}=(-1)^{m_1+s_1}G^{s_1,-s_2,-s_3}_{l_1,l_2,l_3, -m_1,m_2,m_3}$}.
If the condition $s_1 + s_2 + s_3 = 0$ is satisfied (which will always be the case for us), the angular integral can be expressed in terms of the Wigner $3$-$j$ symbols~\cite{Spiers:2023mor}:
\begin{equation}\label{eq:G->3j}
G^{s_1,s_2,s_3}_{l_1,l_2,l_3, m_1,m_2,m_3} = \sqrt{\frac{(2l_1+1)(2l_2+1)(2l_3+1)}{4 \pi}} 
        \begin{pmatrix}
            l_1 & l_2 &l_3\\
            s_1 & s_2 & s_3  
        \end{pmatrix}
        \begin{pmatrix}
            l_1 & l_2 &l_3\\
            m_1 & m_2 & m_3  
        \end{pmatrix} \,.
\end{equation}
This expression is useful because the Wigner $3$-$j$ symbols are efficiently implemented in \texttt{Mathematica}~\cite{Wigner3jMathematica}.  
We also note the useful symmetry, inherited from the $3$-$j$ symbols~\cite{Spiers:2023mor}, 
\begin{equation}
    \label{G_Symmetry}
    G^{s_1, s_2, s_3}_{l_1, l_2, l_3, m_1, m_2, m_3} = (-1)^{(l_1 + l_2 + l_3)} G^{s_1, s_2, s_3}_{l_1, l_2, l_3, -m_1, -m_2, -m_3}\, .
\end{equation}

In equation \eqref{Umemdot}, one could worry that the memory piece of the waveform is sourced by the time-derivative of the complete waveform $h_{\ell m}$, which contains the memory terms $\dot{\cU}_{lm}^\text{(mem)}$ we are trying to compute. However, this cyclic definition is easily solved within perturbation theory, because the non-oscillatory memory terms  evolve on slow radiation-reaction time scales, whilst the oscillatory terms evolve on much faster orbital timescales. Thus, the contributions of time derivatives $\dot{\cU}_{lm}^\text{(mem)}$ on the right-hand side of \eqref{Umemdot} enter at  5PN  and 2SF orders beyond the leading-order oscillatory terms, and can thus be treated hierarchically.

In the case of a non-precessing quasi-circular orbit, there is a clean split between the oscillatory modes, which correspond to $m \neq 0$, and the non-oscillatory memory modes, which correspond to $m = 0$. Since there is no memory contribution to the current-type radiative moments, we will in fact only need to compute   
\begin{equation}
    \label{UtoStrain}
    h_{l0}^\text{(mem)} = -\frac{1}{\sqrt{2}}\, \cU_{l0}^\text{(mem)} \,.
\end{equation}

Once the derivative $\dot \cU_{l0}^{\text{(mem)}}$ is calculated, all that is left to do is to perform the time integral in equation~\eqref{Umem}. In the case of circular orbits, this is achieved by changing variables in the integral, eliminating time in favor of, e.g., the orbital frequency $\Omega$. Crucially, this requires controlling the ``chirp'' or time-evolution of the frequency $\dot{\Omega}$. The chirp can be controlled to low PN or GSF orders using rigorous asymptotic matching, but as we will see, it can be controlled to much higher order by assuming the validity of some flux balance laws.

\section{Post-Newtonian calculation through 3.5PN order for non-spinning, quasi-circular binaries}\label{sec:3.5PN}

The memory contribution to the waveform can be computed fully analytically within the PN approximation. After a review of the modern theoretical framework for PN calculations, we will discuss how the nonlinear memory naturally arises for it, but can \textit{a priori} only be controlled at low PN orders. Then, assuming that the flux balance law for energy holds at 3.5PN  and applying the formalism of section~\ref{sec:theory}, we will be able to derive the memory contribution for non-spinning quasi-circular orbits up to 3.5PN order.

\begingroup
\allowdisplaybreaks
\subsection{Review of the PN-MPM construction}

The first concrete estimates for the memory effects were performed in the PN approximation, and more specifically using the formalism of Blanchet and Damour~\cite{Blanchet:1985sp, Blanchet:1986dk, Blanchet:1987wq, Blanchet:1992br,Blanchet:1998in}. This formalism combines the PN expansion in small orbital velocities ($v/c\to 0$) with the multipolar post-Minkowskian (MPM) method, which combines the PM or non-linearity expansion ($G\to 0$) with multipolar series parametrized by specific multipole moments.  The PN expansion is performed in the ``near zone'', which encompasses the matter source, but whose radius is much less than a gravitational wavelength. The MPM expansion is valid in the ``exterior vacuum zone'', which overlaps with the near zone of the source (but excludes the domain where there is matter) and extends into the far wave zone. The MPM field represents the most general solution of the Einstein field equation (say, in harmonic coordinates) in the exterior zone. The PN and MPM expansions are matched in the ``buffer zone'', which is the region where both expansions are valid: the whole procedure is called the post-Newtonian-multipolar-post-Minkowkian (PN-MPM) formalism.

In this section, we will restore $G$ and $c$, as they are useful for power counting within the PN approach. The starting point of the formalism is the Einstein equations, written in the ``relaxed'' form of Landau and Lifschitz~\cite{LL}. Introducing the \textit{gothic metric deviation}\footnote{At linearized level in $h$, we have $\frak{h}^{\mu\nu} = - h^{\mu\nu} + \frac{1}{2} h \eta^{\mu\nu} + \mathcal{O}(h^2)$, where $h^{\mu\nu} = g^{\mu\nu} - \eta^{\mu\nu}$ is the perturbation of the inverse metric, and $h = h^{\mu\nu}\eta_{\mu\nu}$ its trace. Thus, asymptotically and in the TT gauge, we have  $\frak{h}_{ij}^\text{TT} = - h_{ij}^\text{TT} + \mathcal{O}(1/R^2)$.}, defined by \mbox{$\frak{h}^{\mu\nu}=\sqrt{-g}g^{\mu\nu}-\eta^{\mu\nu}$}, and imposing the harmonic gauge condition~$\partial_\mu \frak{h}^{\mu\nu} = 0$, the Einstein equations read
\begin{equation}
\Box_\eta \frak{h}^{\mu\nu} = \frac{16\pi G}{c^4} \tau^{\mu\nu}\,,
\end{equation}
where $\Box_\eta = \eta^{\alpha\beta} \partial_\alpha \partial_\beta$ is the flat-space d'Alembert operator, and the stress-energy pseudo-tensor is given by 
\begin{equation}
\tau^{\mu\nu}= (-g) T^{\mu\nu} + \frac{c^4}{16\pi G}\Lambda^{\mu\nu}[\frak{h}^{\alpha\beta}] \,.
\end{equation}
Here,  $T^{\alpha\beta}$ is the stress-energy tensor, $g$ is the determinant of the metric and $\Lambda^{\alpha\beta}$ is a functional that is at least quadratic in $\frak{h}^{\mu\nu}$ [see (56) of~\cite{Blanchet:2024mnz}].  We also introduce the harmonic coordinate system $(t,\bm{x} = r \bm{n})$, associated to retarded time $u = t - r/c$.

When specializing to the MPM construction, the metric is first written as a formal expansion thanks to the book-keeping parameter $G$, namely 
\begin{equation}
\frak{h}^{\mu\nu}_\text{MPM} = \sum_{n=1}^{+\infty} G^n \,\frak{h}^{\mu\nu}_n \,. 
\end{equation}
Since the MPM construction is valid only in the exterior vacuum zone, the stress-energy tensor will drop out of the field equations. At linearized order ($n=1$), one needs to solve $\Box_\eta \frak{h}^{\mu\nu}_1 = 0$ along with the harmonic gauge condition $\partial_\mu \frak{h}^{\mu\nu}_1 = 0$. The most general solution (under some usual physical conditions, such at asymptotic flatness and no-incoming radiation) can be parametrized by a set of six moments: a mass-type \textit{source moment}~$\cI_L$, a current-type source moment~$\cJ_L$, and four \textit{gauge moments}, denoted $\cW_L$, $\cX_L$, $\cY_L$ and $\cZ_L$, which are symmetric and trace-free (STF) with respect to their $l$ indices [see (64-66) of~\cite{Blanchet:2024mnz}]. For now, these moments are entirely free and parametrize the possible solutions to the field equations, but we will determine them shortly. At nonlinear orders~($n\ge 2$), one then needs to solve a hierarchy of equations, $\Box_\eta \frak{h}_n^{\mu\nu} = \Lambda_n[\frak{h}^{\alpha\beta}_1, ..., \frak{h}^{\alpha\beta}_{n-1}]$, also under the harmonic gauge condition $\partial_\mu  \frak{h}_n^{\mu\nu} = 0$. Since the source requires the metric at order $n$, one only needs to know the metric at order~$\le n-1$, so the procedure is constructive.

Once the whole MPM construction is performed, one can show~\cite{Blanchet:1998in} that the matching procedure between the exterior PM metric and the interior PN metric entirely determines the source and gauge moments as functionals of the gravitation stress-energy pseudo-tensor $\overline{\tau}^{\alpha\beta}$ (the overbar indicates a PN expansion). For instance,
\begin{equation}\label{eq:ILsource}
	\cI_L = \FP \int \dd^3\mathbf{x} \,\Bigl(\frac{r}{r_0}\Bigr)^B \hat{x}_L \bigl[\overline{\tau}^{00} + \overline{\tau}^{ii}\bigr]+ \cdots\,,
\end{equation}
where we recall the notation for the STF projection $\hat{x}_L = x_{\langle i_1} ... x_{i_{\ell} \rangle}$, the finite part (FP) denotes a particular IR regularization when $B\to 0$ depending on an arbitrary regularization scale $r_0$, and the ellipsis denote other terms, known to all orders for general systems [see (135) and (141) of~\cite{Blanchet:2024mnz}]. In the case of compact binary systems, the multipolar moments can be determined explicitly in terms of the positions, velocities and spins of the two particles, see e.g.~\cite{Marchand:2020fpt, Larrouturou:2021dma,Larrouturou:2021gqo} for $\cI_{ij}$ at 4PN in the non-spinning case (at these high orders, \eqref{eq:ILsource} must be modified to account for the introduction of dimensional regularization). 

Note that the matching procedure also  determines the unspecified homogeneous solution that appears in the  PN near-zone metric in terms of the exterior multipolar moments~\cite{Blanchet:1987wq, Blanchet:1993ng, Blanchet:2005ft, Poujade:2001ie}, hence effectively acting as a boundary condition.  This result has been used to determine the radiation-reaction part of the equations of motion up to 4.5PN order~\cite{Nissanke:2004er,Gopakumar:1997ng,Blanchet:2024loi} as well as the conservative tail contribution at 4PN  (see also \cite{Porto:2024cwd, Almeida:2024klb} for recent discussions of the conservative memory contributions to the equations of motion at 5PN).

An alternative version of the MPM construction consists in fixing the residual gauge freedom of the harmonic gauge by asking that the MPM metric be described by only two \textit{canonical moments}, namely $\cM_L$ and $\cS_L$. This is practical as it reduces the number of moments one needs to work with, but an explicit matching to the source (analogous to \eqref{eq:ILsource} for  the source moments) cannot be obtained. Instead, one usually first determines the source and gauge moments explicitly, then finds relations of the type $\cM_L = \cI_L +  \mathcal{O}(c^{-5})$ and $\cS_L = \cJ_L +  \mathcal{O}(c^{-5})$, where the $\mathcal{O}(c^{-5})$ terms contain at least quadratic combinations of source or gauge moments (see~\cite{Blanchet:2022vsm, Blanchet:2023sbv} for a 4PN-accurate relation between $\cI_L$ and $\cM_L$).

Finally, once the full MPM metric is determined in terms of the canonical moments, which are themselves determined in terms of the orbital variables of the source, one can read off the asymptotic expression of the metric at future null infinity. Due to tail effects, harmonic coordinates do not have the correct peeling properties and the leading fall-off is in fact $\ln(r)/r$. This is corrected by a coordinate transformation towards so-called \textit{radiative coordinates}, which can either be performed at the end or by introducing yet another MPM algorithm in terms of \textit{radiative canonical moments}~\cite{Blanchet:1986dk, Trestini:2022tot}. In radiative coordinates, the dominant $1/R$ can be written as a multipolar expansion in term of \textit{radiative moments} $\cU_L$ and $\cV_L$, as in equation \eqref{eq:metricRadiativeMoments}. The radiative moments are related  to the canonical moments by relations of the type $\cU_L = \cM^{(l)}_L + \bigO(c^{-3})$ and $\cV_L = \cS^{(l)}_L + \bigO(c^{-3})$, where the $\bigO(c^{-3})$ terms are  at least quadratic in the canonical moments. These corrections contain hereditary terms such as tails, which stem from the interaction between the mass-type quadrupole $\cM_{ij}$ and the conserved mass-type monopole $\cM$ (which is in fact the ADM mass of the system), and read~\cite{Blanchet:1992br}
\begin{equation}
\cU_{ij}^{\cM\times \cM_{pq}} =  \frac{2G \cM}{c^3} \int_{0}^{+\infty} \dd \tau \,\cM^{(4)}_{ij}(U-\tau) \left[ \ln\left(\frac{c\tau}{2b_0}\right) + \frac{11}{12} \right] \,.
\end{equation}

Note the arbitrary length scale $b_0$ appearing in the previous tail term. In fact, when decomposing the final spherical harmonic modes in terms of the orbital phase~$\phi$, this scale will appear in complex corrections to the amplitudes of the GW. This is because the orbital phase $\phi$ is not a physical observable for an asymptotic observer. To solve this, we introduce an auxiliary phase~\cite{Blanchet:1993ec}, which corresponds to the half-phase associated to the $(2,2)$ mode, and reads   
\begin{equation}\label{eq:aux_phase}
	\psi = \phi_p - \frac{2 G \cM \Omega}{c^3}\ln\left(\frac{\Omega}{\Omega_0}\right) \,,
\end{equation}
where we recall that the orbital frequency reads $\Omega = \dd \phi_p / \dd t$ and that \mbox{$\Omega_0 = \exp\left(11/12-\gamma_{\mathrm{E}}\right) c  /{(4 b_0)}$}. At 3.5PN order, the orbital frequency and gravitational-wave frequency $\dd \psi / \dd t$ associated to the two phases are identical, so we do not have to worry about distinguishing them (see~\cite{Blanchet:2023sbv,Blanchet:2023bwj} for how they differ at 4PN order). Finally, we introduce the PN parameter 
\begin{equation}\label{eq:PNx}
	x = \left(\frac{G M \Omega}{c^3}\right)^{2/3}\,,
\end{equation}
which is dimensionless, invariant under a large family of gauge transformations, and a small 1PN quantity.
The amplitudes can then be decomposed as 
\begin{equation}\label{eq:hlm_vs_Hlm}
	h_{\ell m} = 2 M \nu x \sqrt{\frac{16 \pi}{5}} H_{\ell m}\mathrm{e}^{-\mathrm{i}m \psi} \,,
\end{equation}
where $H_{\ell m}$ is generically complex-valued to account for the dephasing between different modes.

\subsection{Memory terms in the standard PN-MPM approach}
\label{subsec:memory_standard_PN_MPM}
In the context of the PN-MPM approach, the displacement memory arises from nonlinear interactions. The first memory contribution stems from a quadrupole-quadrupole interaction~\cite{Blanchet:1997ji}, which reads
\begin{align}\label{eq:Uij_Mij_Mij}
\cU_{ij}^{\cM_{pq}\times \cM_{rs}} =  \frac{G}{c^5} \Bigg\{&- \frac{2}{7}  \int_{0}^{+\infty} \!\!\dd \tau \,\Big[ \cM^{(3)}_{a\langle i}  \cM^{(3)}_{j\rangle a}\Big](U-\tau) \nonumber \\*
& + \frac{1}{7} \cM^{(5)}_{a\langle i}  \cM_{j\rangle a} - \frac{5}{7} \cM^{(4)}_{a\langle i}  \cM^{(1)}_{j\rangle a}  - \frac{2}{7}\cM^{(3)}_{a\langle i}  \cM^{(2)}_{j\rangle a}\Bigg\} \,.
\end{align}
The instantaneous terms entering \eqref{eq:Uij_Mij_Mij} are irrelevant for the memory, but the integral is crucial. Note that, unlike the case of tails, taking a time derivative of \eqref{eq:Uij_Mij_Mij} entirely removes the nonlocal-in-time behavior, so we refer to these terms as semi-hereditary. Thus, the energy flux does not exhibit any memory-like integrals, because it only depends on the time-derivatives of the radiative moments. 

Consider now the case of circular orbits~\cite{Arun:2004ff}, where in the center-of-mass frame, the canonical quadrupole moment reads $\cM_{ij} = M \nu x_{\langle i} x_{j \rangle}$, at leading order. Performing the time-derivatives and reducing to the case of quasi-circular orbits, we find at leading order that\footnote{Here, the positions of the two particles are denoted $\bm{y}_1$ and $\bm{y_2}$ and their velocities $\bm{v}_1 = \dd \bm{y}_1 / \dd t$  and $\bm{v}_2 = \dd \bm{y}_2 / \dd t$ .  The relative separation vector is given by  $r_{12} = |\bm{y}_1 - \bm{y}_2|$ and $\bm{n}_{12} = (\bm{y}_1 - \bm{y}_2)/r_{12}$, whereas the relative velocity is defined as $\bm{v}_{12} = \bm{v}_1 - \bm{v}_2$. We then define the right-handed orthonormal triad $(\bm{n}_{12}, \bm{\lambda}_{12}, \bm{l}_{12})$ such that $\bm{\lambda}_{12} \cdot \bm{v}_{12} >0$ and $\bm{l}_{12}$ is orthogonal to the orbital plane.}
%
%
\begin{align}\label{eq:MijMij_expr1}
\Big[\cM^{(3)}_{a\langle i}  \cM^{(3)}_{j\rangle a}\Big](U-\tau) = \frac{16 c^{10} \nu x^5}{G^2}\bigg[n_{12}^{\langle i} (U-\tau) n_{12}^{j \rangle}(U-\tau) + \lambda_{12}^{\langle i}(U-\tau) \lambda_{12}^{j \rangle}(U-\tau) \bigg] \,.
\end{align}
%
In view of the integration, we now project the corotating basis vectors at any time in the remote past onto the corotating basis vectors at retarded time $U$, namely 
\begin{subequations}\label{eq:basis_change}\begin{align}
n_{12}^i(U-\tau) = \cos\Big[\omega(U-\tau)\Big] n_{12}^i(U) + \sin\Big[\omega(U-\tau)\Big] \lambda_{12}^i(U) \,, \\*
\lambda_{12}^i(U-\tau) = -\sin\Big[\omega(U-\tau)\Big] n_{12}^i(U) + \cos\Big[\omega(U-\tau) \Big] \lambda_{12}^i(U) \,.
\end{align}\end{subequations}
Injecting \eqref{eq:basis_change} into \eqref{eq:MijMij_expr1} and doing some trigonometry, we find that
\begin{align}\label{eq:MijMij_expr2}
\Big[\cM^{(3)}_{a\langle i}  \cM^{(3)}_{j\rangle a}\Big](U-\tau)= \frac{16 c^{10} \nu x^5}{G^2}\bigg[n_{12}^{\langle i} (U)\, n_{12}^{j \rangle}(U) + \lambda_{12}^{\langle i}(U) \, \lambda_{12}^{j \rangle}(U) \bigg] \,. 
\end{align}
Remarkably, at leading order, $\cM^{(3)}_{a\langle i}  \cM^{(3)}_{j\rangle a}$ does not have any oscillatory piece (at subleading order, an oscillatory piece will appear though). In this approach, the nonoscillatory memory term can be interpreted as the interference at quadratic order between two quadrupole moments oscillating at the same frequency. One could worry that, for a perfectly circular orbit, injecting \eqref{eq:MijMij_expr2} into the integral appearing in \eqref{eq:Uij_Mij_Mij} will lead to an infinite contribution. However, we are dealing with a \textit{quasi-circular} orbit, and the time dependence of \eqref{eq:MijMij_expr2} is now entirely contained in the frequency variable $x$, which secularly increases due to radiation reaction. At leading order, its evolution~\cite{Favata:2008yd} is dictated by $\dot{x} =  64 c^3 \nu x^5 /(G m)$, such that we can perform a change of variables analogous to \eqref{eq:evolution}, namely
\begin{equation}
\int_0^{+\infty} \dd \tau \, x^5(U-\tau)    =  \int_{0}^{x(U)} \dd x  \, \frac{x^5}{\dot{x}} = \frac{G M}{64 c^3 \nu}   x(U) \,.
\end{equation}
Using this formula to integrate \eqref{eq:MijMij_expr2}, injecting the result into \eqref{eq:Uij_Mij_Mij},  working our way back to the spherical harmonic decomposition using \eqref{eq:UlmVlm_vs_ULVL} and \eqref{eq:strainModeMultipoles}, and normalizing the mode as prescribed by \eqref{eq:hlm_vs_Hlm}, we finally find that 
\begin{equation}
H_{20} = -\frac{5}{14\sqrt{6}} \,.
\end{equation}


Although our computation started at relative 2.5PN order, we have found that the memory contribution enters at Newtonian order! It is in fact  well known~\cite{Arun:2004ff} that when dealing with non-oscillatory effects, one acquires a $-2.5$PN correction to the order one is working at (this corresponds in fact to the order of radiation reaction!).  Do note, however, that numerically, we have $H_{20} \approx -0.146$, which is small compared to $H_{22} \approx 1$.

When going to higher orders, we find that the non-oscillatory memory terms stem from various quadratic interactions between two canonical moments, which should of course be evaluated at the relevant PN order. Cubic interactions contribute to the memory for the first time at 4PN in the PN-MPM counting, which corresponds to $1.5$PN order in the waveform, once the $-2.5$PN correction to the order counting is accounted for. These so-called ``tails of memory'' arise from the interaction between the ADM mass and two mass-type quadrupole moments, and read~\cite{Trestini:2023wwg}
\begin{align}
\mathcal{U}_{ij}^{\cM\times\cM_{ij} \times \cM_{ij}} = \frac{8 G^2\cM}{7 c^8}\Bigg\{&\int_0^{+\infty} \!\dd\rho\,  \cM_{a \langle i}^{(4)}(U-\rho) \int_0^{+\infty} \!\dd \tau\,  \cM_{j \rangle a}^{(4)}(U-\rho-\tau)   \ln\left(\frac{c\tau}{2 r_0}\right) \nonumber\\*
& + \text{(tail-like terms)} + \text{(instantaneous terms)}\Bigg\}\,,
\end{align}
where $r_0$ is another arbitrary length scale. Thanks to this expression and the higher-order memory terms, the memory terms in the $(2,0)$ mode were computed for circular orbits at 1.5PN order. Although both the cubic tails of memory and the higher PN corrections to the quadratic memory have non-vanishing $1.5$PN contributions to the memory, these remarkably cancel out in the final result, which reads~\cite{Blanchet:2023sbv}
\begin{equation}\label{eq:H20_1p5PN_PNMPM}
H_{20} = -\frac{5}{14\sqrt{6}}\bigg[1 + \left(- \frac{4075}{4032} + \frac{67}{48}\nu \right)x + \mathcal{O}(x^2)\bigg] \,.
\end{equation}

\subsection{Favata's approach to memory and its extension at 3.5PN for non-spinning compact binaries}\label{sec:PNresults}

When the complete oscillatory part of the non-spinning 3PN waveform for quasi-circular orbits was available~\cite{Blanchet:2008je}, Favata used the methods presented in section~\ref{sec:theory} to complete it with the corresponding non-oscillatory (memory) piece~\cite{Favata:2008yd}. First, he computed $\dot{\cU}_{l0}$ consistently at 3PN order by injecting the modes given in reference~\cite{Blanchet:2008je} into equation~\eqref{Umemdot}. Then, using the 3PN phasing~\cite{Blanchet:2001aw,Blanchet:2001ax,Blanchet:2004ek}, he was able to perform the time integration in order to obtain $h_{l0}$, thus completing the 3PN waveform. Note, however, that the 3PN phasing obtained in reference~\cite{Blanchet:2001ax} relies on the key assumption that the flux-balance law for energy holds at 3PN order. This law was only recently proven to hold  at 2PN order~\cite{Blanchet:2024loi}, and subtleties in the formulation of the flux-balance laws, due to the difference between orbital and GW frequency, enter at 4PN~\cite{Blanchet:2023bwj}.
 
Since the oscillatory waveform for quasi-circular orbits is now available at 3.5PN~\cite{Favata:2008yd, Faye:2012we, Faye:2014fra, Henry:2021cek, Henry:2022ccf}, we were able to compute the full memory piece of the waveform at 3.5PN order, using exactly the same methodology as Favata~\cite{Favata:2008yd}. In the non-spinning case, we obtain the following expressions for the time-derivatives of the radiative moments:  
\begin{subequations}\begin{align}
\dot \cU^\text{(mem)}_{20} &= \frac{256}{7 }\sqrt{\frac{\pi }{15}} \frac{c^6\nu^2 x^5}{G^2} \Bigg\{1+ x \left(- \frac{1219}{288} + \frac{1}{24} \nu \right) +4\pi x^{3/2}\nonumber\\*
&\qquad\qquad  + x^2 \left(- \frac{793}{1782} - \frac{14023}{6336} \nu - \frac{4201}{1584}\nu^2\right) + x^{5/2} \pi \left(- \frac{2435}{144} - \frac{23}{12} \nu\right) \nonumber\\*
&\qquad\qquad + x^3\Bigg[\frac{174213949439}{1816214400} +\frac{16 }{3}\pi ^2 -\frac{1712}{105} \gamma_\text{E}  -\frac{856}{105} \log (16 x) \nonumber\\*
   &\qquad\qquad\qquad \qquad  +\left(-\frac{126714689}{4447872} + \frac{41}{48} \pi^2\right) \nu  +\frac{4168379 }{123552}\nu ^2+ \frac{142471}{46332} \nu ^3 \Bigg] \nonumber\\*
   &\qquad\qquad + x^{7/2}\pi \left(-\frac{33199  }{8448} +\frac{532909}{38016}\nu -\frac{27949}{864}\nu ^2\right) + \mathcal{O}(x^4)\Bigg\},\\
\dot \cU^\text{(mem)}_{40} &= \frac{64}{315} \sqrt{\frac{\pi }{5}} \frac{c^6\nu^2 x^5}{G^2}  \Bigg\{1+ x \left(- \frac{10133}{704} + \frac{25775}{528} \nu \right) +4\pi x^{3/2}\nonumber\\*
&\qquad\qquad  + x^2 \left( \frac{322533}{4576} - \frac{721593}{2288} \nu - \frac{237865}{5148}\nu^2\right) + x^{5/2} \pi \left(- \frac{1028}{11} + \frac{11114}{33} \nu\right) \nonumber\\*
&\qquad\qquad + x^3\Bigg[\frac{32585924257}{403603200} +\frac{16 }{3}\pi ^2 -\frac{1712}{105} \gamma_\text{E}  -\frac{856}{105} \log (16 x) \nonumber\\*
   &\qquad\qquad \qquad  +\left(\frac{4669843}{164736} + \frac{41}{48} \pi^2\right) \nu  +\frac{16531}{52}\nu ^2 - \frac{1145725 }{92664}\nu ^3 \Bigg] \nonumber\\*
   &\qquad\qquad + x^{7/2}\pi \left(\frac{37621537  }{82368} -\frac{8366815 }{4576}\nu -\frac{8011895 }{10296}   \nu ^2\right) + \mathcal{O}(x^4)\Bigg\},\\
\dot \cU^\text{(mem)}_{60} &= -\frac{839}{693} \sqrt{\frac{\pi }{2730}} \frac{c^6\nu^2 x^6}{G^2} \Bigg\{1 - \frac{3612}{839}\nu   \nonumber\\*
&\qquad\qquad+ x \left(- \frac{982361}{75510} + \frac{56387}{839} \nu  - \frac{62244}{839} \nu^2\right) + x^{3/2} \pi \left(\frac{5540}{839} - \frac{23184}{839}\nu \right) \nonumber\\
&\qquad\qquad  + x^2 \left( \frac{302491414}{4492845} - \frac{1516457957}{3851010} \nu + \frac{27377867}{42789}\nu^2 + \frac{1106868}{14263} \nu^3\right) \nonumber\\*
&\qquad\qquad + x^{5/2} \pi \left(- \frac{488407}{4195} + \frac{7810432}{12585} \nu - \frac{9431984}{12585} \nu^2\right)  + \mathcal{O}(x^3)\Bigg\},\\
\dot \cU^\text{(mem)}_{80} &=  \frac{75601 }{347490}\sqrt{\frac{\pi }{1190}} \frac{c^6\nu^2 x^7}{G^2} \Bigg\{1 - \frac{452070}{75601}\nu + \frac{733320}{75601} \nu^2 \nonumber\\*
&\qquad\qquad+ x \left(- \frac{7655551}{604808} + \frac{369735869}{4309257} \nu  - \frac{248030070}{1436419} \nu^2 + \frac{135873360}{1436419} \nu^3 \right) \nonumber\\*
&\qquad\qquad+ x^{3/2} \pi \left(\frac{708606}{75601} - \frac{4245492}{75601}\nu  + \frac{6707184}{75601}\nu^2 \right) + \mathcal{O}(x^2)\Bigg\},\\
\dot \cU^\text{(mem)}_{10,0} &= - \frac{525221 }{15752880}\sqrt{\frac{\pi }{385}} \frac{c^6\nu^2 x^8}{G^2} \Bigg\{1 - \frac{79841784}{9979199}\nu + \frac{198570240}{9979199} \nu^2 \nonumber\\
& \qquad\qquad\qquad\qquad\qquad\qquad - \frac{172307520}{9979199}\nu^3+ \mathcal{O}(x)\Bigg\}.
\end{align}\end{subequations}
Of course, these expressions agree at 3PN order with equation~(3.13) of~\cite{Favata:2008yd}, and we have also checked that they agree (at 3.5PN, at leading order in the symmetric mass ratio, and in the non-spinning limit) with the PN-GSF results of Eq.~\eqref{eq:U1SFPNl2} and of~\mbox{\ref{sec:further_comparisons}}.  In view of performing the time integration, we recall the 3.5PN-accurate expression of $\mathrm{d}x/\mathrm{d}t$ given by equation~(3.19) of~\cite{Favata:2008yd}, which we reproduce here:
\begin{align}\label{eq:dxdt_3.5PN}
\frac{\dd x}{\dd t} =   \frac{64 c^3 \nu}{5 G M} &  x^5 \Bigg\{ 1 + x \left( - \frac{743}{336} - \frac{11}{4} \nu \right) + 4 \pi x^{3/2}  \nonumber\\
& + x^2 \left( \frac{34\,103}{18\,144} + \frac{13\,661}{2016} \nu + \frac{59}{18} \nu^2 \right)+ \pi x^{5/2} \left( - \frac{4159}{672} - \frac{189}{8} \nu \right)  \nonumber\\ 
&+ x^3 \Bigg[ \frac{16\,447\,322\,263}{139\,708\,800} + \frac{16}{3}\pi^2 - \frac{856}{105} ( 2\gamma_E + \ln 16x )  \nonumber\\ 
&\qquad\quad + \left( - \frac{56\,198\,689}{217\,728} + \frac{451}{48}\pi^2 \right) \nu + \frac{541}{896} \nu^2 - \frac{5605}{2592} \nu^3 \Bigg] \nonumber\\
&  + \pi x^{7/2} \left( -\frac{4415}{4032} + \frac{358\,675}{6048} \nu + \frac{91\,495}{1512} \nu^2 \right) + O\left(\frac{1}{c^8}\right) \Bigg\} .
\end{align}
Note that \eqref{eq:dxdt_3.5PN} was obtained under the assumption that the 3.5PN flux-balance law  for energy  holds. We can now perform the time integration appearing in \eqref{eq:NonlinearMemory} by performing the change of variables
\begin{equation}\label{eq:PN time integration}
\int_{-\infty}^{U} \dd t\, f(x(t)) = \int_0^{x(U)} \frac{\dd x}{\dot{x}} f(x) \,,
\end{equation}
which can then be performed explicitly by using \eqref{eq:dxdt_3.5PN} and then PN-expanding  the integrand in powers of $x$.
Finally, we can work our way back to the $h_{l0}$ modes using \eqref{eq:strainModeMultipoles}. The modes are then normalized following \eqref{eq:hlm_vs_Hlm}, and read
\begin{subequations}\begin{align}
H_{20} &= - \frac{5}{14\sqrt{6}}\Bigg\{ 1 + x\left(-\frac{4075}{4032} + \frac{67}{48}\nu\right) + x^2 \left(-\frac{151877213}{67060224} - \frac{123815}{44352}\nu + \frac{205}{352} \nu^2 \right) \nonumber\\*
&\qquad\qquad\quad\ \ + x^{5/2} \pi \left(-\frac{253}{336} + \frac{253}{84}\nu\right)  \nonumber\\*
&\qquad\qquad\quad\ \  + x^{3} \Bigg[- \frac{4397711103307}{532580106240}+ \left( \frac{700464542023}{13948526592}- \frac{205}{96}\pi^2  \right) \nu \nonumber \\*
&\qquad\qquad\quad\ \ \qquad\ \ + \frac{69527951}{166053888} \nu^2  + \frac{1321981}{5930496}\nu^3 \Bigg]  \nonumber\\*
&\qquad\qquad\quad\ \   + x^{7/2} \pi \left( \frac{38351671}{28740096} - \frac{3486041}{598752} \nu - \frac{652889}{598752} \nu^2 \right) + \mathcal{O}(x^4) \Bigg\}\,, \label{eq:H20PN}\\
H_{40} &= - \frac{1}{504\sqrt{2}}\Bigg\{ 1 + x\left(-\frac{180101}{29568} + \frac{27227}{1056}\nu\right) \nonumber\\*
&\qquad\qquad\ \ + x^2 \left(\frac{2201411267}{158505984} - \frac{34829479}{432432}\nu + \frac{844951}{27456} \nu^2 \right) \nonumber\\*
&\qquad\qquad\ \ + x^{5/2} \pi \left(- \frac{13565}{1232} + \frac{13565}{308}\nu\right)  \nonumber\\*
&\qquad\qquad\ \  + x^{3} \Bigg[ \frac{15240463356751}{781117489152}+ \left(- \frac{1029744557245}{27897053184}- \frac{205}{96}\pi^2  \right) \nu \nonumber \\*
&\qquad\qquad\ \ \qquad\ \ - \frac{4174614175}{36900864} \nu^2  + \frac{221405645}{11860992}\nu^3 \Bigg]  \nonumber\\*
&\qquad\qquad\ \   + x^{7/2} \pi\left(\frac{21681663715}{747242496} - \frac{4531656385}{31135104}\nu + \frac{1502076385}{15567552} \nu^2 \right)  + \mathcal{O}(x^4) \Bigg\}\,, \\
H_{60} &=   \frac{4195}{1419264\sqrt{273}}\, x \Bigg\{ 1 - \frac{3612}{839}\nu  + x\left(-\frac{45661561}{6342840} + \frac{101414}{2517}\nu - \frac{48118}{839} \nu^2  \right) \nonumber\\*
&\qquad\quad\ \   + x^{3/2}\pi\left(\frac{1248}{839} - \frac{4992}{839} \nu\right)  \nonumber\\*
& \qquad\quad\ \ +x^2 \left(\frac{3012132889099}{144921208320} - \frac{27653500031}{191694720} \nu + \frac{1317967427}{4107744} \nu^2 - \frac{24793657}{342312} \nu^3 \right)  \nonumber\\*
&\qquad\quad\ \  + x^{5/2} \pi  \left(- \frac{74044807}{2718360} + \frac{18121333}{113265} \nu - \frac{26966906}{113265} \nu^2 \right)  + \mathcal{O}(x^3) \Bigg\}\,, \\
H_{80} &=  - \frac{75601}{213497856\sqrt{119}}\, x^2  \Bigg\{ 1 - \frac{452070}{75601}\nu  + \frac{733320}{75601} \nu^2 \nonumber \\*
&\qquad\qquad\ \ + x\left(-\frac{265361599}{33869248} + \frac{18177898147}{321757856}\nu - \frac{722521125}{5745676} \nu^2 + \frac{261283995}{2872838} \nu^3 \right) \nonumber\\*
&\qquad\qquad\ \   + x^{3/2}\pi\left(\frac{812404}{226803} - \frac{1624808}{75601} \nu + \frac{2515936}{75601} \nu^2\right)  + \mathcal{O}(x^2) \Bigg\}\,, \\
H_{10,0} &=  \frac{525221}{6452379648\sqrt{154}}\, x^3 \Bigg\{1 - \frac{79841784}{9979199}\nu + \frac{198570240}{9979199}\nu^2 - \frac{172307520}{9979199} \nu^3 + \mathcal{O}(x)\!\Bigg\} \,.
\end{align}\end{subequations}
These expressions are also in perfect agreement with the 3PN results given in (4.3) of~\cite{Favata:2008yd}, and we have also checked that they agree (at 3.5PN, at leading order in the symmetric mass ratio, and in the non-spinning limit) with the PN-GSF results of Eq.~\eqref{eq:H1SFPNl2} and of~\mbox{\ref{sec:further_comparisons}}. Note that with this approach, we have very easily recovered the result of equation~\eqref{eq:H20_1p5PN_PNMPM}, obtained with the PN-MPM approach, which featured a very nontrivial cancellation at 1.5PN order between the tails of memory and the higher-order PN correction to the quadratic memory.  
This result, combined with reference~\cite{Henry:2022ccf}, completes our knowledge of the full 3.5PN waveform, including both oscillatory and memory contributions.

Finally, note that  we were not able to obtain the full memory contribution to the 4PN waveform, despite the amplitude of the $(\ell,m) = (2,2)$ mode being known at 4PN order~\cite{Blanchet:2023bwj,Blanchet:2023sbv}. This is because other necessary modes are only known at 3.5PN order\footnote{When discussing modes, the PN order-counting convention is that \textit{Newtonian order} refers to the leading order of the $(2,2)$ mode. This means that the $(2,1)$, $(3,3)$ and $(3,1)$ modes first enter at 0.5PN order, the $(3,2)$, $(4,4)$ and $(4,2)$ modes first enter at 1PN order, etc. Thus, if the $(3,2)$ mode is known at 3.5PN, it means it is known at relative 2.5PN beyond its leading order expression.}. More specifically: 
\begin{itemize}
\item  $h_{20}^{\mathrm{(mem)}}$ at 4PN requires  $h_{32}$ and $h_{42}$  at 4PN 
\item  $h_{40}^{\mathrm{(mem)}}$ at 4PN  requires  $h_{32}$, $h_{42}$, $h_{52}$ and $h_{62}$  at 4PN 
\item  $h_{60}^{\mathrm{(mem)}}$ at 4PN  requires  $h_{42}$, $h_{52}$, $h_{62}$, $h_{72}$ and  $h_{82}$  at 4PN 
\item  $h_{80}^{\mathrm{(mem)}}$ at 4PN  requires   $h_{62}$, $h_{72}$ and  $h_{82}$ at 4PN 
\item  $h_{10,0}^{\mathrm{(mem)}}$ at 4PN  requires  $h_{82}$ at 4PN 
\end{itemize}
Note that $h_{12,0}^{\mathrm{(mem)}}$ is the only mode that can already be computed at 4PN order, since it only requires the oscillatory modes at leading order, which are known analytically for any $(\ell, m)$ provided $m\neq 0$, see reference~\cite{Kidder:2007rt}. 

\subsection{Including spin effects to the memory for non-precessing binary black holes}
\label{sec:PNspin}
The previous results can be fairly straightforwardly extended to the case of spinning, non-precessing, black hole binaries. The non-precessing condition is imposed by requiring the spin of both objects to be aligned or anti-aligned with the orbital angular momentum.  The main theoretical ingredient making this possible is the oscillatory 3.5PN waveform of Refs.~\cite{Henry:2022ccf,Henry:2022dzx}, which contain all relevant spinning contributions to the waveform (spin-orbit, spin-spin and cubic-in-spin)  at 3.5PN order. This immediately gives us access to $\dot{\mathcal{U}}_{l0}^{\text{(mem)}}$. The second ingredient is the energy and energy flux at infinity, which were obtained at 4PN in the spin-orbit and spin-spin sectors \cite{Cho:2022syn} as well as in the cubic-in-spin sector \cite{Marsat:2014xea}. The third ingredient are the horizon fluxes, which enter in the spinning case at 2.5PN relative to the quadrupolar formula, and were obtained at 4PN in Ref.~\cite{Saketh:2022xjb}. Since horizon fluxes are meaningful only in the case of black holes, we limit the present analysis to the case of a black hole binary. Inserting these expressions for the energy, flux at infinity, and horizon fluxes into the energy flux-balance law, we  can straightforwardly obtain the complete 3.5PN phasing required to access $H_{l0}^{\text{(mem)}}$.

Following standard conventions for spin-aligned systems, we introduce a dimensionless spin parameter for each object, denoted $\chi_1$ and $\chi_2$. Non-spinning black holes satisfy $\chi_A = 0$, whereas maximally spinning black holes satisfy $\chi_A = 1$ if the spin is aligned with the orbital angular momentum and $\chi_A = -1$ if it is anti-aligned.  We then define the dimensionless spin parameters $s = (m_1^2 \chi_1 + m_2^2 \chi_2)/M^2$ and $\sigma = (m_2 \chi_2 - m_1 \chi_1)/M$. Since both objects are black holes, the spin-induced quadrupolar and octupolar deformabilities are entirely fixed, namely $\kappa_1 = \kappa_2= \lambda_1=\lambda_2 = 1$. In view of comparing to PN-GSF results, we identify $\chi_1$ with the primary spin $\tilde{a} = a/m_1$, send the secondary spin $\chi_2$ to zero, and note that $\delta = \sqrt{1-4\nu} = 1 +\mathcal{O}(\nu)$. 

The $\dot{\mathcal{U}}_{l0}^{\text{(mem)}}$ are obtained using the same method as the non-spinning case. The~$(2,0)$ mode reads
 \begin{align}
\dot{\mathcal{U}}_{20}^{\text{(mem)}} &= \frac{256}{7}  \sqrt{\frac{\pi }{15}}  \frac{c^6\nu ^2 x^5}{G^2}    \Bigg\{ 1 + x \left(-\frac{1219}{288}+\frac{\nu }{24}\right) +x^{3/2} \left(4 \pi -\frac{7}{3}s +\frac{19}{64} \delta  \sigma  \right) \nonumber  \\
   & +x^2 \Bigg[-\frac{793}{1782} +8 s^2 +8 \delta s \sigma  +\frac{63}{32} \sigma^2 +\nu\left(  -\frac{14023}{6336} - 8   \sigma^2 \right) -\frac{4201}{1584}\nu^2 \Bigg] \nonumber \\
      & +x^{5/2} \Bigg[-\frac{2435}{144} \pi-\frac{79}{9}s -\frac{1007}{144} \delta  \sigma  +\nu  \left(  -\frac{23}{12} \pi+\frac{67}{9} s + \frac{127}{64} \delta  \sigma\right) \Bigg] \nonumber \\
&+ x^3 \Bigg[\frac{174213949439}{1816214400} +\frac{16}{3}\pi^2  -\frac{1712}{105}\gamma_{\mathrm{E}} -\frac{856}{105} \log (16x)  \nonumber\\
& \qquad\ \  -\frac{28}{3} \pi  s   +\frac{121}{96} \pi  \delta  \sigma   -\frac{691}{24} s^2 -\frac{3841}{96} \delta  s \sigma  -\frac{2257}{192} \sigma^2 \nonumber \\
&\qquad\ \ +\nu  \left( -\frac{126714689}{4447872} + \frac{41}{48} \pi ^2 -\frac{15}{2} s^2-\frac{15}{2} \delta  s  \sigma  +\frac{2129}{48} \sigma^2\right) \nonumber\\
&\qquad\ \    +\nu ^2 \left(\frac{4168379}{123552}+\frac{15}{2} \sigma^2\right) +\frac{142471}{46332} \nu^3 \Bigg] \nonumber \\
& +x^{7/2} \Bigg[-\frac{33199}{8448} \pi +\frac{18728305}{342144}  s +\frac{361003}{19008}  \delta  \sigma   +32 \pi  s^2 +32 \pi  \delta  s   \sigma +\frac{127}{16} \pi  \sigma^2 \nonumber\\
& \qquad \ \ + 8 s^3 +\frac{211}{8} \delta   s^2 \sigma +\frac{41}{2}  s \sigma^2 + \frac{445}{96} \delta  \sigma^3\nonumber\\
&\qquad\ \ +\nu  \left(\frac{532909}{38016} \pi +\frac{29281}{297}s +\frac{70819}{1408}\delta \sigma -32 \pi  \sigma^2 -\frac{163}{2} s \sigma^2  -\frac{147}{8} \delta  \sigma^3  \right)  \nonumber \\
   &\qquad\ \ +\nu ^2 \left(-\frac{27949}{864} \pi +\frac{5167}{264}  s +  \frac{35653}{6336} \delta  \sigma \right) \Bigg]  + \mathcal{O}(x^{4})\Bigg\}\,, 
\end{align}
and the other modes are relegated to the Supplementary Material~\cite{SuppMaterial}. We have checked that they agree (at 3.5PN, at leading order in the symmetric mass ratio, and in the case of a non-spinning secondary) with the PN-GSF results of Eq.~\eqref{eq:U1SFPNl2} and of~\mbox{\ref{sec:further_comparisons}}.

Compiling known results for the non-spinning, spin-orbit, non-spinning and cubic-in-spin sectors (e.g. from \cite{Blanchet:2023sbv, Cho:2022syn, Marsat:2014xea}), and specializing to the case of two black holes, we write down the full 3.5PN-accurate expressions for the energy and flux of energy at infinity in the case of spin-aligned binaries, at at all relevant order in spin: 
 \begin{align}
E &=  - \frac{c^2 M  \nu  x}{2} \Bigg\{1+  x \left(-\frac{3}{4}-\frac{\nu}{12} \right)  +x^{3/2} \left(\frac{14}{3}s + 2 \delta  \sigma \right) \nonumber \\
  & +x^2 \Bigg[-\frac{27}{8} - 4 s^2 -4 \delta  \sigma  s -\sigma^2   +\nu  \left(\frac{19}{8}+4 \sigma^2\right)-\frac{\nu ^2}{24}\Bigg] \nonumber \\
     & +x^{5/2} \Bigg[11 s +3 \delta  \sigma +\nu  \left(-\frac{61}{9} s-\frac{10}{3}\delta  \sigma  \right)\Bigg] \nonumber\\
&+x^3 \Bigg[-\frac{675}{64}  -\frac{25}{9} s^2 +\frac{10}{3} \delta  s \sigma  +\frac{5}{2} \sigma^2  \nonumber\\
    &\qquad\quad +\nu  \left(\frac{34445}{576} -\frac{205}{96} \pi ^2 + \frac{10}{3} s^2 +\frac{10}{3} \delta  \sigma  s  -\frac{15}{2} \sigma^2 \right)   \nonumber\\
    & \qquad\quad +\nu ^2 \left( -\frac{155}{96} -\frac{10}{3} \sigma^2\right) -\frac{35}{5184} \nu ^3 \Bigg] \nonumber\\
   & +x^{7/2} \Bigg[\frac{135}{4} s +\frac{27}{4} \delta  \sigma  - 8 s^3   -16 \delta   s^2 \sigma  -10  s \sigma^2 - 2\delta  \sigma^3  \nonumber\\
    & \qquad\quad+\nu  \left(-\frac{367}{4} s - 39 \delta  \sigma + 8 \delta  \sigma^3 +40 s \sigma^2\right) +\nu ^2 \left(\frac{29}{12}  s + \frac{5}{4} \delta  \sigma \right) \Bigg] + \mathcal{O}(x^4) \Bigg\} \,, \\
\mathcal{F}_{\infty} &=  \frac{32 c^5 \nu ^2 x^5}{5 G} \Bigg\{1 +x \left(-\frac{1247}{336}-\frac{35}{12}\nu\right)  +x^{3/2}\left(4 \pi - 4 s   - \frac{5}{4}\delta  \sigma\right) \nonumber\\
   & +x^2  \Bigg[-\frac{44711}{9072} +8 s^2+8 \delta  \sigma  s+\frac{33}{16} \sigma^2 +\nu  \left(\frac{9271}{504}-8 \sigma^2\right) + \frac{65}{18} \nu ^2   \Bigg] \nonumber\\*
    & +x^{5/2} \Bigg[ -\frac{8191}{672}\pi   -\frac{9}{2} s  -\frac{13}{16} \delta  \sigma +\nu  \left(-\frac{583}{24}\pi +\frac{272}{9} s +\frac{43}{4} \delta  \sigma\right) \Bigg] \nonumber\\
&+ x^3 \Bigg[\frac{6643739519}{69854400} +\frac{16}{3}\pi
   ^2 -\frac{1712}{105} \gamma_\mathrm{E}   -\frac{856}{105}  \log (16x) \nonumber\\
   & \qquad\quad-16 \pi  s  -\frac{31}{6} \pi  \delta  \sigma  -\frac{3839}{252}s^2 - \frac{1375}{56} \delta  s \sigma  -\frac{227}{28} \sigma^2 \nonumber\\
   &\qquad\quad +\nu  \left(-\frac{134543}{7776} +\frac{41}{48}\pi ^2 -43 s^2-43 \delta   s \sigma+\frac{3481}{168}\sigma^2\right) \nonumber\\
   &\qquad\quad +\nu^2   \left(-\frac{94403}{3024}+43 \sigma^2\right) -\frac{775}{324} \nu^3  \Bigg] \nonumber\\
   & +x^{7/2} \Bigg[-\frac{16285}{504}\pi +\frac{476645}{6804}s +\frac{9535}{336} \delta  \sigma  \nonumber\\
   &\qquad\quad +32 \pi  s^2  +32  \pi  \delta s \sigma +\frac{65}{8} \pi  \sigma^2    -\frac{16}{3} s^3 + \frac{2}{3}   \delta s^2 \sigma +\frac{9}{2} s \sigma^2  +   \frac{35}{24}\delta  \sigma^3    \nonumber\\
   &\qquad\quad +\nu  \left(\frac{214745}{1728} \pi  +\frac{6172}{189} s +\frac{1849}{126} \delta  \sigma -32 \pi  \sigma^2-\frac{56}{3} s \sigma^2  -6 \delta  \sigma^3 \right) \nonumber\\
   &\qquad\quad +\nu ^2 \left(\frac{193385}{3024} \pi -\frac{2810}{27} s -\frac{1501}{36} \delta  \sigma \right)  \Bigg] + \mathcal{O}(x^4)\Bigg\} \,.
\end{align}

The horizon fluxes, namely $\dot{m}_1$ and $\dot{m}_2$, are known\footnote{Note, however, that there is a typo in (4.25) of the published version of \cite{Saketh:2022xjb}: the $M^2$ in the prefactor should in fact be a $m_1^2$.} at 4PN from~\cite{Saketh:2022xjb}, see their Eqs. (4.22).
Note that the 3.5PN expression has already been obtained in \cite{Chatziioannou:2016kem}, but the 4PN piece of that work was incorrect. In view of obtaining the phasing, we apply the chain rule on the flux balance law for energy,
\begin{equation}
\frac{\dd E}{\dd t} = \frac{\partial E}{\partial x} \dot{x} + \frac{\partial E}{\partial m_1} \dot{m}_1 + \frac{\partial E}{\partial m_2} \dot{m}_2 = - \mathcal{F}_{\infty} - \dot{M}\,,
\end{equation}
where we can ignore the spin evolutions $\dot{\chi}_1$ and $\dot{\chi}_2$ at this order.

We will need differential equation governing the time evolution of $x$ in terms of $x$ itself, which is straightforwardly obtained from the flux balance law using 
\begin{equation}
\dot{x} = - \left(\frac{\partial E}{\partial x}\right)^{-1}\left(\mathcal{F}_\infty + \delta \mathcal{F}\right) \,,\end{equation}
where we can read off\footnotemark[\value{footnote}] from (4.22) of \cite{Saketh:2022xjb} that
\begin{align}
\delta \mathcal{F} &= \dot{M} +\frac{\partial E}{\partial m_1} \dot{m}_1 + \frac{\partial E}{\partial m_2} \dot{m}_2 \nonumber\\
&= \frac{32 c^5 \nu^2 x^{15/2}}{5 G}\Bigg\{ -\frac{1}{4}s - \frac{3}{2}s^3 - \frac{9}{4}\delta s^2 \sigma   -\frac{9}{4} s \sigma^2 -\frac{3}{4} \delta \sigma^3  \nonumber\\*
&\qquad\qquad\qquad + \nu\left(\frac{1}{2}s + \frac{1}{4} \delta \sigma + \frac{9}{2} s \sigma^2 + \frac{3}{4}\delta \sigma^3\right)\nonumber\\
&\qquad +x \Bigg[ - s - \frac{9}{4}s^3 - \frac{9}{16}\delta s^2 \sigma - \frac{9}{16} s \sigma^2  - \frac{3}{16}\delta \sigma^3    \nonumber\\*
&\qquad\!\qquad+\nu  \left( \frac{13}{4} s + \delta \sigma + \frac{15}{4} s^3 + \frac{45}{8} \delta s^2 \sigma + \frac{9}{8} s \sigma^2 + \frac{3}{16} \delta \sigma^3 \right) \nonumber \\*
&\qquad\!\qquad+\nu ^2 \left( - \frac{5}{4}s - \frac{5}{8} \delta \sigma  - \frac{45}{4} s \sigma^2 - \frac{15}{8} \delta \sigma^3   \right)\Bigg] + \mathcal{O}(x^{3/2})\Bigg\} \,.
\end{align}	
We then find our main result (which reduces to \eqref{eq:dxdt_3.5PN} in the non-spinning limit): 
\begin{align}
\frac{\dd x}{\dd t} &=\frac{64 c^3 \nu  x^5}{5 G M } \Bigg\{ 1+x\left(-\frac{743}{336} -\frac{11}{4}\nu\right)  + x^{3/2} \left(4 \pi -\frac{47}{3}  s -\frac{25}{4} \delta  \sigma    \right)\nonumber\\
& +x^2 \Bigg[\frac{34103}{18144} +20 s^2 +20 \delta  \sigma  s  +\frac{81}{16} \sigma^2 +\nu  \left(\frac{13661}{2016}-20 \sigma^2\right) +\frac{59}{18} \nu ^2  \Bigg] \nonumber\\
& +x^{5/2} \Bigg[ -\frac{4159}{672}\pi -\frac{5897}{144} s -\frac{809}{84} \delta  \sigma  - \frac{3}{2} s^3  - \frac{9}{4}\delta s^2 \sigma -\frac{9}{4}s \sigma^2    -\frac{3}{4} \delta  \sigma^3  \nonumber\\
& \qquad +\nu  \left(-\frac{189}{8}\pi   +\frac{1007}{12} s  + \frac{283}{8} \delta \sigma + \frac{9}{2} s \sigma^2 + \frac{3}{4} \delta \sigma^3 \right) \Bigg] \nonumber\\
& + x^3 \Bigg[ \frac{16447322263}{139708800} +\frac{16}{3}\pi ^2 -\frac{1712 }{105} \gamma_{\mathrm{E}}  -\frac{856}{105}  \log (16x) \nonumber\\
& \qquad  -\frac{188}{3}  \pi  s     -\frac{151}{6} \pi  \delta  \sigma +\frac{7649}{42}s^2 +\frac{19627}{168} \delta  s \sigma   +\frac{3159}{224} \sigma^2\nonumber\\
& \qquad+\nu  \left(-\frac{56198689}{217728} +\frac{451}{48}\pi ^2  -86 s^2 - 86 \delta  \sigma  s -\frac{57221}{672} \sigma^2 \right) \nonumber\\
& \qquad +\nu ^2 \left(\frac{541}{896}+86 \sigma^2\right) -\frac{5605}{2592}\nu ^3 \Bigg]  \nonumber\\
& +x^{7/2} \Bigg[  -\frac{4415}{4032}\pi -\frac{4348507}{18144} s -\frac{1195759}{18144} \delta  \sigma  + 80 \pi  s^2   + 80 \pi  \delta   s  \sigma +\frac{161}{8} \pi  \sigma^2  \nonumber\\
&\qquad   -\frac{2371}{6} s^3 -\frac{25325}{48} \delta s^2 \sigma -\frac{471}{2} s \sigma^2   -\frac{419}{12} \delta \sigma^3  \nonumber\\
&\qquad  +\nu  \left(\frac{358675}{6048} \pi +\frac{146455}{224} s   +\frac{258409}{1008} \delta  \sigma -80 \pi \sigma^2 \right. \nonumber\\*
&\qquad\qquad  \left. + \frac{7}{2} s^3 + \frac{21}{4} \delta s^2 \sigma +\frac{5581}{6} s \sigma^2  + \frac{2147}{16} \delta  \sigma^3 \right)   \nonumber\\ 
& \qquad+\nu^2 \left(\frac{91495}{1512}\pi -\frac{11213}{54} s -\frac{8765}{96} \delta  \sigma - \frac{21}{2} s \sigma^2 - \frac{7}{4} \delta \sigma^3  \right)  \Bigg] + \mathcal{O}(x^4)\Bigg\} \,.
\end{align}

We take this opportunity to present the full time-domain phasing, including all orders in spin and the horizon flux contributions, because it has, to our knowledge, never appeared in the literature (see however (5.10) of \cite{Saketh:2022xjb} for the frequency-domain phasing in the SPA approximation). From the flux balance law, we first obtain 
\begin{equation}
t_0 - t = - \int_{x}^{x_0}\frac{\partial E/\partial x}{\mathcal{F}_\infty(x) + \delta \mathcal{F}(x)} \,  \dd x \,,
\end{equation}
which is then inverted (in the sense of a PN series) to obtain an expression for $x$ in terms of time. Introducing the dimensionless time variable $\tau =  \nu c^3 (t_0-t)/(5GM)$, and choosing the integration constant such that $t_0$ can be interpreted as the time of coalescence~\cite{Blanchet:2023bwj}, we find the 3.5PN accurate expression for the ``chirp'':
\begin{align}
x &=  \frac{\tau^{-1/4}}{4} \Bigg\{1+ \tau^{-1/4}\left(\frac{743}{4032}+\frac{11}{48} \nu \right)  +\tau^{-3/8}\left(-\frac{\pi }{5} +\frac{47}{60} s +\frac{5}{16} \delta  \sigma \right) \nonumber\\
&  + \tau^{-1/2}\Bigg[\frac{19583}{254016}-\frac{5}{8} s^2 -\frac{5}{8} \delta s \sigma -\frac{81}{512} \sigma^2 +\nu  \left(\frac{24401}{193536}+\frac{5}{8} \sigma^2\right)+\frac{31}{288} \nu^2 \Bigg] \nonumber\\
&  + \tau^{-5/8}\Bigg[-\frac{11891}{53760}\pi  +\frac{358763}{161280} s +\frac{96473}{129024}  \delta  \sigma  + \frac{1}{32} s^3+ \frac{3}{64} s^2 \sigma + \frac{3}{64} s \sigma^2  + \frac{1}{64}\delta  \sigma^3 \nonumber\\
&\qquad\quad +\nu  \left(\frac{109}{1920} \pi  -\frac{247}{5760} s  - \frac{29}{512} \delta  \sigma - \frac{3}{32}s \sigma^2 - \frac{1}{64} \delta \sigma^3\right)\Bigg]\nonumber\\
& +\tau^{-3/4}\Bigg[-\frac{10052469856691}{6008596070400}+\frac{\pi^2}{6}+\frac{107}{420}\gamma_{\mathrm{E}}   -\frac{107}{3360} \log \left(\frac{\tau}{256}\right) \nonumber\\
&\qquad\!\qquad -\frac{47}{48} \pi  s  -\frac{149}{384} \pi  \delta  \sigma - \frac{1583}{4032} s^2 - \frac{529}{3584} \delta  s \sigma  +\frac{329}{8192}  \sigma^2     \nonumber\\
&\qquad\!\qquad+\nu  \left(\frac{3147553127}{780337152} -\frac{451}{3072} \pi^2 -\frac{3}{8} s^2 -\frac{3}{8} \delta  \sigma  s -\frac{1175}{7168} \sigma^2     \right)\nonumber\\
&\qquad\!\qquad +\nu ^2 \left(-\frac{15211}{442368}+\frac{3}{8} \sigma^2\right)+\frac{25565}{331776}  \nu^3\Bigg]\nonumber\\
&  + \tau^{-7/8} \Bigg[-\frac{113868647}{433520640}\pi +\frac{24634336547}{2601123840}  s +\frac{281190779}{99090432} \delta  \sigma  +\frac{21}{16} \pi  s^2  +\frac{21}{16} \pi  \delta  s   \sigma   \nonumber\\
&\qquad\quad  +\frac{1711}{5120} \pi  \sigma^2 - \frac{969085}{258048} s^3 -\frac{969701}{172032} \delta  s^2 \sigma  -\frac{2326643}{860160} s \sigma^2   -\frac{427997}{1032192}\delta   \sigma^3 \nonumber\\
&\qquad\quad+\nu  \left(- \frac{31821}{143360} \pi    -\frac{102279787}{15482880}  s  -\frac{13594213}{4128768}\delta  \sigma  -\frac{21}{16} \pi  \sigma^2  \right.\nonumber\\
&\qquad\qquad\quad  \left. + \frac{239}{3072} s^3  + \frac{239}{2048}\delta s^2 \sigma+\frac{326785}{28672} s \sigma^2+ \frac{991009}{516096} \delta  \sigma^3\right)\nonumber\\
&\qquad\quad+\nu ^2 \left(\frac{294941}{3870720}\pi   +\frac{1363}{368640} s -\frac{3331}{98304} \delta  \sigma - \frac{239}{1024} s \sigma^2 - \frac{239}{6144} \delta \sigma^3\!\right)\Bigg] \!\!+ \mathcal{O}\left(\tau^{-1}\right)\!\!\Bigg\}\,,
\end{align}
The phasing in terms of $x$ is obtained using
\begin{equation}
\psi = \int_{t_0}^t \dd t' \Omega(t') = - \frac{c^3 }{G M}\int_{x_0}^x \dd x \, x^{3/2}  \frac{\partial E/ \partial x}{\mathcal{F}_\infty(x) + \delta \mathcal{F}(x) } \,.
\end{equation}
The 3.5PN-accurate expression for the phasing for spin-aligned black holes, including all spin terms, then reads
\begin{align}
\psi &= \psi_0 -\frac{1}{32 \nu  x^{5/2}} \Bigg\{ 1+x\left(\frac{55}{12}\nu+\frac{3715}{1008}\right) +x^{3/2} \left(-10 \pi +\frac{235}{6} s  + \frac{125}{8} \delta  \sigma \right) \nonumber\\
&\qquad+x^2 \Bigg[\frac{15293365}{1016064} -100 s^2-\frac{405}{16} \sigma^2-100 \delta  \sigma s + \nu  \left(\frac{27145}{1008}+100 \sigma^2\right) + \frac{3085}{144} \nu^2 \Bigg]\nonumber\\
&\qquad+x^{5/2} \log (x) \Bigg[ \frac{38645}{1344} \pi - \! \frac{555605}{2016} s - \! \frac{41745}{448} \delta  \sigma - \frac{15}{4} s^3 - \frac{45}{8} \delta s^2 \sigma - \frac{45}{8} s \sigma^2  -\frac{15}{8} \delta  \sigma^3 \nonumber\\
&\quad\qquad\qquad\qquad+\nu  \left(-\frac{65}{16} \pi  -\frac{45}{8} s+\frac{5}{2} \delta  \sigma + \frac{45}{4} s \sigma^2 + \frac{15}{8} \delta \sigma^3\right)\Bigg]\nonumber\\
&\qquad +x^3 \Bigg[\frac{12348611926451}{18776862720}-\frac{160}{3} \pi ^2-\frac{1712}{21}\gamma_{\mathrm{E}} -\frac{856 \log (16x)}{21}  \nonumber\\
&\qquad\qquad  + \frac{940}{3} \pi s +  \frac{745}{6}\pi \delta  \sigma +\frac{7915}{63} s^2 +\frac{2645}{56} \delta  s \sigma -\frac{1645}{128} \sigma^2 \nonumber\\
&\qquad\qquad+\nu  \left(-\frac{15737765635}{12192768}+\frac{2255}{48} \pi^2+120 s^2 +120 \delta s \sigma   +\frac{5875}{112} \sigma^2\right)  \nonumber\\
&\qquad\qquad +\nu ^2 \left(\frac{76055}{6912}-120 \sigma^2\right)-\frac{127825}{5184} \nu^3\Bigg]\nonumber\\
&\qquad+x^{7/2} \Bigg[  \frac{77096675}{2032128}\pi -\frac{9018232555}{6096384} s -\frac{170978035}{387072} \delta  \sigma   - 200 \pi s^2   -200 \pi\delta s \sigma \nonumber\\
&\qquad\ \,\qquad  -\frac{815}{16}\pi \sigma^2    +\frac{125925}{224}s^3 + \frac{379805}{448}   \delta   s^2  \sigma+\frac{182755}{448} s \sigma^2  +\frac{1315}{21} \delta  \sigma^3  \nonumber\\
&\qquad\ \,\qquad +\nu  \left(\frac{378515}{12096}\pi+\frac{3329545}{3024} s  +\frac{2909765}{5376} \delta  \sigma  +  200\pi \sigma^2 \right.  \nonumber\\
&\qquad\ \,\qquad\qquad \left. - \frac{95}{8} s^3 - \frac{285}{16} \delta s^2 \sigma -\frac{385825}{224} s \sigma^2 -\frac{130615}{448} \delta  \sigma^3\right)\nonumber\\
&\qquad\ \,\qquad +\nu ^2 \left(-\frac{74045}{6048}\pi  + \frac{835}{288} s  + \frac{7015}{1152} \delta  \sigma  + \frac{285}{8} s \sigma^2 + \frac{95}{16} \right)\Bigg] + \mathcal{O}(x^4)\Bigg\}\,,
\end{align}
where the integration constant $\psi_0$ is determined by initial conditions, e.g., when the wave frequency enters the detector's band.
Of course, this expression disagrees at 2.5PN with~(14)~of~\cite{Cho:2022syn} and~(6.21)~of~\cite{Marsat:2014xea}, because the latter works discard the contribution of the horizon fluxes.  

 Following the procedure described in Sec.~\ref{sec:PNresults}, we can now obtain the memory amplitudes. The~$(2,0)$~mode reads
\begin{align}
H_{20} &= -\frac{5}{14 \sqrt{6}}   \Bigg\{1+x\left(-\frac{4075}{4032}+\frac{67}{48} \nu \right)+x^{3/2} \left(\frac{16}{3} s+\frac{419}{160} \delta  \sigma \right)  \nonumber\\* 
& \qquad +x^2 \Bigg[-\frac{151877213}{67060224} -4 s^2 - 4 \delta s \sigma -\frac{33}{32} \sigma^2  +\nu  \left(-\frac{123815}{44352}+4 \sigma^2\right)+ \frac{205}{352} \nu^2 \Bigg] \nonumber\\ 
& \qquad +x^{5/2} \Bigg[-\frac{253}{336}\pi  +\frac{181381}{21168} s+\frac{289127}{225792} \delta  \sigma  + \frac{3}{7} s^3 + \frac{9}{14} \delta s^2 \sigma +\frac{9}{14} s \sigma^2  + \frac{3}{14} \delta  \sigma^3 \nonumber\\ 
& \qquad\qquad +\nu  \left(\frac{253}{84} \pi +\frac{283}{252} s+\frac{1583}
{2688} \delta  \sigma  - \frac{9}{7} s \sigma^2 - \frac{3}{14} \delta \sigma^3 \right) \Bigg]\nonumber\\ 
& \qquad + x^3 \Bigg[-\frac{4397711103307}{532580106240}  + \frac{23}{384} \pi  \delta  \sigma   +\frac{997}{336} s^2 +\frac{173191}{16128} \delta  s \sigma +\frac{14171}{3072} \sigma^2    \nonumber\\ 
& \qquad\qquad +\nu  \left(\frac{700464542023}{13948526592}  -\frac{205}{96} \pi ^2 -\frac{31}{12} s^2 - \frac{31}{12} \delta   s\sigma  -\frac{142529}{8064} \sigma^2 \right)\nonumber\\ 
& \qquad\qquad +\nu ^2 \left(\frac{69527951}{166053888}+\frac{31}{12} \sigma^2\right)+\frac{1321981}{5930496} \nu^3 \Bigg] \nonumber\\ 
& \qquad +x^{7/2} \Bigg[\frac{38351671}{28740096} \pi  +\frac{9820292015}{301771008} s +\frac{14713254113}{3218890752} \delta  \sigma   +\frac{\pi}{24} \sigma^2 \nonumber\\ 
& \qquad \qquad -\frac{68833}{6048} s^3  -\frac{95153}{4032} \delta s^2  \sigma 
   -\frac{58669}{4032} s\sigma^2  -\frac{274927}{96768} \delta \sigma^3 \nonumber\\
   & \qquad\qquad +\nu  \left(-\frac{3486041}{598752} \pi -\frac{1670164}{18711} s -\frac{66525761}{1596672} \delta  \sigma \right.  \nonumber\\
   & \qquad\qquad\qquad +\left. \frac{77}{72}s^3 + \frac{77}{48} \delta s^2 \sigma+\frac{41569}{672} s \sigma^2 + \frac{154261}{12096} \delta  \sigma^3 \right)\nonumber\\
   & \qquad\qquad+\nu ^2 \left(-\frac{652889}{598752} \pi -\frac{39821}{14256} s - \frac{226313}{152064} \delta  \sigma - \frac{77}{24} s \sigma^2 - \! \frac{77}{144} \delta \sigma^3 \right) \!\!   \Bigg] \! +  \mathcal{O}(x^4) \!\Bigg\} \,,
\end{align}
and the other modes are relegated to the Supplementary Material~\cite{SuppMaterial}. In the spin sector, they agree with the 2PN results of Ref.~\cite{Mitman:2022kwt} in the spin-aligned case, both at linear and quadratic order in spin. Moreover, all $H_{l0}$ modes are once again in perfect agreement (at 3.5PN, at leading order in the symmetric mass ratio, and in the case of a non-spinning secondary) with the PN-GSF results of Eq.~\eqref{eq:H1SFPNl2} and of~\mbox{\ref{sec:further_comparisons}}. The inclusion of the horizon fluxes was crucial to obtain agreement.

\endgroup

\section{Calculation of gravitational-wave memory \textit{via} the gravitational self-force approach: overview}\label{sec:SF_theory}

In this section, we describe the necessary formalism for calculating GW memory in GSF theory. We begin with the multiscale expansion of the Einstein field equations as our overarching computational framework. We then summarize the calculations at first and second order in the mass ratio. The breakdown of the multiscale expansion on large spatial scales at second order~\cite{Pound:2015wva} necessitates a more thorough analysis of spacetime asymptotics for asymmetric binaries, which we carry out in section \ref{far-zone}. 

Throughout this section, we appeal to standard methods of black hole perturbation theory. We refer to~\cite{PoundAndWardell} for a detailed exposition of those methods. 

\subsection{Multiscale expansion}

The GSF framework applies when the mass ratio $\epsilon$ is small, such that the secondary object can be treated as a point mass that perturbs the spacetime of the primary black hole. One then solves the Einstein field equations using a series expansion in powers of~$\epsilon$~\cite{Barack:2018yvs}. Here, we specifically use a multiscale formulation of that expansion~\cite{Miller:2020bft,Miller:2023ers}, which accurately captures both the periodic behavior on the orbital time scale and the slow, secular evolution on the radiation-reaction timescale~\cite{Hinderer:2008dm,PoundAndWardell}.

We work with the standard Boyer-Lindquist spatial coordinates $y^i=\{r,\theta,\phi\}$ defined on the background spacetime of the primary. 
For our time coordinate, we use a hyperboloidal time $\uptau=t-\upkappa(r_*)$ that interpolates between (i)~advanced time $v=t+r_*$ at the future horizon, (ii)~Boyer-Lindquist time $t$ in a region including the particle, and (iii)~retarded time $u=t-r_*$ at future null infinity. 
Here, $r_*$ is the tortoise coordinate satisfying $\dd r_*/\dd r =(r^2+\mathring{a}^2)/(r^2-2\mathring{m}_1 r+\mathring{a}^2)$ and $\mathring{a}$ is the spin parameter of the background Kerr black hole~\cite{Zenginoglu:2011jz,Miller:2020bft,Miller:2023ers,PanossoMacedo:2022fdi}.\footnote{As explained below, the background parameters $\mathring{m}_1$ and $\mathring{a}$ differ from the physical black hole parameters by ${\cal O}(\epsilon)$ due to the black hole's interaction with the particle~\cite{Miller:2020bft}. 
However, we always re-expand final results to express them in terms of the physical parameters.
} 
Using a hyperboloidal time variable significantly simplifies calculations at second order in $\e$, as we review below. 
However, we note that when solving field equations at first order in $\epsilon$, we use Boyer-Lindquist~$t$ throughout the spacetime and then perform a simple transformation to obtain the solution as a function of $\uptau$.

We specialise to quasi-circular motion with orbital phase $\phi_p$ and define the orbital frequency to be $\Omega \equiv d\phi_p/dt$.
During the inspiral, the orbital phase evolves rapidly, on the fast orbital timescale, whereas the orbital frequency evolves slowly, on the radiation-reaction timescale. 
In our multiscale expansion the evolution of the binary is given by%
\begin{subequations}\begin{align}
	\frac{d\phi_p}{dt} 	&= \Omega, \\*
	\frac{d\Omega}{dt}  &= \epsilon \left[F^{(0)}(\Omega)+ \epsilon F^{(1)}(\Omega) + \mathcal{O}(\epsilon^2)\right],\label{eq:Omegadot}
\end{align}\end{subequations}
where $F^{(0)}(\Omega)$ and $F^{(1)}(\Omega)$ are known as the adiabatic (0PA) and post-adiabatic (1PA) forcing terms, respectively.
Inclusion of these two terms ensures that over a radiation-reaction timescale the error in $\phi_p$ scales linearly with $\epsilon$. This accuracy is expected to be sufficient for EMRI modelling~\cite{Burke:2023lno}.
The mass and spin of the primary also evolve at 1PA order due to the absorption of GWs by the black hole; we include their evolution consistently throughout our calculation, but for brevity we will mostly elide them here.

In order to compute the forcing terms $F^{(n)}(\Omega)$ and the waveform, we expand the metric as a product of slowly evolving amplitudes and a rapidly evolving phase:
\begin{equation}\label{eq:metric_expansion}
	g_{\alpha\beta} = \mathring{g}_{\alpha\beta} + \sum_{m\in \mathbb{Z}}\left[ \epsilon h^{(1),m}_{\alpha\beta}(\Omega;y^i) + \epsilon^2 h^{(2),m}_{\alpha\beta}(\Omega;y^i)\right] e^{-i m \phi_p},
\end{equation}
where $\mathring{g}_{\alpha\beta}$ is the background Kerr metric. We treat the background mass and spin parameters $\mathring{m}_1$ and $\mathring{a}$ as constants and include the evolving $\mathcal{O}(\epsilon)$ corrections to them in~$h^{(1),0}_{\alpha\beta}$. The metric $g_{\alpha\beta}$ is assumed to have no time dependence apart from its dependence on $\phi_p$ and $\Omega$ (and the corrections to the primary's evolving mass and spin), and $\phi_p$ and $\Omega$ are promoted to fields on spacetime by extending them off the particle's worldline as constant on slices of constant~$\uptau$. 
Field equations for the metric amplitudes, $h^{(n),m}_{\alpha\beta}$, can be obtained by substituting equation~\eqref{eq:metric_expansion} into the Einstein field equations $G_{\alpha\beta}[g] = 8\pi T_{\alpha\beta}$, where $G_{\alpha\beta}$ is the Einstein tensor, $T_{\alpha\beta}$ is the stress-energy tensor of the particle~\cite{Upton:2021oxf}, and hereafter we omit indices on tensorial arguments of functionals.
We can expand the stress-energy and Einstein tensors in powers of the mass ratio as
\begin{subequations}\begin{align}
	T_{\alpha\beta}			&= \epsilon T^{(1)}_{\alpha\beta} + \epsilon^2 T^{(2)}_{\alpha\beta} + \mathcal{O}(\epsilon^3), \\
	G_{\alpha\beta}[g] 	&= \epsilon\delta G_{\alpha\beta}[h^{(1)}] + \epsilon^2\delta^2 G_{\alpha\beta}[h^{(1)},h^{(1)}] + \mathcal{O}(\epsilon^3),
\end{align}\end{subequations}
where 
\begin{equation}
h^{(n)}_{\alpha\beta} = \sum_{m\in\mathbb{Z}}h^{(n),m}_{\alpha\beta}e^{-im\phi_p}, 
\end{equation}
 $\delta G_{\alpha\beta}$ is the linearized Einstein tensor, $\delta^2 G_{\alpha\beta}$ is quadratic in its arguments, and we have used that $G_{\alpha\beta}[\mathring{g}] = 0$. 
Time derivatives appearing in $G_{\alpha\beta}$ can be evaluated using $\partial_\uptau = \dot{\phi}_p\partial_{\phi_p} + \dot{\Omega}\partial_\Omega = \Omega\partial_{\phi_p} + \epsilon F^{(0)}\partial_\Omega + \mathcal{O}(\epsilon^2)$.
Using this, we further expand the linearized and quadratic Einstein tensors as \mbox{$\delta^n G_{\alpha\beta}  = \delta^n G^{(0)}_{\alpha\beta} + \epsilon \delta^n G^{(1)}_{\alpha\beta} + \mathcal{O}(\epsilon^2)$}, where $n\in\{1,2\}$, $\partial_\uptau$ is replaced by $\Omega\partial_{\phi_p}$ in $\delta^nG^{(0)}_{\alpha\beta}$, and $\delta^nG^{(1)}_{\alpha\beta}\propto F^{(0)}$. For details, we refer to equations~(26)--(40) in reference~\cite{Miller:2023ers} (noting the differences in notation $\uptau\to s$ and $\delta^n G^{(j)}_{\alpha\beta}\to G^{(n,j)}_{\alpha\beta}$). 

These expansions put the Einstein equation in the form of a series in $\e$ with coefficients that depend on $\Omega(\uptau,\e)$ and $\phi_p(\uptau,\epsilon)$. Within the multiscale framework, $\phi_p$ and $\Omega$ are treated as independent variables, such that we can equate coefficients of explicit powers of $\e$ in the expanded Einstein equation despite the $\e$ depence in $\Omega$ and $\phi_p$. The field equations at each order in the mass ratio then take the following form: 
\begin{subequations}\label{eq:h1h2_field_equation}\begin{align}
	\delta G^{(0),m}_{\alpha\beta}[h^{(1),m}] &= T^{(1),m}_{\alpha\beta}, \label{eq:h1_field_equation}\\*
	\delta G^{(0),m}_{\alpha\beta}[h^{(2),m}] &= T^{(2),m}_{\alpha\beta} - \sum_{m',m''}\delta^2 G^{(0),m}_{\alpha\beta}[h^{(1),m'},h^{(1),m''}] - \delta G^{(1),m}_{\alpha\beta}[h^{(1),m}], \label{eq:h2_field_equation}
\end{align}\end{subequations}
where the exponential $e^{im\phi_p}$ has been factored out of each equation, and the sum over $m'$ and $m''$ runs over all pairs satisfying \mbox{$m'+m''=m$}. The linear operator $\delta G^{(0),m}_{\alpha\beta}$ on the left-hand side is identical to the linearized Einstein operator one would obtain when applying the full linearized Einstein tensor $\delta G_{\alpha\beta}$ to a function of the form $h^{(n),m}_{\alpha\beta}(y^i)e^{-im\Omega\uptau}$; this is the standard linearized Einstein tensor in the frequency domain, with frequencies $\omega=m\Omega$.

Equations~\eqref{eq:h1h2_field_equation} represent an expansion of the Einstein field equations in powers of $\epsilon$ at fixed $(\phi_p,\Omega,x^i)$ (i.e., treating $\phi_p$, $\Omega$, and $x^i$ as independent variables). Because the point-particle source $T^{(1)}_{\alpha\beta}$ only depends on $\phi$ and $\phi_p$ in the combination $(\phi-\phi_p)$, this functional dependence propagates through every order, such that the only $\phi$ dependence in $h^{(n),m}_{\alpha\beta}$, $T^{(n),m}_{\alpha\beta}$, and $\delta^n G^{(k),m}_{\alpha\beta}$ is an overall exponential $e^{im\phi}$. The Fourier mode number $m$ hence becomes identified with the azimuthal mode number $m$ in the spherical (or spheroidal) harmonic expansions we adopt throughout the paper. 

We note that the quadratic source term in equation~\eqref{eq:h2_field_equation} is highly singular. While it admits a distributional interpretation that makes equation~\eqref{eq:h2_field_equation} well defined~\cite{Upton:2021oxf}, in practice we regularize the equation \textit{via} the introduction of puncture fields in the vicinity of the secondary's worldline~\cite{Pound:2012nt,Pound:2014xva}.
This subtlety will not enter into our  analysis of the GW memory, which will involve studying the asymptotic behaviour of the metric perturbation outside the support of the puncture fields near the worldline.

Finally, for the waveform, we decompose the metric perturbation at future null infinity into  $h_+$ and  $h_\times$ polarizations in a TT gauge; see section~\ref{sec:theory}. Expanding equation~\eqref{eq:metric_expansion} on a basis of spin-weight $-2$ spherical harmonics, we write the waveform as%
\begin{align}\label{eq:multiscale waveform}
	h\equiv h_+ - i h_\times = \frac{1}{r}\sum_{lm}\left[\epsilon h^{(1)}_{lm}(\Omega) + \epsilon^2 h^{(2)}_{lm}(\Omega) \right]e^{-im\phi_p} {}_{-2}Y_{lm}(\theta,\phi),
\end{align}
noting that $r$ and $R$ are asymptotically equal.

The distinguishing feature of memory in the multiscale context is that it accumulates on the radiation-reaction timescale, promoting it by one order in $\epsilon$, analogous to the $-2.5$PN promotion described in section~\ref{subsec:memory_standard_PN_MPM}. Consider an integral of the form~\eqref{Umem} with a multiscale integrand, which for illustration purposes we write as
\begin{equation}\label{eq:example integral}
    O_{lm}(u) = \int_{-\infty}^u  K_{lm}(\Omega(u'))e^{-im\phi_p(u')} \dd u'
\end{equation}
for some $K_{lm}$, working with $u$ rather than $U$ for simplicity. If $m\neq0$, the rapidly oscillating phase factor prevents the integral from accumulating. In that case, we can repeatedly integrate by parts to reduce the integral to an instantaneous function of $u$:
 \begin{equation}\label{eq:integration by parts}
    O_{lm} = -\left\{\frac{K_{lm}}{im\Omega}+\frac{1}{im\Omega}\frac{d}{du}\left(\frac{K_{lm}}{im\Omega}\right)+\frac{1}{im\Omega}\frac{d}{du}\left[\frac{1}{im\Omega}\frac{d}{du}\left(\frac{K_{lm}}{im\Omega}\right)\right]+\ldots\right\}e^{-im\phi_p}. 
\end{equation}
Since $d\Omega/du= \mathcal{O}(\epsilon)$, each successive term in curly brackets is suppressed by one additional order in $\epsilon$. Contrast this with an integral for a quasistationary ($m=0$) term. Adopting $\Omega$ as the integration variable and using equation~\eqref{eq:Omegadot}, we can write
\begin{equation}
   O_{l0} = \int_{-\infty}^u  K_{l0}(\Omega(u')) \dd u' = \frac{1}{\epsilon}\int_{0}^{\Omega(u)}  \frac{K_{l0}(\Omega)}{F^{(0)}(\Omega)} d\Omega +\mathcal{O}(\epsilon^0).
\end{equation}
The implication of this is that, while the GW flux is $\mathcal{O}(\epsilon^2)$, it generates an $\mathcal{O}(\epsilon)$ quasistationary contribution to the waveform. Concretely, the memory terms in equation~\eqref{eq:multiscale waveform} are%
\begin{align}\label{eq:memory terms}
	h^{\rm mem}_+ - i h^{\rm mem}_\times = \frac{1}{r}\sum_{l=2}^\infty\left[\epsilon h^{(1)}_{l0}(\Omega) + \epsilon^2 h^{(2)}_{l0}(\Omega) \right] {}_{-2}Y_{l0}(\theta,\phi).
\end{align} 
We emphasize that traditional GSF waveform models have always omitted these terms, making the models technically incomplete even at leading order in  $\epsilon$.

As alluded to below equation~\eqref{UtoStrain}, actually evaluating the time integral in equation~\eqref{Umem} at 1PA order currently requires us to adopt some additional assumptions; cf.~the discussion in the PN context in section~\ref{sec:PNresults}. In principle, $F^{(0)}$ and $F^{(1)}$ can be computed from the local first- and second-order self-forces acting on the particle~\cite{Miller:2020bft}, and  the integral at 1PA order is given by
\begin{equation}
    \label{eq:hl0}
    \cU_{l0}^\text{(mem)} = \int_{0}^{\Omega(U)} \frac{\dot{\cU}_{l0}^\text{(mem)}}{\dot \Omega} \dd \Omega' = \frac{1}{\epsilon}\int_{0}^{\Omega(U)} \frac{\dot{\cU}_{l0}^\text{(mem)}}{F^{(0)}(\Omega')+\epsilon F^{(1)}(\Omega')} \dd \Omega'
\end{equation}
without additional assumptions or approximations. The relationship between orbital frequency $\Omega$ and retarded time $U$ at future null infinity depends on the choice of slicing $\uptau$, as different foliations will connect a time at future null infinity to different points on the particle's worldline. However, the choice of slicing also affects $F^{(1)}$ through the foliation-dependent source term $\delta G^{(1),m}_{\alpha\beta}$ in the field equation~\eqref{eq:h2_field_equation}. These two effects cancel out, such that the function $\cU_{l0}^\text{(mem)}(U)$ is in fact insensitive to the choice of slicing (up to 2PA differences). 

Unfortunately, at the time of writing, the local second-order self force has not yet been computed. 
As a consequence, the forcing functions $F^{(n)}$ are instead obtained from an energy-balance law.  
For a given choice of time variable $\uptau$, we define a binding energy 
\begin{equation}\label{eq:binding E}
\mathcal{E} = M_{\rm B} - m_1 - m_2,
\end{equation} 
where the Bondi mass $M_{\rm B}$ and the black hole mass $m_1$ are evaluated, respectively, on the cuts of future null infinity and of the future horizon defined by $\uptau=\text{constant}$. Note that here we work with the physical black hole mass 
\begin{equation}
m_1\equiv \mathring{m}_1 + \epsilon \delta m_1. 
\end{equation}
$\mathcal{E}$ can be calculated as a function of $\{\Omega(\uptau),m_1(\uptau),m_2,a(\uptau)\}$~\cite{Pound:2019lzj,Bonetto:2021exn}, such that differentiating equation~\eqref{eq:binding E} yields
\begin{equation}\label{dE/dtau}
\frac{\partial\mathcal{E}}{\partial\Omega}\dot\Omega + \frac{\partial\mathcal{E}}{\partial m_1}{\cal F}_{\cal H} +\frac{\partial\mathcal{E}}{\partial a}\dot a = -{\cal F}_\infty - {\cal F}_{\cal H} \equiv -{\cal F}.
\end{equation}
On the left-hand side, we have applied the chain rule. On the right, we have used the balance laws $dM_{\rm B}/d\uptau =-\mathcal{F}_\infty$ and $dm_1/d\uptau = \mathcal{F}_{\cal H}$, where $\mathcal{F}_{\infty}$ and $\mathcal{F}_{\cal H}$ are the flux of energy to future null infinity and the future horizon at time $\uptau$; these flux-balance laws are exact results in GR~\cite{Madler:2016xju,Ashtekar:2004cn}.
We have also used the fact that $m_2$ is exactly constant to high order in $\epsilon$~\cite{Poisson:2004cw}.  Rearranging equation~\eqref{dE/dtau} for $\dot\Omega$, we obtain
\begin{equation}\label{Omegadot balance}
\dot\Omega = -\frac{{\cal F} + (\partial{\cal E}/\partial m_1){\cal F}_{\cal H} + (\partial\mathcal{E}/\partial a)\dot a}{\partial{\cal E}/\partial\Omega}. 
\end{equation}
At this stage we have not made any additional assumptions. We can now expand the binding energy in powers of $\epsilon$, 
\begin{equation}
{\cal E} = m_2 \left[{\cal E}_{(0)}(m_1\Omega,a\Omega)-1 + \epsilon\, {\cal E}_{(1)}(m_1\Omega,a\Omega) + {\cal O}(\epsilon^2)\right],
\end{equation}
where ${\cal E}_{(0)}$ is the specific orbital energy (rather than specific binding energy) of a circular Kerr geodesic, and we have used $M_{\rm B} = m_1 + m_2{\cal E}_{(0)} + m_2 \e\, \mathcal{E}_{(1)}$. Similarly, we expand the fluxes as
\begin{equation}
{\cal F}_{\infty/\cal H} = \e^2 {\cal F}^{(1)}_{\infty/\cal H}(m_1\Omega,a\Omega) + \e^3 {\cal F}^{(2)}_{\infty/\cal H}(m_1\Omega,a\Omega) + {\cal O}(\epsilon^4),
\end{equation}
where ${\cal F}^{(1)}_{\infty/\cal H}$ is calculated from products of $h^{(1),m}_{\alpha\beta}$ with itself, and ${\cal F}^{(2)}_{\infty/\cal H}$ is calculated from products of $h^{(2),m}_{\alpha\beta}$ with $h^{(1),m}_{\alpha\beta}$. Fully expanding equation~\eqref{Omegadot balance} in powers of $\epsilon$ yields the forcing functions $F^{(n)}$ in equation~\eqref{eq:Omegadot} in terms of quantities measurable at infinity and the horizon. However, we now adopt one approximation and one assumption: 
\begin{itemize}
	\item We neglect 1PA horizon flux terms. This includes the explicit ${\cal O}(\epsilon^3)$ terms $(\partial{\cal E}/\partial m_1){\cal F}_{\cal H}$ and $(\partial{\cal E}/\partial a)\dot a$ in equation~\eqref{Omegadot balance}, as well as the contribution of ${\cal F}^{(2)}_{\cal H}$ to the total flux ${\cal F}$. The second-order horizon flux, ${\cal F}^{(2)}_{\cal H}$, cannot presently be included as it has not been calculated, but all 1PA horizon flux terms are expected to be numerically small~\cite{Albertini:2022rfe}. 
	\item We assume that the binding energy ${\cal E}_{(1)}(m_1\Omega,a\Omega)$, as defined by equation~\eqref{eq:binding E}, is equal (at least to a sufficiently good approximation) to the binding energy defined from the first law of binary mechanics~\cite{LeTiec:2011dp}. This assumption is required because the binding energy calculated from equation~\eqref{eq:binding E} was calculated with a different choice of $\uptau$ than the fluxes. Even with these different slicings, the two quantities are numerically very close to one another~\cite{Pound:2019lzj}, but the difference between them has a potentially significant impact on the GW phasing~\cite{Albertini:2022rfe}.
\end{itemize}
Given these assumptions, we write $\dot\Omega$ as
\begin{equation}
    \label{eq:evolution}
    \frac{\dd \Omega}{\dd t} =  -\frac{\mathcal{F}}{\partial \mathcal{E} / \partial \Omega} \, ,
\end{equation}
with the understanding that we neglect ${\cal F}^{(2)}_{\cal H}$ in ${\cal F}$ and use the first-law value of  ${\cal E}_{(1)}$.
We provide more detailed formalism required to calculate the metric perturbation and the associated memory contribution at first and second order in the next sections. 

\subsection{First-order memory for quasi-circular orbits around a Kerr black hole}

To calculate the first-order metric amplitudes, $h^{(1)}_{lm}$, we use the Teukolsky formalism~\cite{TeukolskyI,TeukolskyII,TeukolskyIII,PoundAndWardell}. In this section, we derive a convenient formula for the memory in terms of the asymptotic amplitudes of the (first order) spin-weight $-2$ Weyl scalar $\psi^{(1)}_4$,
\begin{equation}\label{eq:psi4}
    \psi_4^{(1)} = C^{(0)}_{\mu\nu\gamma\delta}[h^{(1)}]n^\mu \bar{m}^\nu n^\gamma \bar{m}^\delta =  \rho^4 \sum_{l=2}^{\infty}\sum_{m=-l}^{l} \tensor*[_{-2}]{R}{_{\ell m \omega}^{(1)}}(r) \tensor*[_{-2}]{S}{_{\ell m}^{a \omega}}(\theta,\phi)e^{-i m\phi_p}.
\end{equation}
Here, $C^{(0)}_{\mu\nu\gamma\delta}[h]$ is the linearized Weyl tensor (neglecting $\dot\Omega$ terms that arise from time derivatives), $n^\mu$ and $\bar{m}^\mu$ are legs of the Kinnersley tetrad, $\rho = -1/(r-i \mathring{a} \cos \theta)$, and $\tensor*[_{-2}]{S}{_{\ell m}^{a \omega}}$ is a spin-weighted spheroidal harmonic~\cite{PoundAndWardell}. The frequencies are harmonics of the orbital frequency, $\omega=m\Omega$. The spin parameter appearing in the spheroidal harmonics is the background spin $\mathring{a}$. However, our calculations in a Kerr background are limited to first perturbative order, meaning they can neglect the evolving correction to the spin and set $\mathring{a}=a$.

As mentioned above, we can use $t$ rather than $\uptau$ as our time coordinate in first-order calculations, in which case $\tensor*[_{-2}]{R}{_{ \ell m \omega}^{(1)}}$ satisfies the radial Teukolsky equation in the form given in equation~(84) of reference~\cite{PoundAndWardell}. After obtaining the solution in the $t$-foliation, one can transform to the time coordinate $\uptau=t-\upkappa(r_\star)$ using $\tensor*[_{-2}]{R}{_{ \ell m \omega}^{(1)}}\to \tensor*[_{-2}]{R}{_{ \ell m \omega}^{(1)}}e^{-i\omega\upkappa}$~\cite{Miller:2020bft,Miller:2023ers}. Note that such a simple transformation only applies at leading order in $\epsilon$. 

We stress that, at first order in our multiscale expansion, the coefficients $\tensor*[_{-2}]{R}{_{ \ell m \omega}^{(1)}}$ are identical to those sourced by a particle on a precisely circular, nondissipating geodesic with orbital frequency $\Omega$; the evolution of the frequency only affects the field equations at second order. In the geodesic case and in the $t$-foliation, the phase factor $e^{-i m\phi_p}$ becomes $e^{-i m\Omega t}$, and the radial Teukolsky equation is insensitive to this replacement. Therefore, when calculating the $m\neq0$, oscillatory waveform amplitudes $h^{(1)}_{lm}$, one can either consider geodesic, circular source orbits or evolving, quasi-circular source orbits. 

The waveform amplitudes $h^{(1)}_{lm}$ themselves are straightforwardly extracted from $\tensor*[_{-2}]{R}{_{ \ell m \omega}^{(1)}}$.  Writing the Weyl tensor in terms of $h$, projecting it onto the relevant tetrad legs, and taking the limit as $r\rightarrow\infty$,
 one finds that most of the terms in the Weyl scalar drop out, leaving
\begin{equation}
    \label{hdoubledotreasoning}
     \psi_4 = -\frac{1}{2}\ddot{h}_{\bar{m}\bar{m}}= -\frac{1}{2}(\ddot h_+ - i \ddot h_\times), \;\;\;\;\; r\rightarrow\infty;
\end{equation}
this holds fully nonlinearly, not restricted to leading order. At first order, the field equation is homogeneous at all points away from the particle. Since the solution to the homogeneous radial Teukolsky equation (in the $t$-foliation) goes as 
\begin{equation}
    \tensor*[_{-2}]{R}{_{\ell m\omega}^{(\rm hom)}}(r) = r^3 e^{+ i \omega r_*}, \;\;\;\;\; r\rightarrow\infty,
\end{equation}
this leaves us with
\begin{equation}
    \label{eq:psi4Limit}
     \psi^{(1)}_4  = \frac{1}{r} \sum_{\ell=2}^{\infty}\sum_{m=-\ell}^\ell\tensor*[_{-2}]{Z}{^{\text{Up}}_{\ell m \omega}}e^{- i m\phi_p(u)}\tensor*[_{-2}]{S}{_{\ell m}^{a \omega}}(\theta,\phi),  \;\;\;\;\; r\rightarrow\infty,
\end{equation}
where the Teukolsky amplitude $\tensor*[_{-2}]{Z}{^{\text{Up}}_{\ell m \omega}}$ is an $\Omega$-dependent constant (determined by solving the inhomogeneous Teukolsky equation and reading off the coefficient of $1/r$ at large $r$), and we have cancelled the factor $e^{+ i \omega r_*}$ by transforming to the $\uptau$-foliation.  
Integrating with respect to $u$ leaves us with
\begin{equation}
    \label{hdotexpression}
    \dot h^{(1)} = \frac{2}{r} \sum_{\ell=2}^{\infty}\sum_{m=-\ell}^\ell \frac{\tensor*[_{-2}]{Z}{^{\text{Up}}_{\ell m\omega}}}{im\Omega}e^{- i m \phi_p(u)}\tensor*[_{-2}]{S}{_{\ell m}^{a\omega}}(\theta,\phi).
\end{equation}
A second integration yields  
\begin{equation}
    \label{hexpression}
    h^{(1)} =  \frac{2}{r} \sum_{\ell=2}^{\infty}\sum_{m=-\ell}^\ell \frac{\tensor*[_{-2}]{Z}{^{\text{Up}}_{\ell m \omega}}}{(m\Omega)^2}e^{- i m \phi_p(u)}\tensor*[_{-2}]{S}{_{\ell m}^{a\omega}}(\theta,\phi).
\end{equation}
These integrations are performed in the manner of equation~\eqref{eq:integration by parts}, neglecting all subleading terms. Such terms are automatically accounted for in the second-order metric amplitudes.

Calculations in the Teukolsky formalism in Kerr spacetime are done in a spheroidal harmonic basis. To facilitate analytical integrations 
over the combinations of harmonics appearing in, e.g., equation~\eqref{Umem}, and in keeping with the expressions shown in section~\ref{sec:theory},
we must project these results onto a basis of spherical harmonics. This has the fortunate side effect of allowing easier comparisons with PN and NR. The expansion
is written as 
\begin{equation}
    \tensor*[_{-2}]{S}{_{\ell m}^{a \omega}} = \sum _{l= l_{\text{min}}}^{\infty}   \tensor*[_{-2\!}]{Y}{_{l m}} \; b_{\ell l m}^{a \omega},
\end{equation}
where the $b$ coefficients can be obtained following appendix A of~\cite{Hughes:1999bq}, and \mbox{$l_{\text{min}} = \ell _{\text{min}} = \text{max}(m,2)$}.
This allows us to write
\begin{equation}
h^{(1)}_{lm} = \sum _{\ell = \ell_{\text{min}}}^\infty h^{(1)}_{\ell m} b_{\ell l m}^{a \omega},  
\end{equation}
and by considering the form of $h_{\ell m}$  we can further define
\begin{equation}\label{eq:spherical_Z}
    Z^{(1)}_{lm} = \sum _{\ell = \ell_{\text{min}}}^\infty {}_{-2}Z^{\rm Up}_{\ell m\omega}\, b_{\ell l m}^{a \omega}.
\end{equation}
Using these expressions, we can express equation~\eqref{hdotexpression} in a spherical basis by making the swap $\ell \rightarrow l$, and replacing each 
spin-weighted spheroidal harmonic with the corresponding spin-weighted spherical harmonic. In the Schwarzschild case, where $\mathring{a} = 0$, we get that
$b_{\ell l m}^{0} = \delta_{\ell l}$ because spheroidal harmonics reduce to spherical ones in this case, and no projection is necessary. In either case, in the expansion~\eqref{eq:multiscale waveform}, we have
\begin{equation}
h^{(1)}_{lm} = \frac{2 Z^{(1)}_{lm}}{(m\Omega)^2}.
\end{equation}

Substituting $h$ in the above form into equation~\eqref{Umemdot}, we can write the time derivative of the radiative mass moment as
\begin{multline}
    \label{eq:UdotExplicit}
    \dot \cU_{lm}^\text{(1,mem)} = \sqrt{\frac{2(l-2)!}{(l+2)!}}
    \sum_{l'=2}^\infty\sum_{l''=2}^\infty\sum_{m' =-l'}^{l'}\sum_{m''=-l''}^{l''}\frac{(-4) (-1)^{m+m''} }{ m' m'' \Omega^2} \\*\times  Z^{(1)}_{l'm'}Z^{(1)*}_{l''m''}e^{- i (m' -m'')\phi_p(u)} G^{2,-2,0}_{l' \!,\, l'' \!,\, l, m' \!,\, -m''\!,\, -m} \,.
\end{multline}
Finally, we reduce the summation using several symmetries. We eliminate the sum over $m''$ using the fact that $G^{2,-2,0}_{l', l'', l, m', -m'', -m}$ vanishes unless $m'-m''=m$, by virtue of the 3-$j$ symbols in equation~\eqref{eq:G->3j}. We restrict to positive $m'$ by using the symmetry $Z^{(1)}_{lm} =(-1)^{l}Z^{(1)*}_{l-m}$. Also restricting ourselves to the non-oscillatory $m=0$ modes, we obtain
\begin{equation}
    \label{eq:InstantaneousMemory}
        \dot{\cU}_{l0}^{\text{(1,mem)}}  
        = -8\sqrt{\frac{2(l-2)!}{(l+2)!}}\sum_{l'=2}^{\infty}\sum_{l''=2}^{\infty}\sum_{m'=1}^{l'} \frac{(-1)^{m'}}{(m'\Omega)^2} \Re\! \left(Z^{(1)}_{l'm'}Z^{(1)*}_{l''m'} \right) G^{2,-2,0}_{l' \!,\, l'' \!,\, l,m' \!,\, -m' \!,\, 0} \ ,
\end{equation}
where we have additionally used the symmetry~\eqref{G_Symmetry}. Note that by virtue of the system's up-down symmetry, $\dot{\cU}_{l0}$ vanishes for odd values of $l$.

We calculate the accumulated memory over the inspiral using equation~\eqref{eq:hl0}. The evolution 
of the frequency is governed by equation~\eqref{eq:Omegadot}, which can equivalently be rewritten in terms of energy as equation~\eqref{eq:evolution}. At leading (0PA) order, we have 
\begin{equation}
\dot\Omega = \epsilon F^{(0)}(\Omega) = -\frac{\epsilon{\cal F}^{(1)}(\Omega)}{\partial {\cal E}_{(0)}/\partial\Omega},\label{eq:Omegadot 0PA}
\end{equation}
where ${\cal E}_{(0)}$ is the specific energy of a circular geodesic~\cite{1972ApJ...178..347B}, 
\begin{equation}
\mathcal{E}_{(0)} = \frac{1 - 2 (m_1 \Omega)^{2/3}(1 - a \Omega)^{1/3}}{\sqrt{1- (a \Omega)^2 -3(m_1 \Omega)^{2/3}(1 - a \Omega)^{4/3} }},
\end{equation}
and $\mathcal{F}^{(1)}$ is the instantaneous flux of energy (per unit mass squared) carried by $h^{(1)}_{\alpha\beta}$ into the black hole and out to infinity. This flux is readily calculated from the Teukolsky mode amplitudes at infinity and the black hole horizon~\cite{PoundAndWardell}. Unlike the 1PA balance law, the 0PA balance law~\eqref{eq:Omegadot 0PA} is known to hold exactly and is independent of one's choice of $\uptau$~\cite{Miller:2020bft}. We also note that if calculations are restricted to 0PA order, as ours are in Kerr, then we can everywhere replace $\mathring{m}_1$ and $\mathring{a}$ with $m_1$ and $a$ and treat them as constants (though the horizon flux ${\cal F}^{(1)}_{\cal H}$ must be included in ${\cal F}^{(1)}$).

\subsection{Second-order memory for quasi-circular orbits around a Schwarzschild black hole}
\label{sec:memoryRicci}

We now turn to the computation of the memory at second order in the mass ratio. In this section, we develop a simple extension of the formula~\eqref{eq:InstantaneousMemory} in terms of a certain effective second-order Teukolsky amplitude. 

Our calculation of the second-order memory begins from the metric amplitudes $h^{(2)}_{lm}$ that were computed in Schwarzschild in reference~\cite{Warburton:2021kwk}.  Rather than involving the Teukolsky formalism, these were computed by directly solving the field equations~\eqref{eq:h1_field_equation} and \eqref{eq:h2_field_equation} in the Lorenz gauge, as described in Refs.~\cite{Akcay:2010dx,Miller:2020bft,Miller:2023ers} and utilizing the ingredients from Refs.~\cite{Pound:2014xva,Pound:2015wva,Miller:2016hjv,Durkan:2022fvm}. At second order and beyond, the relationship between the orbital frequency $\Omega$ and the asymptotic metric amplitudes depends on the choice of time foliation. The calculations in reference~\cite{Warburton:2021kwk} specifically use a ``sharp slicing''~\cite{Miller:2023ers} in which $\uptau=t$ in a shell $r_-<r_{(0)}<r_+$ centered on the particle's leading-order (evolving) orbital radius $r_{(0)}$; $\uptau=v$ in a region extending from the shell to the horizon; and  $\uptau=u$ from the shell to infinity.

The GW memory is calculated from $\dot h$, which we can obtain from a time derivative of equation~\eqref{eq:multiscale waveform},
\begin{align}\label{eq:multiscale hdot}
	\dot h = \frac{1}{r}\sum_{lm}\left[\epsilon (-im\Omega)h^{(1)}_{lm} + \epsilon^2(-im\Omega) h^{(2)}_{lm} +\epsilon^2 F^{(0)}\partial_\Omega h^{(1)}_{lm}\right]e^{-im\phi_p}\,  {}_{-2}Y_{lm}(\theta,\phi).
\end{align}
Here we have accounted for the time derivative of the amplitude $h^{(1)}_{lm}$ and used equation~\eqref{eq:Omegadot}. We can work directly with this formula, but it will be convenient to define an effective second-order Teukolsky amplitude $Z^{(2)}_{lm}$ that allows us to easily extend the first-order expression~\eqref{eq:InstantaneousMemory}:
 \begin{equation}\label{eq:Z2eff}
\tensor*{Z}{^{(2)}_{l m}}\equiv \frac{1}{2}im \Omega \left(-im\Omega h^{(2)}_{lm} + F^{(0)}\partial_\Omega h^{(1)}_{lm}\right)  = \frac{1}{2}(m\Omega)^2\left(h^{(2)}_{lm} + \frac{iF^{(0)}}{m\Omega}\partial_\Omega h^{(1)}_{lm}\right).
\end{equation}
This amplitude is what would arise by ignoring $\dot\Omega$ terms when taking the second time derivative in equation~\eqref{hdoubledotreasoning}. Since the derivation at first order assumed $d/du = -im\Omega$ when relating $\dot h$ to $\psi_4$, we can immediately extend equation~\eqref{eq:InstantaneousMemory} to second order by making the simple substitution
\begin{equation}
    Z^{(1)}_{l m} \rightarrow  Z_{l m}^{(1)} + \epsilon\, Z_{l m}^{(2)} +\mathcal{O}(\epsilon^2),
\end{equation} 
and reading off the terms linear in $\epsilon$. We then write the total $\dot \cU_{l0}^\text{(mem)}$ as
\begin{equation}
    \dot \cU_{l0}^\text{(mem)} = \epsilon^2 \left[\dot \cU_{l0}^\text{(1, mem)} + \epsilon\, \dot \cU_{l0}^\text{(2, mem)} + \mathcal{O}(\epsilon^2)\right], 
\end{equation}
with $\dot \cU_{l0}^\text{(2, mem)}$ given by equation~\eqref{eq:InstantaneousMemory} with the replacement 
\begin{equation}\label{eq:Z1Z1replaceZ1Z2}
	\Re\! \left(Z^{(1)}_{l'm'}Z^{(1)*}_{l''m'} \right)\to \Re\! \left(Z^{(1)}_{l'm'}Z^{(2)*}_{l''m'} +Z^{(2)}_{l'm'}Z^{(1)*}_{l''m'}\right).
\end{equation}
An equivalent formula for $\dot \cU_{l0}^\text{(2, mem)}$ can readily be obtained in terms of the physical first-and second-order Teukolsky amplitudes (as opposed to the effective $Z^{(2)}_{lm}$). To do so, one would (i)~take a time derivative of equation~\eqref{eq:multiscale hdot} and include all $\dot\Omega$ terms, (ii)~use equation~\eqref{hdoubledotreasoning} to equate the physical second-order Teukolsky amplitude to terms in the derivative of equation~\eqref{eq:multiscale hdot}, and (iii)~integrate over $u$ as in equation~\eqref{eq:integration by parts} (including the first subleading term) to obtain the second-order piece of $\dot h$ in terms of the physical Teukolsky amplitudes.

\section{Spacetime asymptotics and hereditary effects in the self-force regime}\label{far-zone}

In the preceding section, we have implicitly assumed that solving equation~\eqref{eq:h1h2_field_equation} gives us direct access to the asymptotic waveform amplitudes $h^{(1)}_{lm}$ and $h^{(2)}_{lm}$. At first order in SF theory, this assumption is correct: we can obtain $h^{(1)}_{lm}$ simply by evaluating $h^{(1)}_{\alpha\beta}$ at $r\to\infty$ (at fixed $u$). If $h^{(1)}_{\alpha\beta}$ is calculated in a gauge that is asymptotically irregular, such as the Regge-Wheeler-Zerilli gauge, then there are standard prescriptions for transforming to a gauge that extends smoothly to future null infinity; see, e.g., reference~\cite{Martel:2003jj} for the case of the Regge-Wheeler-Zerilli gauge. If $h^{(1)}_{\alpha\beta}$ is calculated in the Lorenz gauge, no transformation is required.

At second order in perturbation theory, extracting the waveform becomes significantly more difficult due to an interplay between the multiscale expansion and gauge choice. As explored in reference~\cite{Pound:2015wva}, in a gauge in which the linearized Einstein tensor reduces to a wave operator (as it does in the Lorenz gauge), the field equation~\eqref{eq:h2_field_equation} suffers from logarithmic infrared divergences stemming from the source terms' slow falloff at large distances. 

The infrared breakdown confines the multiscale expansion to the near zone defined by $r\ll M/\epsilon$ (note this definition of the near zone is not related to the wavelength of the emitted radiation, unlike the definition used in the PN-MPM construction). To overcome the breakdown, we introduce a weak-field, PM expansion in the far zone $r\gg M$, inspired by the MPM formalism. Information is transmitted between the multiscale expansion and the PM expansion in a buffer region (the `near far zone' $M\ll r\ll M/\epsilon$) where both approximations are valid, corresponding to the large-$r$ limit of the near zone and the small-$r$ limit of the far zone.  Re-expanding the far-zone PM solution in this buffer region provides physical boundary conditions for the multiscale solution. 

We illustrate the breakdown of the multiscale expansion in section~\ref{sec:breakdown}, and we outline the PM expansion in section~\ref{sec:PM expansion}. We then apply the formalism to show that
\begin{enumerate}
\item matching to the far-zone solution introduces hereditary, nonlocal-in-time effects in the near-zone dynamics, which cannot otherwise be captured by the multiscale expansion, 
\item the hereditary near-zone effects arise from precisely the same piece of the metric that corresponds to memory in the GW, 
\item `most' of the oscillatory part of the waveform at true asymptotic infinity, in the very far zone $r\gg M/\epsilon$, can be directly extracted from a certain piece of the multiscale solution in the large-$r$ limit of the near zone; this part of the metric propagates undistorted from the near zone to the very far zone. 
\item additional oscillatory modes with hereditary amplitudes arise from products between first-order memory terms and first-order oscillatory terms. The impact of these new modes will require further investigation. 
\end{enumerate}
The first three of these are in close analogy with classic results of the PN-MPM construction~\cite{Blanchet:1987wq,Blanchet:1992br}. However, we note that in the PN-MPM calculations, both tails and memory effects give rise to nonlocal-in-time integrals, while in our context, nonlocal-in-time integrals are always associated with memory-type, low-frequency effects. Ordinary tails, which involve oscillatory integrands, are always reduced to local-in-time functions by virtue of equations such as~\eqref{eq:integration by parts}.

The framework we describe in this section underlies the 2SF results in references~\cite{Pound:2019lzj,Warburton:2021kwk,Wardell:2021fyy}. Here we describe it and its links to GW memory for the first time. 
To clearly split between quasistationary and oscillatory effects, in this section we introduce the notation
\beq\label{av + osc}
f(u) = \<f(u)\> + f^{\rm osc.}(u)
\eeq 
for functions $f(u)=\sum_{m\in\mathbb{Z}} f_m(\Omega(u)) e^{-im\phi_p(u)}$, where 
\beq
\<f\> = f_0 \quad\text{and}\quad f^{\rm osc.}=\sum_{m\neq0} f_m e^{-im\phi_p}.
\eeq

\subsection{Breakdown of the multiscale expansion}\label{sec:breakdown}

As context for our analysis in the next sections, we first review the infrared breakdown of the multiscale expansion. This breakdown, as mentioned above, stems from the large-$r$ behavior of the source terms in  equation~\eqref{eq:h2_field_equation}. Since the problem arises from large-$r$ behavior, our illustration in this section will focus on the large-$r$ limit of our multiscale field equations.

If the $t$-foliation is used, then the first-order solution behaves as $h^{(1),m}_{\alpha\beta}\sim\frac{e^{im\Omega r_*}}{r}$ at large $r$, meaning the source $\delta G^{(1),m}_{\alpha\beta}[h^{(1),m}]$ behaves particularly poorly. A second $t$-derivative in the orginal four-dimensional $\delta G_{\alpha\beta}$ gives rise to terms like $\Omega\dot\Omega\partial_\Omega$, leading to  
\begin{equation}
\delta G^{(1),m}_{\alpha\beta}[h^{(1),m}] \sim \Omega F^{(0)}\partial_\Omega h^{(1),m}_{\alpha\beta}\sim \Omega F^{(0)}\partial_\Omega \frac{e^{im\Omega r_*}}{r} = im\Omega F^{(0)}\frac{r_* e^{im\Omega r_*}}{r}.
\end{equation}
This is immediately cured by using hyperboloidal time $\uptau$ in the multiscale expansion~\cite{Pound:2015wva,Miller:2020bft,Miller:2023ers}, which eliminates the exponential $e^{im\Omega r_*}$ (and hence eliminates the $\Omega$ derivative of it) as well as eliminating second time derivatives from $\delta G_{\alpha\beta}$ (and hence replacing  $\Omega\dot\Omega\partial_\Omega$ terms in $\delta G^{(1),m}_{\alpha\beta}$ with $\dot\Omega\partial_r\partial_\Omega$ terms).  

A more pernicious problem arises from the quadratic source term $\delta^2 G^{(0),m}_{\alpha\beta}$. In a generic gauge and the $t$-foliation, this decays as the product of two oscillating GWs,
\begin{equation}\label{eq:d2G falloff}
\delta^2 G^{(0),m}_{\alpha\beta}\sim \sum_{m',m''} m'm''\Omega^2 h^{(1),m'}h^{(1),m''} \sim  \frac{e^{im\Omega r_\star}}{r^2}.
\end{equation}
In  hyperboloidal slicing, the exponential $e^{im\Omega r_\star}$ is removed but the falloff is otherwise unchanged. We will explain the consequences of this by adapting an analysis from reference~\cite{Pound:2015wva} (see also section~VI of~\cite{Pound:2014koa}). First, recall that in the multiscale framework, after $e^{-im\phi_p}$ is factored out, the field equations only involve spatial derivatives and functions of $\Omega$. For example, at large $r$ in $t$ slicing, the field equation~\eqref{eq:h2_field_equation} in the Lorenz gauge reduces to\footnote{cf. the exact form of the Lorenz-gauge linearized Einstein tensor, \eqref{E operator} below.}  
\begin{equation}\label{large-r h2 field eqn}
\Box^{(0)}_{\eta} h^{(2),m}_{\alpha\beta} = \frac{e^{im\Omega r}}{r^2}s^{m}_{\alpha\beta}(\Omega,\theta)e^{im\phi},\qquad r\to\infty ,
\end{equation}
where $s^{m}_{\alpha\beta}e^{im\phi}$ is the coefficient of $e^{im\Omega r_\star}/r^2$ in the large-$r$ expansion of the quadratic source, and $\Box^{(0)}_{\eta}=(m\Omega)^2+\vec{\nabla}^2$ is the flat-space d'Alembert operator with $\partial_t=-im\Omega$.  One solves each such field equation at fixed $\Omega$. We can hence formally define the retarded solution to equation~\eqref{eq:h2_field_equation} as the integral of the source against a retarded Green's function  $G_{\alpha\beta}{}^{\alpha'\beta'}(\omega,x^i,x'^i)$ for the frequency-domain linearised Einstein equation. The contribution from the source's large-$r$ tail, as written in equation~\eqref{large-r h2 field eqn}, is then
\begin{equation}\label{eq:large-r h2 soln}
h^{(2),m}_{\alpha\beta}(x^i) = \int dr' \int d\Omega'_2 \, G_{\alpha\beta}{}^{\alpha'\beta'}(\omega,x^i,x'^i)e^{im\Omega r'}s^{m}_{\alpha'\beta'}(\Omega,\theta')e^{im\phi'} + \ldots,
\end{equation}
where the Green's function is evaluated at $\omega=m\Omega$ and we have cancelled the $r^{-2}$ in the source with the $r^2$ in the volume element. The ellipses indicate that this formula only represents the contribution of the large-$r$ behavior of the source.

Since we are considering the behavior at large $r$ coming from the source at large $r'$, we can approximate $G_{\alpha\beta}{}^{\alpha'\beta'}$ as its flat-spacetime limit, 
\begin{equation}\label{eq:Gret Fourier}
G_{\alpha\beta}{}^{\alpha'\beta'} = -\frac{1}{4\pi}\delta_{\alpha}^{\alpha'}\delta_{\beta}^{\beta'}\frac{e^{i\omega|\vec{x}-\vec{x}'|}}{|\vec{x}-\vec{x}'|}, 
\end{equation}
which satisfies
\beq
(\omega^2+\vec{\nabla}^2)G_{\alpha\beta}{}^{\alpha'\beta'} = \delta_{\alpha}^{\alpha'}\delta_{\beta}^{\beta'} \delta^3(\vec x-\vec x').
\eeq
Here $\vec{x}=(x,y,z)$ are Cartesian coordinates and we work with Cartesian components. Expanded in scalar spherical harmonics, for large $r$ and $r'$, and for $\omega\neq0$ modes, this becomes
\begin{equation}\label{eq:large-r Greens function}
G_{\alpha\beta}{}^{\alpha'\beta'} \approx -\frac{1}{2i \omega rr'} \delta_{\alpha}^{\alpha'}\delta_{\beta}^{\beta'}\sum_{l=0}^\infty\sum_{m'=-l}^l \left[(-1)^l e^{i\omega r}-e^{-i\omega r}\right]e^{i\omega r'}Y^*_{lm'}(\theta',\phi')Y_{lm'}(\theta,\phi)
\end{equation}
for $r'>r$ (and $r\leftrightarrow r'$ if $r>r')$; cf. equation~(14) of~\cite{Miller:2016hjv}. For $\omega=0$, the dominant behavior is confined to $l=0$, and the Green's function reduces to
\begin{equation}\label{eq:large-r omega=0 Greens function}
G_{\alpha\beta}{}^{\alpha'\beta'} \approx -\frac{1}{4\pi} \delta_{\alpha}^{\alpha'}\delta_{\beta}^{\beta'}\left(\frac{\theta(r-r')}{r}+\frac{\theta(r'-r)}{r'}\right);
\end{equation}
cf. equation~(15) of~\cite{Miller:2016hjv}.

First, consider  $\omega\neq0$ modes. The approximations~\eqref{eq:large-r h2 soln} and \eqref{eq:large-r Greens function} imply
\begin{equation}
h^{(2),m}_{\alpha\beta} \approx -\frac{1}{2 i\omega r}\sum_{l\geq|m|}Y_{lm}\left[e^{i\omega r}\int^r_{R_0} \frac{\beta_l(\omega r')e^{i\omega r'}}{r'} dr' +\beta_l(\omega r)\int_r^\infty \frac{e^{2i\omega r'}}{r'} dr'\right] s^{lm}_{\alpha\beta}(\Omega),
\end{equation}
where for convenience we have introduced $\beta_l(\omega r) =\left[(-1)^l e^{i\omega r}- e^{-i\omega r}\right]$. We have set an arbitrary (large) lower limit $R_0<r$ for the radial integral; this is freely done in the present illustration because we are only interested in the contribution of the large-$r$ source. Also note that the integration over $\phi'$ has imposed $m'=m$. The radial integral from $r$ to $\infty$ converges and yields a subdominant term of order $1/r^2$ in $h^{(2,m)}_{\alpha\beta}$. The first term in the integral from~$R_0$ to~$r$, $\int_{R_0}^r dr' \,{e^{2i\omega r'}}/{r'} $, does likewise (at leading order for large $R_0$). However, the second term in that integral contributes a logarithm: $\int_{R_0}^r dr' \,{(-1)^{l+1}}/{r'} = (-1)^{l+1}\ln(r/R_0)$. Therefore, for $m\neq0$ modes, the retarded integral converges but yields the irregular, logarithmic asymptotic behavior
\begin{equation}\label{eq:h2m asymptotic solution}
h^{(2),m}_{\alpha\beta} \approx \frac{\ln(r) e^{im\Omega r}}{2 i m\Omega r}\sum_{l\geq |m|} s^{lm}_{\alpha\beta}(\Omega)Y_{lm},
\end{equation}
mirroring well-known behavior in harmonic coordinates~\cite{Blanchet:1986dk}. Here we have discarded the arbitrary radius $R_0$, which will not appear in the actual retarded integral using the fully relativistic source and Green's function; that integral will run from the horizon to infinity and involve no arbitrary constants.  

Next, consider $\omega=0$ modes. The approximations~\eqref{eq:large-r h2 soln} and \eqref{eq:large-r omega=0 Greens function} imply
\begin{equation}
h^{(2),0}_{\alpha\beta} \approx -\left(\frac{1}{r}\int^r_{R_0} dr' + \int_r^\infty \frac{dr'}{r'}\right) s^{00}_{\alpha\beta}(\Omega)Y_{00}\,.
\end{equation}
The first integral contributes a constant term $\propto r^0$ in $h^{(2,0)}_{\alpha\beta}$, making it already asymptotically irregular. The second integral does not converge at all.  Its logarithmic divergence is the principal signature of the breakdown of the multiscale expansion. Note that here it occurs in the $l=0$ mode, but this is an artefact of having expanded Cartesian components in scalar spherical harmonics. What appears as a pure $l=0$ mode in this expansion would contribute to $0\leq l\leq 2$ modes in an expansion of tetrad components in spin-weighted spherical harmonics (while remaining confined to $m=0$).

We have performed the analysis in this section in $t$ slicing for simplicity, but the results are the same in $\uptau$ slicing; the only effect is to remove the exponential factor from equation~\eqref{eq:h2m asymptotic solution}. We refer to~\cite{Pound:2015wva} for a more detailed analysis in the case of a nonlinear scalar toy model.

Before proceeding, we comment on an underlying cause of this breakdown: in a generic gauge, $u={\rm constant}$ is not a light cone of the perturbed spacetime (cf. historical references given on p. 386 of \cite{Blanchet:1986dk}). As highlighted in reference~\cite{Spiers:2023cip}, this can be cured by transforming $h^{(1)}_{\alpha\beta}$ to a Bondi-Sachs gauge~\cite{Madler:2016xju}, which enforces that $u={\rm constant}$ is an outgoing null cone in both the background and perturbed spacetime.  If we were to transform $h^{(1)}_{\alpha\beta}$ to a Bondi-Sachs gauge before constructing the source for the second-order field equations, then certain components of the quadratic source term in the field equations would decay more rapidly at large $r$. This suffices to eliminate infrared divergences in the retarded integral for the second-order Teukolsky equation~\cite{Spiers:2023cip}, though the impact of this gauge choice on the full second-order Einstein field equations has not been fully investigated. 

In principle, one could follow a three-step procedure: (i) solve for $h^{(1),m}_{\alpha\beta}$ in any convenient gauge, (ii) transform $h^{(1),m}_{\alpha\beta}$ to a Bondi-Sachs gauge and construct the source terms in equation~\eqref{eq:h2_field_equation} from it, and (iii) impose any convenient gauge for $h^{(2),m}_{\alpha\beta}$ when solving equation~\eqref{eq:h2_field_equation}. This strategy, which would mirror an analogous approach in reference~\cite{Trestini:2022tot}, will be further developed in a sequel paper~\cite{Spiers:InPrep}. For the remainder of this section, we describe instead the strategy used in~\cite{Pound:2019lzj} and~\cite{Warburton:2021kwk}, which will also allow us to directly relate our calculations to the derivation of GW memory due to Blanchet and Damour~\cite{Blanchet:1992br}.

\subsection{Post-Minkowskian expansion in the far zone $r\gg M$}\label{sec:PM expansion}

We cure the multiscale expansion's large-$r$ pathology by adopting a far-zone, MPM-like expansion based on reference~\cite{Pound:2015wva}. The essential point in this construction is that we solve the field equations directly on spacetime rather than treating $(\phi_p,\Omega)$ as independent variables, using four-dimensional retarded integrals rather than three-dimensional, fixed-frequency retarded integrals. Evaluating the retarded integrals is made tractable by working in the PM limit outside the source region, at $r\gg M$, meaning we solve flat-spacetime wave equations rather than curved-spacetime ones.

Our starting point is the self-consistent formulation of the small-$\epsilon$ expansion~\cite{Pound:2009sm,Pound:2015fma}, as extended in reference~\cite{Miller:2020bft}. This formulation is valid over the entire spacetime, in both the near zone and the far zone. When the self-consistent formulation is re-expanded in multiscale form, it recovers the multiscale formulation in the near zone~\cite{Miller:2020bft}. In this section, we expand it in a PM form valid in the far zone.

In the self-consistent framework, we treat $h_{\alpha\beta}$ as a function of spacetime coordinates, of the particle's mass and the background black hole parameters, and of a set of matter variables $\Psi(t,\e)=\{\gamma(t,\e),\delta m_1(t,\e),\delta a(t,\e)\}$ comprising the particle's trajectory $\gamma$ and the black hole's small, evolving mass and spin corrections. We then define an expansion in powers of the mass ratio at fixed spacetime coordinates and fixed~$\Psi$:
\begin{equation}\label{eq:self-consistent expansion}
h_{\alpha\beta} = \epsilon h^{[1]}_{\alpha\beta}(x^\mu;\Psi) + \epsilon^2 h^{[2]}_{\alpha\beta}(x^\mu;\Psi)  + \ldots,
\end{equation}
where we use square brackets to distinguish orders in this expansion from orders in the multiscale expansion. As we did when solving the multiscale-expanded field equations through second order, we adopt the Lorenz gauge condition, which we now state explicitly:
\begin{equation}\label{eq:Lorenz gauge}
\mathring{g}^{\alpha\beta}\mathring{\nabla}_\alpha \bar h_{\beta\gamma} = 0 \,,
\end{equation}
where $\bar h_{\alpha\beta}\equiv h_{\alpha\beta}-\frac{1}{2}\mathring{g}_{\alpha\beta}\mathring{g}^{\mu\nu}h_{\mu\nu}$ and $\mathring{\nabla}_\alpha$ is the covariant derivative compatible with $\mathring{g}_{\alpha\beta}$. With this gauge choice, the linearized Einstein tensor reduces to a wave operator,
\begin{equation}\label{E operator}
\delta G_{\alpha\beta}[h] = -\frac{1}{2}\left(\mathring{\Box}\bar h_{\alpha\beta} + 2\mathring{R}_{\alpha}{}^\mu{}_\beta{}^\nu \bar h_{\mu\nu}\right)\equiv -\frac{1}{2}E_{\alpha\beta}[\bar h] \,,
\end{equation} 
where $\mathring{\Box} = \mathring{g}^{\alpha\beta} \mathring{\nabla}_\alpha \mathring{\nabla}_\beta$ and where $\mathring{R}_{\mu\nu\rho\sigma}$ is the Riemann tensor associated with the background metric $\mathring{g}_{\alpha\beta}$. We can write a sequence of field equations for each $h^{[n]}_{\alpha\beta}$,
\begin{subequations}\label{self-consistent EFEs}%
\begin{align}%
E_{\alpha\beta}[\bar h^{[1]}] &= -16\pi T^{[1]}_{\alpha\beta} \,,\label{SC EFE1}\\*
E_{\alpha\beta}[\bar h^{[2]}] &= -16\pi T^{[2]}_{\alpha\beta}+ 2\delta^2 G_{\alpha\beta}[h^{[1]},h^{[1]}]\,.
\end{align}
\end{subequations}
These are coupled to the equation for the particle's self-forced trajectory $x^\alpha_p$,
\begin{equation}\label{eq:EOM}
\mathring{u}^\beta\mathring{\nabla}_{\!\beta} \mathring{u}^\alpha = \epsilon \delta f^{\alpha}[h^{[1]}] + \epsilon^2\left( \delta f^{\alpha}[h^{[2]}] +\delta^2f^{\alpha}[h^{[1]},h^{[1]}]\right),
\end{equation}
where  $\mathring{u}^\beta\equiv\frac{dx^\beta_p}{d\mathring{\tau}}$ is the particle's four-velocity normalized in the background spacetime, and $f^{\alpha}_{[1]}$ and $f^{\alpha}_{[2]}$ are the linear and quadratic self-forces~\cite{Pound:2017psq}. The evolving corrections to the primary's mass and spin are incorporated into $h^{[1]}_{\alpha\beta}$ as described in reference~\cite{Miller:2020bft}. The field equations~\eqref{self-consistent EFEs} and matter equations are to be solved together, self-consistently, as a coupled set. The gauge condition~\eqref{eq:Lorenz gauge} is then automatically enforced by the matter evolution equations, so long as initial data satisfies the gauge condition~\cite{Pound:2012dk}.

To see why the self-consistent expansion evades the infrared divergences of the multiscale expansion, we can formally write the self-consistent solutions as retarded integrals over all spacetime, 
\begin{subequations}\label{eq:h[n] retarded integrals}%
\begin{align}
\bar h^{[1]}_{\alpha\beta}(x^\mu) &= -16\pi \int G_{\alpha\beta}{}^{\alpha'\beta'}(x^\mu,x'^\mu) T^{[1]}_{\alpha'\beta'} \sqrt{-\mathring{g}'}d^4x' + \bar h^{\delta m_1,\delta a_1}_{\alpha\beta}  ,\\
\bar h^{[2]}_{\alpha\beta}(x^\mu) &= -16\pi \int G_{\alpha\beta}{}^{\alpha'\beta'}(x^\mu,x'^\mu)\left(T^{[2]}_{\alpha'\beta'} -\frac{1}{8\pi}\delta^2 G_{\alpha'\beta'}[h^{[1]},h^{[1]}]\right)\sqrt{-\mathring{g}'}d^4x'.\label{eq:h[2] retarded integral}
\end{align}
 \end{subequations}
Here $\bar h^{\delta m_1,\delta a_1}_{\alpha\beta}$ represents the contribution from the primary's evolving mass and spin corrections, and  $G_{\alpha\beta}{}^{\alpha'\beta'}$ is the retarded Green's function for the operator $E_{\alpha\beta}$, satisfying
\begin{equation}
\mathring{\Box} G_{\alpha\beta}{}^{\alpha'\beta'} + 2\mathring{R}_\alpha{}^\mu{}_\beta{}^\nu G_{\mu\nu}{}^{\alpha'\beta'} = \delta^{\alpha'}_{\alpha} \delta^{\beta'}_{\beta}\frac{\delta^4(x^\mu-x'^\mu)}{\sqrt{-\mathring{g}}}.
\end{equation}
Crucially, the integrals~\eqref{eq:h[n] retarded integrals} involve integration over time, and the trajectory (and the primary's mass and spin) and metric perturbations evolve self-consistently within that integration. This is important because the amplitude of GW content in $h^{[1]}_{\alpha\beta}$ decays toward \mbox{$u=-\infty$} (since the binary asymptotes to a Newtonian, nonradiating state in the infinite past). Consequently, the problematic, $1/r^2$ term in $\delta^2 G_{\alpha'\beta'}[h^{[1]},h^{[1]}]$ also has decaying amplitude toward $u=-\infty$; this can be understood from the fact that the $1/r^2$ part of $\delta^2 G_{\alpha'\beta'}[h^{[1]},h^{[1]}]$ arises from the product of two GWs, as in equation~\eqref{eq:d2G falloff}. The decay of $\delta^2 G_{\alpha'\beta'}[h^{[1]},h^{[1]}]$ in turn ensures the convergence of the four-dimensional integral~\eqref{eq:h[2] retarded integral}. In the multiscale solution, on the other hand, the second-order field equation~\eqref{eq:h2_field_equation} is solved at fixed values of $\Omega$ and therefore at fixed values of the GW amplitudes. The solution in that case is given by a three-dimensional integral of the form~\eqref{eq:large-r h2 soln} (see reference~\cite{Miller:2023ers} for the explicit, fully relativistic construction), which does not naturally capture the quadratic source term's decay to zero in the distant past. This is analogous to the discussion in Sec.~\ref{subsec:memory_standard_PN_MPM} in the PN context; see also \cite{Arun:2004ff}.

We take the self-consistent expansion as our starting point for our PM expansion in the far zone. Outside the source region, where $R_{\alpha\beta}[\mathring{g}+h]=0$, the self-consistent field equations~\eqref{self-consistent EFEs} become%
\begin{subequations}\label{self-consistent vacuum EFEs}%
\begin{align}%
E_{\alpha\beta}[h^{[1]}] &= 0,\label{SC EFE1}\\*
E_{\alpha\beta}[h^{[2]}] &= 2\delta^2 R_{\alpha\beta}[h^{[1]},h^{[1]}],
\end{align}
\end{subequations}
where we have used $\delta R_{\alpha\beta}[h] = E_{\alpha\beta}[h]$ in the Lorenz gauge. All information about the matter fields $\Psi$ will now be transmitted to $h^{[n]}_{\alpha\beta}$ through matching to the multiscale near-zone solution. We do not directly impose the gauge condition but instead allow it to be enforced indirectly through this matching; this suffices because the field equations automatically preserve the gauge condition if the boundary data satisfies it~\cite{Pound:2012dk}.

The expansion~\eqref{eq:self-consistent expansion} is, effectively, an expansion in powers of the mass $m_2$ (while holding the orbital trajectory  fixed). We next introduce a post-Minkowskian expansion by formally expanding in powers of the background mass $\mathring{m}_1$, under the assumption that in the weak-field region, the effect of $\mathring{m}_1$ is a small, post-Minkowskian perturbation. Concretely, we expand at fixed Cartesian coordinates $(t,x^i)$ and at fixed $\Psi$. We define $x^i$ from the tortoise polar coordinates $(r_*, \theta, \phi)$, such that $r_*=\sqrt{\delta_{ij}x^ix^j}$. This ensures that the outgoing null surfaces of the Minkowski background $\eta_{\alpha\beta}$, defined by constant retarded time $u=t-r_*$, are also the outgoing null surfaces of $\mathring{g}_{\alpha\beta}$: $\eta^{\alpha\beta}\partial_\alpha u\,\partial_\beta u = 0 = \mathring{g}^{\alpha\beta}\partial_\alpha u\,\partial_\beta u = 0$. The expansion puts the operators $E_{\alpha\beta}$ and $\delta^2R_{\alpha\beta}$ in the forms
\begin{align}
E_{\alpha\beta}[h] &= \Box_\eta h_{\alpha\beta} + \sum_{j\geq1}(\mathring{m}_1)^j E^{[j]}_{\alpha\beta}[h],\label{eq:E PM expansion}\\
\delta^2 R_{\alpha\beta}[h,\ldots,h] &= \sum_{j\geq0}(\mathring{m}_1)^j\delta^2R^{[j]}_{\alpha\beta}[h,\ldots,h],\label{eq:dnR PM expansion}
\end{align}
and the metric perturbations in the form
\beq\label{eq:h PM expansion}
h^{[n]}_{\alpha\beta} = \sum_{j\geq0}(\mathring{m}_1)^j h^{[n,j]}_{\alpha\beta}(t,x^i;\Psi).
\eeq
Here all components are in the coordinate basis $\partial_\alpha t$, $\partial_\alpha x^i$. 

Equations~\eqref{eq:E PM expansion} and \eqref{eq:dnR PM expansion} represent expansions in powers of $\mathring{m}_1/r$ and $\mathring{a}/r$ ($\propto \mathring{m_1}/r$) because they are expansions of operators constructed from the Kerr metric, which contains no length scale other than $\mathring{m}_1$. In contrast, equation~\eqref{eq:h PM expansion} does not correspond simply to an expansion in powers of $\mathring{m}_1/r$; indeed, inverse length scales other than $1/r$, such as $\dot\Omega/\Omega$, arise to form dimensionless combinations with the dimensionful expansion parameter $\mathring{m}_1$. Also note that equation~\eqref{eq:h PM expansion} is not an expansion in powers of $\mathring{m}_1\Omega$; we must avoid such an expansion because $\mathring{m}_1\Omega$ is not small here (unlike in the PN-MPM context).

With these expansions, the field equations become a sequence of Minkowskian wave equations, one at each order in $m_2$ and $\mathring{m}_1$. The equations read%
\begin{subequations}\label{EFEnj}
\begin{align}
\Box_\eta h^{[1,0]}_{\alpha\beta} &= 0,\label{EFE10}\\*
\Box_\eta h^{[1,1]}_{\alpha\beta} &= -E^{[1]}_{\alpha\beta}[h^{[1,0]}]\equiv S^{[1,1]}_{\alpha\beta},\label{EFE11}\\*
&\hspace{6.5pt} \vdots \nonumber\\*
\Box_\eta h^{[2,0]}_{\alpha\beta} &= 2\delta^2 R^{[0]}_{\alpha\beta}[h^{[1,0]},h^{[1,0]}] \equiv S^{[2,0]}_{\alpha\beta},\label{EFE20}\\*
\Box_\eta h^{[2,1]}_{\alpha\beta} &= 2\delta^2 R^{[1]}_{\alpha\beta}[h^{[1,0]},h^{[1,0]}] + 4\delta^2 R^{[0]}_{\alpha\beta}[h^{[1,0]},h^{[1,1]}] - E^{[1]}_{\alpha\beta}[h^{[2,0]}] \equiv S^{[2,1]}_{\alpha\beta},\label{EFE21}\\*
&\hspace{6.5pt}\vdots \nonumber
\end{align}
\end{subequations}
We demand that the solutions to these equations match the multiscale solution when they are re-expanded in a common regime, meaning the multiscale solution is re-expanded for $r\gg M$ and the far-zone solution is expanded for $r\ll M/\epsilon$. This shared regime is the `near far zone' at $M\ll r \ll M/\epsilon$.

\subsection{Algorithm for constructing the GSF-MPM far-zone solution}

We now develop a scheme to solve the field equations~\eqref{EFEnj} in a manner consistent  (i)~with retarded propagation and (ii)~with matching to the multiscale near-zone solution. Since our method draws heavily on the PN-MPM formalism, we refer to it as a GSF-MPM scheme.

The retarded solutions take a form analogous to equations~(135)--(136) in reference~\cite{Pound:2015wva}. Explicitly, the retarded solution to the vacuum equation~\eqref{EFE10}, meaning the solution with no incoming radiation, can be written as~\cite{Damour:1990gj}
\begin{align}\label{h10}
h^{[1,0]}_{\alpha\beta} &= \sum_{l\geq0}\partial_L\frac{F^{[1]}_{\alpha\beta L}(u)}{r_*},
\end{align}
where the multipole moments $F^{[1]}_{\alpha\beta L}$ are STF in their last $l$ indices, and 
\begin{equation}
\partial_i r_* = n_i \equiv \delta_{ij}x^j/r_*, 
\end{equation}
meaning $\partial_i u = -n_i$. As in the PN-MPM scheme, the multipole moments will encode the full freedom in our far-zone solution, and they are what will be determined by matching to the near-zone solution.  Here it will be important that no corrections to the multipole moments appear in the higher-order terms $h^{[1,j]}_{\alpha\beta}$, such that $F^{[1]}_{\alpha\beta L}$ on its own appropriately matches the coefficient of $1/r$ appearing in the near-zone, multiscale solution. In other words, we require that no terms of order $1/r$ appear in $h^{[1,j]}_{\alpha\beta}$ for $j>0$.

Before proceeding to subleading orders $h^{[1,j] }_{\alpha\beta}$, we first explain the importance of avoiding new $1/r$ terms in those subleading orders. Consider re-expanding~\eqref{h10} in multiscale form. For a function $K(u,\epsilon)$, the multiscale expansion reads
\begin{equation}
K(u,\epsilon) = \sum_{m\in\mathbb{Z}}\sum_{n=0}^\infty \epsilon^n K_{(n)}^m(\Omega(u,\epsilon))e^{-im\phi_p(u,\epsilon)}.
\end{equation}
Here for simplicity we assume that $\uptau=u$ in the region where we perform the matching to the near zone, such that we do not need to expand $u$ in terms of $\uptau$ and $r_*$. Starting from a multiscale expansion of $F^{[1]}_{\alpha\beta L}$, we then have
\begin{align}\label{eq:h10 multiscale}
h^{[1,0]}_{\alpha\beta} &= \frac{1}{r}\sum_{l=0}^\infty\sum_{m=-l}^l [im\Omega(u,\epsilon)]^l\hat n^L F^{[1],m}_{\alpha\beta L}(\Omega(u,\epsilon))e^{-im\phi_p(u,\epsilon)} +\mathcal{O}(\epsilon,1/r^2),
\end{align}
observing that the leading term in $\epsilon$ and $1/r$ comes from $\partial_L$ acting on the phase factor $e^{-im\phi_p}$ in the multiscale expansion of $F^{[1]}_{\alpha\beta L}$. Recalling the notation for the STF projection $\hat n^L\equiv n^{\langle i_1}\cdots n^{i_l\rangle}$, we have used the fact the the multipolar moments are intrinsically STF in their indices to write $n^L F^{[1],m}_{\alpha\beta L}=\hat n^L F^{[1],m}_{\alpha\beta L}$. The contraction $\hat n^L F^{[1],m}_{\alpha\beta L}$ is equal to an expansion of each Cartesian component in scalar (spin-weight 0) spherical harmonics, $\sum_{lm'} F^{[1],mlm'}_{\alpha\beta}(\Omega)Y_{lm'}$.  Matching to the near-zone solution will enforce $m'=m$, such that  
\begin{align}\label{eq:h10 multiscale}
h^{[1,0]}_{\alpha\beta} &= \frac{1}{r}\sum_{l=0}^\infty \sum_{m=-l}^l (im\Omega)^l F^{[1],lm}_{\alpha\beta}(\Omega)Y_{lm}e^{-im\phi_p} +\mathcal{O}(\epsilon,1/r^2),
\end{align}
discarding the now-extraneous second azithumal mode number $m'$ (i.e., defining $F^{[1],lm}_{\alpha\beta} \equiv F^{[1],mlm}_{\alpha\beta}$).

Now consider if we allowed additional $1/r$ terms of the same form in each $h^{[1,j]}_{\alpha\beta}$, such that the total coefficient of $1/r$ is proportional to the sum $F^{[1],lm}_{\alpha\beta} + \sum_{j>0}(\mathring{m}_1)^j F^{[1,j],lm}_{\alpha\beta}$. This would need to match the $1/r$ term in the large-$r$ expansion of the multiscale solution $h^{(1)}_{\alpha\beta}$, say
\begin{equation}\label{eq:h(1) large r generic}
h^{(1)}_{\alpha\beta} = \frac{1}{r}\sum_{l=0}^\infty\sum_{m=-l}^l G^{(1),lm}_{\alpha\beta}(\Omega)Y_{lm}e^{-im\phi_p} +\mathcal{O}(1/r^2),
\end{equation}
which would imply $G^{(1),lm}_{\alpha\beta}=(im\Omega)^l\bigl[F^{[1],lm}_{\alpha\beta} + \sum_{j>0}(\mathring{m}_1)^j F^{[1,j],lm}_{\alpha\beta}\bigr]$. Such an equality would not be useful because $G^{(1),lm}_{\alpha\beta}$ is obtained numerically on a grid of $\Omega$ values; there is not a sense in which we can further expand it in powers of $\mathring{m}_1$ in order to match each term $F^{[1,j],lm}_{\alpha\beta}$. More fundamentally, the corrections $F^{[1,j],lm}_{\alpha\beta}$ are ruled out by dimensional analysis. $F^{[1,j],lm}_{\alpha\beta}$ is a function of $\Omega$ only, and each term $(\mathring{m}_1)^j F^{[1,j],lm}_{\alpha\beta}$ must have the same dimension as $F^{[1],lm}_{\alpha\beta}$. Therefore the only possible form these higher-order terms can take is $(\mathring{m}_1\Omega)^j b^{[1,j],lm}_{\alpha\beta}$ for some dimensionless numbers $b^{[1,j],lm}_{\alpha\beta}$. Such an expansion is only sensible if $\mathring{m}_1\Omega\ll1$ --- the slow-velocity, PN limit. Hence, we must exclude such higher-order terms to keep our GSF-MPM expansion relativistic.

To construct the subleading terms without introducing these problematic new $1/r$ outgoing waves, we adopt the solution \beq\label{h1j-gen}
h^{[1,j]}_{\alpha\beta} = \FP \Box^{-1}_{\eta}\left(r^B_* S^{[1,j]}_{\alpha\beta}\right)+k^{[1,j]}_{\alpha\beta}.
\eeq
Here $\Box^{-1}_{\eta}f$ denotes the integral $\int G(t-t',\vec x,\vec{x}') f(t',\vec{x}')dt' d^3x'$ against the flat-spacetime retarded Green's function,
\begin{equation}\label{eq:Gret time}
G(t-t',\vec x,\vec{x}') = -\frac{1}{4\pi}\frac{\delta(t-t'-|\vec{x}-\vec{x}'|)}{|\vec{x}-\vec{x}'|}.
\end{equation}
We have introduced a Hadamard finite part operation $\FP$ because $S^{[1,j]}_{\alpha\beta}$ diverges at $r_*=0$, as is clear from the form~\eqref{h10} (see \cite{Blanchet:1985sp} for the origin of this finite part regularization procedure in the PN-MPM context). However, note that $\FP \Box^{-1}_{\eta}\left(r^B_* S^{[1,j]}_{\alpha\beta}\right)$ is a particular solution to $\Box_\eta h^{[1,j]}_{\alpha\beta} =S^{[i,j]}_{\alpha\beta}$, with causal propagation, for all $r_*>0$. The second term in equation~\eqref{h1j-gen}, $k^{[1,j]}_{\alpha\beta}$, is a homogeneous solution of the form~\eqref{h10}, such that equation~\eqref{h1j-gen} represents the most general causal solution to $\Box_\eta h^{[1,j]}_{\alpha\beta} =S^{[i,j]}_{\alpha\beta}$. We choose $k^{[1,j]}_{\alpha\beta}$ through our requirement that $h^{[1,j]}_{\alpha\beta}$ contains no terms of the form $G_{\alpha\beta}(u,n^i)/r_*$. The integral $\FP \Box^{-1}_{\eta}(r^B_* S^{[1,j]}_{\alpha\beta})$ generically involves outgoing waves of that form, which we precisely cancel  with $k^{[1,j]}_{\alpha\beta}$. Note that this involves no loss of generality: the solution $\sum_{j\geq0}(\mathring{m}_1)^j h^{[1,j]}_{\alpha\beta}$ is the most general solution to equation~\eqref{SC EFE1} containing no incoming radiation; all the freedom in that solution is contained in the multipole moments $F^{[1]}_{\alpha\beta L}(u)$ in $h^{[1,0]}_{\alpha\beta}$. 

We follow the same general procedure in solving the higher-order field equations in the sequence~\eqref{EFEnj}. The most general solution to equation~\eqref{EFE20} with no incoming radiation can be written as
\beq\label{hnj-gen}
h^{[n,j]}_{\alpha\beta} = \FP \Box^{-1}_{\eta}\left(r^B_* S^{[n,j]}_{\alpha\beta}\right)+k^{[n,j]}_{\alpha\beta}.
\eeq
Here $k^{[n,j]}_{\alpha\beta}$ is a homogeneous solution of the form~\eqref{h10}. For each $n$, the leading homogeneous solution, $k^{[n,0]}_{\alpha\beta}$, is left arbitrary, to be determined through matching to the near-zone solution. The subleading homogeneous solutions, $k^{[n,j]}_{\alpha\beta}$ with $j>0$, are chosen to cancel any outgoing radiation in $\FP \Box^{-1}_{\eta}\left(r^B_* S^{[n,j]}_{\alpha\beta}\right)$. This contrasts with the traditional MPM scheme, where $k^{[n,j]}_{\alpha\beta}$ would be chosen to enforce the gauge condition on each $h^{[n,j]}_{\alpha\beta}$; we avoid that choice in order to isolate the outgoing wave entirely in the leading PM term $h^{[n,0]}_{\alpha\beta}$ at each order in $\epsilon$, recalling that the gauge condition will be enforced through matching to the near-zone solution.

As illustrated in section~\ref{sec:breakdown}, the breakdown of the multiscale expansion at second order in the near zone is specifically associated with the presence of $1/r^2$ source terms.\footnote{A practical consequence of our choice of tortoise coordinates $x^i$ is that no source terms $\sim 1/r_*^2$ arise in the field equations for $h^{[1,j]}_{\alpha\beta}$.} This breakdown is cured, and physical boundary conditions are found for our near-zone multiscale field $h^{(2)}_{\alpha\beta}$, by examining the part of $h^{[2,0]}_{\alpha\beta}$ sourced by the most slowly falling piece of $S^{[2,0]}_{\alpha\beta}$. Writing 
\begin{equation}
S^{[2,0]}_{\alpha\beta}=\frac{s_{\alpha\beta}(u,n^i)}{r^2_*}+\O\!\left( \frac{1}{r_*^3}\right), 
\end{equation}
we define
\beq\label{jmunu-soln}
j_{\alpha\beta} = \FP\Box^{-1}_\eta\left(r^{B-2}_*s_{\alpha\beta}\right).
\eeq
The retarded integral converges even without the finite part operation, since $1/r^2_*$ is integrable at $r_*=0$, meaning we can equivalently write
\beq\label{jmunu-soln2}
j_{\alpha\beta} = \Box^{-1}_\eta\left(r_*^{-2}s_{\alpha\beta}\right).
\eeq
We then define
\begin{equation}
\Delta j_{\alpha\beta}\equiv \FP \Box^{-1}_{\eta}\left(r^B_* S^{[2,0]}_{\alpha\beta}-\frac{s_{\alpha\beta}(u,n^i)}{r^2_*}\right) = \FP \Box^{-1}_{\eta}\left(r^B_* S^{[2,0]}_{\alpha\beta}\right) - j_{\alpha\beta}.
\end{equation}
This can be decomposed into an outgoing-wave homogeneous solution plus a particular solution to $\Box_\eta \Delta j_{\alpha\beta} = (S^{[2,0]}_{\alpha\beta}-r_*^{-2}s_{\alpha\beta})$ that decays more rapidly than $1/r$ toward infinity. The particular solution is constructed in the same manner as described previously: calculate the (finite part) retarded integral, read off the large-$r$ coefficient of $1/r$ in the solution, and then subtract an outgoing-wave homogeneous solution with that same coefficient of $1/r$. The homogeneous solution, of the form~\eqref{h10}, we denote $\Delta k^{[2,0]}_{\alpha\beta}$. This allows us to write $h^{[2,0]}_{\alpha\beta}$ in the alternative form
\beq\label{h=j+}
h^{[2,0]}_{\alpha\beta} = j_{\alpha\beta} + \tilde k^{[2,0]}_{\alpha\beta} + \left(\Delta j_{\alpha\beta}-\Delta k^{[2,0]}_{\alpha\beta}\right).
\eeq
The second term, $\tilde k^{[2,0]}_{\alpha\beta}\equiv k^{[2,0]}_{\alpha\beta}+\Delta k^{[2,0]}_{\alpha\beta}$, is a homogeneous, outgoing-wave solution, while the combination of terms in parentheses decays more rapidly than $1/r$ toward future null infinity, meaning it will not contribute to the emitted GWs.

In the next sections we show the following: 
\begin{enumerate}
\item $j_{\alpha\beta}$ is sourced by the emitted GW flux (plus oscillatory source terms).
\item When re-expanded in the near far zone $M\ll r\ll M/\e$, $j_{\alpha\beta}$ provides a boundary condition for $h^{(2)}_{\alpha\beta}$ that introduces nonlocal-in-time terms into the orbital dynamics. We define a certain piece of this re-expansion of $j_{\alpha\beta}$ as a puncture field $h^{(2)\cal P}_{\alpha\beta}$ (so called because it is singular at the large-$r$ limit of the near zone),  and we adopt the residual field $h^{(2)\cal R}_{\alpha\beta}=h^{(2)}_{\alpha\beta}-h^{(2)\cal P}_{\alpha\beta}$ as the numerical variable in the near zone~\cite{Miller:2023ers}.
\item When re-expanded in the very far zone $r\gg M/\epsilon$, the quasistationary piece of $j_{\alpha\beta}$ becomes the first-order GW memory $h^{(1)}_{l0}$ in equation~\eqref{eq:memory terms}.
\item The numerically calculated $h^{(2)\cal R}_{\alpha\beta}$ determines the nonhereditary oscillatory part of the second-order waveform $h^{(2)}_{lm}$ in equation~\eqref{eq:multiscale waveform}.
\item Oscillatory modes of $h^{[3]}_{\alpha\beta}$ can be sourced by products of first-order memory modes and first-order oscillatory modes, promoting them by $1/\e$. These oscillatory terms might contribute to the oscillatory modes of $h^{(2)}_{lm}$ in equation~\eqref{eq:multiscale waveform}, and consequently they could contribute to the second-order GW memory $h^{(2)}_{l0}$.
\end{enumerate}

\subsection{Asymptotic source and GW flux}\label{sec:asymptotic_source_and_GW_flux}

The source for $j_{\alpha\beta}$ is the $1/r^2_*$ term in $2\delta^2 R_{\alpha\beta}$, which is computable entirely from the $1/r_*$ term in $h^{[1,0]}_{\alpha\beta}$. We write that term compactly as 
\begin{equation}\label{h10 large r}
h^{[1,0]}_{\alpha\beta}=\frac{f^{[1]}_{\alpha\beta}(u,n^i)}{r_*}+\O\!\left(\frac{1}{r^2_*}\right). 
\end{equation}
In this section we develop a useful decomposition of $f^{[1]}_{\alpha\beta}$ into a mass piece plus oscillatory pieces. We then derive an expression for the coefficient of $1/r^2_*$ in $2\delta^2 R_{\alpha\beta}$ in terms of those pieces, which will be the starting point for our subsequent derivations.

Equation~\eqref{h10 large r} must match the $1/r$ term in the multiscale solution $h^{(1)}_{\alpha\beta}$.  We see from equations~\eqref{eq:h10 multiscale} and \eqref{eq:h(1) large r generic} that this term can be divided into an oscillatory piece  ($m\neq0$) and a quasistationary monopole ($l=0$). More explicitly, we can write
\beq\label{h(1) large r}
h^{(1)}_{\alpha\beta}=\frac{2M_{\rm B}^{(1)}(u)\delta_{\alpha\beta} + z^{(1)}_{\alpha\beta}(u,n^i)}{r_*} + \O\!\left(\frac{\ln r}{r_*^{2}}\right)
\eeq
in $(t,x^i)$ coordinates. Here $z^{(1)}_{\alpha\beta}$ is an oscillatory function made up of $m\neq0$ modes, and $M^{(1)}_{\rm B}$ represents a contribution to the Bondi mass coming from the particle's leading-order orbital energy $m_2{\cal E}_{(0)}$ as well as the evolving correction $ \delta m_1$ to the primary black hole's mass~\cite{Miller:2020bft}: 
\begin{equation}
\e M^{(1)}_{\rm B} = \e \delta m_1 + m_2 {\cal E}_{(0)}.
\end{equation}
The Kronecker-delta form of the mass term in  equation~\eqref{h(1) large r} follows from the Lorenz gauge condition.  

Given equations~\eqref{h10 large r} and~\eqref{h(1) large r}, matching between the near- and far-zone solutions requires
\beq\label{c-form}
f^{[1]}_{\alpha\beta} = z^{(1)}_{\alpha\beta}(u,n^a) + 2M^{(1)}_{\rm B}(u)\,\delta_{\alpha\beta} + \O(\e).
\eeq
The order-$\epsilon$ differences arise because the matching condition involves a re-expansion of $h^{[1,0]}_{\alpha\beta}$. In the remainder of our calculations, we will work with the leading-order approximation, discarding the $\O(\e)$ remainder.

To evaluate the source for $j_{\alpha\beta}$, we further specify $z^{(1)}_{\alpha\beta}$ using the Lorenz gauge condition~\eqref{eq:Lorenz gauge}. The gauge condition simplifies due to
\begin{equation}
\mathring{\nabla}_\gamma \frac{f^{[1]}_{\alpha\beta}(u,n^i)}{r_*} = -\frac{\dot f^{[1]}_{\alpha\beta}k_\gamma}{r_*} + \O\!\left(\frac{1}{r^2_*}\right),
\end{equation}
where a dot indicates differentiation with respect to $u$, 
\begin{equation}
k_\alpha \equiv -\partial_\alpha u = (-1,n_i) 
\end{equation}
is the outgoing principal null vector, and there is no contribution from a derivative acting on the $n^i$ dependence in $f^{[1]}_{\alpha\beta}$ because $\partial_i n^j\propto 1/r_*$. It immediately follows that the gauge condition imposes 
\begin{equation}
\dot z^{(1)}_{\alpha\beta}k^\beta=\frac{1}{2}\dot z^{(1)} k_\alpha, 
\end{equation}
with $z^{(1)}\equiv\eta^{\alpha\beta} z^{(1)}_{\alpha\beta}$. Integrating this with the initial condition $z^{(1)}_{\alpha\beta}(-\infty)=0$ gives us
\beq\label{gauge-z}
z^{(1)}_{\alpha\beta}k^\beta = \frac{1}{2}z^{(1)} k_\alpha.
\eeq
In retarded polar coordinates $(u,r_*,\theta^A)$, we have $k^\alpha=(0,1,0,0)$ and $k_\alpha=(-1,0,0,0)$. Equation~\eqref{gauge-z} then implies
\begin{equation}\label{gauge z polar}
z^{(1)}_{r_* r_*} = z^{(1)}_{r_*A} = \Omega^{AB}z^{(1)}_{AB} = 0. 
\end{equation}
The last equality shows that $z^{(1)}_{AB}$ is equal (up to a factor of $r_*^2$) to the Bondi shear contained in $h^{[1]}_{\alpha\beta}$: 
\beq
z^{(1)}_{AB} = z^{(1)}_{\langle AB\rangle} = r_*^2 C^{[1]}_{AB} +\O(\e),
\eeq
where we recall that the angular brackets denote the STF combination of indices with respect to the unit-sphere metric $\Omega_{AB}$. Note that $C^{[1]}_{AB}$ accounts for the full oscillatory part of the shear at linear order in the mass ratio, but it does not include the first-order memory contribution, which will emerge from a $1/\e$ promotion of $j_{\alpha\beta}$. We reserve the symbol $C^{(1)}_{AB}$ for the complete linear-in-$\e$ contribution to the waveform~\eqref{eq:multiscale waveform}.
Now substituting equation~\eqref{c-form} into the explicit expression for $2\delta^2R_{\alpha\beta}$ ---  equation~(6) of~\cite{PoundAndWardell} --- using equation~\eqref{gauge-z}, and simplifying, we find 
\beq\label{s-form}
s_{\alpha\beta} = -\Pi\, k_\alpha k_\beta +\frac{\partial}{\partial u}\bigl(\bar z^{\mu\nu}_{(1)}\dot z^{(1)}_{\mu\nu}\bigr)k_\alpha k_\beta + 4 M^{(1)}_{\rm B} \ddot z^{(1)}_{\mu\nu}q^\mu_\alpha q^\nu_\beta,
\eeq
which should be compared to  (2.10) of \cite{Blanchet:1992br} in the PN-MPM context.
Here, \mbox{$q^\mu_\alpha\equiv \delta^\mu_\alpha+n^\mu k_\alpha$} projects orthogonally to the ingoing principal null vector $n^\mu$, whose Cartesian components are $n^\mu=\frac{1}{2}(1,-n^i)$, and we have defined the quantity
\beq
\Pi \equiv\frac{1}{2}\dot z^{\alpha\beta}_{(1)}\dot z^{(1)}_{\alpha\beta} - \frac{1}{4}\dot z^{(1)}\dot z^{(1)}.
\eeq
Using the coordinate form~\eqref{gauge z polar} of the gauge condition, it is straightforward to relate~$\Pi$ to the angular distribution of GW flux~\eqref{modeflux - news} carried by $h^{(1)}_{\alpha\beta}$:
\beq\label{Pi-Edot}
\Pi = \frac{1}{2} \dot C^{AB}_{[1]}\dot C_{AB}^{[1]} = 16\pi\left.\frac{d^2E^{\rm GW}}{du d\Omega_2}\right|_{h^{[1]}}.
\eeq
Equation~\eqref{gauge z polar} also allows us to write the second and third terms in equation~\eqref{s-form} in terms of the Bondi shear and Bondi news, leading to another useful form of equation~\eqref{s-form},
\beq\label{s-form Bondi}
s_{\alpha\beta} = \left[-\frac{1}{2}\dot C^{AB}_{[1]}\dot C_{AB}^{[1]} + \frac{\partial}{\partial u}\left(C^{AB}_{[1]}\dot C^{[1]}_{AB}\right)\right]k_\alpha k_\beta +4 r_*^2M^{[1]}_{\rm B}\ddot C^{[1]}_{AB} \Omega^A_\alpha \Omega^B_\beta,
\eeq
where $\Omega^A_\alpha\equiv r_*^{-2}\Omega^{AB}\eta_{\alpha\beta}({\partial x^\beta}/{\partial\theta^B})$ reduces to $\delta^A_\alpha$ in retarded polar coordinates $(u,r_*,\theta^A)$ and to ${r_*^{-1}}\Omega^{AB}\delta_{\alpha i}({\partial n^i}/{\partial\theta^B})$ in Cartesian coordinates $(t,x^i)$. 
Note that the second and third terms in equation~\eqref{s-form} vanish if averaged over $\phi_p$ (up to terms that are higher order in $\epsilon$, which arise from the $u$ derivative in the second term acting on the $\Omega$ dependence of $z_{\alpha\beta}$). Equations~\eqref{s-form} and~\eqref{s-form Bondi} hence split $s_{\alpha\beta}$ into a flux term plus a purely oscillatory one,
\begin{equation}\label{d2R[h1,h1]}
\delta^2R_{\alpha\beta}[h^{[1]},h^{[1]}] = \frac{8\pi k_\alpha k_\beta}{r^2}\left.\frac{d^2E^{\rm GW}}{du d\Omega_2}\right|_{h^{[1]}} +\text{oscillatory terms} + \O\!\left(\frac{1}{r^3}\right).
\end{equation} 
Moreover, this same equality holds for $\delta^2G_{\alpha\beta}$ since all three terms in equation~\eqref{s-form} have vanishing trace. The fact that $\delta^2G_{\alpha\beta}[h^{[1]},h^{[1]}]=\frac{8\pi k_\alpha k_\beta}{r^2}\left.\frac{d^2E^{\rm GW}}{du\,d\Omega_2}\right|_{h^{[1]}}$ on average is the familiar statement that the averaged $\delta^2G_{\alpha\beta}/(8\pi)$ can be thought of as an effective stress-energy tensor of the emitted GWs~\cite{Isaacson:1968zza}. The oscillatory terms, in particular the final term in equation~\eqref{s-form Bondi}, give rise to GW tails~\cite{Blanchet:1992br}.

\subsection{Puncture fields and memory in the near zone $r\ll M/\epsilon$}\label{sec:near-zone memory}

To obtain physical boundary conditions for the near-zone solution, we must re-expand our far-zone solution $\sum_{n,j}\epsilon^n (\mathring{m}_1)^j h^{[n,j]}_{\alpha\beta}$ for $r\ll M/\epsilon$ --- or equivalently, for $\epsilon\ll M/r$.

Here we focus on the leading second-order term, $h^{[2,0]}_{\alpha\beta}$, which suffices to resolve the infrared divergence in the near zone. More specifically, we focus on 
the dominant piece of  $h^{[2,0]}_{\alpha\beta}$, namely $j_{\alpha\beta}=\Box^{-1}_\eta\left[r_*^{-2}s_{\alpha\beta}(u,n^i)\right]$, which contributes the dominant large-$r$ behavior in the near zone and resolves the infrared divergence. 

We first note that the function $s_{\alpha\beta}(u,n^i)$, as given in equation~\eqref{s-form}, is constructed from the coefficient of $1/r_*$ in $h^{[1,0]}_{\alpha\beta}$, meaning it can be calculated from the multipole moments $F^{[1]}_{\alpha\beta L}(u)$ in equation~\eqref{h10}. As explained below that equation, the multipole moments inherit a multiscale form through matching to the coefficient of $1/r$ in the near-zone solution $h^{(1)}_{\alpha\beta}$. Hence, we can write
\begin{equation}
s_{\alpha\beta}(u,n^i) = \sum_{l\geq0}s_{\alpha\beta L}(u)\hat n^L = \sum_{l\geq0}\sum_{m=-l}^l s^m_{\alpha\beta L}(\Omega(u,\epsilon)) e^{-im\phi_p(u,\epsilon)}\hat n^L 
\end{equation}
up to $\O(\epsilon)$ corrections. $j_{\alpha\beta}$ can then be written as
\beq\label{eq:j multiscale}
j_{\alpha\beta} = \Box^{-1}_\eta\left(r_*^{-2}s_{\alpha\beta}\right) =\sum_{l\geq0}\sum_{m=-l}^l \Box^{-1}_\eta\left[r_*^{-2}s^m_{\alpha\beta L}(\Omega(u,\epsilon))e^{-im\phi_p(u,\epsilon)}\hat n^L\right],
\eeq
again up to $\O(\epsilon)$ corrections. 
Reference~\cite{Pound:2015wva} derived near-zone expansions of such integrals, exploiting methods from reference~\cite{Blanchet:1987wq}. Here we will merely quote the results and discuss the implications.

Since $\nhat^L$ diagonalizes $\Box^{-1}_\eta$, $j_{\alpha\beta}$ naturally decomposes into the $\nhat^L$ basis,
\beq\label{jmunu}
j_{\alpha\beta} = \sum_{l\geq0}j_{\alpha\beta L}(u,r_*,\epsilon)\nhat^{L},
\eeq
with $j_{\alpha\beta L}\nhat^{L} =  \Box^{-1}_\eta\left[r_*^{-2}s_{\alpha\beta L} \hat n^L\right]$. Moreover, as shown in reference~\cite{Pound:2015wva}, $\Box^{-1}_\eta$ also preserves the expansion in  Fourier modes $e^{-im\phi_p}$, such that
\begin{equation}
j_{\alpha\beta} = \sum_{m=-\infty}^\infty  j^m_{\alpha\beta}(\Omega(u,\e),r_*,n^i)e^{-im\phi_p(u,\epsilon)} = \sum_{l\geq0} \sum_{m=-l}^l j^m_{\alpha\beta L}(\Omega(u,\e),r_*) \nhat^L e^{-im\phi_p(u,\epsilon)}.
\end{equation}
with $j^m_{\alpha\beta L}\nhat^L e^{-im\phi_p}=\Box^{-1}_\eta[s^m_{\alpha\beta L}\nhat^L e^{-im\phi_p}]+\O(\e)$.

For oscillatory, $m\neq0$ modes, reference~\cite{Pound:2015wva} found that the harmonic coefficients in the near-zone expansion of equation~\eqref{eq:j multiscale} are given by
\begin{equation}\label{eq:j soln m!=0}
j^{m}_{\alpha\beta L} = \frac{\ln r }{2 i m\Omega r}s^{m}_{\alpha\beta L}(\Omega(u)) + \O\!\left(\frac{1}{r},\epsilon\right),
\end{equation}
where we emphasise the $\O(1/r)$ terms contain no $\ln r$. The choice of length scale in the logarithm is discussed in~\ref{sec: vfz in Bondi-Sachs}. For simplicity we suppress the $\epsilon$ dependence of $\Omega$ here and below. Equation~\eqref{eq:j soln m!=0} agrees with the solution~\eqref{eq:h2m asymptotic solution} found directly from the near-zone retarded integral, after converting~\eqref{eq:h2m asymptotic solution} to $\uptau$-slicing by removing the exponential. In other words, matching to a far-zone solution is not strictly needed for $m\neq0$ modes at first and second order in $\epsilon$; the correct solution is obtained in the near zone without any additional information from the far zone. However, because of its irregular, $\ln(r)/r$ behavior, treating $j^{m}_{\alpha\beta}$ as a puncture is still most practical, subtracting it from the numerical variable and solving for the residual.

For quasistationary, $m=0$ modes, reference~\cite{Pound:2015wva} derived the following near-zone expansion of equation~\eqref{eq:j multiscale}: 
\begin{align}\label{eq:j soln m=0}
\< {j}_{\alpha\beta L}\>  = \begin{dcases*} \<s_{\alpha\beta L}(\Omega)\>[\ln (2 r) - 1] - \int^u_{-\infty}\!\! dz\,\,  \frac{d\<s_{\alpha\beta L}(\Omega(z))\>}{dz}\ln(u-z) + \O(\epsilon) & if $ l=0$,   \\
  -\displaystyle\frac{\<s_{\alpha\beta L}(\Omega)\>}{ l( l+1)} + \O(\epsilon) & if $ l>0$,
\end{dcases*}
\end{align}
where we adopt the $\<\>$ notation for $m=0$ modes, as in equation~\eqref{av + osc}. This result can be compared to equation~(5.22) of~\cite{Blanchet:1987wq} in the PN-MPM case. Note that in this instance any choice of length scale in the logarithms cancels between the terms inside and outside the integral.  

Unlike in the oscillatory case, in equation~\eqref{eq:j soln m=0} there are no omitted terms of order $\epsilon^0/r$; omitted terms all come with powers of $\epsilon$ and higher, rather than lower, powers of~$r$. However, the next order in the GSF-MPM expansion, $h^{[2,1]}_{\alpha\beta}$, involves terms of order $\ln(r_*)/r_*$ sourced by $E^{[1]}_{\alpha\beta}[j]$. These terms must be included in $h^{(2)\cal P}_{\alpha\beta}$ in order to ensure the $1/r$ terms in $h^{(2)\cal R}_{\alpha\beta}$ represent a regular, outgoing homogeneous solution. However, we leave the description of $h^{[2,1]}_{\alpha\beta}$ to future work as it does not directly enter into the calculation of GW memory at the orders in $\e$ we consider.

The most important aspect of $\<{j}_{\alpha\beta L}\>$ is the presence of a hereditary integral over all past times. Such nonlocal-in-time behavior could not be predicted within the multiscale expansion. This term will enter the multiscale solution $\<h^{(2)}_{\alpha\beta}\>=h^{(2),0}_{\alpha\beta}$ and influence the conservative sector of the 2SF dynamics, in analogy with memory effects that enter the 5PN dynamics~\cite{Porto:2024cwd,Almeida:2024klb}. 

At this order in $\epsilon$ the memory contribution is confined to $l=0$ due to a suppression by $\epsilon^l$ that arises when re-expanding $j_{\alpha\beta}$ in the near zone~\cite{Pound:2015wva}.  However, the reader should recall that $l$ here refers to an expansion of Cartesian components in scalar spherical harmonics. One can straightforwardly check, using the tensorial structure $\<{j}_{\alpha\beta L}\> \propto k_\alpha k_\beta$, that the hereditary integral contributes to both $l=0$ and $l=2$ ($m$=0) terms in an expansion in spin-weighted or tensor harmonics. To see this, we recall that the $m=0$ piece of $s_{\alpha\beta}$ is given by $-\<\Pi\> k_\alpha k_\beta$, where $\Pi$ is related to the GW flux by equation~\eqref{Pi-Edot}. Expanding $\Pi$ as 
\beq\label{Pi expansion}
\Pi=\sum_{l\geq0}\Pi_L \nhat^L 
\eeq
and using equations~(2.7) and (2.13) of reference~\cite{Damour:1990gj}, we find
\begin{subequations}\label{s0 source STF}%
\begin{align}
\<s_{tt}\> &= -\sum_{l\geq 0}\<\Pi_L\> \nhat^L ,\\
\<s_{ta}\> &= \sum_{l\geq 0}\left(\nhat_{aL}\<\Pi_L\> + \frac{l}{2l+1}\nhat_{L-1}\<\Pi_{aL-1}  \> \right) ,\\
\<s_{ab}\> &= -\sum_{l\geq 0}\biggl(\nhat_{abL}\<\Pi_L\> +\frac{2l}{2l+3}\nhat_{L-1(a}\<\Pi_{b)L-1}\>+ \frac{1}{2l+3}\delta_{ab}\nhat_L \<\Pi_L\> \nonumber\\*
&\qquad\qquad + \frac{l(l-1)}{(2l+1)(2l-1)}\nhat_{L-2}\<\Pi_{abL-2}\> \biggr). 
\end{align}
\end{subequations}
Equation~\eqref{eq:j soln m=0} then implies that the scalar $l=0$ terms contribute the following to the near-zone re-expansion of $j_{\alpha\beta}$:
\begin{subequations}%
\begin{align}\label{eq:j soln m=0 explicit}
\<{j}^{l_{\rm sc.}=0}_{tt}\> &= -\kappa[\Pi_0],\\
\<{j}^{l_{\rm sc.}=0}_{ti}\> &= \frac{1}{3}\kappa[\<\Pi_i\>],\\
\<{j}^{l_{\rm sc.}=0}_{ij}\> &= -\frac{1}{3}\delta_{ij}\kappa[\Pi_0] -\frac{2}{15}\kappa[\<\Pi_{ij}\>] ,
\end{align}
\end{subequations}
where ``$l_{\rm sc.}=0$'' is a reminder that this corresponds to the $l=0$ term in a scalar-harmonic decomposition of Cartesian components. We have introduced the shorthand
\beq
\kappa[f] \equiv f(\Omega(u))[\ln (2 r) - 1] - \int^u_{-\infty}\!\! dz\,\,  \frac{df(\Omega(z))}{dz}\ln(u-z),
\eeq
and used $\Pi_0$ to denote the $l=0$ term in the expansion~\eqref{Pi expansion}, observing that $\Pi_0=\<\Pi_0\>$. 
The Cartesian quantity $\<\Pi_i\>$ in equation~\eqref{eq:j soln m=0 explicit} corresponds to an $l=1$, $m=0$ even-parity vector harmonic mode in polar coordinates; see equations~(A4b) and (A7) in reference~\cite{Poisson:2009qj}, for example. But note that $l=1$, $m=0$ even-parity vector harmonic modes vanish for up-down symmetric systems~\cite{Miller:2020bft}, implying $\<\Pi_i\> = 0$ for the binaries we consider. On the other hand, the quantity $\<\Pi_{ij}\>$ corresponds to an $l=2$, $m=0$ even-parity vector harmonic mode, which is nonvanishing. This quadrupole mode in the near-zone metric is the counterpart of the GW memory at future null infinity. While the GW memory represents re-emission of soft gravitons toward future null infinity, the memory term in the near zone represents the re-emission back into the binary system.

The physically correct near-zone solution must match the behavior of the re-expanded far-zone solution. To enforce this boundary condition, we adopt a puncture scheme, as outlined in reference~\cite{Miller:2023ers}. We define the puncture field $h^{(2){\cal P},m}_{\alpha\beta}$ as the right-hand side of equation~\eqref{eq:j soln m!=0} for $m\neq0$ and \eqref{eq:j soln m=0} for $m=0$. We then define the residual field
\begin{equation}
h^{(2){\cal R},m}_{\alpha\beta} = h^{(2),m}_{\alpha\beta} - h^{(2){\cal P},m}_{\alpha\beta},
\end{equation}
which we take as our numerical variable. 
Defining the right-hand side of equation~\eqref{eq:h2_field_equation} as $S^{(2),m}_{\alpha\beta}$, we move the puncture to the right-hand side to obtain an equation for the residual field
\begin{align}\label{BoxhRes}
\delta G^{(0),m}_{\alpha\beta}[h^{(2){\cal R},m}] = S^{(2),m}_{\alpha\beta} - \delta G^{(0),m}_{\alpha\beta}[h^{(2){\cal P},m}]\equiv S^{(2),m \; {\rm eff}}_{\alpha\beta}.
\end{align}
The effective source $S^{(2),m\; \rm eff}_{\alpha\beta}$ decays sufficiently rapidly to ensure that $h^{(2){\cal R},m}_{\alpha\beta}$ satisfies regular, outgoing boundary conditions as $r\to\infty$. Solving equation~\eqref{BoxhRes} subject to these boundary conditions then fully determines the physical near-zone solution $h^{(2)}_{\alpha\beta}=h^{(2)\cal R}_{\alpha\beta}+h^{(2)\cal P}_{\alpha\beta}$. Reference~\cite{Miller:2023ers} derives necessary and sufficient conditions on when punctures must be introduced in order to obtain the physical solution in the near zone.

By adding a memory integral to the large-$r$ boundary conditions of the near zone, $h^{(2){\cal P},0}_{\alpha\beta}$ will insert a term proportional to that integral in the metric perturbation at the particle's location. Since this is an $m=0$ mode, it will contribute to the conservative 2SF and therefore to the 2PA phase evolution~\cite{Pound:2015wva, Miller:2020bft, PoundAndWardell}.

We will show in the next section that the logarithmic term in equation~\eqref{eq:j soln m!=0} does not contribute to the waveform. By construction, $h^{(2){\cal R},m}_{\alpha\beta}$ then contains the only outgoing wave content in the near-zone field $h^{(2)}_{\alpha\beta}$. This outgoing wave (i.e., the coefficient of $1/r$ when $r\to\infty$) uniquely determines the outgoing wave content in $h^{[2,0]}_{\alpha\beta}$, which comprises the homogeneous solution $\tilde k^{[2,0]}_{\alpha\beta}$ together with any outgoing wave content in $j_{\alpha\beta}$; recall equation~\eqref{h=j+}. Since $\tilde k^{[2,0]}_{\alpha\beta}$ contains all the unknowns in $h^{[2]}_{\alpha\beta}$, it follows that $h^{(2){\cal R},m}_{\alpha\beta}$ fully fixes the freedom in the far-zone solution $h^{[2]}_{\alpha\beta}$. In the next section, we show that the $1/r$ term in $h^{(2){\cal R},m}_{\alpha\beta}$ represents `most' of the second-order contribution to the oscillatory part of the waveform at future null infinity, but it potentially omits terms arising from interactions between memory and emitted waves.

\subsection{Waveforms and memory in the very far zone $r\gg M/\e$}\label{very-far-zone}

The expressions~\eqref{eq:j soln m!=0} and \eqref{eq:j soln m=0} represent the large-$r$ behavior of the near-zone solution. However, this is the behavior found by taking the limit $\e\to0$ at fixed $r$ and then taking the limit $r\to\infty$. This differs significantly from the metric's genuine large-$r$ behavior obtained by taking the limit $r\to\infty$ at fixed $\e$; the latter limit, which represents the behavior in the very far zone $r\gg M/\e$, is the relevant one for computing asymptotic quantities such as the waveform. Here we show that in this limit, the field $\<j_{\alpha\beta}\>$ --- which is associated with hereditary effects in the near zone --- contributes the first-order GW memory $h^{(1)}_{l0}$ in the far zone. We also show that `most' of the oscillatory part of the waveform~\eqref{eq:multiscale waveform}, $h^{(2)}_{lm}$, can be extracted directly from the near-zone residual field $h^{(2){\cal R},m}_{\alpha\beta}$, but there is potentially an additional contribution from the interaction between the memory $\<j_{\alpha\beta}\>$ and the oscillatory wave modes $h^{(1)}_{lm}$. We conclude by discussing the second-order memory $h^{(2)}_{l0}$ and its relationship to the third-order PM field $h^{[3,0]}_{\alpha\beta}$.

The expansion in the very far zone can be obtained using an alternative formula for the retarded integral~\cite{Blanchet:1992br},
\beq
\Box^{-1}_\eta\left[\frac{\nhat_L}{r^2_*}S(t-r_*)\right] = -\nhat_L \int_1^\infty\!\!\! dx\, Q_l (x)S(t-r_*x),
\eeq
where $Q_l$ is a Legendre function of the second kind, with a branch cut on $(-\infty, 1]$. From this we have \beq
{j}_{\alpha\beta L} =  - \int_1^\infty\!\!\!dx\, Q_{l} (x)s_{\alpha\beta L}(t-r_*x).
\eeq
The large-$r$ expansion is then obtained using the Legendre function's asymptotic behavior as $x \rightarrow 0^+$~\cite{Blanchet:1992br} 
\begin{equation}
Q_l(1+x)=-\frac{1}{2}\ln\left(\frac{x}{2}\right)-H_l+\O(x^{-1}\ln x) \,,
\end{equation}
where $H_l=\sum_{n=1}^{ l}\frac{1}{n}$ is the harmonic number. After a change of integration variable to $z=(x-1)r_*$, the integral becomes
\beq\label{very-large-r}
{j}_{\alpha\beta L} =  \frac{1}{r_*}\int_0^\infty\!\!\!d z\,\left[\frac{1}{2}\ln\left(\frac{z}{2r_*}\right)+H_l\right] s_{\alpha\beta L}(u-z) + \O(r^{-2}_*\ln r_*),
\eeq
decaying as $\ln(r_*)/r_*$.

First consider this large-$r$ expansion for oscillatory modes. Writing \mbox{$s_{\alpha\beta}=\sum_{l,m}s^m_{\alpha\beta L}(\Omega(u)) e^{-im\phi_p(u)}\nhat_L$}, we can immediately evaluate the integral~\eqref{very-large-r} using the expansion~\eqref{eq:integration by parts}, which yields 
\beq\label{very-large-r m!=0}
{j}^m_{\alpha\beta} = \frac{\ln r_*}{2im\Omega r_*}\sum_{l\geq |m|} s^m_{\alpha\beta L}(\Omega)\nhat^L + \O(1/r_*,\e) ,\quad m\neq0.
\eeq
This is identical to the result~\eqref{eq:j soln m!=0} in the near-zone re-expansion of $j_{\alpha\beta}$, which in turn was identical to the result~\eqref{eq:h2m asymptotic solution} that was found directly in the near zone. In other words, the oscillatory part of the solution propagates through the entire far zone $r\gg M$ without any change in the leading behavior.
\setcounter{footnote}{0}
\footnote{In fact, for oscillatory modes, one can derive a uniform approximation to $\Box^{-1}_\eta[r_*^{-2}s_{\alpha\beta}]$ valid for all $r\gg M$:
\beq\label{uniform m!=0}
{j}^m_{\alpha\beta} = \sum_{l\geq |m|}\frac{\ln (-2im\Omega r_*) + \gamma_{\rm E} - 2H_l}{2im\Omega r_*} s^m_{\alpha\beta L}(\Omega)\nhat^L + \O(r_*^{-2}\ln r_*,\e),\quad m\neq0,
\eeq
where $\gamma_{\rm E}$ is the Euler gamma. We omit the lengthy derivation of this formula.}

For quasistationary, $m=0$ modes, equation~\eqref{very-large-r} instead becomes 
\beq\label{very-large-r m=0}
\<{j}_{\alpha\beta L}\> =  \frac{1}{\e r_*}\int_0^{\Omega(u)}\!\!\!\frac{d\Omega}{F^{(0)}(\Omega)}\,\left[\frac{1}{2}\ln\left(\frac{u-u_{(0)}(\Omega)}{2 r_*}\right)+H_l\right] \<s_{\alpha\beta L}(\Omega)\> + \O(r^{-2}_*\ln r_*).
\eeq
Here we have changed integration variable to $\Omega(u-z)$, with $\e F^{(0)}(\Omega)$ the leading term in $\dot\Omega$, as in equation~\eqref{eq:Omegadot}. The quantity $u_{(0)}(\Omega)$ is the inverse function of $\Omega(u)$; i.e., it is the solution to $du/d\Omega = 1/[\e F^{(0)}(\Omega)]$ satisfying $u_{(0)}(\Omega(u))=u$. Unlike $m\neq0$ modes, which have the same radial behavior throughout the far zone, for  $m=0$ modes we see starkly different behavior in the different regimes: the near-zone re-expansion of the retarded integrals went as $\ln r$ at large~$r$, while the very-large-$r$ expansion goes as $\ln(r)/ r$. 
More significantly, the integral enhances the perturbation by one order in $\e$, such that the second-order quantity $\e^2\< j_{\mu\nu}\>$ becomes \textit{first} order near null infinity. And unlike in the near zone where the hereditary integrals are suppressed by $\e^{ l}$, here in the very far zone they enter at the same order in $\e$ for all $l$.


We can now consolidate our findings for the behavior of the metric 
in the very far zone. The first-order metric perturbation $h^{[1]}_{\alpha\beta}$ has the form \mbox{$h^{[1]}_{\alpha\beta}=f^{[1]}(u,n^i)/r_* +\O(r_*^{-2})$} given in equation~\eqref{h10 large r}. 
At second order, equation~\eqref{h=j+} implies \mbox{$h^{[2,0]}_{\alpha\beta} = j_{\alpha\beta} + \tilde k^{[2,0]}_{\alpha\beta} + \O(r_*^{-2}\ln r_*)$,} where $\tilde k^{[2,0]}_{\alpha\beta}$ has the same outgoing-wave form as $h^{[1]}_{\alpha\beta}$. However, $j_{\alpha\beta}$ also contains outgoing wave content in the omitted, non-logarithmic $\O(1/r_*)$ terms in equations~\eqref{eq:j soln m!=0} and~\eqref{very-large-r m!=0}. Since we do not include those terms in our puncture field $h^{(2)\cal P}_{\alpha\beta}$, we define the `puncture' part of $j_{\alpha\beta}$ as the sum of (i) the $m=0$ piece $\<j_{\alpha\beta}\>$, and (ii) the $m\neq0$ terms $\propto\ln(r_*)/r_*$ in equation~\eqref{very-large-r m!=0}. We also recall, as mentioned below equation~\eqref{Pi expansion}, that the subleading term $\langle h^{[2,1]}_{\alpha\beta}\rangle$ in the PM expansion includes $\ln(r_*)/r_*$ terms when re-expanded in the near zone, and we must account for it when matching near-zone and far-zone solutions. Combining these three pieces, we define the total `puncture part' of $h^{[2]}_{\alpha\beta}$ as
\beq\label{jP def}
j^{\cal P}_{\alpha\beta} = \<j_{\alpha\beta}\> + \frac{\ln r_*}{r_*} a^{{\cal P}, \rm osc.}_{\alpha\beta}(u,n^i)  + \mathring{m}_1\<h^{[2,1]}_{\alpha\beta}\>,
\eeq
where 
\beq\label{aPosc}
a^{{\cal P},\rm osc.}_{\alpha\beta}(u,n^i) \equiv \sum_{m\neq0}\frac{1}{2im\Omega(u)}s^m_{\alpha\beta L}(\Omega(u))\nhat^L e^{-im\phi_p(u)}
\eeq
can be read off equation~\eqref{j osc soln}. The total metric $ \mathring{g}_{\alpha\beta} +\e h^{[1]}_{\alpha\beta}+\e^2 h^{[2]}_{\alpha\beta}$ then divides into a well-behaved outgoing-wave term plus the `puncture piece':
\beq\label{total far-zone metric}
g_{\alpha\beta} = \mathring{g}_{\alpha\beta}  + \frac{\e f^{[1]}_{\alpha\beta}(u,n^i)+\e^2 f^{[2]}_{\alpha\beta}(u,n^i)}{r_*}  +\e^2 j^{\cal P}_{\alpha\beta} + \O\!\left(\frac{\ln r_*}{r_*^{2}}\right) + \O(3)
\eeq
for some $f^{[2]}_{\alpha\beta}$. We use `$\O(3)$' to indicate that $h^{[3]}_{\alpha\beta}$ and higher terms in the GSF-MPM expansion are neglected. We stress that this does not necessarily correspond to only omitting $\O(\e^3)$ terms in the very far zone because memory terms are promoted by one order in $\e$.


In this section our aim is to understand how the various terms in the metric contribute to the waveform, but we also want to convey how that waveform can be obtained from the numerically computed near-zone field variables. We first note that equation~\eqref{total far-zone metric} is valid throughout the far zone. If re-expanded in the near zone, it must match the large-$r$ expansion of the near-zone metric $\mathring{g}_{\alpha\beta} + \e h^{(1)}_{\alpha\beta} + \e^2\bigl(h^{(2)\cal R}_{\alpha\beta}+h^{(2)\cal P}_{\alpha\beta}\bigr)$. The puncture terms match by construction; $h^{(2)\cal P}_{\alpha\beta}$ is \textit{defined} from the near-zone re-expansion of $j^{\cal P}_{\alpha\beta}$. Therefore the regular, outgoing field $\frac{1}{r_*} \sum_{n=1,2}\e^n f^{[n]}_{\alpha\beta}$ must match the $1/r$ term in the large-$r$ expansion of $\e h^{(1)}_{\alpha\beta} + \e^2 h^{(2)\cal R}_{\alpha\beta}$. 

Focusing first on the oscillatory terms, we write the matching condition as
\beq
\e f^{[1],\rm osc.}_{\alpha\beta}+\e^2 f^{[2],\rm osc.}_{\alpha\beta} = \e^2 z^{(1)}_{\alpha\beta} + \e^2 z^{(2)\cal R}_{\alpha\beta} +\O(3),
\eeq
where we have introduced
\beq
h^{(2){\cal R},\rm osc.}_{\alpha\beta} = \frac{z^{(2)\cal R}_{\alpha\beta}(u,n^i)}{r_*} + \O(r^{-2}_*\ln r_*)
\eeq
in analogy with the oscillatory part of $h^{(1)}_{\alpha\beta}$ discussed in section~\ref{sec:asymptotic_source_and_GW_flux}. The total oscillatory part of the metric, expanded in the very far zone, is therefore
\beq\label{eq:gosc Lorenz}
g^{\rm osc.}_{\alpha\beta} = \frac{\e z^{(1)}_{\alpha\beta}+\e^2 z^{(2)\cal R}_{\alpha\beta}+\e^2 a^{{\cal P}, \rm osc.}_{\alpha\beta}\ln r_*}{r_*} + \O\!\left(\frac{\ln r_*}{r_*^{2}}\right) + \O(3).
\eeq
This does not yet have the form of a regular outgoing GW due to the presence of logarithms. However, the logarithms are an artefact of the Lorenz gauge condition. Transformed to regular, Bondi-Sachs coordinates $X^\alpha = (U,R,\Theta^A)$~\cite{Madler:2016xju}, the oscillatory part of the metric becomes
\beq\label{eq:gosc Bondi-Sachs}
g^{\rm osc.}_{\alpha\beta} = \frac{\e z^{(1)}_{\alpha\beta}+\e^2 z^{(2)\cal R}_{\alpha\beta}}{R} + \O\!\left(\frac{1}{R^{2}}\right) + \O(3).
\eeq
We detail the coordinate transformation in~\ref{sec: vfz in Bondi-Sachs}. The angular components (or equivalently, the TT components) $z^{(1)}_{AB}$ and $z^{(2)\cal R}_{AB}$ are naturally decomposed into a sum of modes in the form~\eqref{eq:multiscale waveform}, as in $z^{(1)}_{AB}\bar m^A \bar m^B=\sum_{l\geq2}\sum_{m\neq0}h^{(1)}_{lm}(\Omega)e^{-im\phi_p}{}_{-2}Y_{lm}$. Equation~\eqref{eq:gosc Bondi-Sachs} hence establishes our first key conclusion: 
\begin{quote}
\textit{Up to possibly relevant $\O(3)$ contributions, the oscillatory modes of the 1PA waveform~\eqref{eq:multiscale waveform} are given directly by the mode amplitudes $h^{(1)}_{lm}$ and $h^{(2)\cal R}_{lm}$ of the near-zone field variables $h^{(1)}_{\alpha\beta}$ and $h^{(2)\cal R}_{\alpha\beta}$.}
\end{quote}
We comment on the $\O(3)$ terms in equation~\eqref{eq:gosc Bondi-Sachs} at the end of this section.

Next, we turn to the quasistationary terms in the metric~\eqref{total far-zone metric}. In this case it is clear that $\O(3)$ memory terms will be promoted to $\O(\e^2)$. We hence restrict our attention to $\O(\e)$ terms. Recalling equation~\eqref{c-form}, we see that $\langle f^{[1]}_{\alpha\beta}\rangle$ only contributes a mass term, $\langle f^{[1]}_{\alpha\beta}\rangle=2M^{(1)}_{\rm B}\delta_{\alpha\beta}$. Meanwhile,  $\langle h^{[2,1]}_{\alpha\beta}\rangle$ falls off at least as fast as $\ln(r_*)/r_*^2$ in the very far zone, implying $\< j^{\cal P}_{\alpha\beta}\>$ reduces to $\<j_{\alpha\beta}\>$. Appealing to equation~\eqref{very-large-r m=0}, we write
\beq
\<j^{\cal P}_{\alpha\beta}\>=\frac{\<a^{\cal P}_{\alpha\beta}(\Omega)\>\ln r_* + b^{\cal P}_{\alpha\beta}(\Omega)}{\e r_*}+\O\!\left(\frac{\ln r_*}{r_*^{2}}\right)
\eeq
for some hereditary integrals $\<a^{\cal P}_{\alpha\beta}(\Omega)\>$ and $b^{\cal P}_{\alpha\beta}(\Omega)$. The quasistationary part of equation~\eqref{total far-zone metric} is therefore
\beq\label{eq:gstat Lorenz}
\<g_{\alpha\beta}\> = \mathring{g}_{\alpha\beta} + \frac{\e \left[2M^{(1)}_{\rm B}\delta_{\alpha\beta}+\<a^{\cal P}_{\alpha\beta}(\Omega)\>\ln r_* + b^{\cal P}_{\alpha\beta}(\Omega)\right]}{r_*}  + \O\!\left(\e^2,\frac{\ln r_*}{r_*^{2}}\right).
\eeq
Again this is not yet regular at future null infinity due to the logarithms. However, after the transformation to Bondi-Sachs coordinates described in~\ref{sec: vfz in Bondi-Sachs}, we find that this becomes
\beq
\<g_{\alpha\beta}\> = \mathring{g}_{\alpha\beta} + \frac{\e \left[2M^{(1)}_{\rm B}k_{\alpha}k_{\beta}+\big\langle c^{(1)}_{\alpha\beta}(\Omega)\big\rangle\right]}{R}  + \O\!\left(\e^2,\frac{1}{R^{2}}\right),
\eeq
where $\langle c^{(1)}_{AB}\rangle \bar m^A\bar m^A=\sum_{l\geq2}h^{(1)}_{l0}(\Omega)\ {}_{-2}Y_{l0}$ is the first-order GW memory. This is our second key conclusion: 
\begin{quote}
\textit{The field $\langle j_{\alpha\beta}\rangle$, which cures the near-zone infrared divergence by introducing memory effects into the near-zone field $h^{(2)}_{\alpha\beta}$, also becomes the first-order GW memory at future null infinity.}
\end{quote}

We round out our analysis by assessing the impact of the neglected $\O(3)$ terms. By construction, only the leading term in $h^{[3]}_{\alpha\beta}$ can contribute to the waveform. That term satisfies the next equation in the sequence~\eqref{EFEnj},
\beq
\Box_\eta h^{[3,0]}_{\alpha\beta} = 4\delta^2 R^{[0]}_{\alpha\beta}[h^{[2,0]},h^{[1,0]}] +2\delta^3 R^{[0]}_{\alpha\beta}[h^{[1,0]},h^{[1,0]},h^{[1,0]}]\label{EFE30}.
\eeq
The cubic source decays as $1/R^3$, generating a regular outgoing solution. The quadratic source is difficult to analyze due to the logarithms in $h^{[2,0]}_{\alpha\beta}$. However, if we first transform $h^{[2,0]}_{\alpha\beta}$ to the Bondi-Sachs gauge, then the quadratic source falls as $1/R^2$, just as $\delta^2 R^{[0]}_{\alpha\beta}[h^{[1]},h^{[1]}]$.  The `oscillation times oscillation' part of this source creates a quasistationary term proportional to the GW flux, as in equation~\eqref{d2R[h1,h1]}; this ultimately generates the second-order GW memory, in perfect analogy with how $\langle\delta^2 R^{[0]}_{\alpha\beta}[h^{[1]},h^{[1]}]\rangle$ generates the first-order GW memory through $\<j_{\alpha\beta}\>$. However, the `oscillation times stationary perturbation' terms in $\delta^2 R^{[0]}_{\alpha\beta}[h^{[2,0]},h^{[1,0]}] $ have a different structure than the analogous terms in $\delta^2 R^{[0]}_{\alpha\beta}[h^{[1]},h^{[1]}]$  because $\langle h^{[2,0]}_{\alpha\beta}\rangle$ includes memory (in addition to a mass perturbation). In the Bond-Sachs gauge, these terms are 
\beq\label{eq:promoted osc source}
4\delta^2 R^{[0]}_{\alpha\beta}[\<h^{[2,0]}\>,h^{[1,0]}] =  \frac{1}{\e R^2} \<C^{AB}_{(1)}\>\ddot C^{(1),\rm osc.}_{AB}k_{\alpha}k_{\beta} + \text{mass term}+ \O\left(\frac{1}{R^3}\right),
\eeq
where $\langle C_{AB}^{(1)}\rangle =R^{-2}\langle c^{(1)}_{AB}\rangle$ is the first-order memory's contribution to the shear. The mass term in equation~\eqref{eq:promoted osc source} has the form of the final term in equation~\eqref{s-form Bondi}, now involving the second-order mass perturbation.

A source of the form~\eqref{eq:promoted osc source} will give rise to oscillatory terms of order $1/\e$ in $h^{[3,0]}_{\alpha\beta}$, generating oscillatory $\sim \frac{\e^2}{r}$ (and $\sim \frac{\e^2\ln r}{r}$) contributions to the asymptotic metric. These terms are straightforwardly calculated following the same steps as we used to solve the field equation for $\<j_{\alpha\beta}\>$, since the source in equation~\eqref{eq:promoted osc source} has the same functional form as the source for $\<j_{\alpha\beta}\>$, which read $-\<\Pi\> k_\alpha k_\beta/R^2$. Concretely, if we (i) put $\langle h^{[2,0]}_{\alpha\beta}\rangle$ in the Bondi-Sachs gauge, (ii) solve equation~\eqref{EFE30} for the part of $h^{[3,0]}_{\alpha\beta}$ sourced by \eqref{eq:promoted osc source}, and (iii) transform the result to the Bondi-Sachs gauge, then we arrive at the solution~\eqref{<j> BS} with $\int^u_{-\infty}du'$ replaced by $1/(-i m \Omega)$ and with $\<\Pi\>$ replaced by (minus) the coefficient of $k_\alpha k_\beta /R^2$ in equation~\eqref{eq:promoted osc source}. Making those replacements in equation~\eqref{<j> BS}, we find the contribution to the emitted GW in Cartesian Bondi-Sachs coordinates $(U,X^a)$ is 
\beq\label{eq: memory distortion GW}
h^{[3,0],Q}_{ab} = \frac{1}{\e R}\sum_{l\geq2}\frac{2N_{L-2}}{(l+1)(l+2)}\perp^{\rm TT}_{ab,ij}Q_{ijL-2}(U)+ \O\!\left(\frac{\ln R}{R^2}\right),
\eeq 
where $Q\equiv-\langle C^{AB}_{(1)}\rangle\dot C^{(1),\rm osc.}_{AB}$ is expanded as $Q=\sum_{l\geq0}Q_L \hat N^L$, and $\perp^{\rm TT}_{ab,ij}$ is the TT projection operator defined in equation~\eqref{eq:TT projector}. Since $Q$ is oscillatory, it appears that the $\O(3)$ terms that we omitted throughout this section do in fact generically contribute to the oscillatory part of second-order waveform modes $h^{(2)}_{lm}$ ($m\neq0$). They will \textit{also} then modify the $\O(\e^3)$ GW flux, altering the results of reference~\cite{Warburton:2021kwk}. This change in the GW flux will additionally alter the second-order memory modes $h^{(2)}_{l0}$. 

We will further investigate these terms in a followup work. Here we only comment that they might be associated with the asymptotic (BMS) frame in GSF calculations. Over the course of a binary's evolution --- from an initial stationary epoch in the distant past, through a radiative epoch, and into a final stationary epoch --- the accumulated memory at future null infinity is equivalent to a BMS transformation between asymptotic frames~\cite{Strominger:2014pwa,Hollands:2016oma}. Terms in the asymptotic metric sourced by equation~\eqref{eq:promoted osc source} represent the distortion of the oscillatory modes due to their interaction with this slow evolution of the asymptotic frame. We will refer to them as `memory distortion' terms.

We also observe that these terms, even if they contribute to the $\O(\e^3)$ GW flux, cannot contribute to the 1PA evolution of the binary, unlike other $\O(\e^3)$ fluxes. This is clear from the fact that memory terms are not promoted by $1/\e$ in the near zone, as mentioned in section~\ref{sec:near-zone memory}. Like the apparent disagreement between the first-law binding energy and the binding energy defined from the Bondi mass~\cite{Pound:2019lzj}, this might point to a breakdown in the relationship between binary quantities and asymptotic quantities, suggesting further subtlety in the application of balance laws as described around equation~\eqref{eq:evolution}.

Finally, we notice that these terms are in close analogy to potential `cubic memory' terms arising from a cubic  interaction between three quadrupole moments in the MPM expansion (namely $\mathcal{M}_{ij}\times\mathcal{M}_{kl}\times\mathcal{M}_{pq}$), which will arise for the first time in the 5PN flux. Indeed, if terms of the type 
\begin{subequations}\begin{equation}
\mathcal{U}_{ij}(U) \propto \frac{G^2}{c^{10}}\int_0^{+\infty} \!\!\dd \rho\  \mathcal{M}_{ab}^{(4)}(U -\rho) \int_0^{+\infty} \!\!\dd \tau \left[\mathcal{M}_{a\langle i}^{(3)} \mathcal{M}_{j\rangle b}^{(3)}\right](U -\rho-\tau) + \ldots
\end{equation}
or of the type
\begin{equation}\label{eq:MPM memory distortion type 2}
\mathcal{U}_{ij}(U) \propto  \frac{G^2}{c^{10}}\mathcal{M}_{ab}^{(3)} \int_0^{+\infty} \!\!\dd \tau \left[\mathcal{M}_{a\langle i}^{(3)} \mathcal{M}_{j\rangle b}^{(3)}\right](U -\tau) + \ldots
\end{equation} \end{subequations}
were to appear, the memory-like integral in $\tau$ would feature a non-oscillatory DC component and thus acquire a $-2.5$PN promotion. This DC term would multiply an oscillatory term, leading to a 2.5PN correction to the radiative moment and the energy flux. Equation~\eqref{eq:MPM memory distortion type 2}, in particular, has the same form as \eqref{eq: memory distortion GW}: a memory integral multiplying a time derivative of a radiative moment --- meaning, at leading PN-MPM order, a quadrupole memory integral multiplying a third time derivative of a source quadrupole moment. A complete calculation of these terms can be performed using the same methods as for the tails of memory~\cite{Trestini:2023wwg}; such a calculation  will hopefully exclude such terms. This will be the subject of future work.

\section{Gravitational-wave memory at first order in the mass ratio for quasi-circular inspirals into a Kerr black hole}\label{sec:KerrMemoryFirstOrder}

We now present results for GSF calculations made at first order in the mass ratio, along with details of the computational methods used.
In this section we focus on quasi-circular inspirals into a Kerr black hole.
In section \ref{sec:PNSF} we compute a double PN-GSF series expansion for the memory through 5PN order.
We find in the non-spinning case that these results agree with the 3.5PN results from  section \ref{sec:PNresults}, and for the spinning case agreement is found with the 3PN results in the literature \cite{Henry:2023tka}.
In section \ref{sec:Numerical Kerr results}, we numerically compute the memory using the Teukolsky formalism.
We use these results to further validate our PN-GSF series, finding agreement between the numerical and analytic results -- see figure \ref{fig:hdot20}.

\subsection{Post-Newtonian calculation at first order in the mass ratio}\label{sec:PNSF}

To analytically calculate the GW memory at leading order in the mass-ratio using our equation \eqref{eq:InstantaneousMemory}, we solve the Teukolsky equation with the additional assumption that the small body is on a PN-like orbit. Generating PN solutions to the Teukolsky equation has been well described in the literature; see e.g. \cite{Fujita:2011zk, Fujita:2014eta, Kavanagh:2016idg, Akcay:2019bvk, Munna:2020civ}. As such we will provide only a brief overview.

Exact homogeneous solutions for the radial Teukolsky equation were given by Mano, Suzuki and Takasugi (MST) \cite{Mano:1996mf, Mano:1996vt, Sasaki:2003xr} in terms of an infinite sum of hypergeometric functions. In the PN regime, the particle's orbital radius is large, \textit{i.e.}, $r_p\gg M$, and likewise $\omega=m\Omega \sim \sqrt{M r_p^{-3}}$. Introducing the order counting parameter $\eta$, and defining the variables $\bar{r}_p=r_p \eta^2$ and $\bar{\omega}=\omega \eta^{-3}$, one finds that the PN solutions are given by expanding the MST solutions for small $\eta$ and fixed barred variables. In this way, every second power of $\eta$ is a full PN order. Our code to generate these expansions and those for the asymptotic amplitudes is publicly available~\cite{JakobSFPN}.

\begin{figure}[hbt!]
    \centering
    \begin{tabular}{ccc}
        \includegraphics[width = 0.5\textwidth]{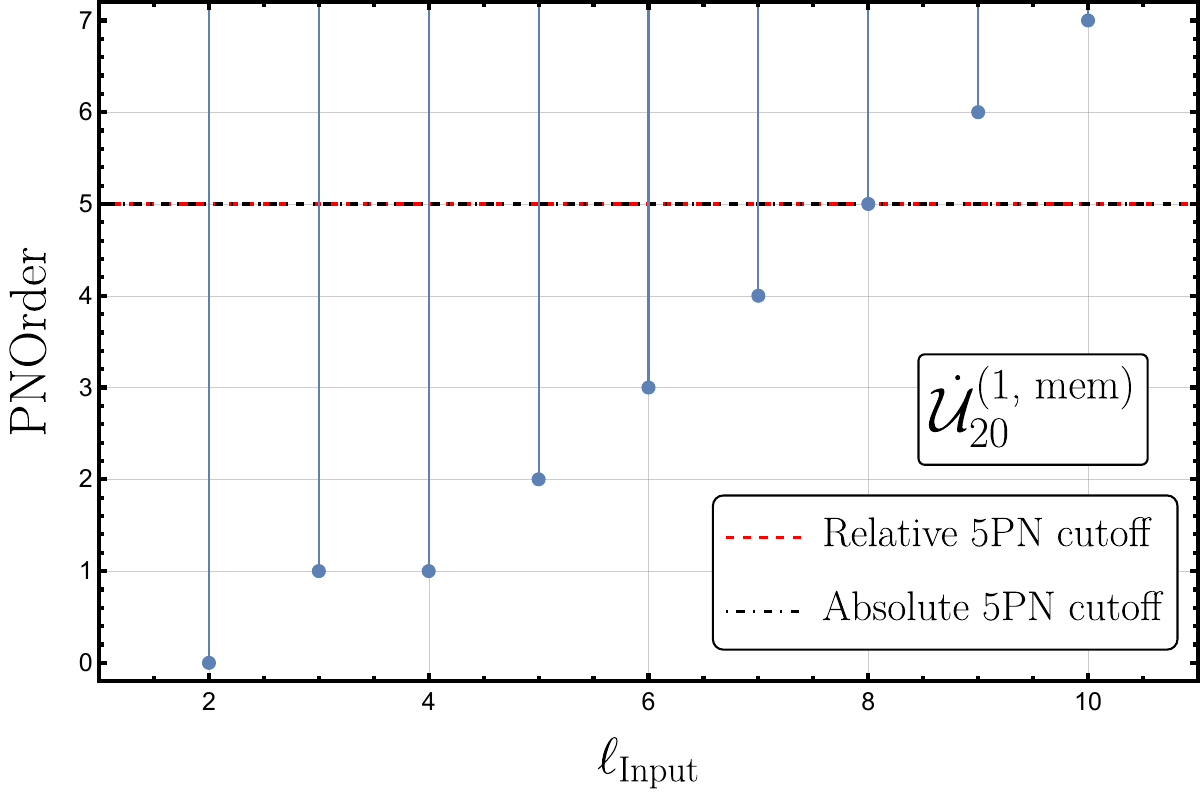}
        \includegraphics[width = 0.5\textwidth]{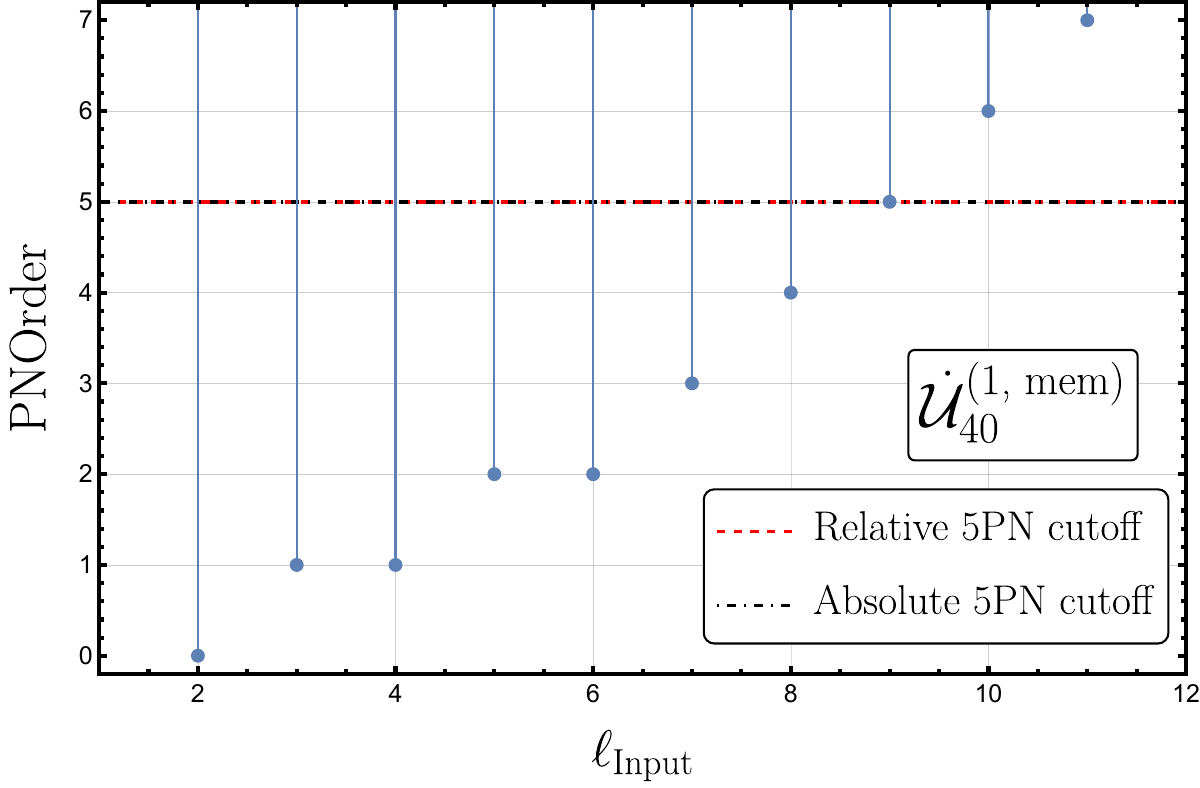}\\
        \includegraphics[width = 0.5\textwidth]{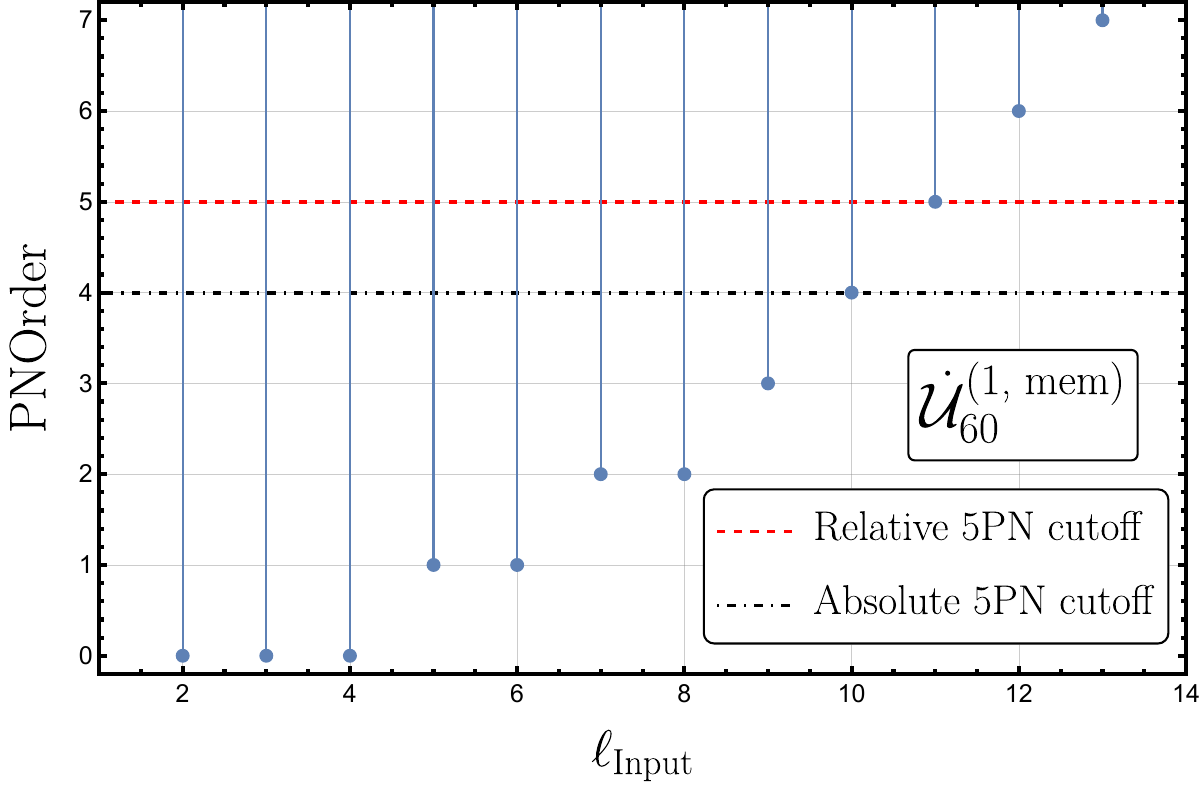}
        \includegraphics[width = 0.5\textwidth]{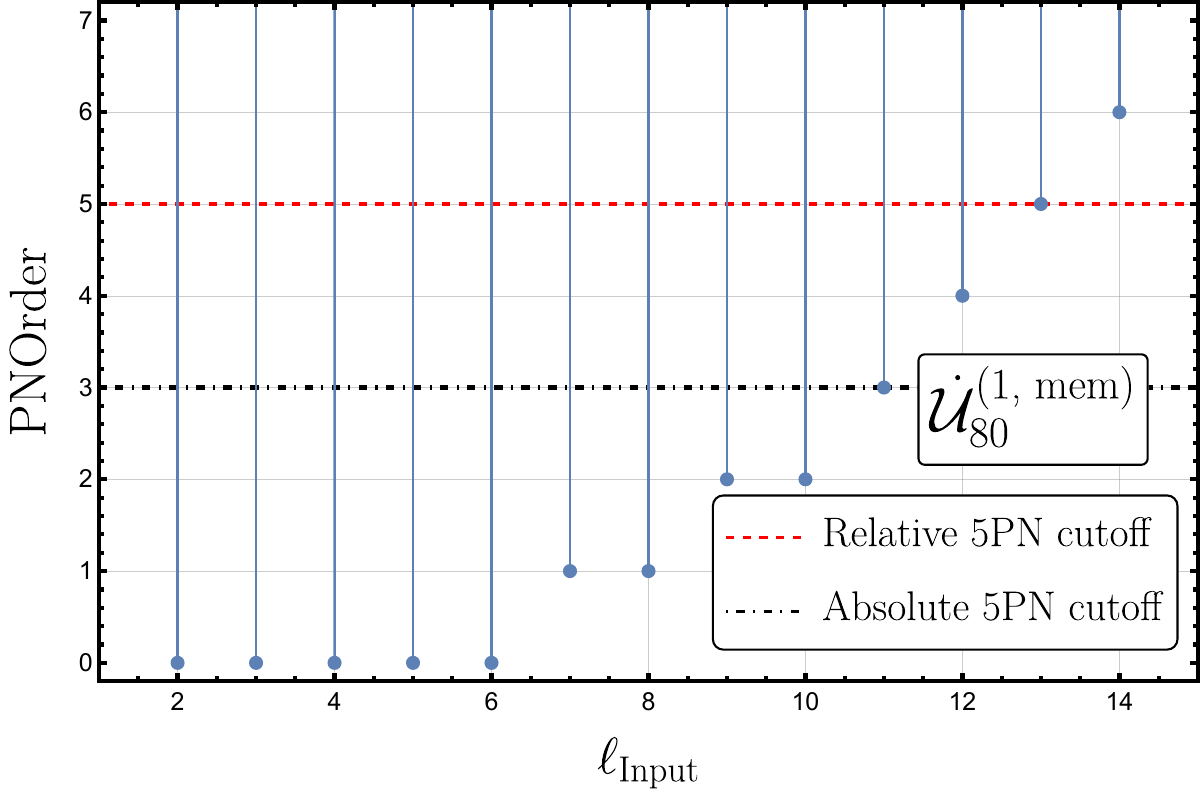}
    \end{tabular}
	\caption{
	The PN order to which each $\tensor*[_{-2}]{Z}{^{\text{Up}}_{\ell m\omega}}$ must be calculated for the memory $\cU_{l0}$ to be accurate to 5PN for $l\in\{2,4,6,8\}$ read left to right, top to bottom. 
    On the plots 0PN refers to the order at which $\cU_{l0}$ appears. The spheroidal $\ell$-number is plotted on the horizontal axis. The (red) dot represents
    the order at which $\tensor*[_{-2}]{Z}{^{\text{Up}}_{\ell m\omega}}$ first appears. The (black) dashed horizontal line represents the 5PN cutoff. We can read off how many orders are needed by counting
     the difference between the cutoff and the order at which a mode appears. For example, to calculate $\cU_{80}$ to 5PN the $\ell=9$ amplitude
    should include 4PN orders beyond its own leading order.
    Note that due to the spheroidal-spherical coupling, equation~\eqref{eq:spherical_Z}, that we are forced to include higher $\ell$ modes in our calculation 
    than Favata~\cite{Favata:2008yd}. 
    }
    \label{fig:PNInputorder}
\end{figure}

The asymptotic amplitudes (in the spin-weighted spheroidal basis) exhibit the following PN scalings as a function of their mode number:
\begin{align}
	{}_{-2}Z^{\rm Up}_{lm\omega}\sim
	\Bigg\{\begin{array}{cc}
	(y \eta^{2})^{\frac{l+6}{2}} &\text{if }  l+m  \text{ is even} \\[0.4cm]
	(y \eta^{2})^{\frac{l+7}{2}} &\text{if }  l+m \text{ is odd}
\end{array} \,,
\end{align}
where $y = (m_1 \Omega)^{2/3}$. 
From this scaling it is a straightforward computation to see which modes, and to what PN order, are required to determine a mode of the memory to a given PN order --- see figure \ref{fig:PNInputorder} for details.

We compute the GW memory for $l\in\{2,4,6,8,10,12,14\}$ for a particle on a circular orbit to all orders in
 the Kerr spin parameter, each computed with 5PN accuracy relative to 
 the leading order of the mode-summed memory, which is set by the dominant
 contributions of  $l=2$ and $l=4$. In addition to the 5PN calculation, we also compute the $l\in\{6,8\}$ modes to 5PN order relative to their respective leading order contributions (i.e.~to 6PN and 7PN overall). While formally, upon computing the total mode-summed memory,
 the higher PN orders for $l\in\{6,8\}$ contribute only as partial contributions beyond 5PN,
 they are useful to have as more stringent checks on the numerically computed modes of the memory in the next subsection.

To present our results we define the adimensionalised spin as $\tilde{a} = a/m_1 $.
We first give the series for the news for the $(l,m)=(2,0)$ mode:
\begin{multline}
\label{eq:U1SFPNl2}
    \dot \cU^\text{(1, mem)}_{20} = \frac{256}{7} \sqrt{\frac{\pi }{15}}\epsilon^2 y^5\bigg\{1-\frac{1219}{288}y+\left(4 \pi -\frac{505 \tilde{a}}{192}\right)y^{3/2}+\left(\frac{63 \tilde{a}^2}{32}-\frac{793}{1782}\right)y^2\\*
    +\left(-\frac{257 \tilde{a}}{144}-\frac{2435 \pi }{144}\right)y^{5/2}+\left(-\frac{103 \tilde{a}^2}{192}-\frac{339 \pi  \tilde{a}}{32}-\frac{856 \log (y)}{105}\right.\\*
    \left.+\frac{16 \pi ^2}{3}-\frac{1712 \gamma_{\rm E} }{105}+\frac{174213949439}{1816214400}-\frac{3424 \log (2)}{105}\right)y^3\\*
    +\left(-\frac{241 \tilde{a}^3}{96}+\frac{127 \pi  \tilde{a}^2}{16}+\frac{1111841 \tilde{a}}{31104}-\frac{33199 \pi }{8448}\right)y^{7/2}\\*+
    \left(\frac{31 \tilde{a}^4}{32}+\frac{118369 \tilde{a}^2}{14256}-\frac{57 \pi  \tilde{a}}{8}+\frac{1789615 \log (y)}{49896}-\frac{4867 \pi ^2}{216}\right.\\*
    \left.+\frac{1789615 \gamma_{\rm E} }{24948}-\frac{4080360541183}{14384418048}+\frac{17890507 \log (2)}{124740}\right)y^4\\*
    +\left(-\frac{3191 \tilde{a}^3}{162}-\frac{565 \pi  \tilde{a}^2}{288}+\frac{153983 \tilde{a} \log (y)}{5040}-\frac{2041 \pi ^2 \tilde{a}}{144}+\frac{153983 \gamma_{\rm E}  \tilde{a}}{2520}-\frac{15694894939 \tilde{a}}{61916400}\right.\\*
    \left.+\frac{308351 \tilde{a} \log (2)}{2520}-\frac{3424}{105} \pi  \log (y)-\frac{6848 \gamma_{\rm E}  \pi }{105}+\frac{7790099187109 \pi }{19372953600}-\frac{13696}{105} \pi  \log (2)\right)y^{9/2}\\*
    +\left(\frac{331 \tilde{a}^4}{96}-\frac{497 \pi  \tilde{a}^3}{48}-\frac{1819}{112} \tilde{a}^2 \log (y)+\frac{85 \pi ^2 \tilde{a}^2}{8}-\frac{1819 \gamma_{\rm E}  \tilde{a}^2}{56}+\frac{69429660961 \tilde{a}^2}{408648240}\right.\\*
    \left.-\frac{7811}{120} \tilde{a}^2 \log (2)+\frac{2932871 \pi  \tilde{a}}{19008}+\frac{3131532667 \log (y)}{1284323040}-\frac{1076725 \pi ^2}{85536}+\frac{3131532667 \gamma_{\rm E} }{642161520}\right.\\*
    \left.-\frac{1328426429865600503}{3933958530902400}-\frac{332418897 \log (3)}{14094080}+\frac{42845174011 \log (2)}{642161520}\right)y^5 +\mathcal{O}(y^{11/2})\bigg\}.
    \end{multline}
    
Following equation~\eqref{eq:hl0}, we integrate the news along the inspiral to compute the memory.
In practice we evaluate the following integral:
\begin{equation}\label{eq:Ul0_PN_Int}
    \cU_{l0}^\text{(mem)}(y) =  \int_{0}^{y} \frac{\dot{\cU}_{l0}^\text{(1, mem)}(y')}{\dot{y}_p(y')} \dd y'\,,
\end{equation}
where 
\begin{equation}
    \label{eq:ydot}
    \dot{y}_p(y') = \frac{\dd y_p}{\dd \mathcal{E}}(y')\left(-\mathcal{F}^\text{PN}(y')\right).
\end{equation}
In this case $\mathcal{F}^\text{PN}$ are PN-GSF series for the fluxes~\cite{2015PTEP.2015c3E01F}. 
The explicit expressions we used are available in a digital format from the \texttt{PostNewtonianSelfForce} package~\cite{niels_warburton_2023_8112975} of the \texttt{Black Hole Pertubation Toolkit}. 
We convert from $\cU_{l0}^\text{(mem)}$ to $H_{l0}^\text{(mem)}$ using equations~\eqref{UtoStrain} and~\eqref{eq:hlm_vs_Hlm} to get our final result for the $(l,m)=(2,0)$ mode:
\begin{multline}\label{eq:H1SFPNl2}
    H_{20}^\text{(1, mem)} = -\frac{5}{14 \sqrt{6}}\bigg\{1-\frac{4075}{4032}y+\left(\frac{1303 \tilde{a}}{480}\right)y^{3/2}+\left(-\frac{33 \tilde{a}^2}{32}-\frac{151877213}{67060224}\right)y^2\\
    +\left(\frac{3 \tilde{a}^3}{14}+\frac{4936811 \tilde{a}}{677376}-\frac{253 \pi }{336}\right)y^{5/2}+\left(-\frac{29107 \tilde{a}^2}{9216}-\frac{23 \pi  \tilde{a}}{384}-\frac{4397711103307}{532580106240}\right)y^3\\
    +\left(\frac{16405 \tilde{a}^3}{32256}+\frac{\pi  \tilde{a}^2}{24}+\frac{38587083163 \tilde{a}}{1379524608}+\frac{38351671 \pi }{28740096}\right)y^{7/2}\\
    +\left(\frac{797 \tilde{a}^4}{512}-\frac{3 \pi  \tilde{a}^3}{5}-\frac{13}{10} \sqrt{1-\tilde{a}^2} \tilde{a}^2-\frac{2966416073 \tilde{a}^2}{162570240}-\frac{\sqrt{1-\tilde{a}^2}}{10}\right.\\
    \left.+\frac{1}{5}  \tilde{a} \Psi_B(\tilde{a})-\frac{3}{5} \sqrt{1-\tilde{a}^2} \tilde{a}^4+\frac{3}{5}  \tilde{a}^3 \Psi_B(\tilde{a})-\frac{6065881 \pi  \tilde{a}}{1935360}+\frac{9160157 \log (y)}{8731800}\right.\\
    \left.+\frac{139 \pi ^2}{420}+\frac{9160157 \gamma_{\rm E} }{4365900}-\frac{21373614805069828883}{405985814986752000}+\frac{9477 \log (3)}{1568}-\frac{1641169 \log (2)}{873180}\right)y^4\\
    +\left(-\frac{3193 \tilde{a}^5}{4928}+\frac{12105358855 \tilde{a}^3}{1967099904}+\frac{266275 \pi  \tilde{a}^2}{354816}-\frac{5279 \tilde{a} \log (y)}{27720}+\frac{23 \pi ^2 \tilde{a}}{264}-\frac{5279 \gamma_{\rm E}  \tilde{a}}{13860}\right.\\
    \left.+\frac{376568705459465749 \tilde{a}}{3866531571302400}-\frac{121}{180} \tilde{a} \log (2)+\frac{222498491599 \pi }{26153487360}\right)y^{9/2}\\
    +\left(\frac{3 \tilde{a}^6}{32}+\frac{1431193 \tilde{a}^4}{114688}-\frac{81583 \pi  \tilde{a}^3}{32256}-\frac{107 \tilde{a}^2 \log (y)}{1120}\right.\\
    \left.-\frac{126443 \sqrt{1-\tilde{a}^2} \tilde{a}^2}{24192}-\frac{\pi ^2 \tilde{a}^2}{16}-\frac{107 \gamma_{\rm E}  \tilde{a}^2}{560}-\frac{1610120285575493 \tilde{a}^2}{21968929382400}-\frac{11471 \sqrt{1-\tilde{a}^2}}{24192}\right.\\
    \left.-\frac{107}{240} \tilde{a}^2 \log (2)+\frac{1}{12}  \tilde{a} \Psi_A(\tilde{a})+\frac{9455  \tilde{a} }{12096}\Psi_B(\tilde{a})-\frac{8951 \sqrt{1-\tilde{a}^2} \tilde{a}^4}{4032}-\frac{1}{16}  \tilde{a}^3 \Psi_A(\tilde{a})\right.\\
    \left.+\frac{9455  \tilde{a}^3 }{4032}\Psi_B(\tilde{a})-\frac{40966887443 \pi \tilde{a}}{4291854336}-\frac{38376467119 \log (y)}{23599435860}-\frac{91903489 \pi ^2}{89413632}\right.\\
    \left.-\frac{38376467119 \gamma_{\rm E} }{11799717930}-\frac{22203886337754680921828309}{238763310807037551575040}-\frac{197306951679 \log (3)}{5524879360}\right.\\
    \left.+\frac{1724214370573 \log (2)}{37759097376}\right)y^5 +\mathcal{O}(y^{11/2})\bigg\}
    \end{multline}
 where $\gamma_{\rm E}$ is Euler's constant and $\Psi_A(\tilde{a})$ and $\Psi_B(\tilde{a})$ are real-valued functions of $\tilde{a}$:
\begin{subequations}\label{eq:PsiAPsiB}\begin{align}
    \Psi_A(\tilde{a}) = i\left[\psi^{(0)}\left(3 + \frac{i \tilde{a}}{\sqrt{1-\tilde{a}^2}}\right) - \psi^{(0)}\left(3 - \frac{i \tilde{a}}{\sqrt{1-\tilde{a}^2}}\right)\right] \,, \\
    \Psi_B(\tilde{a}) = i\left[\psi^{(0)}\left(3 + \frac{2 i \tilde{a}}{\sqrt{1-\tilde{a}^2}}\right) - \psi^{(0)}\left(3 - \frac{2 i \tilde{a}}{\sqrt{1-\tilde{a}^2}}\right)\right]\,,
\end{align}\end{subequations}
with $\psi^{(0)}$ being the digamma function.
We give the PN-GSF series for the \mbox{$l\in\{4,6,8,10,12,14\}$} modes in \ref{sec:further_comparisons}.
The series expressions above are also available in a digital format in the \texttt{PostNewtonianSelfForce} package~\cite{niels_warburton_2023_8112975} of the \texttt{Black Hole Perturbation Toolkit}.

In order to compare the series above to PN results, we note that the PN-GSF inverse separation is linked to the PN inverse separation \eqref{eq:PNx} \textit{via} the relation $y = x + \mathcal{O}(\epsilon)$, whereas the total mass reads $M = m_1 + \mathcal{O}(\epsilon)$ in the EMRI limit.
With these relations, we find that, through 3.5PN, the above series agree perfectly with the results of section~\ref{sec:PNresults}, both in the non-spinning and spinning sector.
Moreover, the spinning terms are found to agree through 2PN with the results of Ref.~\cite{Mitman:2022kwt} and through 3PN with the results of Ref.~\cite{Henry:2023tka}.
We will further test our PN series against the numerical GSF results presented in the next section.

\subsection{Numerical calculation in Kerr spacetime}
\label{sec:Numerical Kerr results}
\begin{figure}
    \centering
    \begin{tabular}{ccc}
        \includegraphics[width = .5\textwidth]{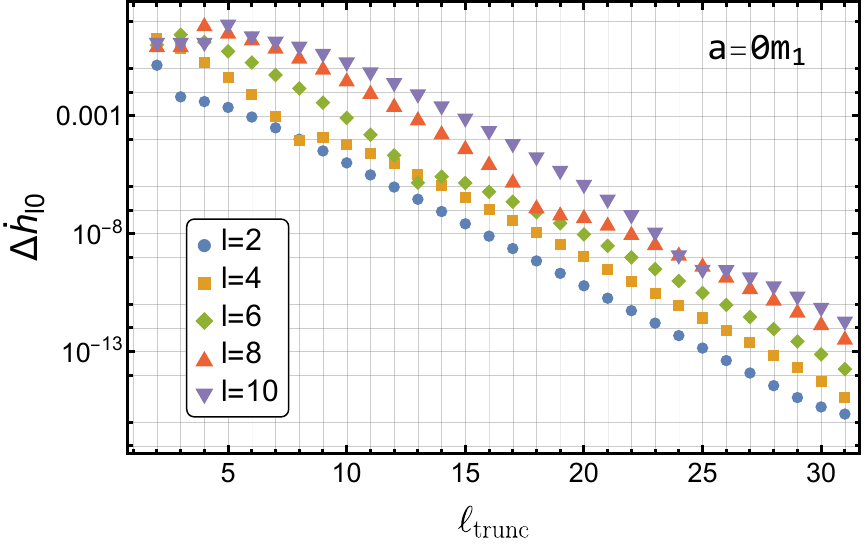}
        \includegraphics[width = .5\textwidth]{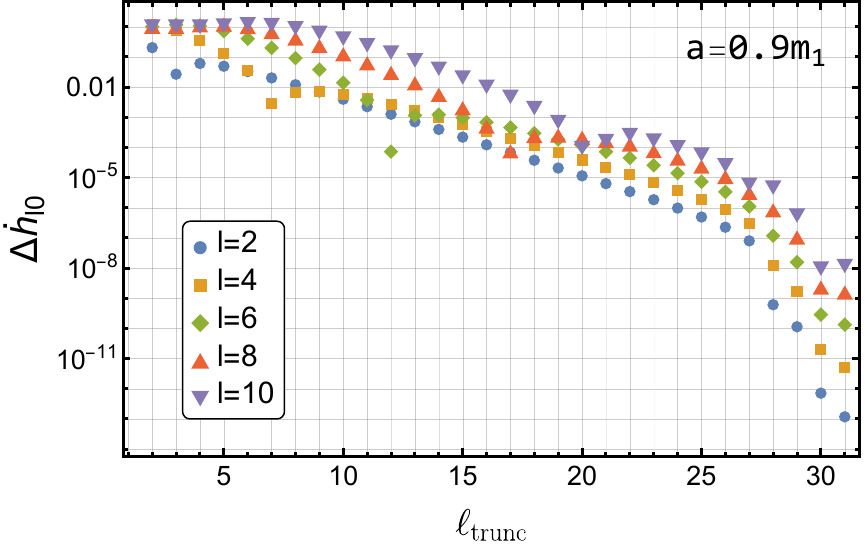}
    \end{tabular}
    \caption{
	The convergence of $\dot{h}_{l0}^{(\rm mem)}$ upon truncating the sum over spheroidal harmonic $\ell$-modes at $\ell_{\rm trunc}$. 
	The relative error is computed with respect to the sum with $\ell_\text{max} = 32$ for an orbit at the ISCO.
	The left plot shows the convergence in the Schwarzschild case ($a=0$) and the right plot shows the Kerr case with $a=0.9m_1$.
    On both plots the convergence for $l=2$ is shown by (blue) circles, $l=4$ by (yellow) squares, $l=6$ by (green) diamonds, $l=8$ by (orange) triangles, and $l=10$ by (purple) upside-down triangles.
	The mode-sum converges more rapidly for larger orbital radii, and thus the figures above show the convergence at the most computationally intensive possible point of each inspiral.
	Note that, in the Kerr plot, the rapid convergence of relative error after $\ell_{\rm trunc} \simeq 27$ is due to the spheroidal-to-spherical harmonic expansion coupling to spheroidal harmonic modes with $\ell > 32$, which we have not computed.
	}
    \label{fig:Convergence}
\end{figure}

In order to numerically compute the memory using equation~\eqref{eq:InstantaneousMemory}, we need to perform calculations within the Teukolsky formalism \cite{TeukolskyI}. 
We do this using tools provided by the Black Hole Perturbation Toolkit~\cite{BHPToolkit}. 
First, we compute the asymptotic amplitudes~$\tensor*[_{-2}]{Z}{^{\text{Up}}_{\ell m\omega}}$ of the spheroidal harmonic modes of the perturbation using the \texttt{GremlinEQ}~\cite{Gremlin} and the \texttt{Teukolsky}~\cite{BHPT_Teukolsky} packages.
For all orbits, we compute spheroidal harmonics modes up to $\ell_{\rm} = 32$.
We then compute the spherical harmonic amplitudes $\tensor*[_{-2}]{Z}{^{(1)}_{l m\omega}}$ using equation \eqref{eq:spherical_Z}.
The spheroidal-to-spherical expansion coefficients $b^{a\omega}_{\ell l m}$ that appear in equation \eqref{eq:spherical_Z} are computed using the \texttt{SpinWeightedSpheroidalHarmonics}~\cite{BHPT_SWSH} package.
Formally, we must sum over infinitely many spheroidal harmonic modes, but, since the mode-sum converges exponentially with $\ell$, in practice we find that computing them up to $\ell_{\rm max} = 32$ is sufficient for this work.
As an estimate of the error in our result from truncating the spheroidal harmonic mode sum, we compute our results with a lower $\ell_{\rm trunc} < 32$ and define the relative error as
\begin{equation}
    \label{eq:convergence}
    \Delta \dot h_{l 0} = \left| \frac{\dot h_{l 0}^\text{(mem)}|_{\ell_\text{max} = \ell_\text{trunc}} - \dot h_{l 0}^\text{(mem)}|_{\ell_\text{max} = 32}}{\dot h_{l 0}^\text{(mem)}|_{\ell_\text{max} = 32}}\right|_{r = r_\text{ISCO}}.
\end{equation}
Details of the convergence rate with spheroidal harmonic $\ell$-modes are shown in figure~\ref{fig:Convergence}.
Due to the coupling between the spheroidal and spherical harmonic modes, the accuracy of our final result varies depending on the spin of the black hole.
The rate of convergence is more rapid for larger orbital radii, so we present results at the ISCO, which is the smallest orbital radius during the inspiral.
For $a=0$, we find that truncating the sum over modes in equation \eqref{eq:InstantaneousMemory} at $\ell = 32$ gives an estimated relative error of $\sim 10^{-15}$ in~$\dot{h}_{l0}$. 
Similarly, for $a=0.9m_1$, we find that truncating the spheroidal harmonic modes at $\ell = 32$ gives an estimated relative error of $\sim 10^{-8}$.

\begin{figure}
    \centering
        \includegraphics[width =0.8\textwidth]{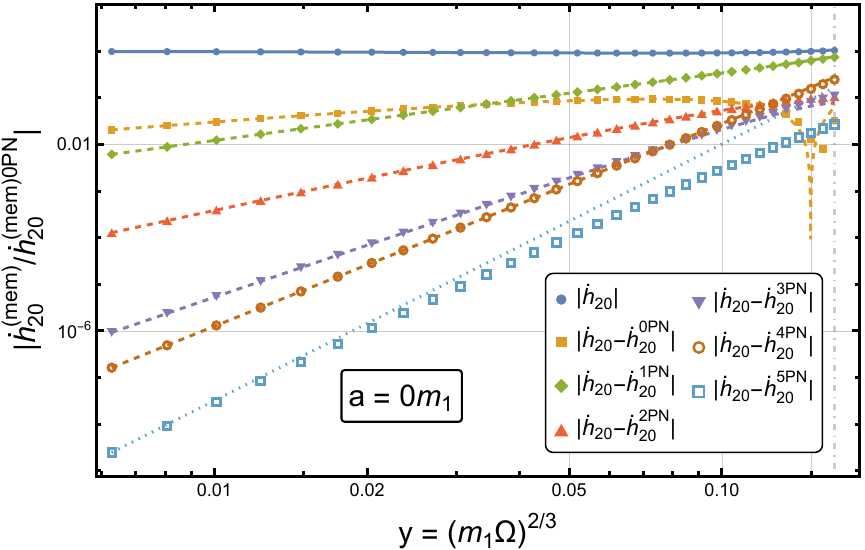}
        \includegraphics[width =0.8\textwidth]{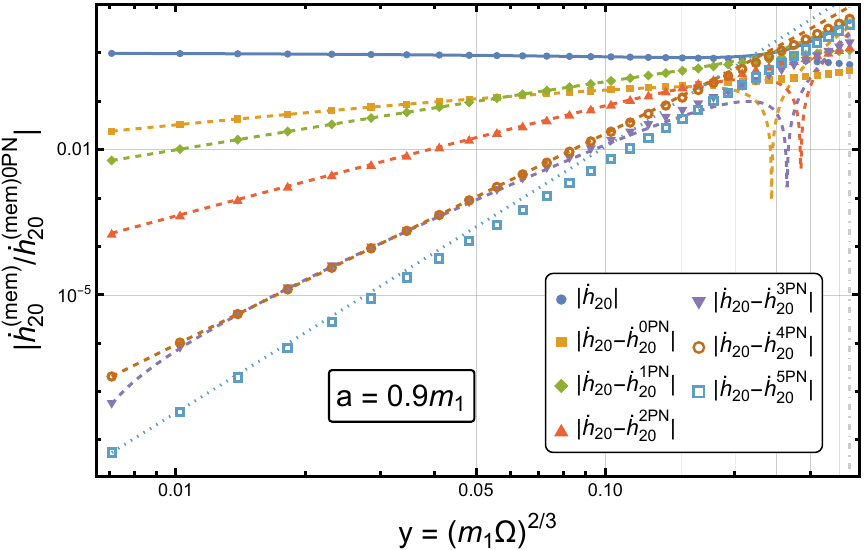}
    \caption[Schwarzschild \& Kerr Memory]
    {The (Newtonian normalised) $l=2,m=0$ Bondi news computed from perturbation theory as compared to the 5PN series at leading order in the mass ratio.
	The top panel shows the comparison for a Schwarzschild black hole ($a=0$), and the bottom panels shows the comparison for a Kerr black hole with spin $a=0.9m_1$.
    In both panels the (dark blue) solid dots show the numerical calculation of $\dot{h}_{l0}^\text{mem}$, while the solid (dark blue) line shows the PN prediction. 
	To get the (yellow) solid squares and (yellow) dashed line, we subtract the leading PN term from both the numerical data and the PN series. 
	We then subtract the next-to-leading term to get the green line, and so on. 
	The (light blue) open squares show the numerical data minus the entire 5PN series. 
	These data points lie close to the (light blue) dotted $y^{5.5}$ reference curve, which shows that the residual has the expected PN scaling.
	The (light gray) dot-dashed vertical line marks the location of the ISCO.
	}
~\label{fig:hdot20}
\end{figure}

In order to make detailed comparisons between our numerical calculations and the PN results of section~\ref{sec:PNSF}, we will consider $\dot{h}_{l0}^\text{(mem)}$ rather than $h_{l0}^\text{(mem)}$, as this avoids having to align the waveforms at a given frequency.
In the terminology of the BMS framework, we are effectively making comparisons of (a scaled version of) the Bondi news rather than the shear.
In figure \ref{fig:hdot20}, we present our results for the dominant $\dot{h}_{20}$ mode, scaled by the leading PN term.
The difference between our numerical data and the leading PN term is consistent with the 1PN term, and repeatedly subtracting PN orders shows consistency through all available PN orders.
The difference between our numerical data, and the full 5PN result shows a residual which scales as $y^{5.5}$, as expected. 
In agreement with Favata~\cite{Favata:2008yd}, we find that the odd-$l$ modes (corresponding to the odd-parity piece of the Bondi news) are zero to the required PN order and within numerical precision. 

\begin{figure}
    \centering
    \begin{tabular}{ccc}
        \includegraphics[width = 0.5\textwidth]{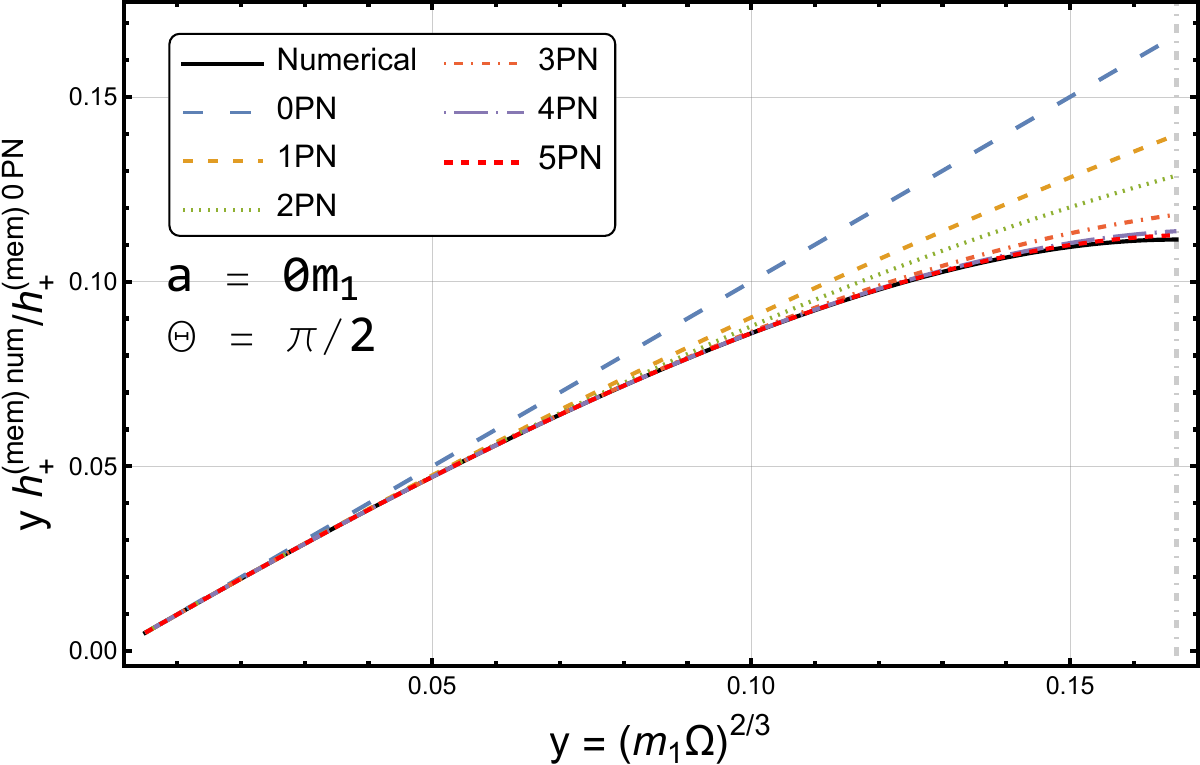}
        \includegraphics[width = 0.5\textwidth]{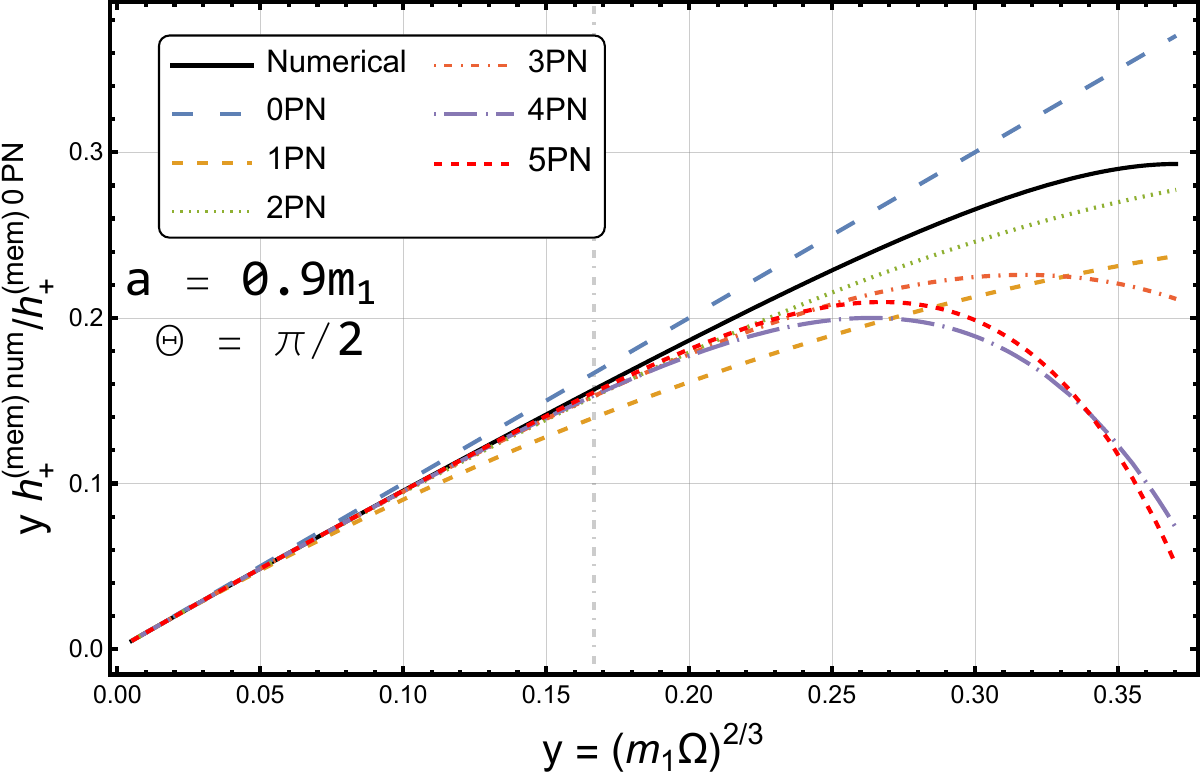}\\
        \includegraphics[width = 0.5\textwidth]{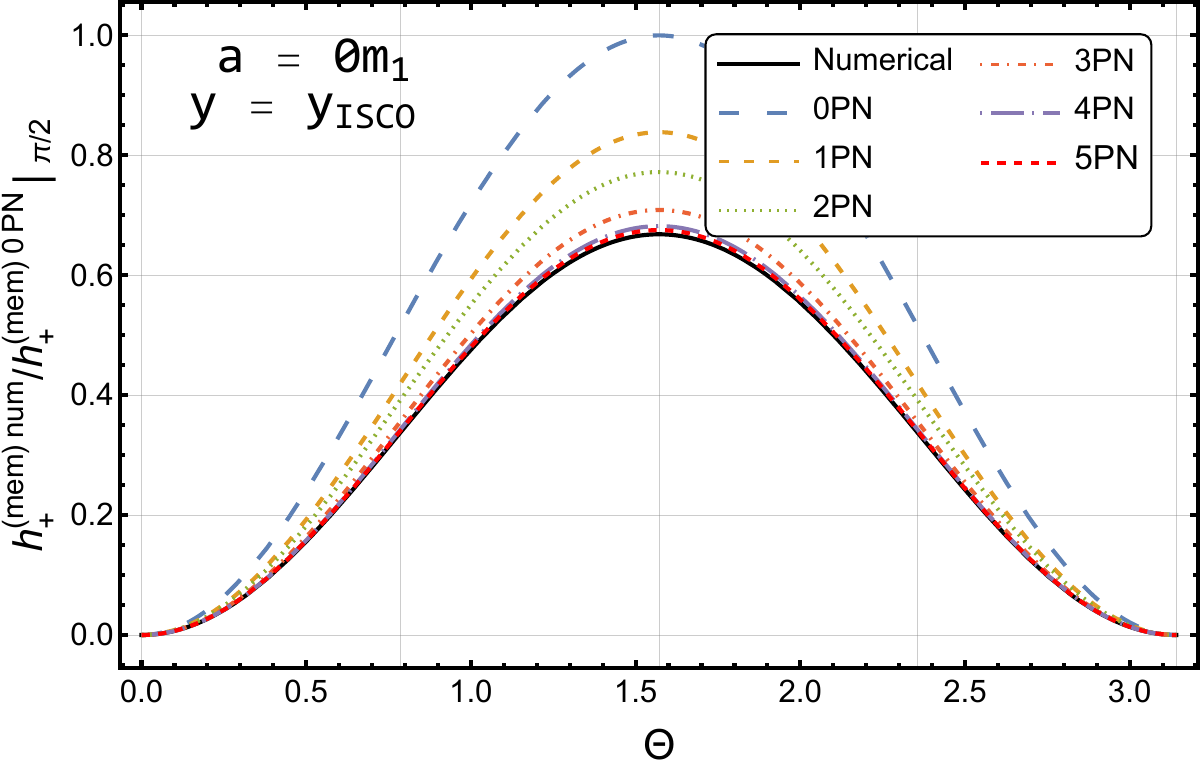}
        \includegraphics[width = 0.5\textwidth]{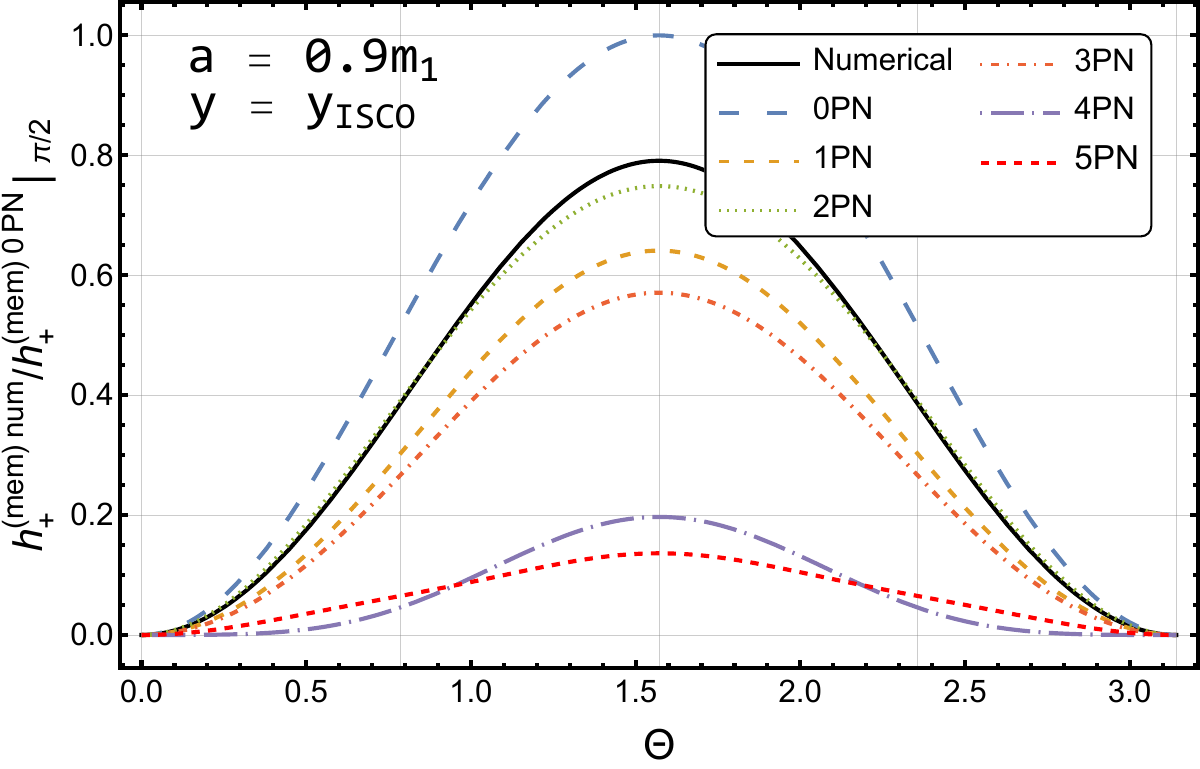}
    \end{tabular}
	\caption{
	(Top row) The (Newtonian normalised) memory accumulated over an inspiral, as measured by an observer in the equatorial plane of the primary black hole (i.e., $\Theta = \pi/2$). 
	The (black) solid curve shows the results of the numerical calculation using equation \eqref{eq:hl0_PN_num}.
	The dotted and dashed curves show the results of the PN-GSF calculation given in section \eqref{sec:PNSF}.
    In both plots, the vertical dot-dashed line shows the Schwarzschild ISCO, i.e., $y = 1/6$. 
   	The top left plot shows the comparison for a Schwarzschild primary.
	Here we see that the 5PN result well approximates the numerical result all the way down to the ISCO.
	The top right plot shows the result for a Kerr primary with $a=0.9m_1$.
	In the strong-field regime past the Schwarzschild ISCO, the PN series converges poorly towards the numerical result.
	(Bottom row) Memory accumulated over an inspiral from infinity to the ISCO, measured as a function of polar position of the observer, $\Theta$.
	The (black) solid curve shows the numerical result.
    This figure is similar to figure~1 of reference~\cite{Favata:2008yd}.
	}
    \label{fig:Inspirals}
\end{figure}

In order to compute the accumulation of memory, we integrate $\dot{h}_{l0}$ over an adiabatic quasi-circular inspiral. 
We do this by integrating equation \eqref{eq:hl0} and using equation \eqref{eq:ydot} to obtain $\dot{y}$, in which we replace
$\mathcal{F}^\text{PN}$ by a numerical flux calculated from the Teukolsky formalism. 
The flux we used was precomputed by Taracchini~\textit{et~al.}~\cite{GremlinFluxes} and is available through the  \texttt{Black Hole Perturbation Toolkit}~\cite{BHPToolkit}. 
As we only have the numerical flux tabulated over a finite range of orbital radii, we truncate the integral in equation~\eqref{eq:hl0} by introducing a finite lower bound, and then estimate the contribution from the rest of the integral using the PN expressions in section \eqref{sec:PNSF}.
The total accumulated memory is thus given by
\begin{equation}\label{eq:hl0_PN_num}
    h_{l0}^\text{(1, mem)}(y) =  h_{l0}^\text{(1, mem),PN}(y_{\rm match})  + \int_{y_{\rm match}}^{y} \frac{\dot{h}_{l0}^\text{(1, mem)}(y')}{\dot y_p(y')} \dd y' \,,
\end{equation} 
where $y_{\rm match}$ is the (inverse) radius at which the PN and numerical results are matched.
In practice, we find that choosing $100 < 1/y_{\rm match} < 200$ is sufficient to ensure a relative accuracy in the memory of $\sim 10^{-6}$.
The comparison of the numerically computed memory with the PN-GSF results are presented in the top row of figure \ref{fig:Inspirals}. 
For inspirals into a Schwarzschild black hole, the 5PN result approximates well the numerical result even close to the ISCO.
For prograde orbits around Kerr black holes, the ISCO moves inward, and we see that the PN expansion is less accurate in the strong-field regime, close to the ISCO.

Finally, we present the dependence of the total accumulated memory at the ISCO on the inclination angle of the binary, $\Theta$, in the bottom row of figure \ref{fig:Inspirals}. 
The memory signal is maximal for $\Theta = \pi/2$ and tends to zero for $\Theta \in\{0,\pi\}$.
This dependence on the inclination angle is opposite to the dominant (2,2) mode, and thus including memory in GW models may help to break the degeneracy between luminosity distance and inclination angle~\cite{Gasparotto:2023fcg,Xu:2024ybt,Goncharov:2023woe}.

\section{Gravitational-wave memory through second order in the mass ratio for quasi-circular inspirals into a Schwarzschild black hole}
\label{Sec:2ndOrderResults}

We now present results for the GW memory through second-order in the mass ratio.
Unlike the previous section, here we focus on inspirals into a non-rotating Schwarzschild black hole.
We first calculate the news before integrating it along a 1PA inspiral to compute the memory.
We compare our results to numerical relativity simulations and find good agreement for mass ratios as small as $q\sim 10$.
We also show that the residual between our second-order results and NR scales in the expected way with the mass ratio.
Finally, we compare with the PN results from section \ref{sec:PNresults}.
Here we do not find agreement on a mode-by-mode basis and we discuss why this might be.

As reviewed in section~\ref{sec:SF_theory}, the calculation of memory at second order in the mass ratio builds upon the framework for computing second-order perturbations that has been developed over the last decade~\cite{Pound:2012nt,Pound:2014xva,Wardell:2015ada,Miller:2020bft,Pound:2019lzj,Warburton:2021kwk,Wardell:2021fyy,Durkan:2022fvm,Miller:2023ers,Spiers:2023mor}.
Specifically, we need the second-order metric amplitudes~\cite{Warburton:2021kwk} and the parametric derivative (with respect to $\Omega$) of the first-order metric perturbation~\cite{Durkan:2022fvm,Miller:2023ers}, from which the second-order effective Teukolsky amplitudes in equation \eqref{eq:Z2eff} can be computed. 
With this in hand, the second-order contribution to the news, $\dot{h}_{l0}^{\text{(2,mem)}}$, can be computed \textit{via} equation~\eqref{eq:InstantaneousMemory} with the replacement given in equation~\eqref{eq:Z1Z1replaceZ1Z2}.

Our computation of waveform amplitudes utilizes the results of the GSF-MPM far-zone formalism in section~\ref{far-zone} and \ref{sec: vfz in Bondi-Sachs}. However, we omit the possible `memory distortion' terms~\eqref{eq: memory distortion GW} in the second-order, oscillatory ($m\neq0$) mode amplitudes. This could limit the accuracy of our second-order memory results and of our frequency evolution that we use to compute the memory $h_{l0}^{(n,\text{mem)}}$ from the news $\dot{h}_{l0}^{(n,\text{mem)}}$. We return to these issues below.

Previous works only computed the second-order metric perturbation for spherical harmonic modes up to $l=5$~\cite{Warburton:2021kwk,vandeMeent:2023ols}.
This proved to be insufficient input for equation~\eqref{Umem}, leading to a large truncation error especially for memory modes with $l \geq 4$. 
To solve this, we computed the second-order metric amplitudes for all modes with $l \le 10$. 
This improved our results for the $l\in\{2,\, 4\}$ memory modes, but was still insufficient for $l\geq6$ (and thus we do not show any results for these modes).
In order to compute the memory along the inspiral, we used the 1PA inspiral as outlined in~\cite{Wardell:2021fyy}.
Finally, in order to improve the comparison with comparable-mass binaries, we re-expand our results in terms of the symmetric mass ratio and the total mass using equations (31) -- (33) of reference~\cite{vandeMeent:2023ols}.

We first compare our second-order memory results with those from numerical relativity.
Until recently it was difficult to extract the $m=0$ modes from NR simulations, as extrapolation techniques were not able to resolve modes of the strain necessary to calculate the modes of the memory~\cite{Mitman:2020pbt}. 
In recent years these issues were resolved by using Cauchy-Characteristic Extraction (CCE) where the metric perturbation at the edge of the computational domain is numerically propagated outward to null infinity~\cite{Mitman:2020pbt}.
NR results including memory have been hybridized with PN results and interpolated across the quasi-circular parameter space for $q \le 8$ in the \texttt{NRHybSur3dq8\_CCE} surrogate model.
Unfortunately, we find the output of this surrogate model to be too noisy to facilitate an accuracy comparison with our results.
Instead, we turn to the BMS balance laws to compute the memory from the radiative modes of the NR simulations~\cite{Mitman:2020bjf}.
Here, we make our comparisons with the ``electric component'' of the memory, $\Delta J^\text{(E)}$, which generally will contain information on BMS charges and the memory~\cite{Moxon:2020gha,Mitman:2020pbt}.
The SXS waveforms we want to extract the memory from involve negligible super-translations and super-Lorenz translations~\cite{Mitman:2020pbt}, meaning the change in the BMS charge in $\Delta J^\text{(E)}$ is dominated by the null memory which we have computed with our GSF methods.
To explicitly extract the electric memory from SXS waveforms, we used the \texttt{waveforms.memory.J\_E function}~\cite{Mitman:2020bjf} in the \texttt{sxs} Python package.

\begin{figure}
    \label{NRq10Comparison}
    \centering
    \begin{tabular}{cc}
         \includegraphics[width =0.5\textwidth]{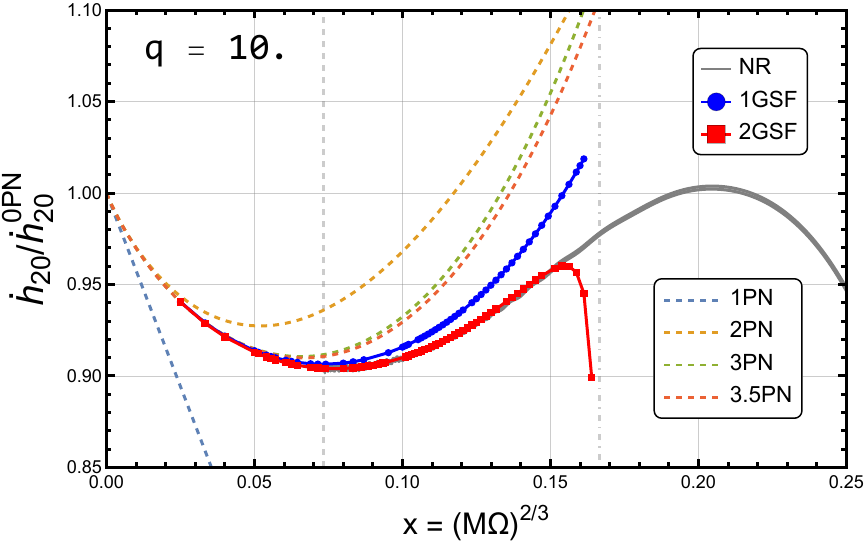}
         \includegraphics[width =0.5\textwidth]{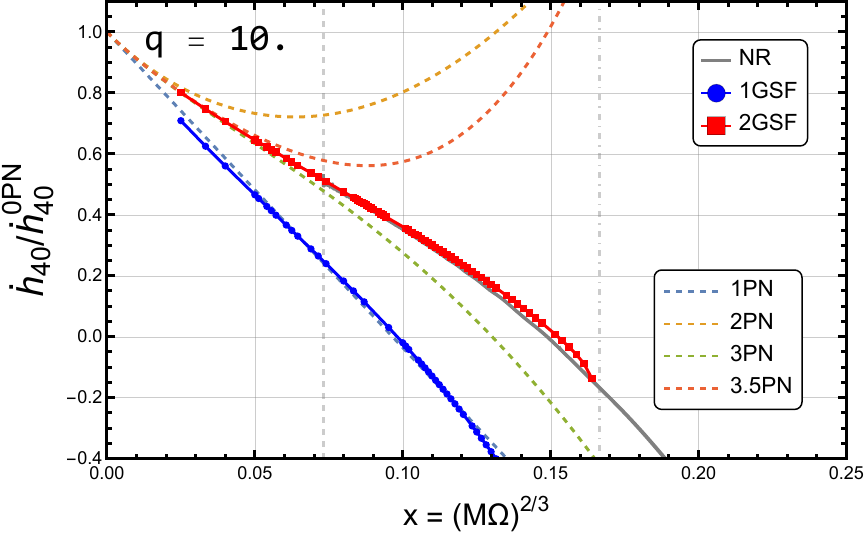}\\
         \includegraphics[width =0.5\textwidth]{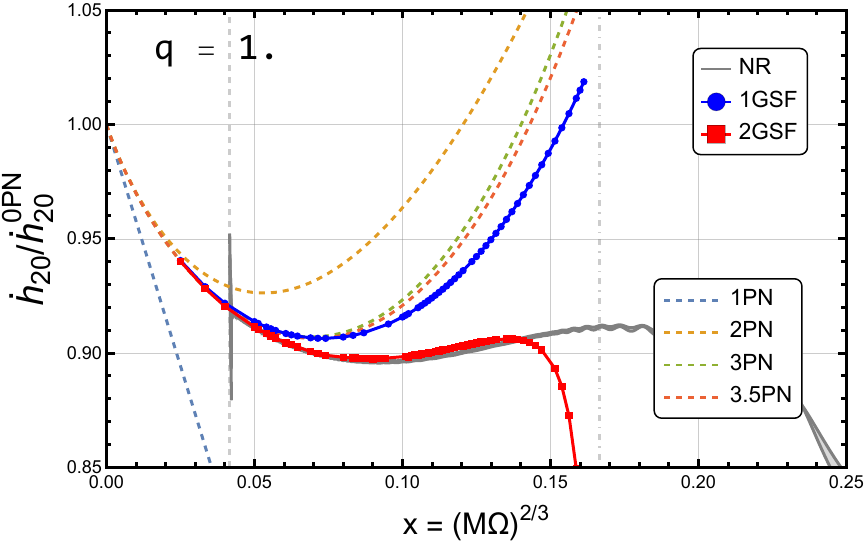}
         \includegraphics[width =0.5\textwidth]{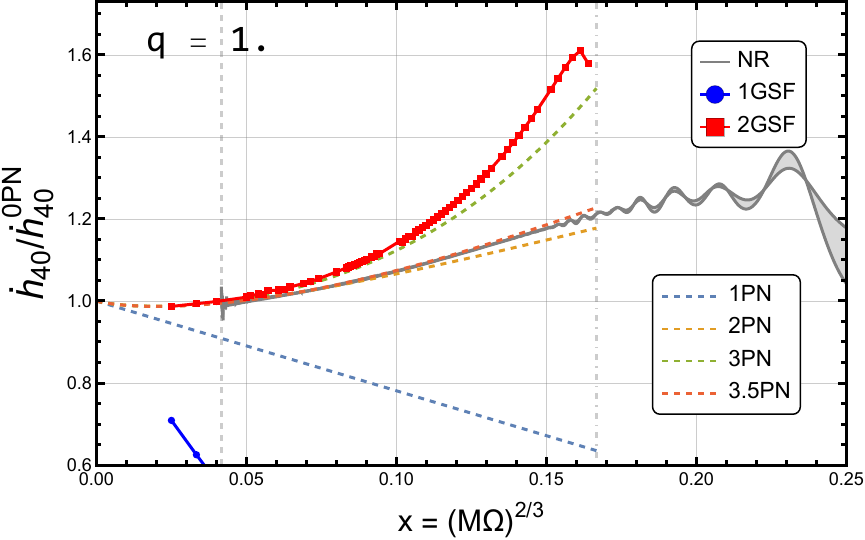}
    \end{tabular}
    \caption{
	Comparisons of $\dot{h}_{l0}$ between NR, PN and GSF for $q=10$ (top row) and $q=1$ (bottom row). 
	 In each plot, the two solid grey lines show the NR results computed by extrapolating the waveform to null infinity with third- and fourth-order polynomials.
	 The shaded grey area between them thus gives an estimate of the error in the NR result.
	 The (blue) circles and (red) squares show our first- and second-order perturbation theory results, respectively.
	 The second-order results agree remarkably well with NR in all cases except $q=1, l=4$.
	 The dashed lines represent PN results up to 3.5PN.
	 The (grey) dot-dashed vertical lines mark the location of the ISCO, while the (grey) dashed vertical lines show where the NR simulations start.
	 The NR data is from simulation SXS:BBH:1107 ($q=10$) and SXS:BBH:1132 ($q=1$).
	 }
~\label{fig:NRMemory}
\end{figure}

The results of the comparison with NR for the $l\in\{2,4\}$ modes and $q\in\{10,1\}$ are shown in figure \ref{fig:NRMemory}.
For $q=10$, we find that the second-order result agrees remarkably well with NR.
For $q=1$, the agreement is still surprisingly good for the $l=2$ mode, although the $l=4$ mode does not agree as well.
It is interesting to note that, for $l >2$, it was possible to resum the flux to improve the agreement with NR~\cite{Warburton:2021kwk}.
The same resummation does not work with the memory modes, but it may be possible to resum the amplitude data before the memory is computed.
We leave this for future work.
In figure \ref{fig:NRMemory}, we also plot the PN results for comparison, including our new 3.5PN results from section \ref{sec:PNresults}.
In the strong-field, we find that the 3.5PN results agree better with the NR results than the 3PN results for the $l=2$ and $q\in\{1,10\}$ cases.
For $l=4$ and $q=1$ the 3.5PN result again performs the best of the PN results, but for $l=4$, $q=10$ it performs worse than 3PN. This is consistent with the well-known alternating nature of the PN series; see, e.g., figure 1 of \cite{Warburton:2024xnr}.

\begin{figure}
    \centering
    \begin{tabular}{cc}
         \includegraphics[width =0.5\textwidth]{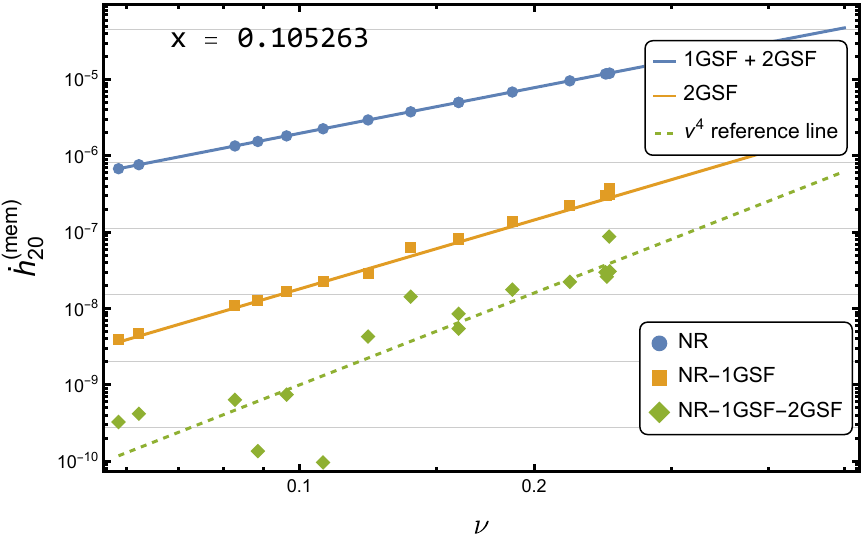}
         \includegraphics[width =0.5\textwidth]{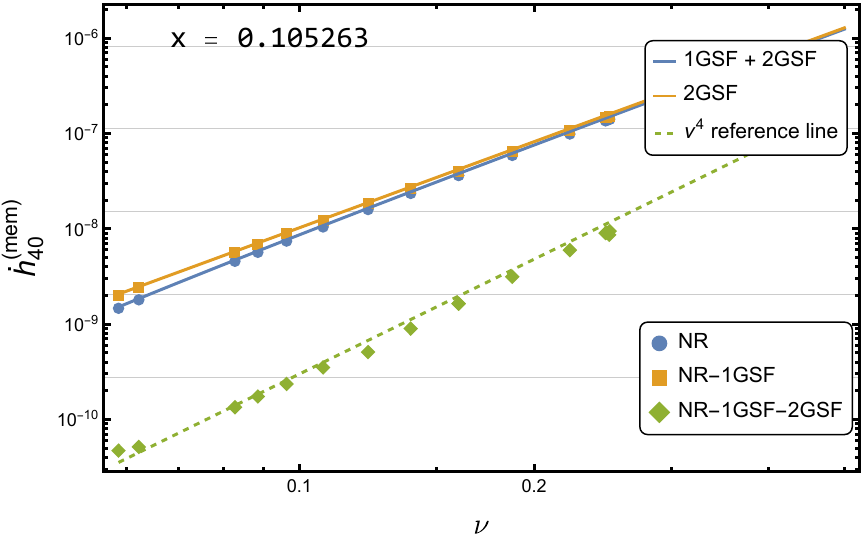}
		 
    \end{tabular}
    \caption[Mass ratio dependence]{
		Behaviour of the Bondi news, $\dot{h}_{l0}^{(\rm mem)}$, as a function of the symmetric mass ratio, $\nu$, for $x=2/19$.
		The (blue) circles show the result from the NR simulation, which closely follows the full GSF result (blue, solid curve) which scales as $\nu^2$.
		After subtracting the first-order result, the residual (orange, squares) follows the 2SF result, which scales as $\nu^3$.
		Finally, after further subtracting the second-order result, the residual (green, diamonds) scales as $\nu^4$, as expected.
		For the $l=2$ mode (left panel) there is significant noise in the residual, which appears to come from small oscillations in the NR waveforms.
		The SXS simulations used to make this figure are listed in table \ref{table:SXSIDs}.
	}
~\label{fig:MassRatio}
\end{figure}

%
In order to check that our results capture the GW memory through second order in the mass ratio, we can compare them with NR simulations for different mass ratios at a fixed frequency.
The time derivative of the memory is compared for the $l\in\{2,4\}$ modes in figure \ref{fig:MassRatio}.
The full memory scales as $\nu^2$ and, after subtracting the first-order self-force (1SF) result, we find that the residual follows closely the 2SF result, which scales as $\nu^3$.
After further subtracting the second-order result, we find that the residual scales as $\nu^4$.
This is most clearly seen for the $l=4$ mode.
In the $l=2$ mode, there is significant noise in the residual which appears to come from small oscillations in the NR waveforms (likely from residual eccentricity and/or centre-of-mass motion in the NR simulation~\cite{Mitman:2021xkq}). This comparison strongly suggests that  there is either no contribution from the omitted `memory distortion' terms or that the numerical magnitude of those terms is small.

\begin{figure}
    \centering
    \includegraphics[width = 1\textwidth]{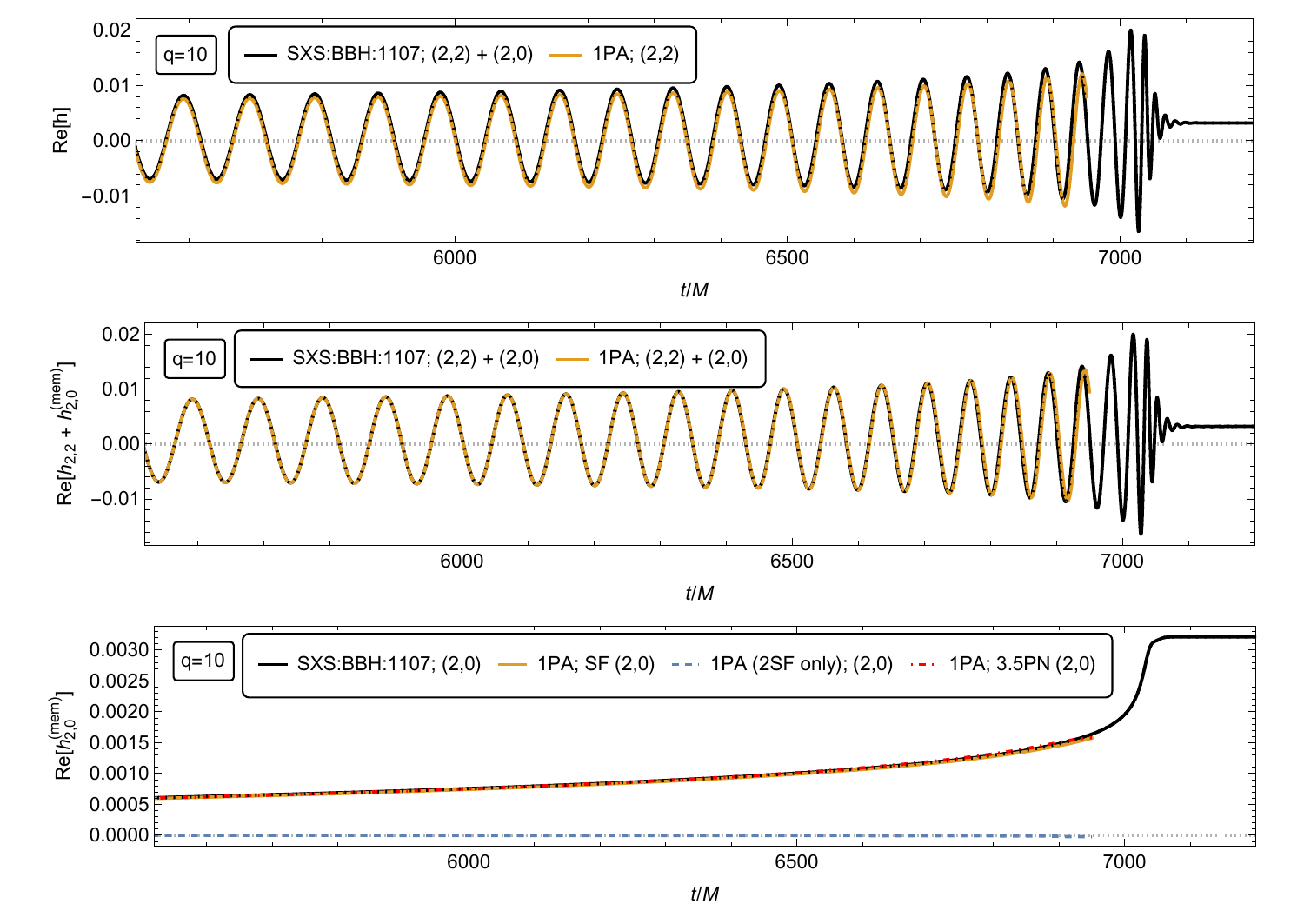}
    \caption{
	Comparison of an NR waveform and the 1PA GSF waveform for a binary with $q=10$.
	In each case, the memory starts to accumulate at the relaxation time $t/M \approx 320$, where the waveforms are also aligned.
	The top panel shows the NR waveform  (in black)  including both the $(2,2)$ and $(2,0)$ modes.
	The 1PA waveform is shown without the contribution from the $(2,0)$ mode, and we see that the amplitudes do not match the NR waveform as the memory accumulates.
	The middle panels shows the same NR waveform, but now the 1PA waveform includes both the $(2,2)$ and $(2,0)$ modes, which brings the amplitudes into close alignment.
	The bottom panel shows just the $(2,0)$ mode for both the NR and 1PA waveforms, which are seen to be in close agreement until near the end of the 1PA waveform, where the transition to plunge sets in and the multiscale expansion breaks down.
	The (blue) dotted curve shows the contribution from just the second-order amplitudes, which is almost negligible.
	The (red) dot-dashed curve shows the result from a hybrid model that combines the 1PA phase with the 3.5PN amplitudes. This hybrid model also compares well with the NR simulation.
	The NR waveform used in this figure is SXS:BBH:1107~\cite{sxs_collaboration_2019_3302023}.
	}
	\label{fig:waveform_q10}
\end{figure}

Next, we compare the NR waveform (including the memory) with the GSF waveform, with and without memory.
The radiative modes, the frequency and the phase evolution of the GSF waveform are computed as outlined in reference~\cite{Warburton:2021kwk}.
The result of the comparison for the $l=2$, $m\in\{0,2\}$ modes for a $q=10$ binary are shown in figure \ref{fig:waveform_q10}.
Both the NR and GSF waveforms are given zero memory contributions at the reference time in the NR simulation.
Without including the $(2,0)$ mode, the amplitude of the GSF waveform does not closely follow the NR waveform as the memory accumulates.
After including the $(2,0)$ mode, we find that the amplitude of the waveforms are closely matched from the reference time to the end of the GSF waveform (where the multiscale expansion breaks down due to the onset of the transition to plunge).
At $q=10$, the first-order contribution to $h^{(1, \rm mem)}_{20}$ dominates the GSF result, with the second-order amplitudes having an almost negligible contribution.
We find that this remains true even for equal mass binaries.
We also consider a model that hybridizes the GSF phasing with the PN amplitudes,  where the frequency evolution is computed using the GSF equation~\eqref{eq:Omegadot}, including the 1PA forcing function $F^{(1)}$, and the memory amplitude at each frequency is computed using our 3.5PN result from equation \eqref{eq:H20PN}.
We find this also agrees well with the NR waveform. 
These results suggest that we only incur small errors, if any, by using the first-law binding energy in the flux-balance law~\eqref{eq:evolution} and in our omission of possible memory distortion contributions to the asymptotic energy flux.

\begin{figure}
    \centering
    \begin{tabular}{cc}
         \includegraphics[width =0.5\textwidth]{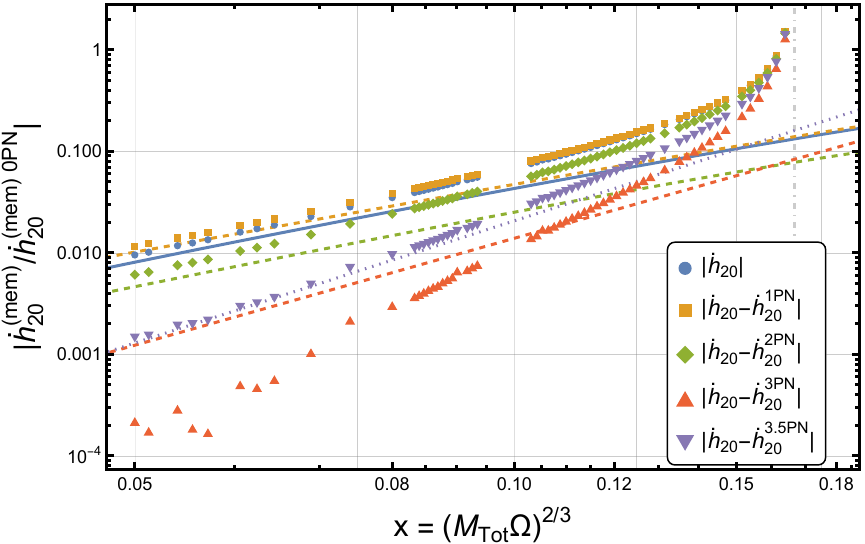}
         \includegraphics[width = 0.5\textwidth]{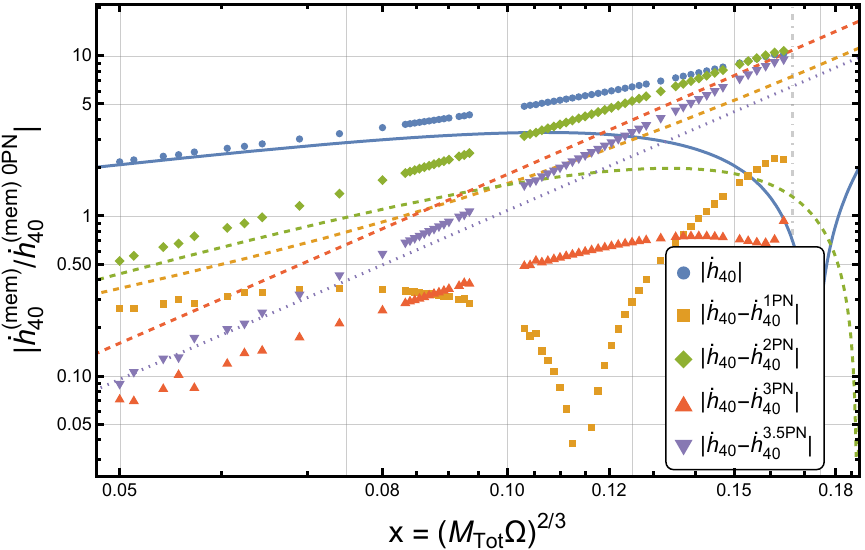}
    \end{tabular}
    \caption[Second Order Memory]
    {The second-order part of the Bondi news computed from Schwarzschild perturbation theory as compared to the PN computation, all normalised by the leading (1SF) PN term. 
    The blue points show the numerical calculation of $\dot{h}_{l0}^\text{mem}$, while the (blue) circles and solid line shows the PN prediction. To get the 
    (yellow) points and dashed line we subtract the leading PN term from the numerical data and the PN series. We then subtract the next-to-leading term 
    to get the (green) diamonds and dashed line. The (orange) triangles and dashed line follow. The (purple) upside-down triangles show the numerical data minus the entire PN series. Since we have no PN expression 
    to  compare at this order, we plot this data with a (purple) dotted $y^{4}$ reference line, showing that our data has approximately the expected subleading behaviour.
    }
~\label{fig:2ndorderMemory3.5PN}
\end{figure}

Finally, we compare our second-order results to the second-order contribution to the PN series from section \ref{sec:PNresults}.
The results of this comparison are presented in figure \ref{fig:2ndorderMemory3.5PN}.
For the $l=2$ and $l=4$ modes, we find agreement with PN up to 2PN order but beyond this there is no clear agreement with either the 3PN terms or our new 3.5PN terms.
This lack of agreement for the individual modes is reminiscent of the lack of agreement  between the modes of the second order and 4.5PN fluxes, as observed in  reference~\cite{Warburton:2024xnr}.
In that work, it was found that comparing the \textit{total} flux  did bring agreement between the PN and GSF results.
The agreement for the total flux but not the individual modes suggests that the GSF and PN waveforms are computed in different frames.
In this work we have only computed the $l=2$ and $l=4$ second-order GSF memory modes (the $l\ge6$ results contain significant numerical error) and so we cannot perform the same check.
Resolving this lack of agreement between the GSF and PN results for the modes of the second-order memory will likely require a combination of (i) understanding the frame the GSF calculation is in, (ii) numerical results for the metric amplitudes for higher $(l,m)$ modes, and (iii) more accurate GSF results at large orbital radii.
As each of these is a significant undertaking, we leave them for future work.

\section{Conclusions}

In this work, we have computed the displacement memory contribution to GWs from quasi-circular compact binaries with a variety of different methods. We have also developed the necessary theoretical framework for memory computations in GSF theory, which has facilitated our calculations and revealed the emergence of memory effects in binary dynamics and waveforms at second order in the mass ratio.

Our explicit computations of  GW memory began by extending the 3PN results of Favata~\cite{Favata:2008yd} to 3.5PN order for spinning, non-precessing, binary  black holes on circular orbits.
We then used GSF theory to compute GW memory contributions at first and second order in the mass ratio.
The first-order memory was computed for inspirals into a spinning black hole both numerically and \textit{via} a 5PN-1SF  expansion.
The second-order results are numerical and restricted to the non-spinning case. 
We find that a 1PA waveform model including memory corrections agrees well with NR simulations with mass ratios as small as $q=10$.
In particular, we observe that including only the first-order memory is sufficient to closely match the NR results.
Similarly, a hybrid model where the waveform phase is computed from the 1PA GSF model but the memory amplitude is computed \textit{via} PN is also in close agreement with NR.

Our 2SF calculations (and all previous ones) rely on a framework of matched asymptotic expansions, in which we match a near-zone multiscale expansion to a far-zone post-Minkowskian expansion. Here we laid out that formalism for the first time, highlighting its  essential role in introducing memory effects into GSF theory. Our analysis demonstrated how first-order GW memory is associated with nonlocal-in-time effects in the second-order near-zone binary dynamics. It also revealed that potential additional `memory distortion' effects might arise in the second-order waveform, potentially contributing to both the oscillatory and memory pieces of the waveform at second order in the mass ratio. While our results for the first-order GW memory are complete, our numerical results at second order might be incomplete through omission of these effects. Our comparisons with NR suggest that the omitted effects are numerically small if they are present, but they bear further investigation. Moreover, we found that analogous effects within the PN-MPM framework  could potentially arise from the `cubic memory' entering formally at 5PN, but in practice contributing to the 2.5PN flux.

There are a number of ways this work could be extended.
In the PN-MPM context, once the missing 4PN waveform amplitudes listed at the end of section \ref{sec:PNresults} are known, the memory can be computed through 4PN order for non-spinning black holes on quasi-circular orbits. A natural extension would be including eccentricity and spin when these are available. Moreover, a full investigation of the `cubic memory', which arises from an interaction between three quadrupole moments, appears to be in order to exclude potential  `memory distortion' terms from the PN-MPM waveform.

In the GSF context, at first order in the mass ratio there are a variety natural of extensions. 
These include:
\begin{itemize}
	\item {\textit{Completing the model with the transition and plunge.}
	Recent work has extended the multiscale expansion to the transition~\cite{Kuchler:2024esj} and plunge regime~\cite{Folacci:2018cic,Silva:2023cer}. 
	The metric amplitudes from these calculations could be used to complete inspiral models developed in this work to include the merger and ringdown phases of the waveform.
	}
	\item{\textit{Extending the PN-GSF expansion beyond 5PN order}. 
	As we see from figure \ref{fig:Inspirals}, 5PN results already capture the numerical result well in the non-spinning case, but for the spinning case, higher-order PN results would be useful in the strong field. 
	These results could be combined with a resummation to capture the behaviour near the ISCO~\cite{Burke:2023lno} to produce a first-order memory amplitude model that is accurate for spinning binaries. 
	As we saw with the comparisons with NR --- see figure \ref{fig:waveform_q10} --- a model including only first-order memory contributions is likely to be sufficient for almost all mass ratios.
	}
	\item {\textit{Exploring near-extremal spins.}
	This work computes the memory for black holes with spins up to $a = 0.9m_1$. For near-extremal spins ($a\gtrsim0.999m_1$) new symmetries appear in the near-horizon regime which allow for new analytic treatments. 
	These were previously exploited to compute the flux and the waveform in this near-extremal, near-horizon regime~\cite{Gralla:2015rpa,Gralla:2016qfw}, and these results could be extended to compute the memory contribution here as well.
	Memory in this regime for an extremal Kerr-Newman black hole was recently explored in reference~\cite{Galoppo:2024vww}.
	}
	\item {\textit{Extending our results beyond quasi-circular inspirals.} 
	This extension is theoretically straightforward for $\dot{\cU}_{lm}$, where one would extend the $Z^{(1)}_{\ell m\omega}$ coefficients of equation~\eqref{eq:psi4Limit} to
	to $Z^{(1)}_{\ell m n k}$, where the change from $\ell m \omega$ indices to $\ell m n k$ corresponds to the change $\omega = m \Omega_\phi\rightarrow \omega = m \Omega_\phi + n \Omega_r + k\Omega_\theta$, where $\Omega_r$ and $\Omega_\theta$ are the frequencies of the radial and azimuthal motion respectively. 
	Many codes now exist that can compute these $Z^{(1)}_{\ell m n k}$ coefficients~\cite{Drasco:2005kz,Fujita:2009us,vandeMeent:2017bcc}, including the open-source \texttt{Teukolsky} package from the Black Hole Perturbation Toolkit~\cite{BHPT_Teukolsky}.
	In practice, the sum over $n$ and $k$ for each $(l,m)$ mode will involve computing hundreds to thousands of modes, which will be computationally burdensome.
	Moreover, the time integration of $\dot{\cU}_{lm}$ requires computing the inspiral, which in practice necessitates an interpolated model of the GW flux across the parameter space.
	Currently, such an interpolated flux model is readily available for eccentric orbits into a non-rotating black hole in the \texttt{FastEMRIWaveforms} code~\cite{Katz:2021yft}.
	The PN-GSF results could also be extended to more generic inspirals using the results of~\cite{Hopper:2015icj, Munna:2020iju, Munna:2023wce} for eccentric orbits, or in~\cite{Isoyama:2021jjd} for generic cases.
	For eccentric orbits the numerical and PN-GSF results could be compared to PN results for eccentric orbits ~\cite{Favata:2011qi, Grilli:2024lfh}.
	}
\end{itemize}

At second order in GSF theory, there are several obvious next steps. One step will be to establish whether the `memory distortion' terms highlighted here modify the waveform and GW flux at second order. Given that these possible corrections cannot enter the local 1PA binary dynamics, as discussed in section~\ref{very-far-zone}, another step will be to further study the validity of the balance law~\eqref{eq:evolution} and of the first-law binding energy at 1PA order. A third step will be to calculate the memory contribution to the second-order binary dynamics (which affects the 2PA orbital evolution) and see if it can shed light on disagreements about the 5PN dynamics, which arise precisely at second order in the mass ratio~\cite{Porto:2024cwd,Almeida:2024klb}. As discussed in reference~\cite{Almeida:2024klb}, disagreements at 5PN might be due to different choices of BMS frame, and analysis of near- and far-zone memory effects might also illuminate the little-understood issue of frame choice in 2SF calculations~\cite{Warburton:2024xnr}.

Once second-order calculations become available for spinning black holes and/or eccentric and precessing orbits, these can also be used to calculate the memory using the framework described in this paper.
In advance of those calculations, the results of this paper can be used as a check on the source for the second-order field equations.
As pointed out in section~\ref{sec:asymptotic_source_and_GW_flux}, the asymptotic behaviour of the second-order Ricci tensor at null infinity can be related to the first-order memory.
Thus, the results for memory from circular orbits around a Kerr black hole presented in this paper can be used as a check on a future source for the second-order field equations in the Kerr spacetime.

Finally, other memory effects could prove to be interesting, for example, centre-of-mass memory or spin memory~\cite{Mitman:2020bjf,Nichols:2017rqr} (Table I of~\cite{Mitman:2020pbt} provides a nice breakdown of the different types of memory). 
These effects are subdominant compared to the displacement memory and are likely to be challenging to detect even with third-generation detectors~\cite{Islam:2021old,Grant:2022bla}.

\section*{Acknowledgements}

We thank David~Nichols and Marc~Favata for useful discussions.
We thank Keefe~Mitman for advice on extracting memory signals from numerical relativity simulations.
We thank Mohammed~Khalil and Quentin~Henry for checks on our \mbox{GSF-PN} results.
AP thanks Luc~Blanchet for numerous discussions of large-$r$ asymptotics that helped initiate our investigations of GW memory in GSF theory. 
DT thanks M.V.S.~Saketh for cross-checks and for pointing out a typo in \cite{Saketh:2022xjb}, Loïc~Honet for cross-checks, and Suprovo~Ghosh for discussions. 
KC acknowledges support from the Irish Research Council under grant number GOIPG/2021/725.
CK and JN acknowledges support from Science Foundation Ireland under Grant number 21/PATH-S/9610.
AP acknowledges the support of a Royal Society University Research Fellowship. AP and DT acknowledge the support of the ERC Consolidator/UKRI Frontier Research Grant GWModels (selected by the ERC and funded by UKRI [grant number EP/Y008251/1]). 
DT also received support from the Czech Academy of Sciences under the Grant No.~LQ100102101.
NW acknowledges support from a Royal Society~-~Science Foundation Ireland University Research Fellowship.
This publication has emanated from research conducted with the financial support of Science Foundation Ireland under Grant numbers 16/RS-URF/3428, 17/RS-URF-RG/3490 and 22/RS-URF-R/3825.
For the purpose of Open Access, the authors have applied a CC BY public copyright licence to any Author Accepted Manuscript version arising from this submission.
This work makes use of the \texttt{Black Hole Perturbation Toolkit}~\cite{BHPToolkit}, in particular the \texttt{SpinWeightedSpheroidalHarmonics}~\cite{BHPT_SWSH}, \texttt{Teukolsky}~\cite{BHPT_Teukolsky}, \texttt{KerrGeodesics}~\cite{BHPT_KerrGeodesics}, and \texttt{PostNewtonianSelfForce}~\cite{niels_warburton_2023_8112975} packages and \texttt{GremlinEQ}~\cite{Gremlin}.
For the PN-GSF calculations we also made use of the \texttt{SFPN} package \cite{JakobSFPN}. For the PN-MPM computations, most computations were performed using the \texttt{xAct} library~\cite{xtensor} for \texttt{Mathematica}.
We thank Ian~Hinder and Barry~Wardell for the \texttt{SimulationTools} analysis package~\cite{SimulationTools}.

\appendix

\addtocontents{toc}{\protect\setcounter{tocdepth}{1}}

\section{Transformation to a Bondi-Sachs gauge}\label{sec: vfz in Bondi-Sachs}

To characterize the contributions to the waveform in section~\eqref{very-far-zone}, we transform the metric~\eqref{total far-zone metric} to a Bondi-Sachs coordinate system $X^\alpha=(U,R,\Theta^A)$. This eliminates all the logarithmic contributions to the very-far-zone metric, and it isolates the waveform content in the Bondi shear $C_{AB}$. In this appendix we provide the transformation to a Bondi-Sachs gauge. 

The Bondi-Sachs gauge conditions read 
\beq\label{Bondi-Sachs gauge}
g_{RR}=g_{RA}=0 \quad \text{and} \quad\Omega^{AB}g_{AB}=\Omega^{AB}\eta_{AB}=2R^2. 
\eeq
These conditions ensure that surfaces of constant $U$ are outgoing null cones of the full spacetime, curves of constant $(U,\Theta^A)$ are outgoing null rays parameterized by $R$, and surfaces of constant $(U,R)$ have the surface element $R^2d\Omega_2$ (and hence surface area $4\pi R^2$). Completing the analysis in section~\eqref{very-far-zone} will only require imposing these conditions at order $1/R$. Like in section~\eqref{very-far-zone}, we limit our attention to $\O(\e)$ for quasistationary terms and to $\O(\e^2)$ for oscillatory terms.

The transformation will combine an $\O(\e^0)$ transformation of the background coordinates along with $\O(\e^n)$ gauge transformations. Under a perturbative gauge transformation generated by a vector $\e\xi^\alpha$, the metric $g_{\alpha\beta}=\mathring{g}_{\alpha\beta}+\sum_{n>0}\e^n h^{[n]}_{\alpha\beta}$ transforms to~\cite{Bruni:1996im,Pound:2015fma}%
\begin{align}\label{second-order gauge transformation}
g_{\alpha\beta}' &= g_{\alpha\beta} + \e{\cal L}_{\xi} \mathring{g}_{\alpha\beta} + \e^2\left(\frac{1}{2}{\cal L}_{\xi}{\cal L}_{\xi}\mathring{g}_{\alpha\beta} + {\cal L}_{\xi} h^{[1]}_{\alpha\beta}\right) + \O(3),
\end{align}
where $\cal L$ denotes a Lie derivative. This corresponds to a coordinate transformation
\beq
x'^\alpha = x^\alpha - \e\xi^\alpha +\frac{1}{2}\e^2\xi^\beta\partial_\beta \xi^\alpha + \O(3) .
\eeq

\subsection{Background coordinates}

We first note that the background metric $\mathring{g}_{\alpha\beta}$ itself contains $\ln(r_*)/r_*$ terms due to our use of $r_*$ as a radial coordinate. These logarithms appear in the angular components of the Schwarzschild metric: $r^2(d\Omega_2)^2 = r^2_*[1+\ln (r_*)/r_*](d\Omega_2)^2$. To eliminate them, we revert to ordinary retarded background coordinates $(u,x^i)$, with $x^i=r n^i$. This puts the background metric in the form 
\beq\label{background in Bondi}
\mathring{g}_{\alpha\beta} = \eta_{\alpha\beta} +\frac{2\mathring{m}_1}{r} k_{\alpha}k_{\beta} + \O(1/r^2),
\eeq
with the Minkowski metric components
\beq
\eta_{uu}=-1,\ \eta_{ui} = -n_i,\ \text{and}\ \eta_{ij}=\delta_{ij}-n_i n_j.
\eeq

In the coordinates $(u,x^i)$, the first-order mass perturbation $\frac{2 M^{(1)}_{\rm B} }{r_*}\delta_{\alpha\beta}$ becomes $\frac{2 M^{(1)}_{\rm B} }{r}p_{\alpha\beta}$, where $p_{\alpha\beta}$ has components 
\beq
p_{uu}=1,\ p_{ui} = n_i,\ \text{and}\ p_{ij}=\delta_{ij}+n_i n_j.
\eeq
Otherwise, the only effect of transforming our expressions to $(u,x^i)$ coordinates is that we simply replace $r_*$ with $r$; this replacement only changes subleading terms of order $\ln(r)/r^2$. The total metric, given by the sum of equations \eqref{eq:gosc Lorenz} and \eqref{eq:gstat Lorenz}, then reads
\begin{equation}\label{g after background transformation}
g_{\alpha\beta} = \eta_{\alpha\beta} +\frac{2\mathring{m}_1  k_{\alpha}k_{\beta} +2\e M^{(1)}_{\rm B}p_{\alpha\beta}}{r} + \frac{\e z^{(1)}_{\alpha\beta}+\e^2 z^{(2)\cal R}_{\alpha\beta}}{r}+ \e^2 j^{\cal P}_{\alpha\beta}  + \ldots,
\end{equation}
where the ellipses indicate the following omitted terms: $\O(3)$ terms (meaning $h^{[3]}_{\alpha\beta}$ and beyond), $\O(\e^2)$ quasistationary terms, $\O(\e^3)$ oscillatory terms, and $\O(r^{-2}\ln r)$ terms. We consistently use ellipses in this way below.

\subsection{Transformation of the mass perturbation}

We next put the mass perturbation $\frac{2\e M^{(1)}_{\rm B}}{r}p_{\alpha\beta}$ into the Bondi-Sachs gauge. An appropriate generator is given by
\beq\label{xi vec dM}
\xi^u =2 M^{(1)}_{\rm B}\ln (r/P) \quad\text{and}\quad \xi^i = - M^{(1)}_{\rm B}\,n^i.
\eeq
Here we have introduced an arbitrary length scale $P$ to adimensionalize the argument of the logarithm. We will return to this scale below. 

Applying equation~\eqref{second-order gauge transformation}, we arrive at
\begin{equation}\label{g after mass transformation}
g''_{\alpha\beta} = \eta_{\alpha\beta} +\frac{2M_{\rm B}}{r} k_{\alpha}k_{\beta} + \frac{\e z^{(1)}_{\alpha\beta}+\e^2 z^{(2)\cal R}_{\alpha\beta}}{r}+ \e^2 \left(j^{\cal P}_{\alpha\beta} + \frac{\xi^u \dot z^{(1)}_{\alpha\beta}}{r}\right)  +\ldots,
\end{equation}
where we have appealed to equation~\eqref{background in Bondi} and identified the total Bondi mass
\beq
M_{\rm B} = \mathring{m}_1+\e M^{(1)}_{\rm B} + \O(\e^2).
\eeq
The manifestly nonlinear terms in equation~\eqref{second-order gauge transformation} have all cancelled, up to $\O(\e^3,1/r^2)$ remainders, except the term $\propto \xi^u \dot z_{\alpha\beta}$. 


\subsection{Transformation of oscillatory terms}


Turning to the oscillatory terms in the metric, we first note that for oscillatory terms of order $1/r$, the Lorenz gauge condition already enforces the Bondi-Sachs gauge condition; compare equation~ \eqref{gauge z polar} to equation~\eqref{Bondi-Sachs gauge}. Hence, in equation~\eqref{g after mass transformation}, the quantities $z^{(1)}_{\alpha\beta}$ and $z^{(2)\cal R}_{\alpha\beta}$ do not need to be transformed, leaving only $j^{{\cal P},\rm osc.}_{\alpha\beta}+\frac{1}{r}\xi^u \dot z^{(1)}_{\alpha\beta}$ to transform, where $j^{{\cal P},\rm osc.}_{\alpha\beta}$ is the oscillatory part of $j^{\cal P}_{\alpha\beta}$.


Transforming these terms into the Bondi-Sachs gauge requires knowledge of the tensorial structure of $j^{{\cal P},\rm osc.}_{\alpha\beta}$. We could start with the expression $j^{{\cal P},\rm osc.}_{\alpha\beta}=r^{-1}\ln (r)a^{{\cal P},\rm osc.}_{\alpha\beta}$, with $a^{{\cal P},\rm osc.}_{\alpha\beta}$ given by equation~\eqref{aPosc}, after putting $s_{\alpha\beta}$ in the form $\sum_{l\geq0}s_{\alpha\beta L}\nhat^L $, starting from equation~\eqref{s-form}. Alternatively, we can obtain $j^{{\cal P},\rm osc.}_{\alpha\beta}$ by noting the tensorial structure of the source in equation~\eqref{s-form Bondi} and seeking a solution to
\beq\label{j osc eqn}
\Box_\eta^{(0)} \left[j^{{\cal P},\rm osc.}_{\alpha\beta} +\O(r^{-2}\ln r)\right] = \frac{s^{\rm osc.}_{uu}k_\alpha k_\beta + s_{AB}\Omega^A_\alpha \Omega^B_\beta}{r^2}
\eeq
with an ansatz 
\beq\label{jP osc ansatz}
j^{{\cal P},\rm osc.}_{\alpha\beta} = -\frac{\ln(r/P)}{2r}\left[{\cal Z}_{uu}(u,\theta^C)k_\alpha k_\beta+ {\cal Z}_{AB}(u,\theta^C)\Omega^A_\alpha \Omega^B_\beta\right],
\eeq
again adimensionalizing the argument of the logarithm using the arbitrary scale $P$. Note that $s_{AB}$ is purely oscillatory, meaning $s_{AB}=s^{\rm osc.}_{AB}$, and recall that $\Box_\eta^{(0)}$ is the flat-space d'Alembert operator neglecting $\dot \Omega$ terms. 

In either approach, we straightforwardly find 
\beq
\dot {\cal Z}_{uu} = s^{\rm osc.}_{uu} \quad \text{and} \quad \dot {\cal Z}_{AB} = s_{AB}, 
\eeq
where overdots should be understood as $\Omega\partial_{\phi_p}$. Substituting $s^{\rm osc.}_{uu}$ and $s^{\rm osc.}_{AB}$ from equation~\eqref{s-form Bondi} then yields
\beq\label{j osc soln}
j^{{\cal P},\rm osc.}_{\alpha\beta} = \frac{\ln(r/P)}{2r}\left\{\left[\partial_u^{-1}\Pi^{\rm osc.}- \left(r^{-4}z^{AB}_{(1)}\dot z^{(1)}_{AB}\right)^{\rm osc.}\right] k_\alpha k_\beta -4M^{(1)}_{\rm B}\dot z^{(1)}_{AB}\Omega^A_\alpha\Omega^B_\beta \right\}.
\eeq
Here we have introduced $\partial_u^{-1}$ as the inverse of $\partial_u $ that acts on oscillatory functions  as 
\beq
\partial_u^{-1} f^{\rm osc.} = -\sum_{m\neq0}\frac{1}{im\Omega}f^m e^{-im\phi_p}.
\eeq

To evaluate $j^{{\cal P},\rm osc.}_{\alpha\beta} + \frac{1}{r}\xi^u \dot z^{(1)}_{\alpha\beta}$ we now combine equation~\eqref{j osc soln} for $j^{{\cal P},\rm osc.}_{\alpha\beta}$ with 
\beq
z^{(1)}_{\alpha\beta} = z^{(1)}_{uu}k_\alpha k_\beta - 2z^{(1)}_{ur}k_{(\alpha} r_{\beta)}  - 2 z^{(1)}_{uA}k_{(\alpha}\Omega^A_{\beta)} + z^{(1)}_{AB} \Omega^A_\alpha\Omega^B_\beta,
\eeq
where $r_\beta = \partial_\beta r$, along with equation~\eqref{xi vec dM} for $\xi^u$. The result is
\begin{multline}
j^{{\cal P},\rm osc.}_{\alpha\beta} + \frac{\xi^u \dot z^{(1)}_{\alpha\beta}}{r} = 
\frac{\ln (r/P)}{r}\Biggl\{\frac{1}{2}\left[\partial_u^{-1}\Pi^{\rm osc.} - \left(r^{-4}z^{AB}_{(1)}\dot z^{(1)}_{AB}\right)^{\rm osc.}\right] k_\alpha k_\beta 
\\*
+ 2M^{(1)}_{\rm B} \left(\dot z^{(1)}_{uu}k_\alpha k_\beta - 2\dot z^{(1)}_{ur}k_{(\alpha} r_{\beta)}- 2\dot z^{(1)}_{uA}k_{(\alpha}\Omega^A_{\beta)}\right)\Biggr\}.
\end{multline}
We eliminate the logarithms using a gauge transformation generated by $\zeta^\alpha$ with 
\begin{subequations}%
\begin{align}
\zeta^u &= \frac{2\ln(r/P)}{r} M^{(1)}_{\rm B} z^{(1)}_{ur},\\
\zeta^r &= \frac{\ln(r/P)}{2r}\partial^{-1}_u\left[\frac{1}{2}\partial^{-1}_u\Pi^{\rm osc.}- \frac{1}{2} \left(r^{-4}z^{AB}_{(1)}\dot z^{(1)}_{AB}\right)^{\rm osc.}+2M^{(1)}_{\rm B}\dot z^{(1)}_{uu}\right]-\zeta^u,\\
\zeta^A &= -\frac{2\ln( r/P)}{r^3} M^{(1)}_{\rm B} \Omega^{AB}z^{(1)}_{uB}.  
\end{align}
\end{subequations}
This leaves the following $\O(\e^2)$ oscillatory contribution to the metric:
\begin{equation}
\e^2\left(\frac{z^{(2)\cal R}_{\alpha\beta}}{r} +j^{{\cal P}, \rm osc.}_{\alpha\beta} + \frac{\xi^u \dot z_{\alpha\beta}}{r} + {\cal L}_\zeta \eta_{\alpha\beta}\right)=  
\frac{\e^2z^{(2)\cal R}_{\alpha\beta}}{r}+\O\!\left(\e^3,\frac{\ln r}{r^2}\right).
\end{equation}

Our total metric~\eqref{g after mass transformation} now reads
\begin{equation}\label{g after osc transformation}
g'''_{\alpha\beta} = \eta_{\alpha\beta} +\frac{2M_{\rm B}}{r} k_{\alpha}k_{\beta} + \frac{\e z^{(1)}_{\alpha\beta}+\e^2 z^{(2)\cal R}_{\alpha\beta}
}{r}+ \e^2 \<j_{\alpha\beta}\> +\ldots
\end{equation}
Although it might not be obvious, the metric now depends on the choice of scale $P$, while it did not previously. In practice, $P$ is chosen when specifying $j^{{\cal P},\rm osc.}_{\alpha\beta}$ because $j^{{\cal P},\rm osc.}_{\alpha\beta}$ is used as input in the field equation~\eqref{BoxhRes} for the near-zone field variable $h^{(2)\cal R}_{\alpha\beta}$. Whichever choice of $P$ is used in $j^{{\cal P},\rm osc.}_{\alpha\beta}$, the same $P$ is used in the gauge generator~\eqref{xi vec dM}. Our numerical results use $P=\mathring{m}_1$, but any other choice is possible. The physical field $h^{(2)\cal P}_{\alpha\beta}+h^{(2)\cal R}_{\alpha\beta}$ is invariant under this choice: through the field equation~\eqref{BoxhRes}, $h^{(2)\cal R}_{\alpha\beta}$ depends on $P$, and a change of $P$ simply exchanges terms between $h^{(2)\cal P}_{\alpha\beta}$ and $h^{(2)\cal R}_{\alpha\beta}$, leaving the sum invariant. This type of invariance under change in puncture is established in section~VC of reference~\cite{Miller:2023ers}. However, after we gauge away the contribution of $j^{{\cal P},\rm osc.}_{\alpha\beta}$ to arrive at the metric in the form~\eqref{g after osc transformation}, we now have no counterbalance to the $P$ dependence of $h^{(2)\cal R}_{\alpha\beta}$, meaning our extracted waveform depends on $P$. 

This dependence on $P$ is precisely analogous to the dependence on the arbitrary scale $b_0$ in PN-MPM calculations, discussed around equation~\eqref{eq:aux_phase}. Like $P$, $b_0$ arises as an arbitrariness in relating the coordinate system used for concrete calculations (the harmonic coordinate system in the PN-MPM context) to a radiative coordinate system used to extract waveforms~\cite{Blanchet:2013haa}. We discuss the implications of this below.


\subsection{Transformation of nonlinear quasistationary term}

Our final task is to transform $\<j_{\alpha\beta}\>=\Box_\eta^{-1}[r^{-2}_*\<s_{\alpha\beta}\>]$ to the Bondi-Sachs gauge. The large-$r$ expansion of this term is given  in Cartesian coordinates $(t,x^i)$ by equation~\eqref{very-large-r}, or somewhat more explicitly by~\eqref{very-large-r m=0}, where the coefficients $s_{\alpha\beta L}$ in the source can be read off equation~\eqref{s0 source STF}. In this section we follow an analogous calculation in reference~\cite{Blanchet:1992br}, omitting many details for that reason.

We transform $\<j_{\alpha\beta}\>$ with a generator
\beq\label{zeta}
\e^2\lambda^\alpha = -\e^2\Box^{-1}_{\eta}\left(\frac{k^\alpha}{2r^2}\int^u_{-\infty}\< \Pi(u',n^a)\>du'\right).
\eeq
We can apply the same large-$r$  expansion~\eqref{very-large-r} to $\lambda^\alpha$, after using $k^\alpha = (1,n^i)$ and equation~(2.7) of \cite{Damour:1990gj} to decompose~$n^i\nhat^L$ into STF tensors. We then find, in $(t,x^i)$ coordinates,
\begin{subequations}
\begin{align}
\lambda^t &= -\frac{1}{2r}\sum_{l\geq0} \nhat_L k_l\!\left[\int^u_{-\infty}\!\< \Pi_L(u')\>du'\right] + \O\left(\frac{\ln r}{r^2}\right),\\
\lambda^a &= -\frac{1}{2r}\sum_{l\geq0}\Biggl\{\nhat_{aL}k_{l+1}\!\left[\int^u_{-\infty}\! \<\Pi_L(u')\>du'\right] \nonumber\\*
&\qquad + \frac{l}{2l+1}\nhat_{L-1}k_{l-1}\!\left[\int^u_{-\infty}\! \<\Pi_{aL-1}(u')\>du'\right]  \Biggr\}+ \left(\frac{\ln r}{r^2}\right),
\end{align}
\end{subequations}
where for convenience we have introduced
\begin{equation}
k_l[f_L(u)] \equiv \int_0^\infty\!\!\!d z\,\left(\frac{1}{2}\ln\frac{z}{2r_*}+H_l\right) f_L(u-z). 
\end{equation}

We see that $\lambda^\alpha$ is of order $1/\e^2$ due to the double integration over $u$, such that $\e^2\lambda^\alpha=\O(\e^0)$. However, because $\e^2\lambda^\alpha$ is of order $\ln(r)/r$, it can still be treated as a perturbative transformation. Moreover, it can only contribute at order $1/r$ when a $u$ derivative acts on it, demoting its impact on the metric to $\O(\e/r)$. Since we limit our analysis to $\O(\e)$ terms for the quasistationary part of the metric, we only require ${\cal L}_\lambda\eta_{\alpha\beta}=2\partial_{(\alpha} \lambda_{\beta)}$ in the transformation law~\eqref{second-order gauge transformation}; all other terms are either $\O(\e^2)$ or $\O(\e r^{-2}\ln r)$.

Calculating $\<j_{\alpha\beta}\>' = \<j_{\alpha\beta}\> + 2\partial_{(\alpha} \lambda_{\beta)}$ involves the following steps: (i) combining equations~\eqref{very-large-r} and~\eqref{s0 source STF} to calculate $ \<j_{\alpha\beta}\>$; (ii) noting that $1/r$ terms in $\partial_{(\alpha} \lambda_{\beta)}$ only arise when derivatives act on $u$, and that these derivatives pass through $k_l$ to collapse the integrals over $u$; (iii) using $\partial_\alpha u = -k_\alpha = (-1,n_a)$ and equation~(2.7) of \cite{Damour:1990gj} to decompose~$\partial_{(\alpha} \lambda_{\beta)}$ into STF tensors $\nhat_L$; (iv) summing $\<j_{\alpha\beta}\> + 2\partial_{(\alpha} \lambda_{\beta)}$ and writing all $\nhat_L$ combinations in terms of $n_L$ using equations~(2.7), (2.13), and (2.14) of \cite{Damour:1990gj}. The result of these straightforward but lengthy manipulations is
\begin{subequations}
\begin{align}
\<j_{tt}\>' &= \O(r^{-2}\ln r),\\
\<j_{ta}\>' &=\frac{1}{2r}\sum_{l\geq0}\frac{1}{l+1}\int_{-\infty}^u du'\<n_{aL}\Pi_L(u') - n_{L-1}\Pi_{aL-1}(u')  \>+ \O(r^{-2}\ln r)  ,\\
\<j_{ab}\>' 
&= -\frac{1}{r}\sum_{l\geq0}\frac{1}{(l+1)(l+2)}\int_{-\infty}^u du'\Bigl\langle(l+1)n_{abL}\Pi_L(u')  -(l-2)n_{L-1(a} \Pi_{b)L-1}(u')  \nonumber\\*
&\qquad\qquad - \delta_{ab}n_L\Pi_L(u') - 2n_{L-2}\Pi_{abL-2}(u') \Bigr\rangle+ \O(r^{-2}\ln r).
\end{align}
\end{subequations}
Converted to $(u,x^a)$ coordinates, the components are
\begin{subequations}\label{<j> BS}%
\begin{align}
\< j_{uu}\>' &= \<j_{tt}\>' = \O(r^{-2}\ln r),\\
\<j_{ua}\>' &= \<j_{ta}\>' +\<j_{tt}\>' n_a  \nonumber\\*
&=-\frac{1}{2r}\sum_{l\geq1}\frac{n_{L-1}}{l+1}\perp_{ai}\int_{-\infty}^u du' \<\Pi_{iL-1}(u')  \>+ \O(r^{-2}\ln r),\\ 
\<j_{ab}\>' &= \<j_{ab}\>' + 2 \<j_{t(a}\>'n_{b)} + \<j_{tt}\>'n_a n_b \nonumber\\
&=\frac{1}{r}\sum_{l\geq2}\frac{2n_{L-2}}{(l+1)(l+2)}\perp^{\rm TT}_{ab,ij}\int_{-\infty}^u du'\<\Pi_{ijL-2}(u')\>+ \O(r^{-2}\ln r),
\end{align} 
\end{subequations}
where we note that  $j_{ab}$ on the left-hand side denotes components in $(u,x^a)$ coordinates, while $j_{ab}$ on the right-hand side denotes components  in $(t,x^a)$ coordinates. Here $\perp_{ai}$ and $\perp^{\rm TT}_{ab,ij}$ are the projectors defined in equation~\eqref{eq:TT projector}; these projectors are not inserted by hand but arise from the transformation to retarded coordinates.

It is now trivial to check that $\<j_{\alpha\beta}\>'$ in equation~\eqref{<j> BS} satisfies the Bondi-Sachs gauge conditions~\eqref{Bondi-Sachs gauge}. Translated to Cartesian coordinates $(u,x^a)$, these conditions read 
\beq
\<j_{ab}\>' n^b=0=\<j_{ab}\>'\delta^{ab}.
\eeq
We immediately see these are satisfied due to the TT projector $\perp^{\rm TT}_{ab,ij}$ in equation~\eqref{<j> BS}. Explicitly, the nonzero polar components are 
\beq
\<j_{uA}\>' = \Omega^a_A \<j_{ua}\>' \quad\text{and}\quad \<j_{AB}\>'=\Omega_A^a\Omega^b_B\<j_{ab}\>', 
\eeq
where $\Omega^a_A \equiv \frac{\partial x^a}{\partial\theta^A} = r\frac{\partial n^a}{\partial\theta^A}$ automatically projects out components along $n^a$, such that 
\beq
\Omega^a_A\perp_{ai}\,=\Omega^i_A \quad \text{and}\quad \Omega^a_A\Omega^b_{B}\perp^{\rm TT}_{ab,ij}\,=\Omega^{(i}_A\Omega^{j)}_B-\frac{1}{2}\Omega_{AB}\delta^{ij}.
\eeq


\subsection{Summary: waveform and phase redefinition}

Having completed our transformation, we now relabel our final, Bondi-Sachs coordinates as $X^\alpha = (U,R,\Theta^A)$. They are related to the original coordinates $(u,r,\theta^A)$ by 
\beq
X^\alpha = x^\alpha - \e\xi^\alpha -\e^2\left(\zeta^\alpha+\lambda^\alpha -\frac{1}{2}\xi^\beta\partial_\beta \xi^\alpha\right) + \O(3) .
\eeq
Our total metric in these coordinates is obtained by combining equations~\eqref{g after osc transformation} and \eqref{<j> BS}. It reads
\begin{equation}\label{g Bondi form}
g^{\rm BS}_{\alpha\beta} = \eta_{\alpha\beta} +\frac{2M_{\rm B}(U)}{R}k_{\alpha}k_{\beta} + \frac{\<c_{\alpha\beta}(U,\Theta^A)\>}{R}  + \frac{z_{\alpha\beta}(U,\Theta^A)}{R} +\ldots
\end{equation}
Here we have defined $z_{\alpha\beta} \equiv \e z^{(1)}_{\alpha\beta}+\e^2 z^{(2)\cal R}_{\alpha\beta}$ as the total oscillatory coefficient of $1/R$. Similarly, we have defined $\<c_{\alpha\beta}(U,\Theta^A)\>$ as the total non-oscillatory, $l>0$ coefficient of $1/R$: 
\beq
\<c_{\alpha\beta}\> =\e \<c^{(1)}_{\alpha\beta}\>+\O(\e^2) =\e^2 R\<j_{\alpha\beta}\>' +\O(\e^2,1/R), 
\eeq
noting that $\<j_{\alpha\beta}\>'=\O(1/\e)$ due to the memory-type integrals in equation~\eqref{<j> BS}. This term's contribution to the shear is $\<C_{AB}\> = R^{-2}\<c_{AB}\>$.

Equation~\eqref{g Bondi form} confirms that the near-zone $1/r$ coefficients $z^{(1)}_{\alpha\beta}$ and $z^{(2)\cal R}_{\alpha\beta}$ directly yield the asymptotic oscillatory part of the waveform, as promised in section~\ref{very-far-zone}, up to possible $\O(\e^2)$ memory distortion terms that arise from the omitted $\O(3)$ pieces of the metric. 

We can likewise now confirm that $\<j_{ab}\>'$ represents the first-order GW memory. Equation~\eqref{<j> BS} is precisely the linear-in-mass-ratio part of the GW memory as written in equation~(224) of reference~\cite{Blanchet:2013haa}. (The sign difference arises from use of the field variable $h^{\alpha\beta} = \sqrt{-g}g^{\alpha\beta}-\eta^{\alpha\beta}$ in reference~\cite{Blanchet:2013haa}.) One can rewrite equation~\eqref{<j> BS} in the form~\eqref{eq:NonlinearMemoryMultipolar}, noting the relationship between $\Pi$ and the GW flux in equation~\eqref{Pi-Edot}. We can  read off the memory contribution to the radiative multipole moment ${\cal U}_{L}$ by comparing equation~\eqref{<j> BS} to equation~\eqref{eq:metricRadiativeMoments}, or read off the shear $\<C^{(1)}_{AB}\> = \lim_{R\to\infty}(\e R^{-1}\<j_{AB}\>')$.

Finally, we address the dependence on the arbitrary scale $P$, as discussed below equation~\eqref{g after osc transformation}. Equation~\eqref{jP osc ansatz} implies that if we change $P$ to a new scale $P'$, the puncture field $j^{{\cal P},\rm osc.}_{\alpha\beta}$ changes by an amount
\beq
\Delta j^{{\cal P},\rm osc.}_{\alpha\beta} = -\frac{\ln(P/P')}{2r}\left[{\cal Z}_{uu}(u,\theta^C)k_\alpha k_\beta+ {\cal Z}_{AB}(u,\theta^C)\Omega^A_\alpha \Omega^B_\beta\right]. 
\eeq
The oscillatory part of the residual field changes by  the opposite amount: $\Delta h^{(2){\cal R},\rm osc.}_{\alpha\beta} = -\Delta j^{{\cal P},\rm osc.}_{\alpha\beta} + \O(1/R^2)$. From the explicit equation~\eqref{j osc soln}, we read off ${\cal Z}_{AB}=-4M^{(1)}_{\rm B}\dot z^{(1)}_{AB}$. This implies that the change in scale induces the following change in the shear: 
\beq
\Delta z^{(2){\cal R}}_{AB} = 2M^{(1)}_{\rm B}\ln(P'/P)\dot z^{(1)}_{AB},
\eeq
which is simply proportional to a time derivative of the lower-order shear. If we write the total shear in terms of $lm$ modes, as in equation~\eqref{eq:multiscale waveform}, we see that the modified waveform is hence given by 
\begin{multline}\label{eq:phase redefinition}
	\left[\epsilon h^{(1)}_{lm}(\Omega) + \epsilon^2 h^{(2)\cal R}_{lm}(\Omega) + \e^2\Delta h^{(2)\cal R}_{lm}(\Omega) +\O(3)\right]e^{-im\phi_p} \\*
	=\left[\epsilon h^{(1)}_{lm}(\Omega) + \epsilon^2 h^{(2)\cal R}_{lm}(\Omega) -2im\e^2\Omega M^{(1)}_{\rm B}\ln(P'/P)h^{(1)}_{lm}(\Omega) + \O(3)\right]e^{-im\phi_p}.
\end{multline}
Here we see that we can absorb the change into a redefinition of the phase:
\begin{multline}
\left[\epsilon h^{(1)}_{lm}(\Omega) + \epsilon^2 h^{(2)\cal R}_{lm}(\Omega) + \e^2\Delta h^{(2)\cal R}_{lm}(\Omega) +\O(3)\right]e^{-im\phi_p} \\*
	= \left[\epsilon h^{(1)}_{lm}(\Omega) + \epsilon^2 h^{(2)\cal R}_{lm}(\Omega) +\O(3)\right]e^{-im\psi},
\end{multline}
where the new phase is
\beq\label{eq:aux_phase_GSF}
\psi = \phi_p + 2\e\, \Omega M^{(1)}_{\rm B}\ln(P'/P).
\eeq
This is a 2PA shift in the waveform phase. Therefore, since we only control the 1PA phase, we can choose $P$ to be any convenient value and safely ignore the waveform's sensitivity to the choice.

Equation~\eqref{eq:aux_phase_GSF} can be compared to the analogous equation~\eqref{eq:aux_phase} in the PN-MPM context. Just as the arbitrary scale $b_0$ drops out of the complete 4PN waveform~\cite{Blanchet:2023sbv,Blanchet:2023bwj}, we expect the arbitrary scale $P$ to drop out of the complete 2PA waveform.

\section{Further PN-GSF results and comparisons with numerical self-force calculations and numerical relativity}
\label{sec:further_comparisons}

\begin{figure}[H]
    \centering
    \begin{tabular}{cc}
         \includegraphics[width = 0.5\textwidth]{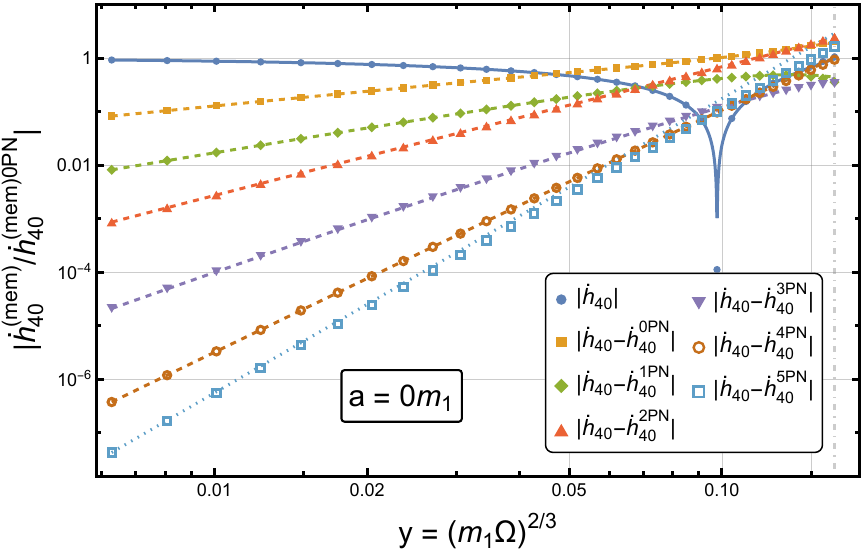}
         \includegraphics[width =0.5\textwidth]{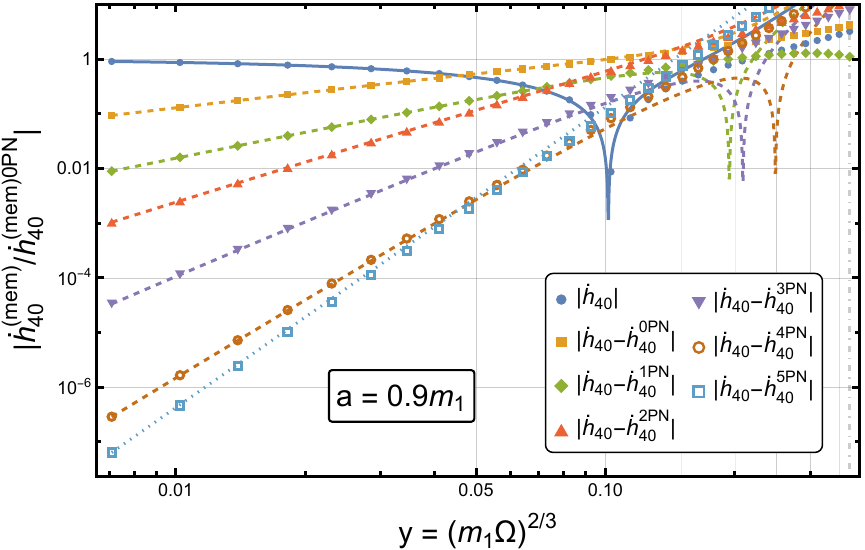}\\
         \includegraphics[width = 0.5\textwidth]{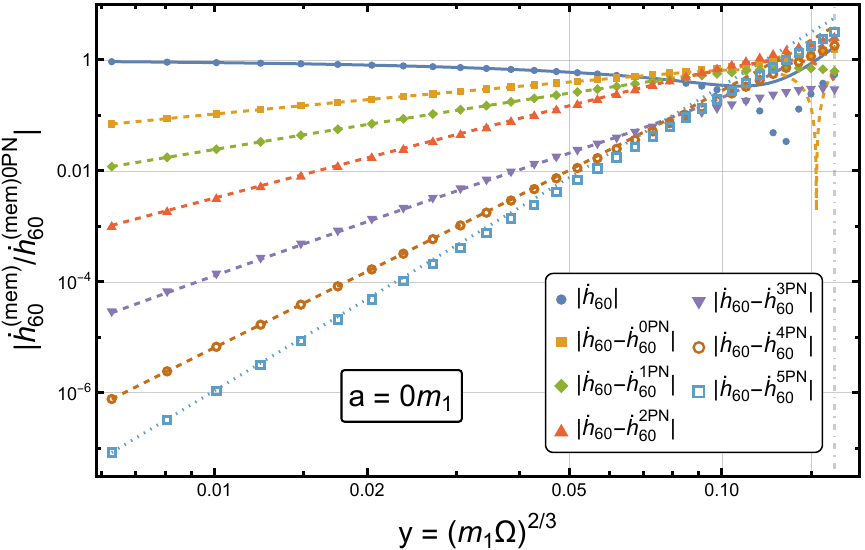}f
         \includegraphics[width = 0.5\textwidth]{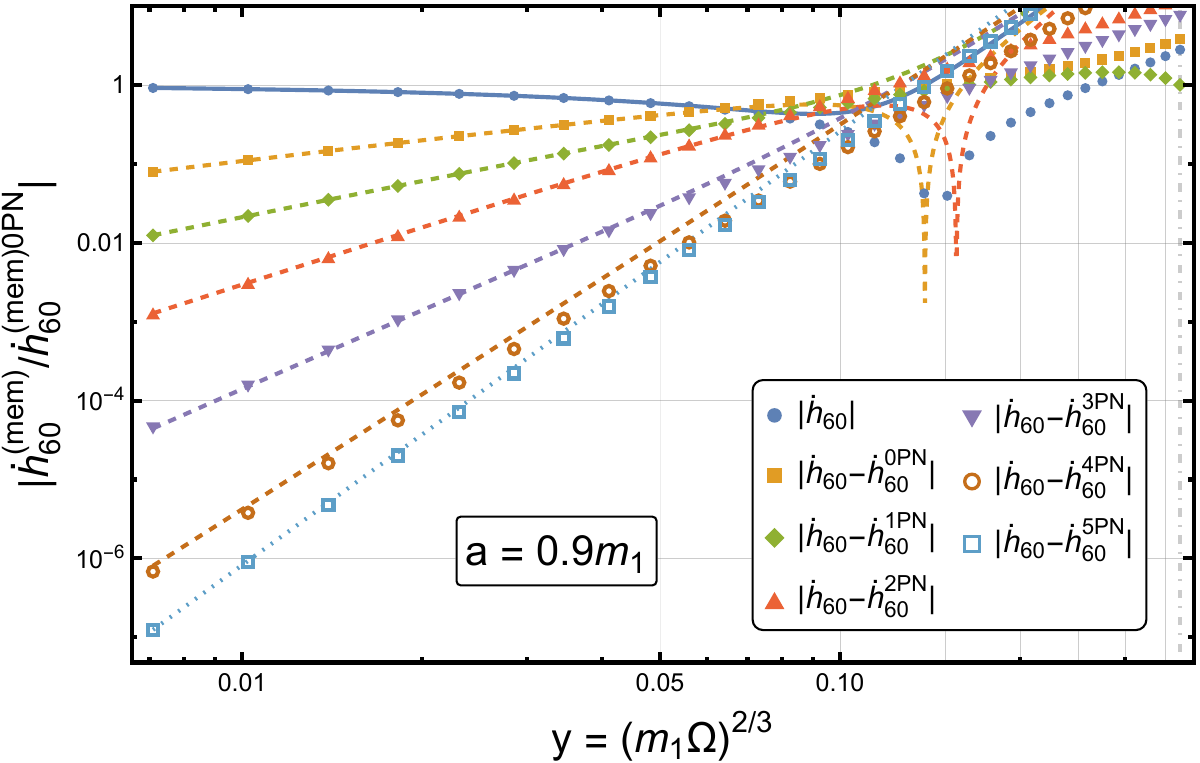}\\
         \includegraphics[width = 0.5\textwidth]{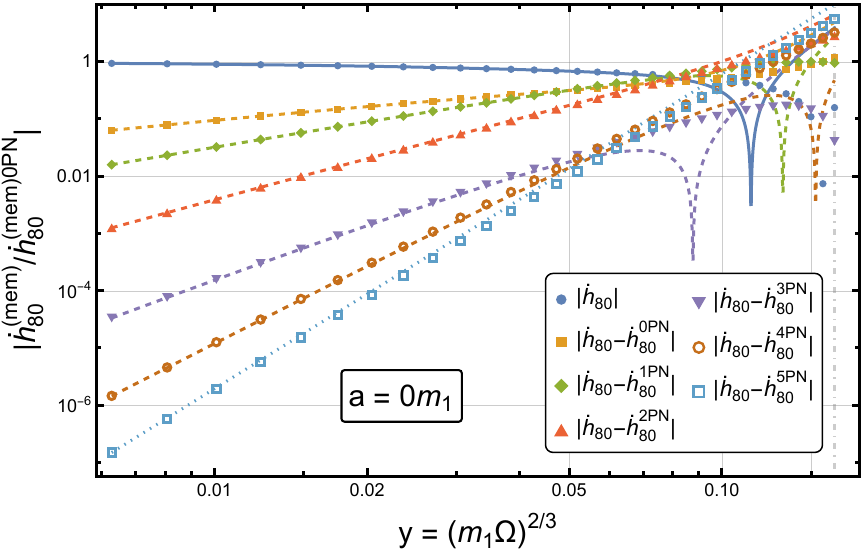}
         \includegraphics[width = 0.5\textwidth]{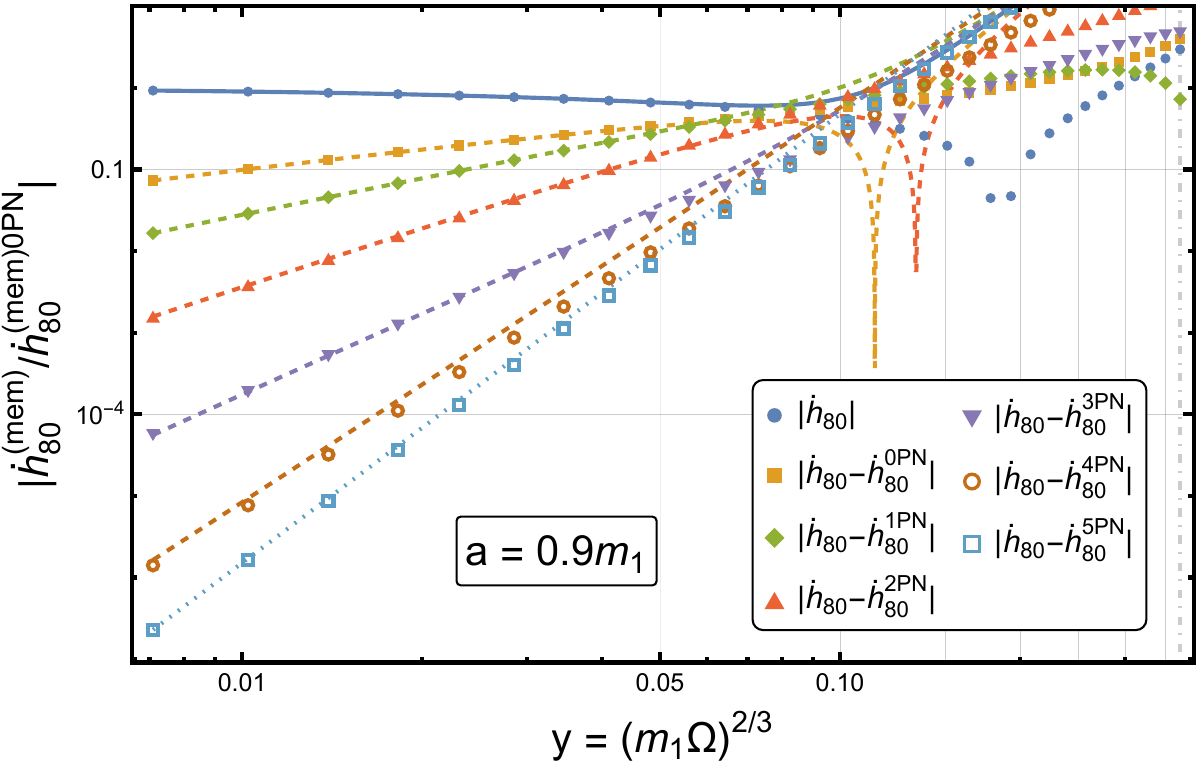}
    \end{tabular}
    \caption[Schwarzschild Memory]
    {The same as figure~\ref{fig:hdot20} but for the $a=0$ and $l\in\{4,6,8\}$ (left column, from top to bottom) and $a=0.9m_1$ and $l\in\{4,6,8\}$ (right column, from top to bottom).}
~\label{fig:hdotlge4}
\end{figure}

\begin{figure}[H]
    \centering
    \includegraphics[width = 1\textwidth]{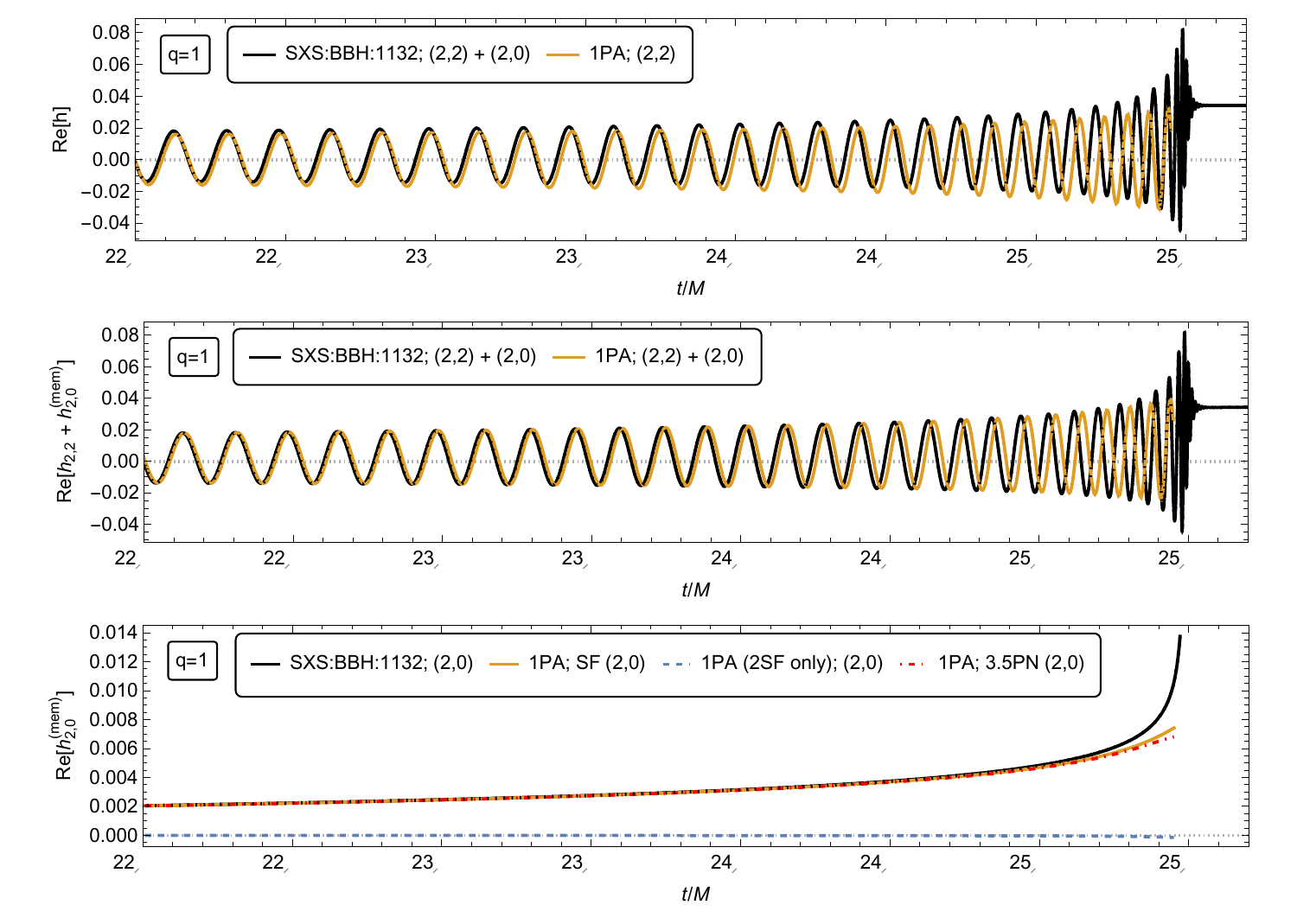}
    \caption{The same as figure \ref{fig:waveform_q10}. The SXS waveform used in this figure had the ID SXS:BBH:1132 \cite{sxs_collaboration_2019_3301955}.}
	\label{fig:waveform_q1}
\end{figure}

In this appendix we present the expressions for the PN-GSF for modes \mbox{$l \in \{4, 6, 8, 10, 12, 14\}$}. 
These results extend the $ l =2$ results given in section~\ref{sec:PNSF}.
We then present a comparison between our numerical 1SF results and these PN-GSF series for modes with $l\ge4$ in figure \ref{fig:hdotlge4}.
These results extend the $l=2$ comparison presented in figure \ref{fig:hdot20} in section \ref{sec:Numerical Kerr results}.
Finally, we present a comparison of the 2SF results with a numerical relativity simulation with $q=1$ in figure~\ref{fig:waveform_q1}.
This complements the $q=10$ waveform shown in figure \ref{fig:waveform_q10}.

The PN-GSF series for $\dot \cU^\text{(1, mem)}_{l>4,0}$ are given below. The results of $l=2$ can be found in equation~\eqref{eq:U1SFPNl2}.
\begin{multline}
    \dot \cU^\text{(1, mem)}_{40} = \frac{64}{315} \sqrt{\frac{\pi }{5}} \epsilon^2 y^5\bigg\{1-\frac{10133}{704}y+\left(4 \pi -\frac{107 \tilde{a}}{48}\right)y^{3/2}+\left(\frac{7 \tilde{a}^2}{4}+\frac{322533}{4576}\right)y^2\\
    +\left(\frac{26749 \tilde{a}}{704}-\frac{1028 \pi }{11}\right)y^{5/2}+\left(-\frac{199511 \tilde{a}^2}{6336}-\frac{235 \pi \tilde{a}}{24}-\frac{856 \log (y)}{105}+\frac{16 \pi ^2}{3}-\frac{1712 \gamma_{\rm E} }{105}\right.\\
    \left.+\frac{32585924257}{403603200}-\frac{3424 \log (2)}{105}\right)y^3+\left(-\frac{41 \tilde{a}^3}{24}+\frac{15 \pi  \tilde{a}^2}{2}-\frac{411053 \tilde{a}}{13728}+\frac{37621537 \pi }{82368}\right)y^{7/2}\\
    +\left(\frac{3 \tilde{a}^4}{4}+\frac{1655447 \tilde{a}^2}{27456}+\frac{313979 \pi \tilde{a}}{1056}+\frac{25254577 \log (y)}{135520}-\frac{103937 \pi ^2}{528}+\frac{25254577 \gamma_{\rm E} }{67760}\right.\\
        \left.-\frac{1809104091184543}{622658836800}+\frac{142155 \log (3)}{352}+\frac{13858871 \log (2)}{40656}\right)y^4\\
    +\left(\frac{3636775 \tilde{a}^3}{57024}-\frac{23021 \pi  \tilde{a}^2}{99}+\frac{21841 \tilde{a} \log (y)}{1260}-\frac{491 \pi^2 \tilde{a}}{36}+\frac{21841 \gamma_{\rm E} \tilde{a}}{630}-\frac{288480723269 \tilde{a}}{227026800}\right.\\
    \left.+\frac{44753}{630} \tilde{a} \log (2)-\frac{3424}{105} \pi  \log (y)-\frac{6848 \gamma_{\rm E}  \pi }{105}+\frac{81330237967 \pi }{146764800}-\frac{13696}{105} \pi  \log (2)\right)y^{9/2}\\
    +\left(-\frac{174959 \tilde{a}^4}{6336}-\frac{35 \pi  \tilde{a}^3}{4}-\frac{3317}{210} \tilde{a}^2 \log (y)+\frac{31 \pi ^2 \tilde{a}^2}{3}-\frac{3317 \gamma_{\rm E}  \tilde{a}^2}{105}+\frac{18136142587 \tilde{a}^2}{67267200}\right.\\
    \left.-\frac{321}{5} \tilde{a}^2 \log (2)-\frac{883835 \pi \tilde{a}}{7722}-\frac{284084196569 \log (y)}{309188880}+\frac{4639615 \pi ^2}{5148}-\frac{284084196569 \gamma_{\rm E} }{154594440}\right.\\
    \left.+\frac{365952109095738585071}{35988435449366400}-\frac{2471692725 \log (3)}{832832}-\frac{508082123 \log (2)}{7135128}\right)y^5+\mathcal{O}(y^{11/2})\bigg\}
    \end{multline} 
    \begin{multline}
        \dot \cU^\text{(1, mem)}_{60} =-\frac{839}{693} \sqrt{\frac{\pi }{2730}} \epsilon^2 y^6\bigg\{1-\frac{982361}{75510}y+\left(\frac{5540 \pi }{839}-\frac{3367 \tilde{a}}{839}\right)y^{3/2}\\*
        +\left(\frac{2505 \tilde{a}^2}{839}+\frac{302491414}{4492845}\right)y^2+\left(\frac{9331007 \tilde{a}}{226530}-\frac{488407 \pi }{4195}\right)y^{5/2}\\
        +\left(-\frac{1205749 \tilde{a}^2}{37755}-\frac{22182 \pi \tilde{a}}{839}-\frac{37002022 \log (y)}{2907135}+\frac{35276 \pi ^2}{2517}-\frac{74004044 \gamma_{\rm E} }{2907135}\right.\\
	\left.+\frac{528802122454577}{3253650956325}-\frac{170586 \log (3)}{5873}-\frac{63555148 \log (2)}{2907135}\right)y^3\\
        +\left(-\frac{6733 \tilde{a}^3}{839}+\frac{16596 \pi  \tilde{a}^2}{839}-\frac{5562048773 \tilde{a}}{134785350}+\frac{14987687671 \pi }{23961840}\right)y^{7/2}\\
        +\left(\frac{2549 \tilde{a}^4}{839}+\frac{15183740813 \tilde{a}^2}{161742420}+\frac{5541347 \pi \tilde{a}}{12585}+\frac{80540281223 \log (y)}{340134795}\right.\\
        \left.-\frac{37935202 \pi ^2}{113265}+\frac{161080562446 \gamma_{\rm E} }{340134795}-\frac{431027561847772107467}{98976062091406500}\right.\\
	\left.-\frac{144066438 \log (3)}{1908725}+\frac{383063021834 \log (2)}{242953425}\right)y^4\\
        +\left(\frac{612422 \tilde{a}^3}{7551}-\frac{4196302 \pi  \tilde{a}^2}{12585}+\frac{89067944 \tilde{a} \log (y)}{1799655}-\frac{141148 \pi^2 \tilde{a}}{2517}+\frac{178135888 \gamma_{\rm E} \tilde{a}}{1799655}\right.\\
        \left.-\frac{453238526769788047 \tilde{a}}{209473147283400}+\frac{34957008 \tilde{a} \log (3)}{323015}+\frac{1130043184 \tilde{a} \log (2)}{12597585}\right.\\
        \left.-\frac{232435288 \pi  \log (y)}{2907135}-\frac{464870576 \gamma_{\rm E}  \pi }{2907135}+\frac{17927953323583441 \pi }{13699582974000}\right.\\
	\left.-\frac{1023516 \pi  \log (3)}{5873}-\frac{423074992 \pi  \log (2)}{2907135}\right)y^{9/2}\\
        +\left(-\frac{148392 \tilde{a}^4}{4195}-\frac{44362 \pi  \tilde{a}^3}{839}-\frac{35656262 \tilde{a}^2 \log (y)}{969045}+\frac{35260 \pi ^2 \tilde{a}^2}{839}\right.\\
        \left.-\frac{71312524 \gamma_{\rm E}  \tilde{a}^2}{969045}+\frac{3957492264770609 \tilde{a}^2}{6507301912650}-\frac{511758 \tilde{a}^2 \log (3)}{5873}-\frac{8317972 \tilde{a}^2 \log (2)}{138435}\right.\\
        \left.-\frac{14082083177 \pi \tilde{a}}{29952300}-\frac{33933906394855169 \log (y)}{26147522230830}+\frac{4691517389 \pi ^2}{2695707}\right.\\
        \left.-\frac{33933906394855169 \gamma_{\rm E} }{13073761115415}+\frac{3157231256292518266521536467}{174999773335940015463000}\right.\\
        \left.+\frac{52841796875 \log (5)}{32633744}+\frac{178771047551451 \log (3)}{97085388400}\right.\\
	\left.-\frac{957529329289620229 \log (2)}{65368805577075}\right)y^5+\mathcal{O}(y^{11/2})\bigg\}
        \end{multline}
\begin{multline}
    \dot \cU^{\text{(1, mem)}}_{80} = \frac{75601 \sqrt{\frac{\pi }{1190}}\epsilon^2 y^7}{347490}\bigg\{1-\frac{7655551}{604808}y+\left(\frac{708606 \pi }{75601}-\frac{1209571 \tilde{a}}{226803}\right)y^{3/2}\\
    +\left(\frac{302428 \tilde{a}^2}{75601}+\frac{142621440727}{2212085260}\right)y^2 +\left(\frac{152578393 \tilde{a}}{3342360}-\frac{418714809 \pi }{2872838}\right)y^{5/2}\\
    +\left(-\frac{3489086921 \tilde{a}^2}{103422168}-\frac{3779202 \pi \tilde{a}}{75601}-\frac{12582713294 \log (y)}{681089409}\right.\\
    \left.+\frac{6320164 \pi ^2}{226803}-\frac{25165426588 \gamma_{\rm E} }{681089409}+\frac{44818614162291479957}{153223049728148400}\right.\\
	\left.+\frac{274223556 \log (3)}{19914895}-\frac{36143658812 \log (2)}{261957465}\right)y^3\\
    +\left(-\frac{1209580 \tilde{a}^3}{75601}+\frac{2834472 \pi  \tilde{a}^2}{75601}-\frac{1743643782493 \tilde{a}}{86271325140}+\frac{17808266307551 \pi }{22120852600}\right)y^{7/2}\\
    +\left(\frac{453726 \tilde{a}^4}{75601}+\frac{705105537139 \tilde{a}^2}{5688219240}+\frac{125613322879 \pi \tilde{a}}{201098660}+\frac{25628500645526629 \log (y)}{83596914060660}\right.\\
    \left.-\frac{4617709549 \pi ^2}{8618514}+\frac{25628500645526629 \gamma_{\rm E} }{41798457030330}-\frac{176751896257744272214175563}{27974813221403990241300}\right.\\
    \left.+\frac{641650390625 \log (5)}{821631668}-\frac{2000899516041 \log (3)}{69838691780}+\frac{13935285756640357 \log (2)}{41798457030330}\right)y^4\\
    +\left(\frac{460185898021 \tilde{a}^3}{4653997560}-\frac{1297746909 \pi  \tilde{a}^2}{2872838}+\frac{17775726179714 \tilde{a} \log (y)}{173677799295}-\frac{4396628 \pi^2 \tilde{a}}{29583}\right.\\
    \left.+\frac{35551452359428 \gamma_{\rm E} \tilde{a}}{173677799295}-\frac{41785663645227228969937 \tilde{a}}{11623883610001657995}-\frac{61529058 \tilde{a} \log (3)}{982813}\right.\\
    \left.+\frac{128154386221828 \tilde{a} \log (2)}{173677799295}-\frac{60822627732 \pi  \log (y)}{378383005}-\frac{121645255464 \gamma_{\rm E}  \pi }{378383005}\right.\\
    \left.+\frac{442964140185376922207 \pi }{158897977495857600}+\frac{1645341336 \pi  \log (3)}{19914895}-\frac{3841533355112 \pi  \log (2)}{3405447045}\right)y^{9/2}\\
    +\left(-\frac{3799784483924486131 \tilde{a}^4}{96674007796706340}-\frac{3551537807176 \pi  \tilde{a}^3}{23838129315}+\frac{46080 \tilde{a}^2 \log ^2(y)}{36157}\right.\\
    \left.+\frac{184320 \tilde{a}^2 \log (2) \log (y)}{36157}+\frac{61440 \gamma_{\rm E}  \tilde{a}^2 \log (y)}{36157}-\frac{2856325144856 \tilde{a}^2 \log (y)}{37459917495}\right.\\
    \left.+\frac{2360838845952 \tilde{a}^2}{9828292164145 \sqrt{1-\tilde{a}^2}}+\frac{277970512 \pi ^2 \tilde{a}^2}{2494833}+\frac{20480 \gamma_{\rm E} ^2 \tilde{a}^2}{36157}-\frac{41837568 \gamma_{\rm E}  \tilde{a}^2}{440829431 \sqrt{1-\tilde{a}^2}}\right.\\
    \left.-\frac{1874214949404496 \gamma_{\rm E}  \tilde{a}^2}{12848751700785}+\frac{1028029855916686188334863531630091 \tilde{a}^2}{908380825509749627396531795625}\right.\\
    \left.+\frac{746496 \pi \tilde{a}}{181518001 \sqrt{1-\tilde{a}^2}}-\frac{3744462336}{15429030085 \sqrt{1-\tilde{a}^2}}+\frac{41837568 \gamma_{\rm E} }{440829431 \sqrt{1-\tilde{a}^2}}\right.\\
    \left.+\frac{184320 \tilde{a}^2 \log ^2(2)}{36157}+\frac{22927073328 \tilde{a}^2 \log (3)}{378383005}+\frac{122880 \gamma_{\rm E}  \tilde{a}^2 \log (2)}{36157}\right.\\
    \left.-\frac{20864132919856 \tilde{a}^2 \log (2)}{37459917495}+\frac{4876732416 \tilde{a}^4}{1965658432829 \sqrt{1-\tilde{a}^2}}-\frac{746496 \pi  \tilde{a}^3}{181518001 \sqrt{1-\tilde{a}^2}}\right.\\
    \left.-\frac{2985984 \gamma_{\rm E}  \pi \tilde{a}}{25931143}-\frac{63571548924052591372 \pi \tilde{a}}{84889366350413625}+\frac{3385240704 \log ^2(y)}{774229841}\right.\\
	\left.+\frac{13540962816 \log (2) \log (y)}{774229841}+\frac{4513654272 \gamma_{\rm E}  \log (y)}{774229841}-\frac{5429502347176427617 \log (y)}{3054220588268850}\right.\\
    \left.+\frac{589843187668048 \pi ^2}{197428609455}+\frac{1155610112 \gamma_{\rm E} ^2}{1880272471}-\frac{2307884947929261061 \gamma_{\rm E} }{654475840343325}\right.\\
	\left.+\frac{2878135067296721788315164758443}{98117976477597995246328000}+\frac{13540962816 \log ^2(2)}{774229841}\right.\\
    \left.-\frac{47967151220703125 \log (5)}{4106515076664}+\frac{169931543044402071 \log (3)}{59108929133800}+\frac{9027308544 \gamma_{\rm E}  \log (2)}{774229841}\right.\\
    \left.+\frac{23054066096723125697 \log (2)}{2231930429888775}\right)y^5+\mathcal{O}(y^{11/2})\bigg\}
\end{multline}

\begin{align}
    \dot{\cU}_{10,0}^\text{(1, mem)} =&-\frac{525221 \sqrt{\frac{\pi }{385}} \epsilon^2 y^8}{15752880}\bigg\{1+-\frac{29957556301}{2295215770}y\nonumber\\
    &+\left(\frac{124075236 \pi }{9979199}-\frac{10504423 \tilde{a}}{1575663}\right)y^{3/2}\nonumber\\
    &+\left(\frac{49895955 \tilde{a}^2}{9979199}+\frac{791778139491934}{12308094566625}\right)y^2+\mathcal{O}(y^{5/2})\bigg\}
\end{align}

\begin{align}
    \dot{\cU}_{12,0}^\text{(1, mem)} =\frac{1816214401 \sqrt{\frac{\pi }{3003}} \epsilon^2 y^9}{73012212000}\bigg\{1-\frac{8129375658025}{588453465924}y+\mathcal{O}(y^{3/2})\bigg\}
\end{align}

\begin{align}
    \dot{\cU}_{14,0}^\text{(1, mem)} =-\frac{435891455999 \sqrt{\frac{\pi }{39585}}\epsilon^2  y^{10}}{18196889760000}\bigg\{1+\mathcal{O}(y)\bigg\}
\end{align}
Integrating over an inspiral as in section~\ref{sec:PNSF} yields the following series for $H_{l>4,0}^\text{(1, mem)}$. The series for $l=2$ is given in equation~\eqref{eq:H1SFPNl2}.  Recall that $\Psi_A(\tilde{a})$ and $\Psi_B(\tilde{a})$ were defined in Eq.~\eqref{eq:PsiAPsiB}.
\begin{multline}
    H_{40}^\text{(1, mem)} = -\frac{1}{504 \sqrt{2}}\bigg\{1-\frac{180101}{29568}y+\left(\frac{23 \tilde{a}}{8}\right)y^{3/2}+\left(\frac{2201411267}{158505984}-\frac{53 \tilde{a}^2}{48}\right)y^2\\
    \left(\frac{3 \tilde{a}^3}{14}-\frac{436207 \tilde{a}}{51744}-\frac{13565 \pi }{1232}\right)y^{5/2}+\left(\frac{1130005 \tilde{a}^2}{405504}-\frac{25 \pi \tilde{a}}{96}+\frac{15240463356751}{781117489152}\right)y^3\\
    +\left(-\frac{127439 \tilde{a}^3}{66528}+\frac{5 \pi  \tilde{a}^2}{36}+\frac{75585906215 \tilde{a}}{4483454976}+\frac{21681663715 \pi }{747242496}\right)y^{7/2}\\+\left(\frac{2297 \tilde{a}^4}{1280}-\frac{3 \pi  \tilde{a}^3}{5}-\frac{13}{10} \sqrt{1-\tilde{a}^2} \tilde{a}^2+\frac{64768014665 \tilde{a}^2}{4649508864}-\frac{\sqrt{1-\tilde{a}^2}}{10}\right.\\
    \left.+\frac{1}{5}  \tilde{a} \Psi_B(\tilde{a})-\frac{3}{5} \sqrt{1-\tilde{a}^2} \tilde{a}^4+\frac{3}{5}  \tilde{a}^3 \Psi_B(\tilde{a})-\frac{74354351 \pi \tilde{a}}{1774080}\right.\\
    \left.+\frac{124474711 \log (y)}{8537760}+\frac{563 \pi ^2}{112}+\frac{124474711 \gamma_{\rm E} }{4268880}-\frac{9400760268126512286283}{30367738961009049600}\right.\\
	\left.+\frac{748683 \log (3)}{8624}-\frac{122327693 \log (2)}{4268880}\right)y^4\\
    +\left(-\frac{835 \tilde{a}^5}{1232}-\frac{115228859183 \tilde{a}^3}{6393074688}+\frac{25465885 \pi  \tilde{a}^2}{1951488}\right.\\
    \left.-\frac{2771 \tilde{a} \log (y)}{1386}+\frac{25 \pi^2 \tilde{a}}{66}-\frac{2771 \gamma_{\rm E} \tilde{a}}{693}+\frac{1546783983423461 \tilde{a}}{32221096427520}\right.\\
 	\left.-\frac{755}{99} \tilde{a} \log (2)+\frac{10775512302887 \pi }{153433792512}\right)y^{9/2}\\
    +\left(\frac{3 \tilde{a}^6}{32}+\frac{151945523 \tilde{a}^4}{22708224}-\frac{36637 \pi  \tilde{a}^3}{44352}-\frac{107}{336} \tilde{a}^2 \log (y)+\frac{1025581 \sqrt{1-\tilde{a}^2} \tilde{a}^2}{177408}\right.\\
    \left.-\frac{5 \pi ^2 \tilde{a}^2}{24}-\frac{107 \gamma_{\rm E}  \tilde{a}^2}{168}+\frac{12969093846952979 \tilde{a}^2}{93734098698240}+\frac{66097 \sqrt{1-\tilde{a}^2}}{177408}-\frac{107}{72} \tilde{a}^2 \log (2)\right.\\
    \left.+\frac{1}{12}  \tilde{a} \Psi_A(\tilde{a})-\frac{80881  \tilde{a} }{88704}\Psi_B(\tilde{a})+\frac{84577 \sqrt{1-\tilde{a}^2} \tilde{a}^4}{29568}-\frac{1}{16} \tilde{a}^3 \Psi_A(\tilde{a})-\frac{80881 \tilde{a}^3 }{29568}\Psi_B(\tilde{a})\right.\\
    \left.-\frac{84035703263 \pi \tilde{a}}{871782912}-\frac{3176039106761267 \log (y)}{78538922542080}-\frac{9382716995 \pi ^2}{1162377216}-\frac{3176039106761267 \gamma_{\rm E} }{39269461271040}\right.\\
    \left.+\frac{31124458733744129163312921839}{65527264188153639154483200}-\frac{1119805664085 \log (3)}{2611761152}\right.\\
    \left.+\frac{1159599747804217 \log (2)}{3020727790080}\right)y^5+\mathcal{O}(y^{11/2})\bigg\}
    \end{multline}
\begin{multline}
        H_{60}^\text{(1, mem)} = \frac{4195 y}{1419264 \sqrt{273}}\bigg\{1-\frac{45661561}{6342840}y+\left(\frac{54403 \tilde{a}}{17619}+\frac{1248 \pi }{839}\right)y^{3/2}\\
        +\left(\frac{3012132889099}{144921208320}-\frac{27879 \tilde{a}^2}{26848}\right)y^2+\left(\frac{\tilde{a}^3}{3}-\frac{87400525 \tilde{a}}{11417112}-\frac{74044807 \pi }{2718360}\right)y^{5/2}\\
        +\left(-\frac{38988079 \tilde{a}^2}{9396800}+\frac{70273 \pi \tilde{a}}{12585}-\frac{1064156 \log (y)}{581427}-\frac{2904 \pi ^2}{4195}-\frac{2128312 \gamma_{\rm E} }{581427}\right.\\
        \left.+\frac{5304042570514164241}{83959812197775360}-\frac{341172 \log (3)}{29365}+\frac{12498056 \log (2)}{2907135}\right)y^3\\
	+\left(\frac{3180087 \tilde{a}^3}{1292060}-\frac{35005 \pi  \tilde{a}^2}{18458}+\frac{597739627499177 \tilde{a}}{11955999686400}+\frac{523891620001 \pi }{5693333184}\right)y^{7/2}\\
        +\left(\frac{787639 \tilde{a}^4}{644352}-\frac{293 \pi  \tilde{a}^3}{839}-\frac{13}{6} \sqrt{1-\tilde{a}^2} \tilde{a}^2+\frac{4311729612143 \tilde{a}^2}{3478108999680}\right.\\
        \left.-\frac{\sqrt{1-\tilde{a}^2}}{6}+\frac{1}{3}  \tilde{a} \Psi_B(\tilde{a})-\sqrt{1-\tilde{a}^2} \tilde{a}^4\right.\\
        \left.+ \tilde{a}^3 \Psi_B(\tilde{a})-\frac{421346027 \pi \tilde{a}}{7248960}\right.\\
        \left.+\frac{11856697105103 \log (y)}{285713227800}-\frac{12005813 \pi ^2}{2114280}+\frac{11856697105103 \gamma_{\rm E} }{142856613900}\right.\\
        \left.-\frac{2139326974527903383632711567}{2145408289127314219008000}-\frac{15603893379 \log (3)}{427554400}+\frac{11061146032747 \log (2)}{28571322780}\right)y^4\\
	+\left(-\frac{526391 \tilde{a}^5}{610792}-\frac{11020949178887 \tilde{a}^3}{941987854080}+\frac{90652033 \pi  \tilde{a}^2}{43977024}\right.\\
        \left.-\frac{16539158096 \tilde{a} \log (y)}{1473917445}-\frac{95920 \pi^2 \tilde{a}}{32721}-\frac{33078316192 \gamma_{\rm E} \tilde{a}}{1473917445}+\frac{1133352424655125030901 \tilde{a}}{5425285511720954880}\right.\\
        \left.-\frac{213569298 \tilde{a} \log (3)}{4199195}+\frac{8959854016 \tilde{a} \log (2)}{1473917445}-\frac{70656 \pi  \log (y)}{29365}-\frac{23360 \pi ^3}{10907}-\frac{141312 \gamma_{\rm E}  \pi }{29365}\right.\\
        \left.+\frac{90952730246335904087 \pi }{354375830704896000}-\frac{104976 \pi  \log (3)}{5873}+\frac{242176 \pi  \log (2)}{29365}\right)y^{9/2}\\
        +\left(\frac{9 \tilde{a}^6}{56}+\frac{37867229509 \tilde{a}^4}{3788789760}+\frac{29937167 \pi  \tilde{a}^3}{19733280}+\frac{4003018 \tilde{a}^2 \log (y)}{1356663}+\frac{108635519 \sqrt{1-\tilde{a}^2} \tilde{a}^2}{14799960}\right.\\
        \left.+\frac{860 \pi ^2 \tilde{a}^2}{839}+\frac{8006036 \gamma_{\rm E}  \tilde{a}^2}{1356663}+\frac{1825224796674054313 \tilde{a}^2}{36391249869004800}+\frac{6526913 \sqrt{1-\tilde{a}^2}}{14799960}\right.\\
        \left.+\frac{2814669 \tilde{a}^2 \log (3)}{164444}-\frac{12221192 \tilde{a}^2 \log (2)}{2261105} +\frac{1}{7}  \tilde{a} \Psi_A(\tilde{a})-\frac{8641193  \tilde{a} }{7399980}\Psi_B(\tilde{a})\right.\\
        \left.+\frac{9169763 \sqrt{1-\tilde{a}^2} \tilde{a}^4}{2466660}-\frac{3}{28} \tilde{a}^3 \Psi_A(\tilde{a})-\frac{8641193  \tilde{a}^3 }{2466660}\Psi_B(\tilde{a})+\frac{27227996458709 \pi \tilde{a}}{2173818124800}\right.\\
        \left.-\frac{18248323045616428823 \log (y)}{122997944573824320}+\frac{7467887568649 \pi ^2}{338149486080}-\frac{18248323045616428823 \gamma_{\rm E} }{61498972286912160}\right.\\
        \left.+\frac{657699594819963548893392282039548513}{318637257693096932121544556544000}+\frac{7548828125 \log (5)}{16316872}\right.\\
       \left.+\frac{17218817988906483 \log (3)}{53280461153920}-\frac{2512350026982328137893 \log (2)}{922484584303682400}\right)y^5+\mathcal{O}(y^{11/2})\bigg\}
\end{multline}

\begin{multline}
    H_{80}^\text{(1, mem)} =-\frac{75601 y^2}{213497856 \sqrt{119}}\bigg\{1-\frac{265361599}{33869248}y+\left(\frac{3704629 \tilde{a}}{1360818}+\frac{812404 \pi }{226803}\right)y^{3/2}\\
    +\left(\frac{3170245055781731}{133786916524800}-\frac{3854499 \tilde{a}^2}{6048080}\right)y^2\\
	+\left(\frac{9 \tilde{a}^3}{22}-\frac{2356859509 \tilde{a}}{349276620}-\frac{82910438887 \pi }{1769668208}\right)y^{5/2}\\
    +\left(-\frac{4993386893 \tilde{a}^2}{413688672}+\frac{823014 \pi \tilde{a}}{75601}-\frac{17575532423 \log (y)}{3405447045}+\frac{39354 \pi ^2}{75601}\right.\\
    \left.-\frac{35151064846 \gamma_{\rm E} }{3405447045}+\frac{86026447950251875497799}{606956821828387430400}+\frac{137111778 \log (3)}{19914895}\right.\\
	\left.-\frac{2760135062 \log (2)}{52391493}\right)y^3 \\
    +\left(\frac{777035577 \tilde{a}^3}{110075056}-\frac{2533125 \pi  \tilde{a}^2}{982813}+\frac{927001137765503 \tilde{a}}{12304755859968}+\frac{6996005362465267 \pi }{39527952609600}\right)y^{7/2}\\
    +\left(\frac{4180725 \tilde{a}^4}{19353856}+\frac{467091 \pi  \tilde{a}^3}{1058414}-\frac{39}{14} \sqrt{1-\tilde{a}^2} \tilde{a}^2-\frac{14926147136200067 \tilde{a}^2}{2996826930155520}-\frac{3 \sqrt{1-\tilde{a}^2}}{14}\right.\\
    \left.+\frac{3}{7}  \tilde{a} \Psi_B(\tilde{a})-\frac{9}{7} \sqrt{1-\tilde{a}^2} \tilde{a}^4+\frac{9}{7}  \tilde{a}^3 \Psi_B(\tilde{a})-\frac{3825867463217 \pi \tilde{a}}{67569149760}\right.\\
    \left.+\frac{647582734396976083 \log (y)}{8192497577944680}-\frac{47337723957 \pi ^2}{1126152496}+\frac{647582734396976083 \gamma_{\rm E} }{4096248788972340}\right.\\
    \left.-\frac{8590362236330250121849682119776517}{4244680626068231841786745651200}+\frac{1924951171875 \log (5)}{5751421676}\right.\\
    \left.+\frac{375675168843693 \log (3)}{27376767177760}-\frac{1674489722549147 \log (2)}{11942416294380}\right)y^4\\
    +\left(-\frac{6909125 \tilde{a}^5}{8467312}+\frac{19087278308687 \tilde{a}^3}{14082833318400}-\frac{2998064169563 \pi  \tilde{a}^2}{57916414080}\right.\\
    \left.-\frac{1295501914846 \tilde{a} \log (y)}{66799153575}+\frac{557902 \pi^2 \tilde{a}}{378005}-\frac{2591003829692 \gamma_{\rm E} \tilde{a}}{66799153575}\right.\\
    \left.+\frac{31920604299579818432637263 \tilde{a}}{128184611871525976166400}+\frac{50748954462 \tilde{a} \log (3)}{1891915025}-\frac{162739019312444 \tilde{a} \log (2)}{868388996475}\right.\\
    \left.-\frac{58632904184 \pi  \log (y)}{3405447045}-\frac{14887456 \pi ^3}{1134015}-\frac{117265808368 \gamma_{\rm E}  \pi }{3405447045}\right.\\
	\left.+\frac{20930253386689200645325901 \pi }{28830449036848402944000}+\frac{1096894224 \pi  \log (3)}{99574475}\right.\\
	\left.-\frac{303421023088 \pi  \log (2)}{1891915025}\right)y^{9/2}\\
    +\left(\frac{27 \tilde{a}^6}{128}-\frac{21060167143990287013 \tilde{a}^4}{263984490623539445760}+\frac{90204503545543 \pi  \tilde{a}^3}{3360830753280}+\frac{17280 \tilde{a}^2 \log ^2(y)}{36157}\right.\\
    \left.+\frac{69120 \tilde{a}^2 \log (2) \log (y)}{36157}+\frac{23040 \gamma_{\rm E}  \tilde{a}^2 \log (y)}{36157}+\frac{1186186536991 \tilde{a}^2 \log (y)}{399572453280}\right.\\
    \left.+\frac{5254539034461063 \tilde{a}^2}{629010698505280 \sqrt{1-\tilde{a}^2}}-\frac{7289665 \pi ^2 \tilde{a}^2}{26611552}+\frac{7680 \gamma_{\rm E} ^2 \tilde{a}^2}{36157}-\frac{15689088 \gamma_{\rm E}  \tilde{a}^2}{440829431 \sqrt{1-\tilde{a}^2}}\right.\\
    \left.+\frac{588226841474953 \gamma_{\rm E}  \tilde{a}^2}{68526675737520}+\frac{457657484408474099689113151445874563 \tilde{a}^2}{79375527707475935441412143677440000}\right.\\
    \left.+\frac{279936 \pi \tilde{a}}{181518001 \sqrt{1-\tilde{a}^2}}+\frac{104137734749}{246864481360 \sqrt{1-\tilde{a}^2}}+\frac{15689088 \gamma_{\rm E} }{440829431 \sqrt{1-\tilde{a}^2}}\right.\\
    \left.+\frac{69120 \tilde{a}^2 \log ^2(2)}{36157}-\frac{41397659577 \tilde{a}^2 \log (3)}{12108256160}+\frac{46080 \gamma_{\rm E}  \tilde{a}^2 \log (2)}{36157}\right.\\
    \left.+\frac{7826868306091 \tilde{a}^2 \log (2)}{199786226640}+\frac{3}{16}  \tilde{a} \Psi_A(\tilde{a})-\frac{1482521 \tilde{a} }{1058414}\Psi_B(\tilde{a})-\frac{75923871 \tilde{a}^6}{16934624 \sqrt{1-\tilde{a}^2}}\right.\\
	\left.-\frac{539961191565829 \tilde{a}^4}{125802139701056 \sqrt{1-\tilde{a}^2}}-\frac{279936 \pi  \tilde{a}^3}{181518001 \sqrt{1-\tilde{a}^2}}-\frac{9}{64}  \tilde{a}^3 \Psi_A\right.\\
    \left.-\frac{4447563 \tilde{a}^3 }{1058414}\Psi_B(\tilde{a})-\frac{1119744 \gamma_{\rm E}  \pi \tilde{a}}{25931143}+\frac{178416578495276962594781 \pi \tilde{a}}{695413689142588416000}\right.\\
    \left.+\frac{1269465264 \log ^2(y)}{774229841}+\frac{5077861056 \log (2) \log (y)}{774229841}+\frac{1692620352 \gamma_{\rm E}  \log (y)}{774229841}\right.\\
    \left.-\frac{5546318829923799001283 \log (y)}{16829567052426777600}+\frac{571071527292260719 \pi ^2}{3164351416934400}+\frac{433353792 \gamma_{\rm E} ^2}{1880272471}\right.\\
  \left.-\frac{11989772046221334444461 \gamma_{\rm E} }{18512523757669455360}+\frac{84815339968154539532300638421823357073}{15394041737207454146213264228352000}\right.\\
    \left.+\frac{5077861056 \log ^2(2)}{774229841}-\frac{327004407177734375 \log (5)}{87605654968832}+\frac{5717381105176736823 \log (3)}{6332365277638400}\right.\\
    \left.+\frac{3385240704 \gamma_{\rm E}  \log (2)}{774229841}+\frac{291437899731825630526109 \log (2)}{63332318118342873600}\right)y^5+\mathcal{O}(y^{11/2})\bigg\}
\end{multline}

\begin{align}
 H_{10,0}^\text{(1, mem)} =& \frac{525221 y^3}{6452379648 \sqrt{154}}\bigg\{1-\frac{4180196800013}{481995311700}y\nonumber\\
 &+\left(\frac{11554854 \tilde{a}}{5777431}+\frac{673267520 \pi }{109771189}\right)y^{3/2}\nonumber\\
 &+\left(\frac{8442034925654838593}{329100310466928000}-\frac{3326613 \tilde{a}^2}{79833592}\right)y^2+\mathcal{O}(y^{5/2})\bigg\}
\end{align}

\begin{align}
    H_{12,0}^\text{(1, mem)} = -\frac{1816214401 y^4}{7476450508800 \sqrt{30030}}\bigg\{1-\frac{955937206631195}{98860182275232}y+\mathcal{O}(y^{3/2})\bigg\}
\end{align}

\begin{align}
    H_{14,0}^\text{(1, mem)} = \frac{435891455999 y^5}{11180169068544000 \sqrt{15834}}\bigg\{1+\mathcal{O}(y)\bigg\}
\end{align}   
The series expressions above are also available in a digital format in the \texttt{PostNewtonianSelfForce}~\cite{niels_warburton_2023_8112975} package of the \texttt{Black Hole Perturbation Toolkit}.

The comparison between the above series and our numerical results is given in figure~\ref{fig:hdotlge4} below.
As in the case for $l=2$, presented in figure \ref{fig:hdot20}, we find excellent agreement between numerical and series results.


Finally, we present a comparison of a 1PA waveform with memory to an numerical relativity simulation with $q=1$ in figure~\ref{fig:waveform_q1}.
Although the oscillatory part of the waveform dephases significantly close to the merger (as expected for a 1PA waveform at this mass ratio) the memory contribution compares well to the NR waveform until closer to the merger.


%

\newpage
\section{List of SXS data used in this work}

In this section we list in table \ref{table:SXSIDs} the data from the SXS Collaboration used in creating figure \ref{fig:MassRatio}.
\begin{table}[H]
    \label{SXSDatasets}
    \begin{center}
        \begin{tabular}{|c|c c| c c | c | c |||}
            \hline
            SXS ID & $q$ & $\nu$ & $|\chi_1|$ & $|\chi_2|$ 	& dataset reference \\
            \hline
            SXS:BBH:2477 & 15.00, & 0.05859 & $6.43\times 10^{-6}$, & $4.52\times 10^{-6}$ 		&	\cite{Yoo:2022erv}								\\
            SXS:BBH:2480 & 14.00, & 0.06222 & $7.62\times 10^{-6}$, & $4.14\times 10^{-6}$		& 	\cite{sxs_collaboration_2024_13147588}			\\
            SXS:BBH:1107 & 10.00, & 0.08264 & $3.659\times 10^{-6}$, & $1.058\times 10^{-7}$ 	& 	\cite{sxs_collaboration_2019_3302023}			\\
            SXS:BBH:1108 & 9.200, & 0.08843 & $2.254\times 10^{-6}$, & $1.464\times 10^{-6}$	&	\cite{kevin_barkett_2019_3302018}				\\
            SXS:BBH:0186 & 8.267, & 0.09626 & $1.370\times 10^{-6}$, & $9.153\times 10^{-8}$	& 	\cite{jonathan_blackman_2019_3302117}			\\
            SXS:BBH:0188 & 7.187, & 0.1072 & $1.553\times 10^{-6}$, & $2.445\times 10^{-5}$		&	\cite{jonathan_blackman_2019_3302249}		\\
            SXS:BBH:0166 & 6.000, & 0.1224 & $4.527\times 10^{-7}$, & $4.672\times 10^{-7}$		&	\cite{kidder_2019_3312642}	\\
            SXS:BBH:0113 & 5.000, & 0.1389 & $3.160\times 10^{-7}$, & $3.877\times 10^{-7}$		& 	\cite{kidder_2019_3312259}		\\
            SXS:BBH:1220 & 4.000, & 0.1600 & $5.627\times 10^{-5}$, & $3.311\times 10^{-5}$		& 	\cite{jonathan_blackman_2019_3302203}		\\
            SXS:BBH:1906 & 4.000, & 0.1600 & $5.767\times 10^{-5}$, & $8.544\times 10^{-5}$		& 	\cite{sxs_collaboration_2019_3310379}		\\
            SXS:BBH:2265 & 3.000, & 0.1875 & $2.238\times 10^{-6}$, & $5.413\times 10^{-6}$		& 	\cite{sxs_collaboration_2019_3387428}		\\
            SXS:BBH:1165 & 2.000, & 0.2222 & $7.908\times 10^{-5}$, & $1.954\times 10^{-5}$		& 	\cite{sxs_collaboration_2019_3315116}		\\
            SXS:BBH:1143 & 1.250, & 0.2469 & $1.365\times 10^{-4}$, & $2.545\times 10^{-5}$		& 	\cite{sxs_collaboration_2019_3301890}		\\
            SXS:BBH:0198 & 1.203, & 0.2479 & $5.042\times 10^{-5}$, & $8.544\times 10^{-5}$		& 	\cite{jonathan_blackman_2019_3302177}		\\
            SXS:BBH:0116 & 1.080, & 0.2496 & $1.067\times 10^{-1}$, & $2.119\times 10^{-1}$		& 	\cite{kidder_2019_3312910}		\\
            SXS:BBH:1132 & 1.000, & 0.2500 & $1.135\times 10^{-7}$, & $1.137\times 10^{-7}$		& 	\cite{sxs_collaboration_2019_3301955}		\\
            \hline
        \end{tabular}
    \end{center}
    \caption{
		List of numerical relativity datasets from the SXS Collaboration \cite{Boyle:2019kee} used in creating figure~\ref{fig:MassRatio}. 
    }\label{table:SXSIDs}
\end{table}

\bibliography{emri_memory_references}

\end{document}